\documentclass[draft,tightenlines,nofootinbib,preprint,aps,eqsecnum,amsmath,amssymb]{revtex4}

\newcommand{\beq}{\begin{equation}}
\newcommand{\eeq}{\end{equation}}
\newcommand{\bea}{\begin{eqnarray}}
\newcommand{\eea}{\end{eqnarray}}
\newcommand{\cir}{{\buildrel \circ \over =}}
\newcommand{\sgn}{\epsilon}
\newcommand{\eo}{{}^4{\buildrel \circ \over E}}

\newcommand{\byd}{{\buildrel {def}\over =}}

\begin{document}

\title{The Einstein-Maxwell-Particle System in the York Canonical Basis
of ADM Tetrad Gravity: I) The Equations of Motion in Arbitrary
Schwinger Time Gauges.}

\medskip

\author{David Alba}

\affiliation{Dipartimento di Fisica\\
Universita' di Firenze\\Polo Scientifico, via Sansone 1\\
 50019 Sesto Fiorentino, Italy\\
 E-mail ALBA@FI.INFN.IT}

\author{Luca Lusanna}

\affiliation{ Sezione INFN di Firenze\\ Polo Scientifico\\ Via Sansone 1\\
50019 Sesto Fiorentino (FI), Italy\\ Phone: 0039-055-4572334\\
FAX: 0039-055-4572364\\ E-mail: lusanna@fi.infn.it}

\today

\begin{abstract}

We study the coupling of N charged scalar particles plus the
electro-magnetic field to ADM tetrad gravity and its canonical
formulation in asymptotically Minkowskian space-times without
super-translations. To regularize the self-energies both the
electric charge and the sign of the energy of the particles are
Grassmann-valued. The introduction of the non-covariant radiation
gauge allows to reformulate the theory in terms of transverse
electro-magnetic fields and to extract the generalization of the
Coulomb interaction among the particles in the Riemannian
instantaneous 3-spaces of global non-inertial frames, the only ones
allowed by the equivalence principle.

Then we make the canonical transformation to the York canonical
basis, where there is a separation between the {\it inertial}
(gauge) variables and the {\it tidal} ones inside the gravitational
field and a special role of the Eulerian observers associated to the
3+1 splitting of space-time. The Dirac Hamiltonian is weakly equal
to the weak ADM energy. The Hamilton equations in Schwinger time
gauges are given explicitly. In the York basis they are naturally
divided in four sets: a) the contracted Bianchi identities; b) the
equations for the inertial gauge variables; c) the equations for the
tidal ones; d) the equations for matter.

Finally we give the restriction of the Hamilton equations and of the
constraints to the family of {\it non-harmonic 3-orthogonal} gauges,
in which the instantaneous Riemannian 3-spaces have a non-fixed
trace ${}^3K$ of the extrinsic curvature but a diagonal 3-metric.
The inertial gauge variable ${}^3K$ (the general-relativistic
remnant of the freedom in the clock synchronization convention)
gives rise to a negative kinetic term in the weak ADM energy
vanishing only in the gauges with ${}^3K = 0$: is it relevant for
dark energy and back-reaction?

In the second paper there will be the linearization of the theory in
these non-harmonic 3-orthogonal gauges to obtain Hamiltonian
Post-Minkowskian gravity (without Post-Newtonian approximatins) with
asymptotic Minkowski background, non-flat instantaneous 3-spaces and
no post-Newtonian expansion. This will allow to explore the inertial
effects induced by the York time ${}^3K$ in non-flat 3-spaces (they
do not exist in Newtonian gravity) and to check how much dark matter
can be explained as an inertial aspect of Einstein's general
relativity: this will be done in a third paper on the
Post-Minkowskian 2-body problem in absence of the electro-magnetic
field and on its 0.5 Post-Newtonian limit.

\end{abstract}

\maketitle

\vfill\eject

\section{Introduction}

By re-expressing the 4-metric in terms of tetrads inside the ADM
action, an ADM formulation of tetrad gravity was presented in a
series of papers \cite{1,2,3,4} for globally hyperbolic space-times
asymptotically Minkowskian, parallelizable (so to admit ortho-normal
tetrads and a spinor structure \cite{5}) and without
super-translations (see Refs.\cite{1,3} for the needed boundary
conditions on the 4-metric and on the tetrads). This allowed the
development of canonical tetrad gravity with its fourteen Dirac
constraints and the identification of the York map as a
Shanmugadhasan canonical transformation adapted to ten of the
constraints \cite{6}. As a consequence we now have a York canonical
basis in which the {\it tidal} effects of the gravitational field
(the polarization of the gravitational waves in the linearized
theory) are separated from the {\it inertial} effects, which are
described by the {\it gauge} variables conjugated to Dirac
first-class constraints. Even if this separation is only 3-covariant
on the instantaneous non-flat 3-spaces, it allows to give a physical
interpretation of all the quantities appearing in tetrad gravity and
to rewrite the ADM Hamilton equations in a new form.\medskip

ADM tetrad gravity is formulated in an arbitrary admissible 3+1
splitting of the globally hyperbolic space-time, i.e. in a foliation
with instantaneous space-like 3-spaces tending to a Minkowski
space-like hyper-plane at spatial infinity in a direction
independent way: they correspond to a clock synchronization
convention and each one of them can be used as a Cauchy surface for
field equations. As shown in Refs.\cite{1,3} the absence of
super-translations implies that  the SPI group of asymptotic
symmetries is reduced to the asymptotic ADM Poincare' group and the
allowed 3+1 splittings  must have the instantaneous 3-spaces tending
to asymptotic special-relativistic Wigner hyper-planes orthogonal to
the ADM 4-momentum $P^{\mu}_{ADM}$ (assumed time-like as a boundary
condition) in a direction-independent way. As a consequence each
3-space is a non-inertial rest frame of the 3-universe and the 3+1
splitting defines a global non-inertial frame. At spatial infinity
there are asymptotic inertial observers, carrying a flat tetrad
$\epsilon^{\mu}_A$ (${}^4\eta_{\mu\nu}\, \epsilon^{\mu}_A\,
\epsilon^{\nu}_B = {}^4\eta_{AB}$, $\epsilon^{\mu}_{\tau} =
P^{\mu}_{ADM}/\sqrt{\sgn\, P^2_{ADM}}$), whose spatial axes can be
identified with the fixed stars of star catalogues.\medskip

Radar 4-coordinates $\sigma^A = (\tau; \sigma^r)$ adapted to the 3+1
splitting and centered on a time-like observer $x^{\mu}(\tau)$
carrying an atomic clock are used instead of local world
4-coordinates $x^{\mu}$. The time variable $\tau$ is an arbitrary
monotonically increasing function of the proper time of the
observer. If $x^{\mu} \mapsto \sigma^A$ is the coordinate
transformation to these adapted 4-coordinates, its inverse $\sigma^A
\mapsto x^{\mu} = z^{\mu}(\tau ,\sigma^r)$ defines the embedding of
the instantaneous 3-spaces $\Sigma_{\tau}$ into the space-time: each
point of $\Sigma_{\tau}$ has its clock synchronized with the atomic
clock carried by the observer. This is a mathematical idealization
of a procedure of clock synchronization. The gradients
$z^{\mu}_A(\tau, \sigma^r) = {{\partial\, z^{\mu}(\tau,
\sigma^r)}\over {\partial\, \sigma^A}}$ are the transition functions
for transforming tensors: for the 4-metric we have ${}^4g_{AB}(\tau,
\sigma^r) = z^{\mu}_A(\tau, \sigma^r)\, z^{\nu}_B(\tau, \sigma^r)\,
{}^4g_{\mu\nu}(z(\tau, \sigma^r))$, so that we get $\sqrt{|det\,
{}^4g_{\mu\nu}|} = \sqrt{|det\, {}^4g_{AB}|} / |det\, z^{\mu}_A|$.
Then we put the decomposition ${}^4g_{AB}(\tau, \sigma^r) =
{}^4E^{(\alpha)}_A(\tau, \sigma^r)\, {}^4\eta_{(\alpha)(\beta)}\,
{}^4E^{(\beta)}_B(\tau, \sigma^r)$, with ${}^4E^{(\alpha)}_A(\tau,
\sigma^r)$ arbitrary cotetrads, inside the ADM action to obtain ADM
tetrad gravity.\medskip

The three independent space-like 4-vectors $z^{\mu}_r(\tau ,\vec
\sigma )$ are tangent to $\Sigma_{\tau}$ in the point $(\tau
,\sigma^r)$. Instead the $\tau$-gradient of the embedding has the
standard decomposition $z^{\mu}_{\tau}(\tau, \sigma^r) = \Big((1+
n)\, l^{\mu} + n^r\, z^{\mu}_r \Big)(\tau, \sigma^r)$ along the unit
normal $l^{\mu}(\tau, \sigma^r) = {{\sqrt{|det\,
{}^4g_{\mu\nu}(z(\tau, \sigma^r))|}}\over {\sqrt{|det\,
{}^4g_{rs}(\tau ,\sigma^u )|}}}\, \epsilon^{\mu}{}_{\alpha\beta
\gamma}\, \Big(z^{\alpha}_1\, z^{\beta}_2\, z^{\gamma}_3\Big)(\tau
,\sigma^u) = {{1 + n(\tau, \sigma^r)}\over {|det\, z^{\mu}_A(\tau,
\sigma^r)|}}\, \epsilon^{\mu}{}_{\alpha\beta\gamma}\,
\Big(z^{\alpha}_1\, z^{\beta}_2\, z^{\gamma}_3\Big)(\tau,\sigma^u) =
z^{\mu}_A(\tau, \sigma^r)\, l^A(\tau, \sigma^r)$ and the tangents to
$\Sigma_{\tau}$ defining the lapse ($N = 1 + n$) and shift ($N^r =
n^r$) functions.\medskip

In the York canonical basis the inertial degrees of freedom of the
gravitational field are the following gauge variables: i) three
angles (the cosine directors of the 3-coordinates lines in each
point of $\Sigma_{\tau}$) describing the freedom in the choice of
3-coordinates; their fixation determines the inertial shift
functions; ii) the York time ${}^3K(\tau, \sigma^r)$ (the trace of
the extrinsic curvature of the 3-space $\Sigma_{\tau}$), describing
the remnant of the special relativistic gauge freedom in choosing
the convention for clock synchronization; it is the only gauge
variable in the set of the canonical momenta (a reflex of the
Lorentz signature of space-time); the fixation of the York time
determines the inertial lapse function. Moreover in ADM tetrad
gravity there are three inertial rotation angles and three inertial
boost parameters describing the O(3,1) gauge freedom of the tetrad
in the tangent space at each point of space-time: these six gauge
variables describe the freedom in the orientation and in the choice
of the transport law along time-like curves of three gyroscopes
located in the chosen point of space-time.\medskip

Let us remark that if we fix the 3-coordinates and the York time and
we find a solution ${}^4g_{AB}(\tau, \sigma^r)$ of the ADM Hamilton
equations in some gauge, then it has an associate preferred 3+1
splitting of space-time, whose instantaneous 3-spaces are {\it
dynamically determined} (dynamical clock synchronization convention)
by the given extrinsic curvature and by the given lapse and shift
functions, as noted in Refs. \cite{7}. As a consequence, there is a
preferred world 4-coordinate system adapted to it. Let us take the
world-line of the time-like observer as origin of the spatial world
coordinates, i.e. $x^{\mu}(\tau) = (x^o(\tau); 0)$. Then the
space-like surfaces of constant coordinate time $x^o(\tau) = const.$
coincide with the dynamical instantaneous 3-spaces $\Sigma_{\tau}$
with $\tau = const.$ of the solution. Then the preferred embedding
is

\beq
 x^{\mu} = z^{\mu}(\tau, \sigma^r) = x^{\mu}(\tau) +
 \epsilon^{\mu}_r\, \sigma^r = \delta^{\mu}_o\, x^o(\tau) +
 \epsilon^{\mu}_r\, \sigma^r.
 \label{1.1}
 \eeq

\noindent If we choose the asymptotic flat tetrads $\epsilon^{\mu}_A
= \delta^{\mu}_o\, \delta^{\tau}_A + \delta^{\mu}_i\, \delta^i_A$
and $x^o(\tau) = x^o_o + \epsilon^o_{\tau}\, \tau = x^o_o + \tau$,
we get $z^{\mu}(\tau, \sigma^r) = \delta^{\mu}_o\, x^o_o +
\epsilon^{\mu}_A\, \sigma^A$, $z^{\mu}_A(\tau, \sigma^r) =
\epsilon^{\mu}_A$ and ${}^4g_{\mu\nu}(x = z(\tau, \sigma^r)) =
\epsilon^A_{\mu}\, \epsilon_{\nu}^B\, {}^4g_{AB}(\tau, \sigma^r)$
($\epsilon^A_{\mu}$ are the inverse flat cotetrads). Then by means
of 4-diffeomorphisms we can write the solution in an arbitrary world
4-coordinate system in general not adapted to the dynamical 3+1
splitting. This gives rise to the 4-geometry containing the given
solution.

\bigskip

In this paper we are going to study ADM tetrad gravity in presence
of the following matter: N positive-energy charged scalar particles
plus the electro-magnetic field.

\bigskip

The isolated system of N positive-energy charged scalar particles
(with Grassmann-valued electric charges for a UV regularization of
self-energies) with mutual Coulomb interaction plus a transverse
electro-magnetic field in the radiation gauge have been studied in
special relativity in Refs.\cite{8,9}, where its {\it inertial
rest-frame instant form} was developed: in it the instantaneous
3-spaces (the Wigner hyper-planes) are orthogonal to the conserved
4-momentum of the isolated system. The extension to non-inertial
rest frames, in which the instantaneous 3-spaces $\Sigma_{\tau}$
tend to Wigner hyper-planes at spatial infinity (asymptotically
orthogonal to the conserved 4-momentum in the non-inertial
rest-frame instant form), was done in Ref.\cite{10}.\medskip

The starting point of these developments was the {\it parametrized
Minkowski theory} for isolated systems \cite{11,12} (positive-energy
particles, strings, fields, fluids; see also
Refs.\cite{13,14,15,16,17,18} ) admitting a Lagrangian formulation,
the precursor of the described formulation of ADM tetrad gravity.
After an admissible 3+1 splitting of Minkowski space-time with the
instantaneous 3-spaces $\Sigma_{\tau}$ described by embedding
functions $x^{\mu} = z^{\mu}(\tau, \sigma^r)$ [$\sigma^A = (\tau;
\sigma^r)$ are radar 4-coordinates centered on an arbitrary
time-like observer] one defines a Lagrangian on $\Sigma_{\tau}$
depending on the given matter and on the embedding through the
induced 4-metric ${}^4g_{AB}(\tau, \sigma^r) = (z^{\mu}_A\,
\eta_{\mu\nu}\, z^{\nu}_B)(\tau, \sigma^r)$, This Lagrangian is
obtained by the matter Lagrangian by coupling it to an external
gravitational field and by replacing the external 4-metric with
${}^4g_{AB}[z(\tau, \sigma^r)]$ (a functional of the embedding)
after having redefined the matter fields so that they know the clock
synchronization convention (for a Klein-Gordon field $\phi(x)$ we
use $\tilde \phi(\tau, \sigma^r) = \phi(z(\tau, \sigma^r))$). Each
admissible  3+1 splitting corresponds to a convention for clock
synchronization and defines a global non-inertial frame centered on
the observer. In special relativity the four embedding functions
$z^{\mu}(\tau, \sigma^r)$ play the role of the {\it inertial
potentials} in the non-inertial frame: the components of the
4-metric ${}^4g_{AB}(\tau, \sigma^r)$ and the extrinsic curvature
${}^3K_{rs}(\tau, \sigma^r)$ of the 3-spaces $\Sigma_{\tau}$ are
derived inertial potentials.\medskip

As said, the Lagrangian of parametrized Minkowski theories depends
on the embeddings $z^{\mu}(\tau ,\sigma^u )$ and on matter variables
adapted to the foliation. Due to the invariance of the action under
frame-preserving diffeomorphisms, the embeddings $z^{\mu}(\tau ,
\sigma^u )$ are {\it gauge variables}. This implies the {\it gauge
equivalence} of the description of the isolated system in any
non-inertial or inertial frame, namely its independence from the
clock synchronization convention and from the choice of
3-coordinates on $\Sigma_{\tau}$.\bigskip

The transition to globally hyperbolic space-times is done in accord
with the equivalence principle, according to which only a
description of tetrad gravity plus matter in global non-inertial
frames is possible (inertial frames exist only locally near a free
falling particle). In particular the absence of super-translations
identifies the non-inertial rest frames (orthogonal to the ADM
4-momentum at spatial infinity) as the {\it only} relevant ones.
\medskip

The real difference with parametrized Minkowski theories is that the
independent variables are now the cotetrads ${}^4E^{(\alpha
)}_A(\tau ,\vec \sigma )$ (or the 4-metric ${}^4g_{AB}(\tau ,\vec
\sigma )$ in metric gravity) and not the embeddings $z^{\mu}(\tau
,\vec \sigma )$ ($z^{\mu}_A$ are only transition coefficients, while
in special relativity they are flat cotetrads): they are given in
Eq.(\ref{1.1}).

\bigskip

Following Ref.\cite{11,13} we will describe N scalar particles of
masses $m_i$, i=1,..,N, with 3-coordinates $\eta^r_i(\tau )$ on the
instantaneous 3-spaces $\Sigma_{\tau}$ (diffeomorphic to $R^3$)
identified by the intersection with $\Sigma_{\tau}$ of their
world-lines $x_i^{\mu }(\tau )=z^{\mu}(\tau ,\eta^r_i(\tau ))$ in
$M^4$. Therefore the world-lines $x^{\mu}_i(\tau)$ are derived
quantities (as shown in Ref.\cite{8} they are covariant
non-canonical predictive coordinates) and, as shown in Ref.\cite{6},
describe particles with a definite sign of the energy (due to clock
synchronization each particle is described by 3, not 4, position
variables). As a consequence, the momenta $p^{\mu}_i(\tau)$ are not
defined: as shown in Ref.\cite{8} it is possible to define them so
that the mass shell constraints $\sgn\, p^2_i - m^2_i\, c^2 \approx
0$ are satisfied also in presence of interactions. However other
definitions are possible.

\bigskip

To regularize the gravitational self-energies we will assume that
the signs $\eta_i$ of the energies of the particles (a topological
number with the two values $\pm 1$) are described by two complex
conjugate Grassmann variables $\theta_i$, $\theta_i^*$: $\eta_i =
\theta_i^*\, \theta_i$, $\eta_i^2 = 0$, $\eta_i\, \eta_j = \eta_j\,
\eta_i \not= 0$ for $i \not= j$ (after a formal quantization they
describe a 2-level system). In this way we are implementing a $i
\not= j$ rule like it was done in classical electrodynamics by using
Grassmann variables to describe the electric charges of charged
relativistic scalar particles \cite{14,15}, \cite{8,9} : $Q_i =
\theta_i^{(Q)*}\, \theta^{(Q)}_i$, $Q^2_i = 0$, $Q_i\, Q_j = Q_j\,
Q_i \not= 0$ for $i \not= j$, to avoid divergencies in the
electro-magnetic self-energies. In electrodynamics this
semi-classical method turns out to be equivalent to a simultaneous
UV and IR regularization of QED (the physics contained in the
discarded, often diverging, quantities is recovered in QED as
radiative corrections) such that only the 1-photon exchange Feynman
diagram survives: it is reformulated as a classical Cauchy problem
with inter-particle action-at-a-distance potentials. In this way it
is possible to arrive at the identification of the Darwin potential
starting from the regularized classical theory (no essential
singularities on the world-lines of the charged point particles so
that the action principle of Section III is well posed) instead of
descending from the Bethe-Salpeter equation of QFT towards the
theory of relativistic bound states. Moreover one gets the
phase-space expression of the Lienard-Wiechert solution and, even if
electro-magnetic self-energies are absent, the asymptotic Larmor
formula for the emission of electro-magnetic waves is recovered if
at least two particles are present (it is an interference effect).
This charged point particle model can also be considered as a
consistent point-like approximation of the classical models for an
extended electron (all suffering of causality problems). In general
relativity there is not an accepted quantum gravity theory replacing
QED. Therefore our point-like approximation of extended objects
(they will be studied elsewhere) is only a UV and IR regularization
killing the gravitational self-energies and corresponding to a
"1-graviton exchange" in an unspecified quantum gravity theory. As
it will be shown in the second paper on the Post-Minkowskian
approximation (weak field limit) the energy, 3-momentum and angular
momentum balance equations for the emission of gravitational waves
turn out to be satisfied in absence of self-energies and of
Post-Newtonian approximations due to the conservation of the ADM
Poincare' generators.
\bigskip

The electro-magnetic field is described by the vector potential
$A_A(\tau ,\sigma^u) = z^{\mu}_A(\tau ,\sigma^u)\, A_{\mu}(z(\tau
,\sigma^u))$ and by the field strength $F_{AB}(\tau ,\sigma^u ) =
\partial_A\, A_B(\tau ,\sigma^u) - \partial_B\,  A_A(\tau ,\sigma^u)
= z^{\mu}_A(\tau ,\sigma^u)\, z^{\nu}_B(\tau ,\sigma^u)\,
F_{\mu\nu}(z(\tau ,\sigma^u))$. Following Ref.\cite{10}, we shall
restrict the formulation to an electro-magnetic field in the {\it
radiation gauge} in the instantaneous 3-spaces $\Sigma_{\tau}$. Even
if this gauge is not covariant, it allows to extract the
non-inertial analogue of the Coulomb potential among the particles
and to work with transverse electro-magnetic fields (the
non-covariant electro-magnetic Dirac observables).
\bigskip

The main aim of this paper is to write explicitly the Hamilton
equations of ADM tetrad gravity coupled to this type of matter in
the York canonical basis of Ref.\cite{6} in the class of Schwinger
time gauges, in which the tetrads are adapted to the 3+1 splitting
of space-time (the time-like tetrad coincides with the unit normal
to the 3-space $\Sigma_{\tau}$ and the three spatial tetrads have a
conventional orientation). In these gauges the gravitational field
is described by 10 configuration variables and 10 conjugate momenta
like in canonical metric gravity. Since there are still 8
first-class constraints there are 8 gauge variables ({\it inertial}
effects): 7 configuration variables (the lapse and shift functions
and 3 variables describing the freedom in the choice of the
3-coordinates in $\Sigma_{\tau}$) and a momentum variable (the trace
${}^3K(\tau, \sigma^r)$ of the extrinsic curvature of the 3-space
$\Sigma_{\tau}$, describing the general relativistic remnant of the
freedom in the clock synchronization convention as already said).
One configuration variable (the element of 3-volume in the
instantaneous 3-space $\Sigma_{\tau}$, canonically conjugate to
${}^3K$) and 7 momenta are determined by the 8 constraints. The
remaining two pairs of conjugate variables describe the physical
{\it tidal} effects contained in the gravitational field (they are
only 3-scalars and partial Dirac observables with respect to 10 of
the 14 constraints of tetrad gravity).

\bigskip

As we shall see, the extrinsic curvature ${}^3K_{rs}(\tau,
\sigma^r)$ of the instantaneous 3-spaces $\Sigma_{\tau}$, viewed  as
sub-manifolds of the space-time, depends upon the matter, the tidal
variables, the three inertial gauge variables for the choice of the
3-coordinates, the shift functions and, finally, upon the inertial
gauge variable ${}^3K(\tau, \sigma^r)$ (its trace, i.e. the York
time). Once the gauge freedoms are fixed the extrinsic curvature
and, as a consequence, the instantaneous 3-spaces are dynamically
determined for every solution of Einstein equations with the allowed
boundary conditions and with Cauchy data compatible with the
constraints.

\bigskip

In canonical ADM tetrad gravity the Dirac Hamiltonian is equal to
the weak ADM energy ${\hat E}_{ADM}$ (its form as a volume integral
over the 3-space $\Sigma_{\tau}$) plus constraints. As a
consequence, there is not a frozen picture like in spatially compact
without boundary space-times (the ones used in loop quantum
gravity). We will see that the gauge momentum (the inertial York
time ${}^3K(\tau, \sigma^r)$; its time nature is a reflex of Lorentz
signature) gives rise to a negative kinetic term in ${\hat E}_{ADM}$
vanishing only in the gauges ${}^3K(\tau, \sigma^r) = 0$.
\bigskip

In Schwinger time gauges the ADM Hamilton equations in the York
canonical basis are naturally separated in four groups. Besides the
Hamilton equations for matter and for the tidal variables, there are
four contracted Bianchi identities (implying the $\tau$-preservation
of the super-Hamiltonian and super-momentum constraints) and the
Hamilton equations for the four inertial gauge variables describing
the 3-coordinates in $\Sigma_{\tau}$ and the clock synchronization
convention (if we add four gauge fixings for these variables, these
Hamilton equations determine the shift and lapse functions).
\bigskip

Then we will define a family of non-harmonic gauges, the {\it
3-orthogonal} ones, in which the 3-metric in the 3-spaces
$\Sigma_{\tau}$ is diagonal but the inertial gauge variable
${}^3K(\tau, \sigma^r)$, fixing the clock synchronization
convention, is equal to an arbitrary numerical function $F(\tau,
\sigma^r)$. The restriction of the Hamilton equations to the
3-orthogonal gauges is given explicitly.\medskip

It is in this family of 3-orthogonal gauges that we will define a
linearization of ADM canonical tetrad gravity plus matter in the
second paper \cite{19}, to obtain a formulation of Hamiltonian
post-Minkowskian gravity (without post-Newtonian expansions) with
non-flat Riemannian 3-spaces and asymptotic Minkowski background.
This will allow to study the dynamical effects of the inertial
potential ${}^3K(\tau, \sigma^r)$. In the third paper \cite{19a} on
the Post-Minkowskian 2-body problem in absence of the
electro-magnetic field, it will be shown that at the 0.5
Post-Newtonian order the inertial York time can, at least partially,
mimic the effects attributed to dark matter, which therefore could
be explained as an inertial relativistic effect.

\bigskip

\bigskip

In Section II there is a review of ADM tetrad gravity, of the York
canonical basis with new results beyond Ref.\cite{6}, showing the
relevance of the Eulerian observers associated to a 3+1 splitting of
space-time, and of the ADM Poincare' charges.

In Section III we give the action of our system in Subsection A,
where we evaluate the Dirac Hamiltonian and the constraints of the
Hamiltonian formulation. In Subsection B we give the Hamilton
equations of the particles. In Subsection C, after having evaluated
the Hamilton equations of the electro-magnetic field, we define the
non-covariant electro-magnetic radiation gauge and give the
restriction to it of the Hamilton equations of the particles and of
the transverse electro-magnetic field. In Subsections D and E we
give the Dirac Hamiltonian, the constraints and the ADM Poincare'
charges in the York canonical basis, while in Subsection F we define
the Schwinger time gauges. Then in Section IV we give the Hamilton
equations in the York canonical basis in these gauges.

In Section V we discuss the formulation in the York canonical basis
and in the Schwinger time gauges of some of the most used gauges for
canonical gravity, included the Hamiltonian formulation of the
harmonic gauges for Einstein equations.

In Section VI we give the restriction of the Dirac Hamiltonian, of
the constraints and of the Hamilton equations to the above defined
family of 3-orthogonal Schwinger time gauges.

In the Conclusions there are some final remarks.

In Appendix A there is a discussion of the contracted Bianchi
identities and a comparison with the standard ADM Hamilton
equations.

In Appendix B there are the calculations needed for the Hamilton
equations in the Schwinger time gauges, while in Appendix C these
calculations are restricted to the 3-orthogonal gauges.

\vfill\eject

\section{Review of Tetrad Gravity and of the York Map}

\subsection{Tetrads and Cotetrads}

We use radar 4-coordinates $\sigma^A = (\sigma^{\tau} = \tau ;
\sigma^r)$, $A = \tau ,r$, adapted to an admissible 3+1 splitting of
the space-time and centered on an arbitrary time-like observer: they
define a non-inertial frame centered on the observer, so that they
are {\it observer and frame- dependent}. The instantaneous 3-spaces
identified by this convention for clock synchronization are denoted
$\Sigma_{\tau}$.\medskip

The 4-metric ${}^4g_{AB}$ has signature $\sgn\, (+---)$ with $\sgn =
\pm$ (the particle physics, $\sgn = +$, and general relativity,
$\sgn = -$, conventions). Flat indices $(\alpha )$, $\alpha = o, a$,
are raised and lowered by the flat Minkowski metric
${}^4\eta_{(\alpha )(\beta )} = \sgn\, (+---)$. We define
${}^4\eta_{(a)(b)} = - \sgn\, \delta_{(a)(b)}$ with a
positive-definite Euclidean 3-metric. From now on we shall denote
the curvilinear 3-coordinates $\sigma^r$ with the notation $\vec
\sigma$ for the sake of simplicity. Usually the convention of sum
over repeated indices is used, except when there are too many
summations.
\bigskip

We shall work with the tetrads ${}^4E^A_{(\alpha)}(\tau, \vec
\sigma)$ and the cotetrads ${}^4E_A^{(\alpha)}(\tau, \vec \sigma)$.
To rebuild the original world tetrads ${}^4E^{\mu}_{(\alpha)}(\tau,
\vec \sigma) = z^{\mu}_A(\tau, \vec \sigma)\,
{}^4E^A_{(\alpha)}(\tau, \vec \sigma)$ we must use Eq.(\ref{1.1})
giving the embedding.

\bigskip

General tetrads ${}^4E^A_{(\alpha )}(\tau, \vec \sigma)$ and
cotetrads ${}^4E_A^{(\alpha )}(\tau, \vec \sigma)$ are connected to
the tetrads and cotetrads adapted to the 3+1 splitting (the
time-like tetrad is the unit normal $l^A$ to $\Sigma_{\tau}$) by a
point-dependent standard Lorentz boost for time-like orbits acting
on the flat indices \footnote{In each tangent plane to a point of
$\Sigma_{\tau}$ the point-dependent standard Wigner boost for
time-like Poincare' orbits $L^{(\alpha )}{}_{(\beta )}(V(z(\sigma
));\,\, {\buildrel \circ \over V}) = \delta^{(\alpha )}_{(\beta )} +
2 \sgn\, V^{(\alpha )}(z(\sigma ))\, {\buildrel \circ \over
V}_{(\beta )} - \sgn\, {{(V^{(\alpha )}(z(\sigma )) + {\buildrel
\circ \over V}^{(\alpha )})\, (V_{(\beta )}(z(\sigma )) + {\buildrel
\circ \over V}_{(\beta )})}\over {1 + V^{(o)}(z(\sigma ))}}\,
{\buildrel {def}\over =}\, L^{(\alpha )}{}_{(\beta
)}(\varphi_{(a)})$ sends the unit future-pointing time-like vector
${\buildrel o\over V}^{(\alpha )} = (1; 0)$ into the unit time-like
vector $V^{(\alpha )} = {}^4E^{(\alpha )}_A\, l^A = \Big(\sqrt{1 +
\sum_a\, \varphi^2_{(a)}}; \varphi^{(a)} = - \sgn\,
\varphi_{(a)}\Big)$, where $l^A$ is the unit future-pointing normal
to $\Sigma_{\tau}$. We have $L^{-1}(\varphi_{(a)}) = {}^4\eta\,
L^T(\varphi_{(a)})\, {}^4\eta = L(- \varphi_{(a)})$. As a
consequence, the flat indices $(a)$ of the adapted tetrads and
cotetrads and of the triads and cotriads on $\Sigma_{\tau}$
transform as Wigner spin-1 indices under point-dependent SO(3)
Wigner rotations $R_{(a)(b)}(V(z(\sigma ));\,\, \Lambda (z(\sigma
))\, )$ associated with Lorentz transformations $\Lambda^{(\alpha
)}{}_{(\beta )}(z)$ in the tangent plane to the space-time in the
given point of $\Sigma_{\tau}$. Instead the index $(o)$ of the
adapted tetrads and cotetrads is a local Lorentz scalar index.}

\medskip

\bea
  {}^4E^A_{(\alpha )} &=& \eo^A_{(\beta )}\, L^{(\beta )}{}_{(\alpha
 )}(\varphi_{(a)}),\qquad
 {}^4E^{(\alpha )}_A = L^{(\alpha )}{}_{(\beta )}(\varphi_{(a)})\, \eo^{(\beta
 )}_A, \nonumber \\
 &&{}\nonumber \\
 {}^4g_{AB} &=& {}^4E^{(\alpha )}_A\, {}^4\eta_{(\alpha
 )(\beta )}\, {}^4E^{(\beta )}_B = \eo^{(\alpha )}_A\, {}^4\eta_{(\alpha
 )(\beta )}\, \eo^{(\beta )}_B,
 \label{2.1}
 \eea

\noindent where the last line gives the resolution of the 4-metric
in terms of cotetrads.
\bigskip

The adapted tetrads and cotetrads   have the expression

\bea
 \eo^A_{(o)} &=& {1\over {1 + n}}\, (1; - n_{(a)}\,
 {}^3e^r_{(a)}) = l^A,\qquad \eo^A_{(a)} = (0; {}^3e^r_{(a)}), \nonumber \\
 &&{}\nonumber  \\
 \eo^{(o)}_A &=& (1 + n)\, (1; \vec 0) = \sgn\, l_A,\qquad \eo^{(a)}_A
= (n_{(a)}; {}^3e_{(a)r}),
 \label{2.2}
 \eea

\noindent where ${}^3e^r_{(a)}$ and ${}^3e_{(a)r}$ are triads and
cotriads on $\Sigma_{\tau}$ and $n_{(a)} = n_r\, {}^3e^r_{(a)} =
n^r\, {}^3e_{(a)r}$ \footnote{Since we use the positive-definite
3-metric $\delta_{(a)(b)} $, we shall use only lower flat spatial
indices. Therefore for the cotriads we use the notation
${}^3e^{(a)}_r\,\, {\buildrel {def}\over =}\, {}^3e_{(a)r}$ with
$\delta_{(a)(b)} = {}^3e^r_{(a)}\, {}^3e_{(b)r}$.} are adapted shift
functions. In Eqs.(\ref{2.2}) $N(\tau, \vec \sigma) = 1 + n(\tau,
\vec \sigma) > 0$, with $n(\tau ,\vec \sigma)$ vanishing at spatial
infinity (absence of super-translations), so that $N(\tau, \vec
\sigma)\, d\tau$ is positive from $\Sigma_{\tau}$ to $\Sigma_{\tau +
d\tau}$, is the lapse function; $N^r(\tau, \vec \sigma) = n^r(\tau,
\vec \sigma)$, vanishing at spatial infinity (absence of
super-translations), are the shift functions.

\bigskip

The adapted tetrads $\eo^A_{(a)}$ are defined modulo SO(3) rotations
$\eo^A_{(a)} = R_{(a)(b)}(\alpha_{(e)})\, {}^4{\buildrel \circ \over
{\bar E}}^A_{(b)}$, ${}^3e^r_{(a)} = R_{(a)(b)}(\alpha_{(e)})\,
{}^3{\bar e}^r_{(b)}$, where $\alpha_{(a)}(\tau ,\vec \sigma )$ are
three point-dependent Euler angles. After having chosen an arbitrary
point-dependent origin $\alpha_{(a)}(\tau ,\vec \sigma ) = 0$, we
arrive at the following adapted tetrads and cotetrads [${\bar
n}_{(a)} = \sum_b\, n_{(b)}\, R_{(b)(a)}(\alpha_{(e)})\,$]

\bea
 {}^4{\buildrel \circ \over {\bar E}}^A_{(o)}
 &=& \eo^A_{(o)} = {1\over {1 + n}}\, (1; - {\bar n}_{(a)}\,
 {}^3{\bar e}^r_{(a)}) = l^A,\qquad {}^4{\buildrel \circ \over
 {\bar E}}^A_{(a)} = (0; {}^3{\bar e}^r_{(a)}), \nonumber \\
 &&{}\nonumber  \\
 {}^4{\buildrel \circ \over {\bar E}}^{(o)}_A
 &=& \eo^{(o)}_A = (1 + n)\, (1; \vec 0) = \sgn\, l_A,\qquad
 {}^4{\buildrel \circ \over {\bar E}}^{(a)}_A
= ({\bar n}_{(a)}; {}^3{\bar e}_{(a)r}),
 \label{2.3}
 \eea

\noindent which we shall use as a reference standard.\medskip

Then Eqs.(\ref{2.1}), namely

\beq
 {}^4E^A_{(\alpha )} = {}^4{\buildrel \circ \over {\bar E}}^A_{(o)}\,
 L^{(o)}{}_{(\alpha )}(\varphi_{(c)}) + {}^4{\buildrel \circ \over
 {\bar E}}^A_{(b)}\, R^T_{(b)(a)}(\alpha_{(c)})\,
 L^{(a)}{}_{(\alpha )}(\varphi_{(c)}),
 \label{2.4}
 \eeq
\medskip

\noindent show that every point-dependent Lorentz transformation
 $\Lambda$ in the tangent planes may be parametrized with the
 (Wigner) boost parameters $\varphi_{(a)}$ and the Euler angles
 $\alpha_{(a)}$, being the product $\Lambda = R\, L$ of a rotation
 and a boost.

\bigskip

The future-oriented unit normal to $\Sigma_{\tau}$ and the projector
on $\Sigma_{\tau}$ are\medskip

\bea
 l_A &=& \sgn\, (1 + n)\, \Big(1;\, 0\Big),\qquad {}^4g^{AB}\, l_A\, l_B =
\sgn ,\nonumber \\
 &&{}\nonumber \\
 l^A &=& \sgn\, (1 + n)\, {}^4g^{A\tau} = {1\over {1 + n}}\, \Big(1;\, - n^r\Big) =
{1\over {1 + n}}\, \Big(1;\, - n_{(a)}\, {}^3e_{(a)}^r\Big),\nonumber \\
 &&{}\nonumber \\
 {}^3h^B_A &=& \delta^B_A - \sgn\, l_A\, l^B,\qquad
 {}^3h^{\tau}_{\tau} = {}^3h^{\tau}_r = 0,\quad {}^3h^r_{\tau} =
 n_{(a)}\, {}^3e^r_{(a)},\quad {}^3h^r_s = \delta^r_s,\nonumber \\
 &&{}\nonumber \\
 &&{}^3h_{\tau\tau} = - \sgn\, n_{(a)}\, n_{(a)},\quad {}^3h_{\tau
 r} = - \sgn\, n_{(a)}\, {}^3e_{(a)r},\quad {}^3h_{rs} = - \sgn\,
 {}^3e_{(a)r}\, {}^3e_{(a)s},\nonumber \\
 &&{}^3h^{\tau\tau} = {}^3h^{\tau r} = 0,\qquad {}^3h^{rs} = -
 \sgn\, {}^3e^r_{(a)}\, {}^3e^s_{(a)}.\nonumber \\
 &&{}
 \label{2.5}
 \eea

\bigskip

\subsection{The 4-metric and the Canonical Variables.}

The 4-metric has the following expression

 \bea
 {}^4g_{\tau\tau} &=& \sgn\, [(1 + n)^2 - {}^3g^{rs}\, n_r\,
 n_s] = \sgn\, [(1 + n)^2 - n_{(a)}\, n_{(a)}],\nonumber \\
 {}^4g_{\tau r} &=& - \sgn\, n_r = -\sgn\, n_{(a)}\,
 {}^3e_{(a)r},\nonumber \\
  {}^4g_{rs} &=& -\sgn\, {}^3g_{rs},\nonumber \\
 &&{}\nonumber \\
 &&{}^3g_{rs} = h_{rs} = {}^3e_{(a)r}\, {}^3e_{(a)s},\qquad {}^3g^{rs} =
 h^{rs} = {}^3e^r_{(a)}\, {}^3e^s_{(a)},\nonumber \\
 &&{}\nonumber \\
 {}^4g^{\tau\tau} &=& {{\sgn}\over {(1 + n)^2}},\qquad
  {}^4g^{\tau r} = -\sgn\, {{n^r}\over {(1 + n)^2}} = -\sgn\, {{{}^3e^r_{(a)}\,
 n_{(a)}}\over {(1 + n)^2}},\nonumber \\
 {}^4g^{rs} &=& -\sgn\, ({}^3g^{rs} - {{n^r\, n^s}\over
 {(1 + n)^2}}) = -\sgn\, {}^3e^r_{(a)}\, {}^3e^s_{(b)}\, (\delta_{(a)(b)} -
 {{n_{(a)}\, n_{(b)}}\over {(1 + n)^2}}),\qquad {}^3g^{rs} = h^{rs},\nonumber \\
 &&{}\nonumber \\
 &&{}^4g^{\tau\tau}\, {}^4g^{rs} - {}^4g^{\tau r}\, {}^4g^{\tau s} = -
{{{}^3g^{rs}}\over {(1 + n)^2}},\qquad {}^3g = \gamma =
({}^3e)^2,\quad {}^3e = det\, {}^3e_{(a)r},\nonumber \\
 &&\sqrt{- g } = \sqrt{|{}^4g|} = {{\sqrt{{}^3g}}\over {\sqrt{\sgn\,
{}^4g^{\tau\tau}}}} = \sqrt{\gamma}\, (1 + n) = {}^3e\, (1 + n).
 \label{2.6}
 \eea

\bigskip

The 3-metric ${}^3g_{rs} = h_{rs}$ has signature $(+++)$, so that we
may put all the flat 3-indices {\it down}. We have ${}^3g^{ru}\,
{}^3g_{us} = \delta^r_s$ ($h^{ru}\, h_{us} = \delta^r_s$),
$\partial_A\, {}^3g^{rs} = - {}^3g^{ru}\, {}^3g^{sv}\, \partial_A\,
{}^3g_{uv}$.\bigskip

The conditions for having an admissible 3+1 splitting of space-time
are:\medskip

a) $1 + n(\tau ,\vec \sigma) > 0$ everywhere (the instantaneous
3-spaces never intersect each other);\hfill\break

b) the M$\o$ller conditions \cite{10,16}, which imply\hfill\break

i) $\sgn\, {}^4g_{\tau\tau} > 0$, i.e. $(1 + n)^2 > {}^3g^{rs}\,
n_r\, n_s$ (the rotational velocity never exceeds the velocity of
light $c$, so that the coordinate singularity of the rotating disk
named "horizon problem" is avoided); \hfill\break

ii) $\sgn\, {}^4g_{rr} = - {}^3g_{rr} < 0$ (satisfied by the
signature of ${}^3g_{rs} = h_{rs}$), ${}^4g_{rr}\, {}^4g_{ss} -
({}^4g_{rs})^2 > 0$ and $ det\, \sgn\, {}^4g_{rs} = - det\,
{}^3g_{rs} < 0$ (satisfied by the signature of ${}^3g_{rs}$) so that
$det\, {}^4g_{AB} < 0$; these conditions imply that ${}^3g_{rs} =
h_{rs}$ has three definite positive eigenvalues $\lambda_r =
\Lambda_r^2$ in the non-degenerate case without Killing symmetries,
the only one we consider;\hfill\break

c) the space-time is asymptotically Minkowskian with the
instantaneous 3-spaces orthogonal to the ADM 4-momentum at spatial
infinity: they are non-inertial rest frames of the 3-universe
(isolated system), there is an asymptotic Minkowski background
4-metric and there are asymptotic inertial observers whose spatial
axes $\epsilon^{\mu}_r$ are  identified by the fixed stars of star
catalogues.

\bigskip

As said in Ref.\cite{6}, in ADM canonical tetrad gravity the 16
configuration variables are: the 3 boost variables
$\varphi_{(a)}(\tau, \vec \sigma)$; the lapse and shift functions
$n(\tau, \vec \sigma)$ and $n_{(a)}(\tau, \vec \sigma)$; the
cotriads ${}^3e_{(a)r}(\tau, \vec \sigma)$. Their conjugate momenta
are $\pi_{\varphi_{(a)}}(\tau, \vec \sigma)$, $\pi_n(\tau, \vec
\sigma)$, $\pi_{n_{(a)}}(\tau, \vec \sigma)$, ${}^3\pi^r_{(a)}(\tau,
\vec \sigma)$. There are 14 first-class constraints: A) the 10
primary constraints $\pi_{\varphi_{(a)}}(\tau, \vec \sigma) \approx
0$, $\pi_n(\tau, \vec \sigma) \approx 0$, $\pi_{n_{(a)}}(\tau, \vec
\sigma) \approx 0$ and the 3 rotation constraints $M_{(a)}(\tau,
\vec \sigma) \approx 0$ implying the gauge nature of the  Euler
angles $\alpha_{(a)}(\tau, \vec \sigma)$, of the boost variables
$\varphi_{(a)}(\tau, \vec \sigma)$ and of the lapse and shift
functions $1 + n(\tau, \vec \sigma)$, $n_{(a)}(\tau, \vec \sigma)$;
B) the 4 secondary super-Hamiltonian and super-momentum constraints
${\cal H}(\tau, \vec \sigma) \approx 0$, ${\cal H}_{(a)}(\tau, \vec
\sigma) \approx 0$. As a consequence there are 14 gauge variables
(the inertial effects) and two pairs of canonically conjugate
physical degrees of freedom (the tidal effects).

\bigskip

At this stage the basis of canonical variables for this formulation
of tetrad gravity, naturally adapted to 7 of the 14 first-class
constraints, is

\beq
 \begin{minipage}[t]{3cm}
\begin{tabular}{|l|l|l|l|} \hline
$\varphi_{(a)}$ & $n$ & $n_{(a)}$ & ${}^3e_{(a)r}$ \\ \hline $
\pi_{\varphi_{(a)}}\, \approx 0$ & $\pi_n\, \approx 0$ &
$\pi_{n_{(a)}}\, \approx 0 $ & ${}^3{ \pi}^r_{(a)}$
\\ \hline
\end{tabular}
\end{minipage}
 \label{2.7}
 \eeq

From Eqs.(5.5) of Ref.\cite{3} we assume the following
(direction-independent, so to kill super-translations) boundary
conditions at spatial infinity ($r = \sqrt{\sum_r\, (\sigma^r)^2}$;
$\epsilon > 0$; $M = const.$): $n(\tau, \vec \sigma)
\rightarrow_{r\, \rightarrow\, \infty}\, O(r^{- (2 + \epsilon)})$,
$\pi_n(\tau, \vec \sigma) \rightarrow_{r\, \rightarrow\, \infty}\,
O(r^{-3})$, $n_{(a)}(\tau, \vec \sigma) \rightarrow_{r\,
\rightarrow\, \infty}\, O(r^{- \epsilon})$, $\pi_{n_{(a)}}(\tau,
\vec \sigma) \rightarrow_{r\, \rightarrow\, \infty}\, O(r^{-3})$,
$\varphi_{(a)}(\tau, \vec \sigma) \rightarrow_{r\, \rightarrow\,
\infty}\, O(r^{- (1 + \epsilon)})$, $\pi_{\varphi_{(a)}}(\tau, \vec
\sigma) \rightarrow_{r\, \rightarrow\, \infty}\, O(r^{-2})$,
${}^3e_{(a)r}(\tau, \vec \sigma) \rightarrow_{r\, \rightarrow\,
\infty}\, \Big(1 + {M\over {2 r}}\Big)\, \delta_{ar} + O(r^{-
3/2})$, ${}^3\pi^r_{(a)}(\tau, \vec \sigma) \rightarrow_{r\,
\rightarrow\, \infty}\, O(r^{- 5/2})$.

\subsection{The York Canonical Basis.}

In Ref.\cite{6} we studied the following point canonical
transformation (it is a Shanmugadhasan canonical transformation
\cite{20}) on the canonical variables (\ref{2.7}), implementing the
York map of Refs.\cite{17,18} and identifying a canonical basis
adapted to the 10 primary first-class constraints . It is realized
in two steps and leads to the following York canonical basis

\bea
 &&\begin{minipage}[t]{3cm}
\begin{tabular}{|l|l|l|l|} \hline
$\varphi_{(a)}$ & $n$ & $n_{(a)}$ & ${}^3e_{(a)r}$ \\ \hline
$\pi_{\varphi_{(a)}} \approx 0$ & $\pi_n \approx 0$ & $
\pi_{n_{(a)}} \approx 0 $ & ${}^3{ \pi}^r_{(a)}$
\\ \hline
\end{tabular}
\end{minipage} \hspace{1cm}\nonumber \\
 && {\longrightarrow \hspace{.2cm}} \
\begin{minipage}[t]{4 cm}
\begin{tabular}{|ll|ll|l|} \hline
$\varphi_{(a)}$ & $\alpha_{(a)}$ & $n$ & ${\bar n}_{(a)}$ &  ${}^3{\bar e}_{(a)r}$\\
\hline
 $\pi_{\varphi_{(a)}} \approx 0$ & $ \pi^{(\alpha)}_{(a)} \approx 0$ & $ \pi_n \approx 0$
 & $\pi_{{\bar n}_{(a)}} \approx 0$ & ${}^3{\bar \pi}^r_{(a)}$ \\ \hline
\end{tabular}
\end{minipage} \nonumber \\
 &&{}\nonumber \\
 &&\hspace{1cm} {\longrightarrow \hspace{.2cm}} \
\begin{minipage}[t]{4 cm}
\begin{tabular}{|ll|ll|l|l|l|} \hline
$\varphi_{(a)}$ & $\alpha_{(a)}$ & $n$ & ${\bar n}_{(a)}$ &
$\theta^r$ & $\tilde \phi$ & $R_{\bar a}$\\ \hline
$\pi_{\varphi_{(a)}} \approx0$ &
 $\pi^{(\alpha)}_{(a)} \approx 0$ & $\pi_n \approx 0$ & $\pi_{{\bar n}_{(a)}} \approx 0$
& $\pi^{(\theta )}_r$ & $\pi_{\tilde \phi}$ & $\Pi_{\bar a}$ \\
\hline
\end{tabular}
\end{minipage}\nonumber \\
 &&{}
 \label{2.8}
 \eea

\noindent where ${}^3{\bar e}_{(a)r} = \sum_b\, {}^3e_{(b)r}\,
R_{(b)(a)}(\alpha_{(e)})$ (with conjugate momenta  ${}^3{\bar
\pi}^r_{(a)}$), ${\bar n}_{(a)} = \sum_b\, n_{(b)}\,
R_{(b)(a)}(\alpha_{(e)})$ are the cotriads and the shift functions
at $\alpha_{(a)}(\tau ,\vec \sigma ) = 0$ after the extraction of
the rotation matrix $R_{(a)(b)}(\alpha_{(e)}(\tau ,\vec \sigma))$,
see after Eq.(\ref{2.2}).
\bigskip

With the first canonical transformation we extract the 3 angles
$\alpha_{(a)}(\tau ,\vec \sigma )$ from the cotriads and we
Abelianize the rotation constraints ${}^3M_{(a)}(\tau ,\vec \sigma )
\approx 0$, replacing them with the momenta conjugate to the angles,
$\pi^{(\alpha )}_{(a)} = - \sum_b\, {}^3M_{(b)}\,
A_{(b)(a)}(\alpha_{(e)}) \approx 0$. The o(3) Lie algebra-valued
Cartan matrix $A_{(l)(m)}(\alpha_{(a)})$ [with $A(0) = B(0) = 1$,
where $B(\alpha_{(a)}) = A^{-1}(\alpha_{(a)})$] is defined in
Ref.\cite{3}.

\bigskip
The second canonical transformation is based on the fact that the
3-metric $ {}^3g_{rs}$ is a real symmetric $3 \times 3$ matrix,
which may be diagonalized with an {\it orthogonal} matrix
$V(\theta^r)$, $V^{-1} = V^T$ ($\sum_u\, V_{ua}\, V_{ub} =
\delta_{ab}$, $\sum_a\, V_{ua}\, V_{va} = \delta_{uv}$, $\sum_{uv}\,
\epsilon_{wuv}\, V_{ua}\, V_{vb} = \sum_c\, \epsilon_{abc}\,
V_{cw}$), $det\, V = 1$, depending on three parameters $\theta^r$
\footnote{Due to the positive signature of the 3-metric, we define
the matrix $V$ with the following indices: $V_{ru}$. Since the
choice of Shanmugadhasan canonical bases breaks manifest covariance,
we will use the notation $V_{ua} = \sum_v\, V_{uv}\, \delta_{v(a)}$
instead of $V_{u(a)}$. We use the following types of indices: $a =
1,2,3$ and $\bar a = 1,2$.} (their conjugate momenta are determined
by the super-momentum constraints). If we choose these three gauge
parameters to be Euler angles ${\hat \theta}^i(\tau, \vec \sigma)$,
we get a description of the 3-coordinate systems on $\Sigma_{\tau}$
from a local point of view, because they give the orientation of the
tangents to the three 3-coordinate lines through each point.
However, as shown in Appendix A of Ref.\cite{21} (see the
bibliography therein), it is more convenient to choose the three
gauge parameters as first kind coordinates $\theta^i(\tau, \vec
\sigma)$ ($- \infty < \theta^i < + \infty$) on the O(3) group
manifold, so that by definition we have $A_{ri}(\theta^n)\, \theta^i
= \delta_{ri}\, \theta^i$ \footnote{A similar diagonalization of the
3-metric using o(3) first kind coordinates has been considered also
in Ref.\cite{22} for numerical gravity purposes. However no attempt
to find a phase space canonical basis like the York one is done in
this paper.}. In this case we have $V_{ru}(\theta^i) = \Big(e^{-
\sum_i\, {\hat T}_i\, \theta^i}\Big)_{ru}$, where $({\hat T}_i)_{ru}
= \epsilon_{rui}$ are the generators of the o(3) Lie algebra in the
adjoint representation, and the Euler angles may be expressed as
${\hat \theta}^i = f^i(\theta^n)$. Since the Cartan matrix has the
form $A(\theta^n) = {{1 - e^{- \sum_i\, {\hat T}_i\, \theta^i}
}\over {\sum_i\, {\hat T}_i\, \theta^i}}$, we get the following
expansions around $\theta^i = 0$: $V_{ru}(\theta^i)\,
\rightarrow_{\theta^i \rightarrow 0}\, \delta_{ru} -
\epsilon_{rui}\, \theta^i + O(\theta^2)$, $A_{ru}(\theta^i)\,
\rightarrow_{\theta^i \rightarrow 0}\, \delta_{ru} - {1 \over 2}\,
\epsilon_{rui}\, \theta^i + O(\theta^2)$, $B_{ru}(\theta^i)\,
\rightarrow_{\theta^i \rightarrow 0}\, \delta_{ru} + {1\over 2}\,
\epsilon_{rui}\, \theta^i + O(\theta^2)$.

\bigskip

In the York canonical basis we have (from now on we will use
$V_{ra}$ for $V_{ra}(\theta^n)$, $A_{ra}$ for $A_{ra}(\theta^n)$,
$B_{ra}$ for $B_{ra}(\theta^n)$, to simplify the notation; the
positive eigenvalues of the 3-metric ${}^3g_{rs}(\tau, \vec \sigma)$
are denoted as $\lambda_a(\tau, \vec \sigma) = \Lambda^2_a(\tau,
\vec \sigma)$)

\begin{eqnarray*}
 {}^4g_{\tau\tau} &=& \sgn\, \Big[(1 + n)^2 - \sum_a\,
 {\bar n}_{(a)}^2\Big],\nonumber \\
 {}^4g_{\tau r} &=& - \sgn\, \sum_a\, {\bar n}_{(a)}\, {}^3{\bar e}_{(a)r} =
 - \sgn\, {\tilde \phi}^{1/3}\, \sum_a\, Q_a\,
 V_{ra}\, {\bar n}_{(a)},\nonumber \\
 {}^4g_{rs} &=& - \sgn\, {}^3g_{rs}
 = - \sgn\, \sum_{uv}\, V_{ru}\, \lambda_u\, \delta_{uv}\,
 V^T_{vs} = - \sgn\, \sum_a\, \Big(V_{ra}\, \Lambda^a\Big)\,
 \Big(V_{sa}\, \Lambda^a\Big) =\nonumber \\
 &=& - \sgn\, \sum_a\, {}^3{\bar e}_{(a)r}\, {}^3{\bar e}_{(a)s} =
 - \sgn\, \sum_a\, {}^3e_{(a)r}\, {}^3e_{(a)s} = - \sgn\, \phi^4\,
 {}^3{\hat g}_{rs}\, =\, - \sgn\, {\tilde \phi}^{2/3}\, \sum_a\, Q^2_a\,
 V_{ra}\, V_{sa},\nonumber \\
 &&{}\nonumber \\
 \Lambda_a &=& \sum_u\, \delta_{au}\, \sqrt{\lambda_u}\,\,
 =\,\, \phi^2\, Q_a\,  =\, {\tilde \phi}^{1/3}\, Q_a,\qquad
 Q_a\, =\, e^{\sum_{\bar a}^{1,2}\, \gamma_{\bar aa}\, R_{\bar a}}
 = e^{\Gamma_a^{(1)}},
 \nonumber \\
 \tilde \phi &=& \phi^6 = \sqrt{\gamma} =
 \sqrt{det\, {}^3g} = {}^3\bar e = \sqrt{\lambda_1\,
 \lambda_2\, \lambda_3} = \Lambda_1\, \Lambda_2\, \Lambda_3,
 \end{eqnarray*}

 \begin{eqnarray*}
 {}^3e_{(a)r} &=& \sum_b\, R_{(a)(b)}(\alpha_{(e)})\, {}^3{\bar
 e}_{(b)r},\qquad {}^3{\bar e}_{(a)r} = {\tilde \phi}^{1/3}\, Q_a\,
 V_{ra},\nonumber \\
 {}^3e^r_{(a)} &=& \sum_b\, R_{(a)(b)}(\alpha_{(e)})\, {}^3{\bar
 e}^r_{(b)},\qquad {}^3{\bar e}^r_{(a)} = {\tilde \phi}^{- 1/3}\, Q^{-1}_a\,
 V_{ra},\nonumber \\
 &&{}\nonumber \\
 {}^3\pi^r_{(a)} &=& \sum_b\,
 R_{(a)(b)}(\alpha_{(e)})\, {\bar \pi}^r_{(b)},\nonumber \\
 &&{}\nonumber \\
 {}^3{\bar \pi}^r_{(a)} &\approx& {\tilde \phi}^{-1/3}\, \Big[
 V_{ra}\, Q^{-1}_a\, (\tilde \phi\, \pi_{\tilde \phi} +  \sum_{\bar b}\,
 \gamma_{\bar ba}\, \Pi_{\bar b}) +\nonumber \\
 &+& \sum_{l}^{l \not= a}\, \sum_{twi}\, Q^{-1}_l\, {{V_{rl}\,
 \epsilon_{alt}\, V_{wt}}\over {Q_l\, Q^{-1}_a - Q_a\, Q^{-1}_l
}}\, B_{iw}\, \pi^{(\theta )}_i \Big],
 \end{eqnarray*}

\bea
 \pi^{(\theta)}_i &=& - \sum_{lmra}\, A_{ml}\,
 \epsilon_{mir}\, {}^3e_{(a)l}\, {}^3{\bar \pi}^r_{(a)},\nonumber \\
  \pi_{\tilde \phi} &=&   {{c^3}\over {12\pi\, G}}\, {}^3K
  \approx {1\over {3\,\, {}^3e}}\, \sum_{ra}\, {}^3{\bar
  \pi}^r_{(a)}\,  {}^3{\bar e}_{(a)r},
  \nonumber \\
  \Pi_{\bar a} &=& \sum_{ra}\, \gamma_{\bar aa}\, {}^3{\bar
  \pi}^r_{(a)}\,  {}^3{\bar e}_{(a)r}.
 \label{2.9}
 \eea

\noindent The set of numerical parameters $\gamma_{\bar aa}$
satisfies \cite{1} $\sum_u\, \gamma_{\bar au} = 0$, $\sum_u\,
\gamma_{\bar a u}\, \gamma_{\bar b u} = \delta_{\bar a\bar b}$,
$\sum_{\bar a}\, \gamma_{\bar au}\, \gamma_{\bar av} = \delta_{uv} -
{1\over 3}$. Each solution of these equations defines a different
York canonical basis.
\medskip

{\it We only consider 3-metrics with 3 distinct positive eigenvalues
$\lambda_r = \Lambda_r^2$}: the degenerate cases should be treated
by adding by hand the constraints $\Lambda_a - \Lambda_b \approx 0$
or $\Lambda_1 \approx \Lambda_2 \approx \Lambda_3$ and by studying
the resulting constraint algebra.

 \bigskip

The assumed boundary conditions given after Eqs.(\ref{2.2}) imply
$\Lambda_a(\tau ,\vec \sigma )\, = \Big({\tilde \phi}^{1/3}\,
Q_a\Big)(\tau, \vec \sigma)\, \rightarrow_{r \rightarrow \infty}\,\,
1 + {M\over {4r}} + {{a_a}\over {r^{3/2}}} + O(r^{-3})$ and $\tilde
\phi (\tau ,\vec \sigma )\, \rightarrow_{r \rightarrow \infty}\,\, 1
+ O(r^{-1})$. Moreover we must have $\pi_i^{(\theta )}(\tau ,\vec
\sigma )\, \rightarrow_{r \rightarrow \infty}\,\, O(r^{-4})$, since
the requirement $\Lambda_a(\tau ,\vec \sigma ) \not= \Lambda_b(\tau
,\vec \sigma )$ for $a \not= b$, needed to avoid singularities,
implies $a_a \not= a_b$ for $a \not= b$ in their asymptotic
behavior, so that we get $\Big({{\Lambda_b}\over {\Lambda_a}} -
{{\Lambda_a}\over {\Lambda_b}}\Big)^{-1}(\tau ,\vec \sigma )\, =
\Big(Q_b\, Q_a^{-1} - Q_a\, Q_b^{-1}\Big)^{-1}(\tau, \vec \sigma)\,
\rightarrow_{r \rightarrow \infty}\,\, {{r^{3/2}}\over { 2\, (a_b -
a_a)}}$. As a consequence, we have $\pi^{(\alpha )}_{(a)}(\tau ,\vec
\sigma )\, \rightarrow_{r \rightarrow \infty}\, O(r^{-5/2})$. Also
the angles $\alpha_{(a)}(\tau ,\vec \sigma )$ and $\theta^i(\tau
,\vec \sigma )$ must tend to zero in a direction-independent way at
spatial infinity. We also have $\pi_{\tilde \phi}(\tau ,\vec \sigma
)\, \rightarrow_{r \rightarrow \infty}\,\, O(r^{-5/2})$ at spatial
infinity.\bigskip

In Eq.(\ref{2.9}) the quantity ${}^3K(\tau, \vec \sigma)$ is the
trace of the extrinsic curvature ${}^3K_{rs}(\tau, \vec \sigma)$ of
the instantaneous 3-spaces $\Sigma_{\tau}$. In the York canonical
basis the extrinsic curvature ${}^3K_{rs}$, the 3-spin connection
${}^3{\bar \omega}_{r(a)} = {}^3\omega_{r(a)}{|}_{\alpha_{(e)} =0}$
(see Eq.(B17) of Ref.\cite{6}) and the 3-Christoffel symbols have
the following expression \cite{6} \footnote{Since in Section III the
action (\ref{3.1}) differs of a factor $- \sgn$ from the action used
in Ref.\cite{6}, also the extrinsic curvature differs from Eq.(A10)
of Ref.\cite{6} by the same factor.}

\begin{eqnarray*}
  {}^3K_{rs} &=& - {{4\pi\, G}\over {c^3\,\, {}^3{\bar e}}}\,
  \sum_{abu}\, \Big[\Big({}^3{\bar
 e}_{(a)r}\, {}^3{\bar e}_{(b)s} + {}^3{\bar e}_{(a)s}\, {}^3{\bar
 e}_{(b)r}\Big)\, {}^3{\bar e}_{(a)u}\, {\bar \pi}^u_{(b)} -
 {}^3{\bar e}_{(a)r}\, {}^3{\bar e}_{(a)s}\, {}^3{\bar e}_{(b)u}\,
 {\bar \pi}^u_{(b)}\Big] \approx \nonumber \\
 &\approx& {}^3{\tilde K}_{rs} =
  - {{4\pi\, G}\over {c^3}}\, {\tilde \phi}^{-1/3}\,
 \Big(\sum_a\, Q^2_a\, V_{ra}\, V_{sa}\, [2\, \sum_{\bar b}\, \gamma_{\bar ba}\,
 \Pi_{\bar b} -  \tilde \phi\, \pi_{\tilde \phi}] +\nonumber \\
 &+& \sum_{ab}\, Q_a\, Q_b\, (V_{ra}\, V_{sb} +
 V_{rb}\, V_{sa})\, \sum_{twi}\, {{\epsilon_{abt}\,
 V_{wt}\, B_{iw}\, \pi_i^{(\theta )}}\over {
 Q_b\, Q^{-1}_a  - Q_a\, Q^{-1}_b}} \Big),
 \end{eqnarray*}

 \begin{eqnarray*}
 {}^3{\bar \omega}_{r(a)} &=& {1\over 2}\, \sum_{bc}\,
 \epsilon_{(a)(b)(c)}\, {}^3{\bar \omega}_{r(b)(c)} =\nonumber \\
 &=& {1\over 2}\, \sum_{bcu}\, \epsilon_{(a)(b)(c)}\, {}^3{\bar e}^u_{(b)}\,
 \Big[\partial_r\, {}^3{\bar e}_{(c)u} - \partial_u\, {}^3{\bar e}_{(c)r}  +
 \sum_{dv}\, {}^3{\bar e}^v_{(c)}\, {}^3{\bar e}_{(d)r}\, \partial_v\,
 {}^3{\bar e}_{(d)u}\Big] =\nonumber \\
 &&{}\nonumber \\
 &=& \sum_{bcu}\, \epsilon_{(a)(b)(c)}\, V_{ub}\, \Big[ - Q_b^{-1}\, Q_c\,
 V_{rc}\,  \partial_u\, \Big({1\over 3}\, ln\, \tilde \phi + \sum_{\bar a}\,
  \gamma_{\bar ac}\, R_{\bar a}\Big) + \nonumber \\
  &+& {1\over 2}\, Q_b^{-1}\, Q_c\,  \Big(\partial_r\, V_{uc} - \partial_u\,
  V_{rc}\Big)   + {1\over 2}\, \sum_{vd}\, Q_b^{-1}\, Q_c^{-1}\,
  Q_d^2\, V_{rd}\, V_{vc}\, \partial_v\, V_{ud} \Big],
 \end{eqnarray*}

 \bea
   {}^3\Gamma^r_{uv}&=& {1\over 3}\,  (\delta_{ru}\, {\tilde \phi}^{-1}\, \partial_v\,
  \tilde \phi + \delta_{rv}\, {\tilde \phi}^{-1}\, \partial_u\, \tilde \phi ) -
   {1\over 3}\, \sum_{abs}\, Q^2_b\, Q^{-2}_a\, V_{ra}\, V_{sa}\,
  V_{ub}\, V_{vb}\, {\tilde \phi}^{-1}\, \partial_s\, \tilde \phi +\nonumber \\
  &+& \sum_{\bar aa}\, \gamma_{\bar aa}\, V_{ra}\, \Big(
  V_{ua}\, \partial_v\, R_{\bar a} + V_{va}\,
  \partial_u\, R_{\bar a}\Big) -
 \sum_{\bar babs}\, \gamma_{\bar bb}\, Q^2_b\, Q^{-2}_a\,
  V_{ra}\, V_{sa}\,  V_{ub}\, V_{vb}\,
  \partial_s\, R_{\bar b} +\nonumber \\
  &+& {1\over 2}\, \sum_a\, V_{ra}\, \Big(\partial_u\, V_{va}
 + \partial_v\, V_{ua}\Big) +\nonumber \\
 &+& {1\over 2}\, \sum_{abs}\, Q_a^{-2}\, Q_b^2\, V_{ra}\,
 V_{sa}\, \Big[V_{ub}\, \Big(\partial_v\, V_{sb} -
 \partial_s\, V_{vb}\Big) +
  V_{vb}\, \Big(\partial_u\, V_{sb} - \partial_s\,
 V_{ub}\Big)\Big],\nonumber \\
 \sum_v\, {}^3\Gamma^v_{uv} &=&  {\tilde \phi}^{-1}\, \partial_u\,  \tilde
 \phi.
 \label{2.10}
 \eea

\bigskip

The previous sequence of canonical transformations  realizes a {\it
York map} because the gauge variable $\pi_{\tilde \phi}$ (describing
the freedom in the choice of the instantaneous 3-spaces
$\Sigma_{\tau}$) is proportional to {\it York internal extrinsic
time} ${}^3K$. It is the only gauge variable among the momenta: this
is a reflex of the Lorentz signature of space-time, because
$\pi_{\tilde \phi}$ and $\theta^n$ can be used as a set of
4-coordinates \cite{7}.

Its conjugate variable, to be determined by the super-hamiltonian
constraint, is $\tilde \phi = \phi^6 = {}^3\bar e$, which is
proportional to {\it Misner's internal intrinsic time}; moreover
$\tilde \phi$ is the {\it 3-volume density} on $\Sigma_{\tau}$: $V_R
= \int_R d^3\sigma\, \phi^6$, $R \subset \Sigma_{\tau}$. Since we
have ${}^3g_{rs} = {\tilde \phi}^{2/3}\, {}^3{\hat g}_{rs}$ with
$det\, {}^3{\hat g}_{rs} = 1$, $\tilde \phi$ is also called the
conformal factor of the 3-metric.

\medskip

The two pairs of canonical variables $R_{\bar a}$, $\Pi_{\bar a}$,
$\bar a = 1,2$, describe the generalized {\it tidal effects}, namely
the independent degrees of freedom of the gravitational field. In
particular the configuration tidal variables $R_{\bar a}$ depend
{\it only on the eigenvalues of the 3-metric} \footnote{If we
consider the eigenvalue equation for the 3-metric ${}^3{\hat
g}_{rs}$ of determinant one, we identify the following {\it two
3-scalars depending only on the tidal variables} $R_{\bar a}$: i)
$Tr\, {}^3{\hat g}_{rs} = \sum_r\, {}^3{\hat g}_{rr} = \sum_a\,
Q_a^2$ (the sum of the eigenvalues); ii) ${}^3{\hat g}_{11}\,
{}^3{\hat g}_{22} - {}^3{\hat g}^2_{12} + {}^3{\hat g}_{22}\,
{}^3{\hat g}_{33} - {}^3{\hat g}^2_{23} + {}^3{\hat g}_{33}\,
{}^3{\hat g}_{11} - {}^3{\hat g}^2_{31} = Q^2_1\, Q^2_2 + Q^2_2\,
Q^2_3 + Q^2_3\, Q^2_1$ (the sum of the possible products of two
eigenvalues). As a consequence of this result and of Eqs.(\ref{2.9})
the tidal variables $R_{\bar a}$, $\Pi_{\bar a}$, are 3-scalars on
$\Sigma_{\tau}$.\hfill\break
 This suggest the possibility of a point canonical transformation
 from $R_{\bar a}$, $\Pi_{\bar a}$ to new 3-scalar tidal variables
  $X(R_{\bar a})$, $\Pi_X$, $Y(R_{\bar a})$, $\Pi_Y$ with
 \begin{eqnarray*}
 X &=& e^{2\, \Big[\gamma_{\bar 11}\, R_{\bar 1} + \gamma_{\bar 21}\, R_{\bar
 2}\Big]} + e^{2\, \Big[\gamma_{\bar 12}\, R_{\bar 1} + \gamma_{\bar 22}\, R_{\bar
 2}\Big]} + e^{2\, \Big[\gamma_{\bar 13}\, R_{\bar 1} + \gamma_{\bar 23}\, R_{\bar
 2}\Big]} = Tr\, {}^3{\hat g}_{rs} = {\tilde \phi}^{-2/3}\, Tr\, {}^3g_{rs},\nonumber \\
 Y &=& e^{2\, \Big[(\gamma_{\bar 11} + \gamma_{\bar 12})\, R_{\bar 1} + (\gamma_{\bar 21} +
 \gamma_{\bar 22})\, R_{\bar 2}\Big]} + e^{2\, \Big[(\gamma_{\bar 12} + \gamma_{\bar 13})\,
 R_{\bar 1} + (\gamma_{\bar 22} + \gamma_{\bar 23})\, R_{\bar
 2}\Big]} + e^{2\, \Big[(\gamma_{\bar 13} + \gamma_{\bar 11})\, R_{\bar 1} + (\gamma_{\bar 23} +
 \gamma_{\bar 21})\, R_{\bar 2}\Big]} =\nonumber \\
 &=& {\tilde \phi}^{-4/3}\, \Big[{}^3g_{11}\, {}^3g_{22} - {}^3g^2_{12} +
 {}^3g_{22}\, {}^3g_{33} - {}^3g^2_{23} + {}^3g_{33}\, {}^3g_{11} -
 {}^3g^2_{31}\Big].
 \end{eqnarray*}  }. They are Dirac
observables {\it only} with respect to the gauge transformations
generated by 10 of the 14 first class constraints. Let us remark
that, if we fix completely the gauge and we go to Dirac brackets,
then the only surviving dynamical variables $R_{\bar a}$ and
$\Pi_{\bar a}$ become two pairs of {\it non canonical} Dirac
observables for that gauge: the two pairs of canonical Dirac
observables have to be found as a Darboux basis of the copy of the
reduced phase space identified by the gauge and they will be (in
general non-local) functionals of the $R_{\bar a}$, $\Pi_{\bar a}$
variables. \medskip

Since the variables $\tilde \phi$  and $\pi_i^{(\theta )}$ are
determined by the super-Hamiltonian and super-momentum constraints,
the {\it arbitrary gauge variables} are $\alpha_{(a)}$,
$\varphi_{(a)}$, $\theta^i$, $\pi_{\tilde \phi}$, $n$ and ${\bar
n}_{(a)}$. As shown in Refs.\cite{6}, they describe the following
generalized {\it inertial effects}:

a) $\alpha_{(a)}(\tau ,\vec \sigma )$ and $\varphi_{(a)}(\tau ,\vec
\sigma )$ are the 6 configuration variables parametrizing the O(3,1)
gauge freedom in the choice of the tetrads in the tangent plane to
each point of $\Sigma_{\tau}$ and describe the arbitrariness in the
choice of a tetrad to be associated to a time-like observer, whose
world-line goes through the point $(\tau ,\vec \sigma )$. They fix
{\it the unit 4-velocity of the observer and the conventions for the
orientation of three gyroscopes and their transport along the
world-line of the observer}. The  {\it Schwinger time gauges} are
defined by the gauge fixings $\alpha_{(a)}(\tau, \vec \sigma)
\approx 0$, $\varphi_{(a)}(\tau, \vec \sigma) \approx 0$.

b) $\theta^i(\tau ,\vec \sigma )$ [depending only on the 3-metric,
as shown in Eq.(\ref{2.9})] describe the arbitrariness in the choice
of the 3-coordinates in the instantaneous 3-spaces $\Sigma_{\tau}$
of the chosen non-inertial frame  centered on an arbitrary time-like
observer. Their choice will induce a pattern of {\it relativistic
inertial forces} for the gravitational field, whose potentials are
the functions $V_{ra}(\theta^i)$ present in the weak ADM energy
$E_{ADM}$ given in Eqs.(\ref{3.14}).

c) ${\bar n}_{(a)}(\tau ,\vec \sigma )$, the shift functions
appearing in the Dirac Hamiltonian, describe which points on
different instantaneous 3-spaces have the same numerical value of
the 3-coordinates. They are the inertial potentials describing the
effects of the non-vanishing off-diagonal components ${}^4g_{\tau
r}(\tau ,\vec \sigma )$ of the 4-metric, namely they are the {\it
gravito-magnetic potentials} \footnote{In the post-Newtonian
approximation in harmonic gauges they are the counterpart of the
electro-magnetic vector potentials describing magnetic fields
\cite{23}: A) $N = 1 + n$, $n\, {\buildrel {def}\over =}\, - {{4\,
\sgn}\over {c^2}}\, \Phi_G$ with $\Phi_G$ the {\it gravito-electric
potential}; B) $n_r\, {\buildrel {def}\over =}\, {{2\, \sgn}\over
{c^2}}\, A_{G\, r}$ with $A_{G\, r}$ the {\it gravito-magnetic}
potential; C) $E_{G\, r} = \partial_r\, \Phi_G -
\partial_{\tau}\, ({1\over 2}\, A_{G\, r})$ (the {\it
gravito-electric field}) and $B_{G\, r} = \epsilon_{ruv}\,
\partial_u\, A_{G\, v} = c\, \Omega_{G\, r}$ (the {\it
gravito-magnetic field}). Let us remark that in arbitrary gauges the
analogy with electro-magnetism  breaks down.} responsible of effects
like the dragging of inertial frames (Lens-Thirring effect)
\cite{23} in the post-Newtonian approximation. The shift functions
are determined by the $\tau$-preservation of the gauge fixings
determining the gauge variables $\theta^i(\tau, \vec \sigma)$.

d) $\pi_{\tilde \phi}(\tau ,\vec \sigma )$, i.e. the York time
${}^3K(\tau ,\vec \sigma )$, describes the non-dynamical
arbitrariness in the choice of the convention for the
synchronization of distant clocks which remains in the transition
from special to general relativity. As said in the Introduction, the
choice of the shape of the instantaneous 3-space as a sub-manifold
of space-time (a pure gauge choice in special relativity) is
dynamically determined by the chosen solution of Einstein's
equations after the fixation of the gauge variables ${}^3K(\tau,
\vec \sigma)$ and $\theta^i(\tau, \vec \sigma)$. Since the York time
is present in the Dirac Hamiltonian \footnote{See Eqs.(\ref{3.43})
and (\ref{3.44}) for its presence in the super-Hamiltonian
constraint and in the weak ADM energy, and Eqs.(\ref{3.41}) for its
presence in the super-momentum constraints.}, it is a  {\it new
inertial potential} connected to the problem of the relativistic
freedom in the choice of the {\it instantaneous 3-space}, which has
no non-relativistic analogue (in Galilei space-time time is absolute
and there is an absolute notion of Euclidean 3-space). Its effects
are completely unexplored.

e) $n(\tau ,\vec \sigma )$, the lapse function appearing in the
Dirac Hamiltonian, describes the arbitrariness in the choice of the
unit of proper time in each point of the simultaneity surfaces
$\Sigma_{\tau}$, namely how these surfaces are packed in the 3+1
splitting. The lapse function is determined by the
$\tau$-preservation of the gauge fixing for the gauge variable
${}^3K(\tau, \vec \sigma)$.

\bigskip

The gauge variables $\theta^i(\tau, \vec \sigma)$, $n(\tau, \vec
\sigma)$, ${\bar n}_{(a)}(\tau, \vec \sigma)$ describe inertial
effects, which are the the relativistic counterpart of the
non-relativistic ones (the centrifugal, Coriolis,... forces in
Newton mechanics in accelerated frames) and which are present also
in the non-inertial frames of Minkowski space-time \cite{10}.

\subsection{The Expansion and the Shear of the Eulerian Observers.}

Let us now consider the geometrical interpretation of the extrinsic
curvature ${}^3K_{rs}$ of the instantaneous 3-spaces $\Sigma_{\tau}$
in terms of the properties of the surface-forming (i.e.
irrotational) congruence of Eulerian (non geodesic) time-like
observers, whose world-lines have the tangent unit 4-velocity equal
to the unit normal orthogonal to the instantaneous 3-spaces
$\Sigma_{\tau}$. If we use radar 4-coordinates, the covariant unit
normal $\sgn\, l_A = (1 + n)\, (1; 0)$  of Eqs.(\ref{2.2}) has the
following covariant derivative (see for instance Ref.\cite{24};
${}^3h_{AB}$ is defined in Eq.(\ref{2.5}); ${\hat b}^r_A$ is defined
in Refs.\cite{1,2})\medskip

\bea
 {}^4\nabla_A\,\, \sgn\, l_B &=& \sgn\, l_A\, {}^3a_B + \sigma_{AB} +
 {1\over 3}\, \theta\, {}^3h_{AB} - \omega_{AB} = \sgn\, l_A\, {}^3a_B
 + {}^3K_{AB},\nonumber \\
 &&{}\nonumber \\
 &&{}\nonumber \\
 {}^3K_{AB} &=& {}^3K_{rs}\, {\hat b}^r_A\, {\hat b}^s_B,\qquad
 {\hat b}^r_A = \delta^r_A + {}^3{\bar e}^r_{(a)}\, {\bar n}_{(a)}\,
 \delta^{\tau}_A,\qquad {}^3h_{AB} = {}^4g_{AB} - \sgn\, l_A\, l_B.
 \nonumber \\
 &&{}
 \label{2.11}
 \eea
\medskip

The quantities appearing in Eqs.(\ref{2.11}) are:\medskip

a) the {\it acceleration} of the Eulerian observers

\bea
 {}^3a^A &=& l^B\, {}^4\nabla_B\, l^A =
 {}^4g^{AB}\, {}^3a_B,\qquad {}^3a_A = {}^3a_r\,
  {\hat b}^r_A,\nonumber \\
 &&{}\nonumber \\
 &&{}^3a_r = - \partial_r\, ln\, (1 + n),\qquad {}^3a_{\tau}
 = {}^3a_r\, {}^3{\bar e}^r_{(a)}\, {\bar n}_{(a)} = -
 {\tilde \phi}^{-1/3}\, Q_a^{-1}\, V_{ra}\, {\bar n}_{(a)}\,
 \partial_r\, ln (1 + n),\nonumber \\
 &&{}^3a^r = - \sgn\, {}^3{\bar e}^r_{(a)}\, {}^3{\bar e}^s_{(a)}\,
 {}^3a_s =  \sgn\, {\tilde \phi}^{-2/3}\, Q_a^{-2}\, V_{ra}\,
 V_{sa}\, \partial_s\, ln (1 + n),\qquad {}^3a^{\tau} = 0;\nonumber \\
 &&{}
 \label{2.12}
 \eea

\medskip

b) the {\it vorticity} or {\it twist} (a measure of the rotation of
the nearby world-lines infinitesimally surrounding the given one),
which is vanishing because the congruence is surface-forming

\bea
 \omega_{AB} &=& - \omega_{BA} =  {{\sgn}\over 2}\,
 (l_A\, {}^3a_B -  l_B\, {}^3a_A) -
 {{\sgn}\over 2}\, ({}^4\nabla_A\,\, l_B -
 {}^4\nabla_B\,\, l_A) = 0,\nonumber \\
 &&{}\nonumber \\
 &&{}\nonumber \\
 &&\omega_{AB}\, l^B = 0,\qquad \omega^A = {1{\sgn}\over 2}\,
{}^4\eta^{ABCD}\, \omega_{BC}\, l_D = 0;
 \label{2.13}
 \eea

 \medskip

c) the {\it expansion} \footnote{It measures the average expansion
of the infinitesimally nearby world-lines surrounding a given
world-line in the congruence.}, which coincides with the {\it York
external time} multiplied by $- \sgn$, is proportional to what is
called the {\it Hubble parameter} $H$ in cosmology \footnote{In
Eqs.(\ref{2.14}) $l$ is a representative length along the integral
curves of ${}^4{\buildrel \circ \over {\bar E}}^A_{(o)}$, describing
the volume expansion (contraction) behavior of the congruence.} and
to the dimensionless parameter (the {\it cosmological deceleration
parameter}) $q = 3\, l^A\, {}^4\nabla_A\, {1\over {\theta}} - 1 = -
3\, \theta^{-2}\, l^A\, \partial_A\, \theta - 1$,

\bea
 \theta &=& {}^4\nabla_A\,\, l^A
 = - \sgn\, {}^3K = - {{4\pi\, G}\over {c^3}}\,
{{{}^3{\bar e}_{(a)r}\, {}^3{\bar \pi}^r_{(a)}}\over {{}^3\bar e}}
=  - \sgn\, {{12\pi\, G}\over {c^3}}\, \pi_{\tilde \phi},\nonumber \\
 &&{}\nonumber \\
 H &=& {1\over 3}\, \theta = {1\over {l}}\, l^A\,
{}^4\nabla_A\, l =  - \sgn\, {{4\pi\, G}\over {c^3}}\, \pi_{\tilde
\phi},\qquad q = 3\, l^A\, {}^4\nabla_A\, {1\over {\theta}} - 1;
 \label{2.14}
 \eea

\medskip

d) the {\it shear} \footnote{It measures how an initial sphere in
the tangent space to the given world-line, which is Lie-transported
along the world-line tangent $l^{\mu}$ (i.e. it has zero Lie
derivative with respect to $l^{\mu}\, \partial_{\mu}$), is distorted
towards an ellipsoid with principal axes given by the eigenvectors
of $\sigma^{\mu}{}_{\nu}$, with rate given by the eigenvalues of
$\sigma^{\mu}{}_{\nu}$.}

\bea
 \sigma_{AB} &=& \sigma_{BA} = - {{\sgn}\over 2}\, ({}^3a_A\,
 l_B + {}^3a_B\, l_A) + {{\sgn}\over
 2}\, ({}^4\nabla_A\, l_B + {}^4\nabla_B\, l_A) -
 {1\over 3}\, \theta\, {}^3h_{AB}
 =\nonumber \\
 &&{}\nonumber \\
 &=& ({}^3K_{rs} - {1\over 3}\, {}^3g_{rs}\, {}^3K)\,
{\hat {\bar b}}^r_A\, {\hat {\bar b}}^s_B,\qquad
 {}^4g^{AB}\, \sigma_{AB} = 0,\qquad \sigma_{AB}\,
 l^B = 0.
 \label{2.15}
 \eea

\bigskip

By explicit calculation we get the following components of the shear
along the tetrads (\ref{2.2})
\medskip

\begin{eqnarray*}
  \sigma_{AB} &=& \sigma_{(\alpha )(\beta )}\, {}^4{\buildrel \circ \over
 {\bar E}}^{(\alpha )}_A\, {}^4{\buildrel \circ \over {\bar E}}^{(\beta )}_B
 = {}^4g_{AC}\, {}^4g_{BD}\, \sigma^{CD}, \nonumber  \\
 &&{}\nonumber \\
 \sigma_{\tau\tau} &=& {\bar n}_{(a)}\, {\bar n}_{(b)}\, \sigma_{(a)(b)} = ({}^3K_{rs}
- {1\over 3}\, {}^3g_{rs}\, {}^3K)\, {\bar n}_{(a)}\, {}^3{\bar
e}^r_{(a)}\, {\bar n}_{(b)}\, {}^3{\bar e}^s_{(b)},\nonumber \\
 \sigma_{\tau r} &=& \sigma_{r\tau} = {\bar n}_{(a)}\, \sigma_{(a)(b)}\,
{}^3{\bar e}_{(b)r} = ({}^3K_{rs} - {1\over 3}\, {}^3g_{rs}\,
{}^3K)\, {\bar n}_{(a)}\, {}^3{\bar e}^s_{(a)},\nonumber \\
 \sigma_{rs} &=& \sigma_{(a)(b)}\, {}^3{\bar e}_{(a)r}\, {}^3{\bar e}_{(b)s} =
 {}^3K_{rs} - {1\over 3}\, {}^3g_{rs}\, {}^3K,\nonumber \\
 &&{}\nonumber \\
 &&\sigma^{\tau\tau} = \sigma^{\tau r} = 0, \qquad \sigma^{rs} =
 {}^3{\bar e}^r_{(a)}\, {}^3{\bar e}^s_{(b)}\, \sigma_{(a)(b)},
 \end{eqnarray*}

\bea
  \sigma_{(\alpha )(\beta )} &=& \sigma_{AB}\,
 {}^4{\buildrel \circ \over {\bar E}}^A_{(\alpha
 )}\, {}^4{\buildrel \circ \over {\bar E}}^B_{(\beta )},\nonumber \\
 &&{}\nonumber \\
 \sigma_{(o)(o)} &=& \sigma_{AB}\, {}^4{\buildrel \circ \over {\bar E}}^A_{(o)}\,
 {}^4{\buildrel \circ \over {\bar E}}^B_{(o )} =
 {1\over {(1 + n)^2}}\, [\sigma_{\tau\tau} - 2\, {\bar n}_{(a)}\,
 {}^3{\bar e}^r_{(a)}\, \sigma_{\tau r} + {\bar n}_{(a)}\, {\bar
 n}_{(b)}\, {}^3{\bar e}^r_{(a)}\, {}^3{\bar e}^s_{(b)}\, \sigma_{rs}] = 0,\nonumber \\
 \sigma_{(o)(a)} &=& \sigma_{AB}\, {}^4{\buildrel \circ \over {\bar E}}^A_{(o)}\,
 {}^4{\buildrel \circ \over {\bar E}}^B_{(a )} =
 {1\over {1 + n}}\, (\sigma_{\tau s} - {\bar n}_{(b)}\, {}^3{\bar
 e}^r_{(b)}\, \sigma_{rs})\, {}^3{\bar e}^s_{(a)} = 0,\nonumber \\
 \sigma_{(a)(b)} &=& \sigma_{AB}\, {}^4{\buildrel \circ \over {\bar E}}^A_{(a)}\,
 {}^4{\buildrel \circ \over {\bar E}}^B_{(b )} =
 \sigma_{rs}\, {}^3{\bar e}^r_{(a)}\, {}^3{\bar e}^s_{(b)}  =
 \sigma_{(b)(a)} = ({}^3K_{rs} - {1\over 3}\, {}^3g_{rs}\, {}^3K)\,
 {}^3{\bar e}^r_{(a)}\, {}^3{\bar e}^s_{(b)},\qquad \sum_a\,
 \sigma_{(a)(a)} = 0. \nonumber \\
 &&{}
 \label{2.16}
 \eea

\noindent $\sigma_{(a)(b)}$ depends upon $\theta^r$, $\tilde \phi$,
$R_{\bar a}$, $\pi^{(\theta )}_r$ and $\Pi_{\bar a}$.

\bigskip

As a consequence we have \footnote{The 3-scalars associated to the
symmetric matrix ${}^3K_{rs}$ are $I =  {}^3K = - \sgn\, \theta$,
$II = det {}^3K_{rs}$, $III = {}^3K_{11}\, {}^3K_{22} - {}^3K^2_{12}
+ {}^3K_{22}\, {}^3K_{33} - {}^3K_{23}^2 + {}^3K_{33}\, {}^3K_{11} -
{}^3K^2_{31}$. If ${}^3{\tilde K}_{rs} = {}^3K_{rs} - {1\over 3}\,
{}^3g_{rs}\, {}^3K$ is the traceless extrinsic curvature, the
3-scalars $II$ and $III$ may be replaced by $ II^{'} = det\,
{}^3{\tilde K}_{rs}$ and $III^{'} = {}^3{\tilde K}_{11}\,
{}^3{\tilde K}_{22} - {}^3{\tilde K}^2_{12} + {}^3{\tilde K}_{22}\,
{}^3{\tilde K}_{33} - {}^3{\tilde K}_{23}^2 + {}^3{\tilde K}_{33}\,
{}^3{\tilde K}_{11} - {}^3{\tilde K}^2_{31}$. } ($n_{r|s}$ is the
covariant derivative on $\Sigma_{\tau}$; $Q_a = e^{\Gamma_a^{(1)}}$)

\medskip

\begin{eqnarray*}
 {}^3K_{rs} &=& - {{\sgn}\over 3}\, {}^3g_{rs}\, \theta +
 \sigma_{(a)(b)}\, {}^3{\bar e}_{(a)r}\, {}^3{\bar e}_{(b)s}
 = {\tilde \phi}^{2/3}\, \sum_{ab}\, \Big(- {{\sgn}\over 3}\,
 \theta\, \delta_{ab} + \sigma_{(a)(b)}\Big)\, Q_a\, Q_b\,
 V_{ra}\, V_{sb},\nonumber \\
 &&{}\nonumber \\
 \Rightarrow&& \partial_{\tau}\, {}^3g_{rs}\, =\, n_{r|s} +
 n_{s|r} - 2\, (1 +n)\, {}^3K_{rs} =\nonumber \\
 &{\buildrel {(\ref{2.9})}\over
 =}&\, {\tilde \phi}^{2/3}\, \sum_a\, Q_a^2\, \Big[2\, ({1\over 3}\,
 {\tilde \phi}^{-1}\, \partial_{\tau}\, \tilde \phi + \partial_{\tau}\,
 \Gamma_a^{(1)})\, V_{ra}\, V_{sa} + \partial_{\tau}\, (V_{ra}\,
 V_{sa})\Big],\nonumber \\
 &&{}\nonumber \\
 &&\Downarrow
 \end{eqnarray*}

\begin{eqnarray*}
 &&\partial_{\tau}\, \tilde \phi\, =\, {1\over 2}\, {\tilde
 \phi}^{1/3}\, \sum_a\, Q_a^{-2}\, {\cal A}_{aa},\nonumber \\
 &&{}\nonumber \\
 &&\partial_{\tau}\, \Gamma_a^{(1)}\, =\, {1\over 2}\, {\tilde
 \phi}^{-2/3}\, \Big(Q_a^{-2}\, {\cal A}_{aa} - {1\over 3}\,
 \sum_b\, Q_b^{-2}\, {\cal A}_{bb}\Big),\nonumber \\
 &&{}\nonumber \\
 &&\sum_r\, V_{ra}(\theta^i)\, \partial_{\tau}\,
 V_{rb}(\theta^i){|}_{a \not= b}\, =\, {\tilde \phi}^{-2/3}\,
 {{{\cal A}_{ba}}\over {Q_b^2 - Q_a^2}}{|}_{a \not= b},\nonumber \\
 &&{}\nonumber \\
 \Rightarrow&& \sum_a\, Q_a^2\, \partial_{\tau}\, \Big(V_{ra}(\theta^i)\,
 V_{sa}(\theta^i)\Big) = {\tilde \phi}^{-2/3}\, \sum_{c \not= d}\,
 V_{rc}(\theta^i)\, V_{sd}(\theta^i)\, {\cal A}_{cd},
 \end{eqnarray*}

 \bea
  &&with\nonumber \\
  &&{}\nonumber \\
  {\cal A}_{ab} &=& - 2\, (1 + n)\, {\tilde \phi}^{2/3}\, Q_a\,
  Q_b\, \Big(\sigma_{(a)(b)} - {{\sgn}\over 3}\, \theta\,
  \delta_{(a)(b)}\Big) +\nonumber \\
  &+& {\tilde \phi}^{1/3}\, \Big(Q_a\, \sum_v\, V_{vb}\, [\partial_v\, {\bar n}_{(a)}
  + (\partial_v\, \Gamma_a^{(1)}  - {1\over 3}\, {\tilde \phi}^{-1}\,
  \partial_v\, \tilde \phi )\, {\bar n}_{(a)}]  +\nonumber \\
 &+& Q_b\, \sum_v\, V_{va}\, [\partial_v\, {\bar n}_{(b)}
  + (\partial_v\, \Gamma_b^{(1)}  - {1\over 3}\, {\tilde \phi}^{-1}\,
  \partial_v\, \tilde \phi )\, {\bar n}_{(b)}] +\nonumber \\
  &+& {2\over 3}\, Q^2_a\, \delta_{ab}\, \sum_{cv}\, V_{vc}\, {\tilde
  \phi}^{-1}\, \partial_v\, \tilde \phi\, {\bar n}_{(c)} +\nonumber \\
 &+& 2\, \sum_{\bar a v}\, \Big[Q_a^2\, \delta_{ab}\, \gamma_{\bar aa}\,
 \sum_c\, Q_c^{-1}\, V_{vc}\, {\bar n}_{(c)} -\nonumber \\
 &-& \gamma_{\bar aa}\, Q_a\, V_{vb}\, {\bar n}_{(a)} + \gamma_{\bar
 ab}\, Q_b\, V_{va}\, {\bar n}_{(b)}
 \Big]\, \partial_v\, R_{\bar a} -\nonumber \\
 &-&\sum_{uvc}\, Q_c^{-1}\, V_{vc}\, {\bar n}_{(c)}\, \Big[Q_a^2\,
 V_{ub}\, (\partial_u\, V_{va} - \partial_v\, V_{ua}) + Q_b^2\,
 V_{ua}\, (\partial_u\, V_{vb} - \partial_v\, V_{ub})
 \Big] \Big),\nonumber \\
 &&{}
 \label{2.17}
 \eea

\noindent where we used the definition of ${}^3K_{rs}$ and the
expression $n_{r|s} + n_{s|r} = \partial_s\, ({\tilde \phi}^{1/3}\,
Q_a\, V_{ra}\, {\bar n}_{(a)}) + \partial_r\, ({\tilde \phi}^{1/3}\,
Q_a\, V_{sa}\, {\bar n}_{(a)}) - 2\, {\tilde \phi}^{1/3}\, Q_a\,
V_{ua}\, {\bar n}_{(a)}\, {}^3\Gamma^u_{rs}$ with
${}^3\Gamma^u_{rs}$ given in Eqs.(\ref{2.10}). These equations give
the first half of the standard ADM Hamilton equations in the York
canonical basis: they are expressed in terms of the expansion and
the shear of the Eulerian observers. We get that; 1)
$\partial_{\tau}\, \tilde \phi$ depends upon $\theta = - \sgn\,
{}^3K$ and $\sigma_{(a)(a)}$; 2) $\partial_{\tau}\, \Gamma_a^{(1)}$
depends upon $\sigma_{(a)(a)}$; 3) the expressions involving the
$\tau$-derivatives of the angles $\theta^i$ depend upon
$\sigma_{(a)(b)}{|}_{a \not= b}$. The diagonal elements
$\sigma_{(a)(a)}$ can be expressed in terms of the tidal velocities
$\partial_{\tau}\, \Gamma_a^{(1)}$ by means of the functions ${\cal
A}_{aa}$.

\bigskip

By using Eqs.(\ref{2.9}) and (\ref{2.10}) we get that the diagonal
elements $\sigma_{(a)(a)}$ of the shear are also connected with the
tidal momenta $\Pi_{\bar a}$, while the non-diagonal elements
$\sigma_{(a)(b)}{|}_{a \not= b}$ are connected with the momenta
$\pi_i^{(\theta)}$ (the unknowns in the super-momentum constraints)

\begin{eqnarray*}
 \tilde \phi\, \sigma_{(a)(a)} &=& - {{8\pi\, G}\over {c^3}}\,
 \sum_{\bar a}\, \gamma_{\bar aa}\, \Pi_{\bar a},\,
 \Rightarrow\, \Pi_{\bar a} = - {{c^3}\over {8\pi\, G}}\, \tilde \phi\,
 \sum_a\, \gamma_{\bar aa}\, \sigma_{(a)(a)},\nonumber \\
 &&{}\nonumber \\
 \tilde \phi\, \sigma_{(a)(b)}{|}_{a \not= b} &=& - {{8\pi\, G}\over
 {c^3}}\, \sum_{tw}\, {{\epsilon_{abt}\, V_{wt}}\over
 {Q_b\, Q_a^{-1} - Q_a\, Q_b^{-1}}}\, \sum_i\, B_{iw}\,
 \pi_i^{(\theta )},\nonumber \\
 &&{}\nonumber \\
 \Rightarrow&& \pi_i^{(\theta )} =  {{c^3}\over {8\pi\, G}}\,
 \tilde \phi\,  \sum_{wtab}\, A_{wi}\, V_{wt}\, Q_a\, Q_b^{-1}\, \epsilon_{tab}\,
 \sigma_{(a)(b)}{|}_{a\not= b},
 \end{eqnarray*}

 \bea
 {}^3K_{rs} &=&  {\tilde \phi}^{2/3}\, \Big[ {{4\pi\, G}\over
 {c^3}}\, \pi_{\tilde \phi}\, \sum_a\, Q^2_a\, V_{ra}\, V_{sa} +
 \sum_{ab}^{a \not= b}\, \sigma_{(a)(b)}\, Q_a\, Q_b\, V_{ra}\,
 V_{sb} -\nonumber \\
 &-& {{8\pi\, G}\over {c^3}}\, {\tilde \phi}^{-1}\, \sum_{a\bar a}\,
 \gamma_{\bar aa}\, \Pi_{\bar a}\, Q^2_a\, V_{ra}\, V_{sa}
 \Big],\nonumber \\
 &&{}\nonumber \\
 &&{}\nonumber \\
 \sigma^2 &{\buildrel {def}\over =}& {1\over 2}\, \sum_{ab}\,
 \sigma^2_{(a)(b)} = {1\over 2}\, \Big({{8\pi\,
 G}\over {c^3}}\Big)^2\, {\tilde \phi}^{-2}\, \Big[\sum_{\bar a}\,
 \Pi^2_{\bar a} +\nonumber \\
 &+&2\, \sum_{ww^{'}}\, \Big({{V_{w1}\, V_{w^{'}1}}\over {(Q_2\, Q_3^{-1}
  - Q_3\, Q_2^{-1})^2}} + {{V_{w2}\, V_{w^{'}2}}\over {(Q_3\, Q_1^{-1}
   - Q_1\, Q_3^{-1})^2}} + {{V_{w3}\, V_{w^{'}3}}\over {(Q_1\, Q_2^{-1}
    - Q_2\, Q_1^{-1})^2}}\Big)\nonumber \\
    && \sum_{ii^{'}}\, B_{iw}\, B_{i^{'}w^{'}}\,
 \pi_i^{(\theta )}\, \pi_{i^{'}}^{(\theta )}\Big].
 \label{2.18}
 \eea

\bigskip

Therefore the Eulerian observers associated to the 3+1 splitting of
space-time allow a physical interpretation of some of the variables
of the York canonical basis:\medskip

1) {\it their expansion $\theta$  is the gauge variable $\pi_{\tilde
\phi}$ determining the non-dynamical part of the shape of the
instantaneous 3-spaces $\Sigma_{\tau}$};

2) {\it the diagonal elements of their shear describe the tidal
momenta $\Pi_{\bar a}$, while the non-diagonal elements are
connected to the variables $\pi_i^{(\theta )}$}, determined by the
super-momentum constraints.

\subsection{The Asymptotic ADM Poincare' Algebra, the Rest-Frame Conditions
and the Center of Mass}

As explained in Ref.\cite{1}, following suggestions of Dirac in
Ref.\cite{25}, the limit of the embedding $z^{\mu}(\tau ,\vec \sigma
)$ for the non-inertial rest-frame instant form of metric and tetrad
gravity at spatial infinity in asymptotically Minkowskian
space-times has the form

\beq
 z^{\mu}(\tau ,\vec \sigma )\, \rightarrow x^{\mu}_{(\infty )}(\tau )
+ b^{\mu}_{(\infty )\, r}\, \sigma^r,
 \label{2.19}
 \eeq

\noindent where $x^{\mu}_{(\infty )}(\tau )$ is an asymptotic
inertial observer and $l^{\mu}_{(\infty )} = b^{\mu}_{(\infty )\,
\tau} = \epsilon^{\mu}_{\tau}$ and $b^{\mu}_{(\infty )\, r} =
\epsilon^{\mu}_r$ are the asymptotic tetrad denoted
$\epsilon^{\mu}_A$ in the Introduction. Eq.(\ref{2.19}) is
Eq.(\ref{1.1}) rewritten in Dirac notation \cite{25}.\medskip

The  generators of the asymptotic ADM Poincare' group are \cite{1}

\bea
 p^{\mu}_{(\infty )} &=& P^{\mu}_{ADM} = b^{\mu}_{(\infty )\, A}(\tau )\, P^A_{ADM}
\approx  l^{\mu}_{(\infty )}\, {\hat P}^{\tau}_{ADM} + b^{\mu
}_{(\infty ) r}\, {\hat P}^{ r}_{ADM},\nonumber \\
 &&{}\nonumber \\
 J^{\mu\nu}_{(\infty )} &=& x^{\mu}_{(\infty )}\, p^{\nu}_{(\infty )} -
x^{\nu}_{(\infty )}\, p^{\mu}_{(\infty )} + S^{\mu\nu}_{(\infty )},
\nonumber \\
 &&{}\nonumber \\
 S^{\mu\nu}_{(\infty )} &=& b^{\mu}_{(\infty )\, A}\,
b^{\nu}_{(\infty )\, B}\, J^{AB}_{ADM} \approx \nonumber \\
 &\approx&    [l^{\mu}_{(\infty )}\, b^{\nu}_{(\infty )
r} - l^{\nu} _{(\infty )}\, b^{\mu}_{(\infty ) r}]\, {\hat J}^{\tau
r} _{ADM} + [b^{\mu}_{(\infty ) r}\, b^{\nu }_{(\infty ) s} - b^{\nu
}_{(\infty ) r}\, b^{\mu}_{(\infty ) s}]\, {\hat J}_{ADM}^{rs},
 \label{2.20}
 \eea

\noindent where $p^{\mu}_{(\infty )}$ is the variable canonically
conjugate to $x^{\mu}_{(\infty )}$ (it is assumed time-like as a
boundary condition). $P^A_{ADM}$, $J^{AB}_{ADM}$ are the {\it
strong} (surface integrals at spatial infinity) ADM Poincare'
charges in adapted radar coordinates: they are weakly equal to the
{\it weak} (volume integrals over $\Sigma_{\tau}$) ADM Poincare'
charges ${\hat P}^A_{ADM}$, ${\hat J}^{AB}_{ADM}$.\medskip

Since $p^{\mu}_{(\infty )}$ is orthogonal to the asymptotic
hyper-plane (\ref{2.19}) due to the requirement of absence of
super-translations \cite{1}, we can make the identifications
$l^{\mu}_{(\infty )} = b^{\mu}_{(\infty )\, \tau}  =
p^{\mu}_{(\infty )}/\sqrt{\sgn\, p^2_{(\infty )}} = h^{\mu} =
(\sqrt{1 + {\vec h}^2}; \vec h) = \epsilon^{\mu}_{\tau}$ and
$b^{\mu}_{(\infty )\, r} = {\tilde \epsilon}^{\mu}_r(\vec h) =
\epsilon^{\mu}_r$, where $h^{\mu}$ and the space-like 4-vectors
${\tilde \epsilon}^{\mu}_r(\vec h)$, orthogonal to $h^{\mu}$, are
the column of the standard Wigner boost sending $P^{\mu} = Mc\,
h^{\mu}$ to its rest-frame form $Mc\, (1; \vec 0)$ (see Ref.\cite{8}
for this notation).\medskip

Then for consistency the first of Eqs.(\ref{2.20}) implies the
rest-frame conditions

\beq
 P^r_{ADM} \approx {\hat P}^r_{ADM} \approx 0,
 \label{2.21}
 \eeq

\noindent and $Mc \approx {\hat P}^{\tau}_{ADM} = {\hat E}_{ADM}/c$.

\medskip

From Eqs.(25) of Ref.\cite{3} the weak or volume form of the ADM
Poincar\'e charges appearing in Eqs.(\ref{2.20}) is

\begin{eqnarray*}
 {\hat P}^{\tau}_{ADM}&=& {1\over c}\, {\hat E}_{ADM} = \int
 d^3\sigma\, \Big[ - {{c^3}\over {16\pi\, G}}\, \sqrt{\gamma}\,\,
 {}^3g^{rs}\, \Big({}^3\Gamma^{u} _{rv}\, {}^3\Gamma^{v}_{su} -
 {}^3\Gamma^{u}_{rs}\, {}^3\Gamma^{v}_{vu}\Big) +\nonumber \\
 &+&{{8\pi\, G}\over { c^3\, \sqrt{\gamma}
 }}\, {}^3G_{rsuv}\, {}^3\Pi^{rs}\, {}^3\Pi^{uv}
 + {\cal M}\Big] (\tau ,\vec \sigma ),\nonumber \\
 {\hat P}^{r}_{ADM}&=& \int d^3\sigma \, \Big[{}^3g^{rs}\, {\cal M}_{s}
 - 2\, {}^3\Gamma^{r}_{su}(\tau ,\vec \sigma )\, {}^3\Pi^{su} \Big](\tau
 ,\vec \sigma) \approx 0,
 \end{eqnarray*}

 \bea
 {\hat J}^{\tau r}_{ADM}&=&- {\hat J}^{r\tau}_{ADM} = \nonumber \\
 &=&\int d^3\sigma\,   \Big( \sigma^{r}
 \, \Big[ {{c^3}\over {16\pi\, G}}\, \sqrt{\gamma}\,\,  {}^3g^{ns}\,
 ({}^3\Gamma^{u}_{nv}\, {}^3\Gamma^{v}_{su} - {}^3\Gamma^{u}_{ns}\,
 {}^3\Gamma^{v}_{vu}) - {{8\pi\, G}\over {c^3\, \sqrt{\gamma}}}\,
 {}^3G_{nsuv}\, {}^3\Pi^{ns}\, {}^3\Pi^{uv} - {\cal M}\Big] +\nonumber \\
 &+& {{c^3}\over {16\pi\, G}}\, \delta^{r}_{u}\, ({}^3g_{vs} - \delta_{vs})\,
 \partial_{n}\, \Big[ \sqrt{\gamma}\, ({}^3g^{ns} \,
 {}^3g^{uv} - {}^3g^{nu}\, {}^3g^{sv})\Big] \Big) (\tau
 ,\vec \sigma ) \approx 0,\nonumber \\
 &&{}\nonumber \\
 {\hat J}^{rs}_{ADM}&=&
 \int d^3\sigma\,  \Big[(\sigma^{r}\, {}^3g^{su} - \sigma^s\, {}^3g^{ru})\,
 {\cal M}_{u} - 2\, (\sigma^{r}\, {}^3\Gamma^{s} _{uv} -
 \sigma^{s}\, {}^3\Gamma^{r}_{uv})\, {}^3\Pi^{uv}
 \Big] (\tau ,\vec \sigma ).
 \label{2.22}
\end{eqnarray}

\noindent  They are the same weak Poincar\'e charges of metric
gravity, expressed in terms of cotriads ${}^3e_{(a)r}$ and their
conjugate momenta ${}^3\pi^r_{(a)}$, by using
${}^3g_{rs}={}^3e_{(a)r}\, {}^3e_{(a)s}$, ${}^3\Pi^{rs}= {1\over 4}
[{}^3e^r_{(a)}\, {}^3\pi^s_{(a)}+{}^3e^s_{(a)}\, {}^3\pi^r_{(a)}]$
(see Eq.(5.7) of Ref.\cite{1}). The Christoffel symbols
${}^3\Gamma^r_{uv}$ are built with the 3-metric
${}^3g_{rs}$.\bigskip

In Eqs.(\ref{2.22}) we  use the notations A) ${\cal M}(\tau ,\vec
\sigma )$  for the internal matter mass density (in general
metric-dependent); B) ${\cal M}_r(\tau ,\vec \sigma )$ for the
internal matter momentum density (metric-independent and universal).
See Section III, Eq.(\ref{3.9}), for their explicit form for the
matter considered in this paper: there we will give the
modifications of the relevant formulas of Ref.\cite{3} due to the
presence of matter, because, in absence of derivative couplings to
matter, they are independent from the type of matter: only the
explicit form of the mass density ${\cal M}$ (and also of the matter
stress tensor $T^{rs}$) depends on the type of matter \footnote{We
have:\hfill\break
 A) For the Klein-Gordon field the Lagrangian is ${1\over 2}\,
 \Big[{}^4g^{AB}\, \partial_A\, \varphi\, \partial_B\, \varphi
 - m^2\, \varphi^2\Big](\tau ,\vec \sigma )$ and we have ${\cal
 M}(\tau ,\vec \sigma ) ={1\over 2}\, \Big[{\tilde \phi}^{-1}\, \pi^2 +
 \tilde \phi\, \Big({}^3e^r_{(a)}\, {}^3e^s_{(a)}\, \partial_r\, \varphi\,
  \partial_s\, \varphi \Big)\Big](\tau ,\vec \sigma )$ and ${\cal
  M}_r(\tau ,\vec \sigma ) = \pi (\tau ,\vec \sigma )\, \partial_r\,
  \varphi (\tau ,\vec \sigma )$, where $\pi(\tau ,\vec \sigma )$ is
  the KG momentum.\hfill\break
  B) For perfect fluids \cite{26} with Lagrangian coordinates $\alpha^i(\tau
  ,\vec \sigma )$ and equation of state $\rho = \rho (n,s)$ and
  fluid momenta $\Pi_i(\tau ,\vec \sigma )$ we have ${\cal M}(\tau
  ,\vec \sigma ) = \Big({{c^2}\over X}\, \Big[\tilde \phi\, X\, \rho ({\tilde \phi}^{-1}\, X)
  + ([det\, (\partial_r\, \alpha^i)]^2 - X^2)\, {{\partial\, \rho (x)}\over
  {\partial\, x}}{|}_{x = {\tilde \phi}^{-1}\, X}\Big]\Big)(\tau ,\vec \sigma )$
 ($X = X({}^3g^{rs}, \tilde \phi , \alpha^i, \Pi_i)$ is the solution of a
 trascendental equation depending on the equation of state)
and ${\cal M}_r(\tau ,\vec \sigma ) = - \Pi_i(\tau ,\vec \sigma )\,
\partial_r\, \alpha^i(\tau ,\vec \sigma )$. For a dust we have
${\cal M}(\tau ,\vec \sigma ) = \sqrt{\mu^2\, c^2 [det\,
(\partial_r\, \alpha^i)]^2 + {\tilde \phi}^{-2/3}\, \sum_u\, e^{-
2\, \sum_{\bar a}\, \gamma_{\bar au}\, R_{\bar a}}\, \Pi_m\,
\partial_u\, \alpha^m\, \Pi_n\, \partial_u\, \alpha^n}(\tau ,\vec \sigma )$.}.
\bigskip

In Section III, Eq.(\ref{3.11}), there is the form of the
energy-momentum tensor $T^{AB}(\tau, \vec \sigma)$ for the type of
matter considered in this paper. Since our formulation is equivalent
to the ADM one, and therefore to Einstein equations, we have
${}^4\nabla_A\, T^{AB}(\tau ,\vec \sigma) \equiv 0$ as a consequence
of the Bianchi identities.

\medskip

In Ref.\cite{27} it is said that the ADM Poincare' charges coincide
with those arising from the Landau-Lifshitz energy-momentum
pseudo-tensor for the gravitational field $t_{LL}^{\mu\nu} =
t_{LL}^{\nu\mu} = - {{c^4}\over {8\pi G}}\, G^{\mu\nu} + {{c^4}\over
{16\pi G\, (- {}^4g)}}\, \partial_{\alpha}\,
\partial_{\beta}\,\Big[(- {}^4g)\, ({}^4g^{\mu\nu}\,
{}^4g^{\alpha\beta} - {}^4g^{\mu\alpha}\, {}^4g^{\nu\beta})\Big]$
\cite{28} (see Ref.\cite{29} for a detailed analysis). As a
consequence Einstein's equations imply the conservation law
$\partial_B\, \Big[(- {}^4g)\, \Big(T^{AB} + t_{LL}^{AB}\Big)\Big] =
0$ (instead the identities ${}^4\nabla_A\, T^{AB}(\tau ,\vec \sigma)
\equiv 0$ are not conservation laws).

\bigskip

In Ref.\cite{10} we gave the form of the {\it internal} Poincare'
generators of an isolated system in special relativity in both the
inertial and non-inertial rest frames. By restricting
Eqs.(\ref{2.22}) to Minkowski space-time (a special solution of
Einstein's equations when $G = 0$) and then to inertial rest-frames
(where ${}^4g_{AB} = {}^4\eta_{AB}$ and the embedding identifies the
Wigner 3-spaces), it turns out that the ADM Poincare' generators
become equal to the special relativistic ones for the given matter
\cite{8} in the inertial rest-frames. Also the non-inertial
rest-frame Poincare' generators \cite{10} could be recovered without
the restriction to inertial rest-frames (now ${}^4g_{AB} =
{}^4g_{AB}[z]$).

\bigskip

As a consequence, this approach leads to the following visualization
of the {\it 3-universe} (with its content of gravitational field and
matter) contained in the instantaneous 3-spaces $\Sigma_{\tau}$: it
is an isolated system described by a decoupled  {\it external}
canonical non-covariant 4-center of mass ${\tilde x}^{\mu}(\tau)$ (a
non-observable \footnote{Let us remark that the non-observability of
this decoupled pseudo-particle is in accord with the viewpoint of
Ref.\cite{30}.} decoupled pseudo-particle; it replaces the
$x^{\mu}_{(\infty)}(\tau)$ of Dirac proposal) carrying a pole-dipole
structure with mass $Mc \approx {\hat E}_{ADM}/c$ and a spin $S^r =
{1\over 2}\, \epsilon^{ruv}\, J^{uv}_{ADM} \approx {1\over 2}\,
\epsilon^{ruv}\, {\hat J}^{uv}_{ADM}$. It is the same structure
present in special relativity \cite{8,10} in the inertial and
non-inertial rest-frame instant forms of dynamics. There, in
non-inertial rest frames the embeddings tend at spatial infinity to
the Wigner hyper-planes $z_W^{\mu}(\tau ,\vec \sigma ) =
Y^{\mu}(\tau ) + \epsilon^{\mu}_r(\vec h)\, \sigma^r$ (the analogue
of Eq.(\ref{2.19})), where $Y^{\mu}(\tau ) = Y^{\mu}(0) + h^{\mu}\,
\tau$ is the covariant non-canonical Fokker-Pryce center of inertia
of the isolated system. Both ${\tilde x}^{\mu}(\tau)$ and
$Y^{\mu}(\tau)$ are well defined functions (given in Ref.\cite{8})
of $\tau$, $Mc$, $S^r$, and of frozen (non-evolving) Jacobi data
$\vec z$, $\vec h$, for the decoupled external center of mass
(${\vec x}_{NW} = \vec z/Mc$ is the non-covariant Newton-Wigner
3-position).\medskip

In special relativity ${\tilde x}^{\mu}(\tau)$  carries also a
universal {\it external} realization of the Poincare' algebra with
generators $P^{\mu} = Mc\, h^{\mu}$, $J^{ij} = z^i\, h^j - z^j\, h^i
+ \epsilon^{iju}\,  S^u$, $K^i = J^{oi} = - \sqrt{1 + {\vec h}^2}\,
z^i + {{({\vec S} \times \vec h)^i}\over {1 + \sqrt{1 + {\vec
h}^2}}}$. In tetrad gravity these generators are
$p^{\mu}_{(\infty)}$, $J^{\mu\nu}_{(\infty)}$.

\medskip

In special relativity inside each instantaneous 3-space
$\Sigma_{\tau}$ there is the isolated system and an {\it internal}
realization of the Poincare' algebra, whose generators are
determined by the energy-momentum tensor: these generators
correspond to the weak ADM Poincare' charges in canonical ADM
gravity.
\medskip

If, like in special relativity \cite{8,10}, we eliminate the {\it
internal} 3-center of mass inside the instantaneous 3-spaces
$\Sigma_{\tau}$ with the following gauge fixings to the constraints
(\ref{2.21}) (together they form three pairs of second class
constraints)

\beq
 {\hat J}^{\tau r}_{ADM} \approx 0,
 \label{2.23}
 \eeq

\noindent then we can choose the world-line of the non-canonical
covariant Fokker-Pryce 4-center of inertia as origin of the
3-coordinates $\sigma^r$ in the 3-spaces $\Sigma_{\tau}$ (instead of
$x^{\mu}_{(\infty)}$ or of $x^{\mu}(\tau)$ of Eq.(\ref{1.1})). As a
consequence the internal realization of the Poincare' algebra is
unfaithful: only the invariant mass (the weak ADM energy) and the
rest spin of the 3-universe are non zero (their value is given as a
boundary condition). Then it should be possible to express both the
gravitational field and the matter only in terms of relative
variables, like it happens in special relativity where the isolated
system inside the Wigner hyper-plane is described by relative
Wigner-covariant variables.

\bigskip

Let us add a final remark. In Eqs. (12.2) and (12.3) of Ref.\cite{1}
and in Section 7 of Ref.\cite{3} it is shown how to make a so-called
Frauendiener-Sen-Witten transport (depending on the extrinsic
curvature of $\Sigma_{\tau}$) of the asymptotic flat tetrads
$\epsilon^{\mu}_A$ to each point of $\Sigma_{\tau}$: this allows to
define a {\it local dynamical compass of inertia} ${}^4{\check
E}^{\mu}_{(\alpha)}$ to be compared in each point of $\Sigma_{\tau}$
with the tetrad ${}^4E^{\mu}_{(\alpha )}$. These special tetrads
have the following form (${\check \alpha}_{(a)}$ are suitable Euler
angles)

\bea
 {}^4{\check E}^{\mu}_{(o)} &=& l^{\mu} = {1\over {1 + n}}\,
 \Big(z^{\mu}_{\tau} - n^r\, z^{\mu}_r\Big) \,
 \rightarrow \, \epsilon^{\mu}_{\tau},\nonumber \\
 {}^4{\check E}^{\mu}_{(c)} &=& z^{\mu}_r\, {}^3{\check e}^r_{(c)}\, \rightarrow\,
 \epsilon^{\mu}_{c},\qquad  {}^3{\check e}^r_{(c)} =
 R_{(c)(d)}({\check \alpha}_{(e)})\, {}^3e^r_{(d)}\, \rightarrow\,
\delta^a_c,
 \label{2.24}
 \eea

\noindent with the special triads ${}^3{\check e}^r_{(a)}$ solution
of the following Frauendiener equations \cite{31} ($\alpha$ is a
proportionality constant)

\bea &&{}^3\nabla_r\, {}^3{\check e}^r_{(1)} = {}^3\nabla_r\,
{}^3{\check e}^r_{(2)} = 0,\qquad
 {}^3\nabla_r\, {}^3{\check e}^r_{(3)} = - \alpha\,
{}^3K = - \alpha\, {{12\pi\, G}\over {c^3}}\, \pi_{\tilde \phi},\nonumber \\
 &&{}\nonumber \\
 &&{}^3{\check e}^r_{(1)}\, {}^3{\check e}^s_{(3)}\,
{}^3\nabla_r\, {}^3{\check e}_{(2)s} + {}^3{\check e}^r_{(3)}\,
 {}^3{\check e}^s_{(2)}\, {}^3\nabla_r\, {}^3{\check e}_{(1)s} +
 {}^3{\check e}^r_{(2)}\, {}^3{\check e}^s_{(1)}\, {}^3\nabla_r\,
{}^3{\check e}_{(3)s} = 0.
 \label{2.25}
\eea

Therefore, these triads are formed by 3 vector fields with the
properties: i) two vector fields are divergence free; ii) the third
one has a non-vanishing divergence proportional to the trace of the
extrinsic curvature (the inertial York time) of $\Sigma_{\tau}$ (on
a maximal slicing hyper-surface, ${}^3K=0$, all three vectors would
be divergence free); iii) the vectors satisfy a cyclic condition. In
Ref.\cite{31} it is shown that these triads do not exist for compact
$\Sigma_{\tau}$ and in general when there is  nontrivial topology
for $\Sigma_{\tau}$.

\bigskip

In conclusion, there are {\it preferred ADM geometrical and
dynamical  Eulerian observers} with unit 4-velocity ${}^4{\check
E}^{\mu}_{(o)}$ and gyroscopes along the spatial axes ${}^4{\check
E}^{\mu}_{(a)}$. The asymptotic world-lines of the congruence of
these observers may replace the static concept of {\it fixed stars}
in the study of the precessional effects of gravito-magnetism on
gyroscopes ({\it dragging of inertial frames}) and seem to be
naturally connected with the definition of post-Newtonian
coordinates \cite{32}. This congruence of time-like preferred
observers is a non-Machian element of these noncompact space-times.
\medskip

These preferred tetrads  correspond to the {\it non-flat preferred
observers} of Bergmann \cite{33}: they are a set of {\it privileged
observers} (privileged tetrads adapted to the instantaneous 3-spaces
$\Sigma_{\tau}$) of {\it geometrical nature}  and not of {\it static
nature}. Since they depend on the intrinsic and extrinsic geometry
of $\Sigma_{\tau}$, on the solutions of Einstein's equations they
also acquire both a {\it dynamical nature} depending on the
configuration of the gravitational field itself and an {\it inertial
nature} due to the dependence on the inertial effect described by
the York time.

\vfill\eject

\section{The ADM Action in Presence of the Electro-magnetic Field and of
Charged  Scalar Particles with Grassmann-valued Electric Charges and
Sign of the Energy.}

Let us now describe N charged scalar particles and the
electro-magnetic field coupled to the gravitational field in ADM
tetrad gravity.

\subsection{The ADM Action and the Constraints}

The tetrad ADM action \footnote{Dimensions of the quantities
appearing in this Section: $[\tau ] = [x^{\mu}] = [\vec \sigma ] =
[{\vec \eta}_i] = [l]$, $[{\vec \kappa}_i] = [P^{\mu}] = [E/c] =
[m\, l\, t^{-1}]$, $[{}^4g] = [{}^3g] = [n] = [n_{(a)}] =
[{}^3e_{(a)r}] = [{\dot {\vec \eta}}_i] = [\theta_i] = [0]$, $[G =
6.7\, 10^{-8}\, cm^3\, s^{-2}\, g^{-1}] = [m^{-1}\, l^3\, t^{-2}]$,
$[G/c^3] = [m^{-1}\, t] \approx 2.5\, 10^{-39}\, sec/g$, $[S] =
[\hbar ] = [J^{AB}] = [m\, l^2\, t^{-1}]$, $[{}^3R] =
[{}^3\Omega_{rs(a)}] = [l^{-2}]$, $[{}^3\omega_{r(a)}] =
[{}^3K_{rs}] = [l^{-1}]$, $[{}^3\pi^r_{(a)}] = [{}^3{\tilde
\Pi}^{rs}] = [m\, l^{-1}\, t^{-1}]$, $[T^{AB}] = [{\cal M}] = [{\cal
M}_r] = [{\cal H}] = [{\cal H}_{(a)}] = [m\, l^{-2}\, t^{-1}]$.} for
tetrad gravity (see Eq.(4) of Ref.\cite{3}) plus the
electro-magnetic field and N charged scalar particles with
Grassmann-valued electric charges $Q_i$ and sign of the energy
$\eta_i$ (whose world-line is $x^{\mu}_i(\tau) = z^{\mu}(\tau, {\vec
\eta}_i(\tau))$ with the embedding of Eq.(\ref{1.1})), depending on
the configuration variables $1 + n(\tau ,\vec \sigma )$,
$n_{(a)}(\tau ,\vec \sigma )$, $\varphi_{(a)}(\tau ,\vec \sigma )$,
${}^3e _{(a)r}(\tau ,\vec \sigma )$, $A_A(\tau ,\vec \sigma)$,
$\eta^r_i(\tau )$, $\theta_i(\tau)$, $\theta_i^{(Q)}(\tau)$, is
\footnote{The use of a positive 3-metric changes the sign definition
of the particle momentum and the Poisson bracket sign with respect
to Ref.\cite{15}. $G$ is the Newton constant. In $S_{Grassmann}$ we
use the Planck constant $\hbar$ for dimensional reasons, since our
regularization of the self-energies is considered a semi-classical
approximation of a quantum theory, which exists for the
electro-magnetic field (QED) but not yet for the gravitational
field.}

\begin{eqnarray*}
 S &=& S_{grav} + S_{em} + S_{part} + S_{Grassmann} =\nonumber \\
 &&{}\nonumber \\
 &=&{{c^3}\over {16\pi\, G}}\, \int d\tau d^3\sigma \,
\lbrace (1 + n)\, {}^3e\, \epsilon_{(a)(b)(c)}\, {}^3e^r_{(a)}\,
{}^3e^s_{(b)}\, {}^3\Omega_{rs(c)}+\nonumber \\
 &+&{{{}^3e}\over
{2 (1 + n)}} ({}^3G_o^{-1})_{(a)(b)(c)(d)} {}^3e^r_{(b)}(n_{(a) |
r}- \partial_{\tau}\, {}^3e_{(a)r})\, {}^3e^s_{(d)}(n_{(c) |
s}-\partial_{\tau} \, {}^3e_{(c) \ s})\rbrace  (\tau ,\vec \sigma )
 -\nonumber \\
 &&{}\nonumber \\
 &-& {1\over 4}\, \int d\tau d^3\sigma\, {}^3e(\tau ,\vec \sigma )\,
 \Big[ - {{2\, {}^3e^r_{(a)}\, {}^3e^s_{(a)}}\over {1 + n}}\,
 F_{\tau r}\, F_{\tau s} + {{4\, {}^3e^r_{(a)}\, n_{(a)}\, {}^3e^s_{(b)}\,
 {}^3e^u_{(b)}}\over {1 + n}}\, F_{\tau u}\, F_{rs} +\nonumber \\
 &+& {}^3e^r_{(a)}\, {}^3e^s_{(b)}\, {}^3e^u_{(c)}\, {}^3e^v_{(d)}\,
 [(1 + n)\, \delta_{(a)(b)}\, \delta_{(c)(d)} - {{\delta_{(a)(b)}\,
 n_{(c)}\, n_{(d)} + \delta_{(c)(d)}\, n_{(a)}\, n_{(b)}}\over
 {1 + n}}]\, F_{ru}\, F_{sv} \Big](\tau ,\vec \sigma ) -
 \end{eqnarray*}

\bea
 &-& \sum_{i=1}^N\, \int d\tau d^3\sigma\,
\delta^3(\vec \sigma , {\vec \eta}_i(\tau ))\, \Big(m_i\, c\, \nonumber \\
 &&\eta_i\, \sqrt{\Big(1 + n(\tau ,\vec \sigma )\Big)^2 -
 \Big({}^3e_{(a)r}(\tau ,\vec \sigma )\, {\dot \eta}
^r_i(\tau ) + n_{(a)}(\tau ,\vec \sigma )\Big)\,
\Big({}^3e_{(a)s}(\tau ,\vec \sigma )\, {\dot \eta}^s_i(\tau ) +
n_{(a)}(\tau ,\vec \sigma )\Big)} +\nonumber \\
 &-& {{\eta_i\,Q_i}\over c}\, \Big[A_{\tau}(\tau ,\vec \sigma ) + A_r(\tau ,\vec \sigma )\,
  {\dot \eta}^r_i(\tau )\Big]\Big) + \nonumber \\
 &&{}\nonumber \\
 &+& {{i\hbar}\over 2}\, \int d\tau \sum_i\, [\theta_i^*(\tau
)\, {\dot \theta}_i(\tau ) - {\dot \theta}_i^*(\tau )\,
\theta_i(\tau )] + {{i\hbar}\over 2}\, \int d\tau \sum_i\,
[\theta_i^{(Q) *}(\tau )\, {\dot \theta}^{(Q)}_i(\tau ) - {\dot
\theta}_i^{(Q) *}(\tau )\,
\theta^{(Q)}_i(\tau )],\nonumber \\
 &&{}\nonumber \\
 &&{}\nonumber \\
 &&\eta_i = \theta^*_i\, \theta_i,\qquad Q_i = \theta^{(Q) *}_i\,
 \theta^{(Q)}_i.
 \label{3.1}
\end{eqnarray}

\noindent While $Q_i$ is the Grassmann-valued electric charge of
particle "i", $\eta_i$ is its Grassmann-valued {\it sign of the
energy} (particles with negative energy have the opposite electric
charge: $\eta_i\, Q_i$).
\medskip

In the action $S_{grav}$ for the gravitational field
${}^3\Omega_{rs(a)} = \partial_r\, {}^3\omega_{s(a)} -
\partial_s\, {}^3\omega_{r(a)} - \epsilon_{(a)(b)(c)}\,
{}^3\omega_{r(b)}\, {}^3\omega_{s(c)}$ is the field strength
associated with the 3-spin connection ${}^3\omega_{r(a)}(\tau ,\vec
\sigma ) = {1\over 2}\, \epsilon_{(a)(b)(c)}\, \Big[{}^3e^u_{(b)}\,
(\partial_r\, {}^3e_{(c)u} - \partial_u\, {}^3e_{(c)r}) + {1\over
2}\, {}^3e^u_{(b)}\, {}^3e^v_{(c)}\, {}^3e_{(d)r}\, (\partial_v\,
{}^3e_{(d)u} - \partial_u\, {}^3e_{(d)v})\Big]$ and
$({}^3G_o^{-1})_{(a)(b)(c)(d)}=\delta_{(a)(c)}\delta_{(b)(d)}+
\delta_{(a)(d)}\delta_{(b)(c)}-2\delta_{(a)(b)}\delta_{(c)(d)}$ is
the flat (with lower indices) inverse  of the flat Wheeler-DeWitt
super-metric ${}^3G_{o(a)(b)(c)(d)}\, = \delta_{(a)(c)}\,
\delta_{(b)(d)} + \delta_{(a)(d)}\, \delta_{(b)(c)} -
\delta_{(a)(b)}\, \delta_{(c)(d)}$, ${}^3G_{o(a)(b)(e)(f)}\,
{}^3G^{-1}_{o(e)(f)(c)(d)} = 2\, (\delta_{(a)(c)}\, \delta_{(b)(d)}
+ \delta_{(a)(d)}\, \delta_{(b)(c)})$.
\medskip

The action $S_{em} = - {1\over 4}\, \int d\tau d^3\sigma\,
\Big[\sqrt{- {}^4g}\, {}^4g^{AC}\, {}^4g^{BD}\, F_{AB}\,
F_{CD}\Big](\tau ,\vec \sigma)$ has the previous form because
${}^4g^{AC}\, {}^4g^{BD}\, F_{AB}\, F_{CD} = 2\, ({}^4g^{\tau\tau}\,
{}^4g^{rs} - {}^4g^{\tau r}\, {}^4g^{\tau s})\, F_{\tau r}\, F_{\tau
s} + 4\, {}^4g^{\tau r}\, {}^4g^{su}\, F_{\tau u}\, F_{rs} +
{}^4g^{rs}\, {}^4g^{uv}\, F_{ru}\, F_{sv}$.

\bigskip

The canonical momenta for the tetrad gravity variables and the
electro-magnetic field are (we are in the canonical basis
(\ref{2.7}))

\begin{eqnarray*}
&&\pi_{\varphi_{(a)}}(\tau ,\vec \sigma ) = {{\delta {S}
_{ADM}}\over {\delta \partial_{\tau} \varphi_{(a)}(\tau
,\vec \sigma )}} = 0, \nonumber \\
 &&\pi_n(\tau ,\vec \sigma ) = {{\delta { S}_{ADM}}\over {\delta
\partial_{\tau} n(\tau ,\vec \sigma )}} = 0,\nonumber \\
 &&\pi_{n_{(a)}}(\tau ,\vec \sigma ) = {{\delta {
S}_{ADM}} \over {\delta \partial_{\tau} n_{(a)}(\tau ,\vec \sigma
)}} = 0,
 \end{eqnarray*}

\begin{eqnarray*}
 &&{}^3\pi^r_{(a)}(\tau ,\vec \sigma
) = {{\delta {S}_{ADM}} \over {\delta \partial_{\tau}
{}^3e_{(a)r}(\tau ,\vec \sigma )}} =\nonumber \\
 &=& - {{c^3}\over {16\pi\, G}}\, [{{
{}^3e}\over {1 + n}}\,  ({}^3G^{-1}_o)_{(a)(b)(c)(d)}\,
{}^3e_{(b)}^r\, {}^3e^s_{(d)}\, (n_{(c) | s}-\partial_{\tau}\,
{}^3e_{(c)s})](\tau ,\vec \sigma ) =\nonumber \\
 &=& - {{c^3}\over {8\pi\, G}}\, [{}^3e ({}^3K^{rs}-{}^3e^r_{(c)}\,
 {}^3e^s_{(c)}\, {}^3K) {}^3e_{(a)s}](\tau ,\vec \sigma ),
 \end{eqnarray*}

 \begin{eqnarray*}
   \pi^\tau(\tau ,\vec \sigma)&=&\frac{\delta\, S_{ADM}}{\delta\,
  \partial_{\tau}\, A_\tau(\tau,\vec{\sigma})} = 0,\nonumber\\
&&\nonumber\\
 \pi^r(\tau ,\vec \sigma)&=&\frac{\delta\,S_{ADM}}{\delta\,
 \partial_{\tau}\, A_r(\tau,\vec{\sigma})} = \Big(\frac{\sqrt{\gamma}}{1 + n}\,
 {}^3e^r_{(a)}\, {}^3e^s_{(a)}\, ( F_{\tau
s} - n^u\, F_{us})\Big)(\tau ,\vec \sigma),\nonumber \\
 &&{}\nonumber \\
 && \{ A_A(\tau ,\vec \sigma), \pi^B(\tau ,{\vec \sigma}^{'}) \} =
 c\, \eta^B_A\, \delta^3(\vec \sigma ,{\vec \sigma}^{'}),
 \end{eqnarray*}

\bea
  &&\lbrace n(\tau ,\vec \sigma ),
\pi_n(\tau ,{\vec \sigma}^{'} ) \rbrace = \delta^3(\vec \sigma
,{\vec \sigma}^{'}),\nonumber \\
 &&\lbrace n_{(a)}(\tau ,\vec
\sigma ),\pi_{n_{(b)}}(\tau ,{\vec \sigma}^{'} )\rbrace
=\delta_{(a)(b)} \delta^3(\vec \sigma ,{\vec
\sigma}^{'}), \nonumber \\
 &&\lbrace \varphi_{(a)}(\tau ,\vec
\sigma ),\pi_{\varphi_{(b)}} (\tau ,{\vec \sigma}^{'} )\rbrace =
\delta_{(a)(b)} \delta^3(\vec \sigma ,
{\vec \sigma}^{'}),\nonumber \\
 &&\lbrace {}^3e_{(a)r}(\tau ,\vec
\sigma ),{}^3\pi^s_{(b)}(\tau , {\vec \sigma}^{'} )\rbrace
=\delta_{(a)(b)} \delta^s_r \delta^3(\vec \sigma , {\vec
\sigma}^{'}),\nonumber \\
 &&{}\nonumber \\
  &&\lbrace
{}^3e^r_{(a)}(\tau ,\vec \sigma),{}^3\pi^s_{(b)}(\tau , {\vec
\sigma}^{'})\rbrace =-{}^3e^r_{(b)}(\tau ,\vec \sigma )\, {}^3e^s
_{(a)}(\tau ,\vec \sigma ) \delta^3(\vec \sigma ,{\vec
\sigma}^{'}), \nonumber \\
 &&\lbrace {}^3e(\tau ,\vec \sigma ),
{}^3\pi^r_{(a)}(\tau ,{\vec \sigma}^{'})\rbrace ={}^3e(\tau ,\vec
\sigma )\, {}^3e^r_{(a)}(\tau ,\vec \sigma )\, \delta^3(\vec \sigma
,{\vec \sigma}^{'}),
 \label{3.2}
\end{eqnarray}

\medskip

\noindent while the particle momenta are

\begin{eqnarray*}
 &&{\tilde \kappa}_{ir}(\tau ) = {{\partial { L}_{ADM}(\tau )}\over
{\partial {\dot \eta}_i^r(\tau )}}\,\, {\buildrel {def}\over =}\,\,
\eta_i\, \kappa_{ir}(\tau) =  {{\eta_i\, Q_i}\over c}\,
A_r(\tau ,{\vec \eta}_i(\tau)) +\nonumber \\
 &&{}\nonumber \\
 &+& {{ \eta_i\, m_i\, c\,\,\, {}^3e_{(a)r}(\tau ,{\vec \eta}_i(\tau ))\,
({}^3e_{(a)s}(\tau ,\vec \sigma )\, {\dot \eta}^s_i(\tau ) +
n_{(a)}(\tau ,\vec \sigma )) }\over {\sqrt{(1 + n(\tau ,\vec \sigma
))^2 - ({}^3e_{(a)r}(\tau ,\vec \sigma )\, {\dot \eta} ^r_i(\tau ) +
n_{(a)}(\tau ,\vec \sigma ))\, ({}^3e_{(a)s}(\tau ,\vec \sigma )\,
{\dot \eta}^s_i(\tau ) + n_{(a)}(\tau ,\vec \sigma ))}}}
{|}_{\vec \sigma = {\vec \eta}_i(\tau )},\nonumber \\
 &&{}\nonumber \\
 &&{}\nonumber \\
   &&\kappa_{ir}(\tau) = \int d\theta_i\, d\theta^*_i\, {\tilde
  \kappa}_{ir}(\tau),\qquad
   \lbrace \eta_i^r(\tau ), \kappa_{is}(\tau ) \rbrace = \delta_{ij}\,
\delta^r_s,
 \end{eqnarray*}

\bea
 &&\Downarrow\nonumber \\
 &&{}\nonumber \\
 &&\sqrt{m^2_i\, c^2 + {}^3e^r_{(a)}(\tau ,\vec \sigma )\, \Big(\kappa_{ir}(\tau
 ) - {{Q_i}\over c}\, A_r(\tau, \vec \sigma)\Big)\, {}^3e^s_{(a)}(\tau ,\vec \sigma )\,
 \Big(\kappa_{is}(\tau ) - {{Q_i}\over c}\, A_s(\tau ,\vec \sigma)\Big)}{|}_{\vec \sigma
 = {\vec \eta}_i(\tau )} =\nonumber \\
 &&{{m_i\, [1 + n(\tau ,\vec \sigma )]}\over {\sqrt{(1 + n(\tau
,\vec \sigma ))^2 - ({}^3e_{(a)r}(\tau ,\vec \sigma )\, {\dot \eta}
^r_i(\tau ) + n_{(a)}(\tau ,\vec \sigma ))\, ({}^3e_{(a)s}(\tau
,\vec \sigma )\, {\dot \eta}^s_i(\tau ) + n_{(a)}(\tau ,\vec \sigma
))}}}{|}_{\vec \sigma = {\vec \eta}_i(\tau )},\nonumber \\
 &&{}\nonumber \\
 &&{}\nonumber \\
&& {\dot \eta}_i^r(\tau ) = {}^3e^r_{(a)}(\tau ,{\vec \eta}_i(\tau
))\, \Big[{{(1 + n)\, {}^3e^s_{(a)}\, \Big(\kappa_{is}(\tau ) -
{{Q_i}\over c}\, A_r\Big)}\over {\sqrt{m^2_i\, c^2 + {}^3e^u_{(b)}\,
{}^3e^v_{(b)}\, \Big(\kappa_{iu}(\tau ) - {{Q_i}\over c}\,
A_u\Big)\, \Big(\kappa_{iv}(\tau ) - {{Q_i}\over c}\, A_v\Big)}}} -
n_{(a)}\Big](\tau
,{\vec \eta}_i(\tau )).\nonumber \\
 &&{}
 \label{3.3}
  \eea

\medskip

To define $\kappa_{ir}$ from ${\tilde \kappa}_{ir}$ we have used the
following property of Grassmann variables \footnote{Following
Ref.\cite{34}, the classical regulated theory can be obtained by
taking the solutions of the mean value   of the final Hamilton
equations, which do not contain singular self-energies, with the
Berezin-Marinov distribution function (namely by putting into them
$< Q_i > = \pm e$). For the U(1) group behind the charge $\eta_i =
\theta^*_i\, \theta_i$ (topological two-levels for the sign of
energy) the {\it positive definite} distribution function is $\rho_i
= a_i + \theta^*_i\, \theta_i$ with $a_i > 0$, $\int \rho_i\,
d\theta_i\, d\theta_i^* = 1$. Therefore the classical regulated
value of a quantity $A$ is $< A >_{12} = \int\, A\, \rho_1\,
d\theta_1\, d\theta^*_1\, \rho_2\, d\theta_2\, d\theta^*_2$ where
Eqs.(\ref{3.4}) have to be used.}

\beq
 \int d\theta_i\, d\theta^*_i = 0,\qquad \int d\theta_i\, d\theta^*_i \,
 \theta^*_i\, \theta_i = 1.
 \label{3.4}
 \eeq

\bigskip

The momenta of the Grassmann variables, implying second-class
constraints, are

 \bea
 &&\pi_{\theta_i^{(Q)}}(\tau ) = {{\partial L}\over {\partial
 {\dot \theta}^{(Q)}_i}} = - {{i\, \hbar}\over 2}\,
 \theta^{(Q) *}_i(\tau ),\qquad
 \pi_{\theta^{(Q) *}_i}(\tau ) = {{\partial L}\over {\partial
 {\dot \theta}^{(Q) *}_i}}
 = - {{i\, \hbar}\over 2}\, \theta^{(Q)}_i(\tau ),\nonumber \\
 &&{}\nonumber \\
 &&\{ \theta^{(Q)}_i(\tau ), \pi_{\theta^{(Q)}_j}(\tau )\} = \{
 \theta^{(Q) *}_i(\tau ), \pi_{\theta^{(Q) *}_j}(\tau )\} = -
 \delta_{ij},\nonumber \\
 &&{}\nonumber \\
 &&\pi_{\theta_i}(\tau ) = {{\partial L}\over {\partial {\dot \theta}_i}}
 = - {{i\, \hbar}\over 2}\, \theta^*_i(\tau ),\qquad
 \pi_{\theta^*_i}(\tau ) = {{\partial L}\over {\partial {\dot \theta}^*_i}}
 = - {{i\, \hbar}\over 2}\, \theta_i(\tau ),\nonumber \\
 &&{}\nonumber \\
 &&\{ \theta_i(\tau ), \pi_{\theta_j}(\tau )\} = \{
 \theta^*_i(\tau ), \pi_{\theta^*_j}(\tau )\} = -
 \delta_{ij}.
 \label{3.5}
 \eea

By going to Dirac brackets for the  second-class constraints the
Grassmann momenta are eliminated and we get (to simplify the
notation we denote $\{.,.\}^*$ with $\{.,.\}$)

\bea
  &&\{ \theta^{(Q)}_i(\tau ), \theta^{(Q) *}_j(\tau )\} = -i\, \delta_{ij},\quad
 \{ \theta^{(Q)}_i(\tau ), \theta^{(Q)}_j(\tau )\} = \{ \theta^{(Q) *}_i(\tau ),
 \theta^{(Q) *}_j(\tau )\} = 0,\nonumber \\
 &&{}\nonumber \\
 &&\{ \theta_i(\tau ), \theta^*_j(\tau )\} = -i\, \delta_{ij},\quad
 \{ \theta_i(\tau ), \theta_j(\tau )\} = \{ \theta^*_i(\tau ),
 \theta^*_j(\tau )\} = 0.
 \label{3.6}
\end{eqnarray}

\bigskip

The primary constraints are

\bea
 &&\pi_{\varphi_{(a)}}(\tau ,\vec \sigma )\,
\approx\, 0,\qquad \pi_n(\tau ,\vec \sigma )\, \approx\, 0,\qquad
 \pi_{n_{(a)}}(\tau ,\vec \sigma
)\, \approx\, 0,\nonumber \\
 &&{}^3M_{(a)}(\tau ,\vec \sigma )
= \epsilon_{(a)(b)(c)}\, {}^3e_{(b)r} (\tau ,\vec \sigma )\,
{}^3\pi^r_{(c)}(\tau ,\vec \sigma )\, \approx\, 0,\nonumber \\
 &&{}\nonumber \\
 &&\pi^{\tau}(\tau ,\vec \sigma) \approx 0.
 \label{3.7}
\eea
\medskip

By evaluating the canonical Hamiltonian by Legendre transformation
(see Ref.\cite{3}) and by asking that the primary constraints are
constants of the motion under the $\tau$-evolution generated by it,
we get the following secondary constraints

\bea
 \Gamma (\tau ,\vec \sigma) &=& \partial_r\, \pi^r(\tau ,\vec \sigma
 ) + \sum_{i=1}^N\, Q_i\, \eta_i\, \delta^3(\vec \sigma, {\vec \eta}_i(\tau))
  \approx 0,\nonumber \\
 &&{}\nonumber \\
 {\cal H}(\tau ,\vec \sigma )&=&   \Big[{{c^3}\over {16\pi\,
G}}\,\,\, {}^3e\,\, \epsilon_{(a)(b)(c)} \, {}^3e^r_{(a)}\,
{}^3e^s_{(b)}\,{}^3\Omega_{rs(c)} -\nonumber \\
 &-&{{2\pi\, G}\over {c^3\, {}^3e}}\,\,\,
{}^3G_{o(a)(b)(c)(d)}\, {}^3e_{(a)r}\, {}^3\pi^r_{(b)}\,
{}^3e_{(c)s}\, {}^3\pi^s_{(d)}\Big] (\tau ,\vec \sigma
) +  {\cal M}(\tau ,\vec \sigma )\approx 0,\nonumber \\
 &&{}\nonumber \\
 {\cal H}_{(a)}(\tau ,\vec \sigma )&=&\Big[\partial_r\,
 {}^3\pi^r_{(a)} -\epsilon_{(a)(b)(c)}\, {}^3\omega_{r(b)}\, {}^3
\pi^r_{(c)} + {}^3e^r_{(a)} {\cal M}_r\Big](\tau ,\vec \sigma
)=\nonumber \\
 &=&-{}^3e^r_{(a)}(\tau ,\vec \sigma ) [{}^3
\Theta_r  - {}^3\omega_{r(b)} \, {}^3M_{(b)}](\tau ,\vec
\sigma )\approx 0,\nonumber \\
 &&{}\nonumber \\
 &\Downarrow& \nonumber \\
  &&{}\nonumber \\
{}^3\Theta_r(\tau ,\vec \sigma )&=&[{}^3\pi^s_{(a)}\partial _r\,
{}^3e_{(a)s}-\partial_s({}^3e_{(a)r}\, {}^3\pi^s_{(a)})] (\tau ,\vec
\sigma ) - {\cal M}_r(\tau ,\vec \sigma )\approx 0.
 \label{3.8}
\eea

\noindent In Eqs.(\ref{3.8}) the following notation has been
introduced for the matter terms (the mass density ${\cal M}(\tau,
\vec \sigma)$ and the mass current density ${\cal M}_r(\tau, \vec
\sigma)$) to conform with the treatment of the same matter in the
non-inertial rest frames of Minkowski space-time given in
Ref.\cite{10}

\bea
 {\cal M}(\tau ,\vec \sigma )&=& \sum_{i=1}^N\, \delta^3(\vec
\sigma ,{\vec \eta}_i(\tau ))\, M_i(\tau ,\vec \sigma )\, c +
{}^3e\, T_{\perp\perp}^{(em)}(\tau ,\vec \sigma),\nonumber \\
 M_i(\tau ,\vec \sigma )\, c &=& \eta_i\, \sqrt{m_i^2\, c^2 +
 {}^3e^r_{(a)}\, \Big(\kappa_{ir}(\tau ) - {{Q_i}\over c}\,
 A_r\Big)\, {}^3e^s_{(a)}\, \Big( \kappa_{is}(\tau ) -
 {{Q_i}\over c}\, A_s\Big)}(\tau ,\vec \sigma),\nonumber \\
 &&{}\nonumber \\
 {\cal M}_r(\tau ,\vec \sigma )&=& \sum_{i=1}^N\, \eta_i\,
 \Big(\kappa_{ir}(\tau ) - {{Q_i}\over c}\, A_r(\tau ,\vec \sigma)\Big)\, \delta^3(\vec
\sigma ,{\vec \eta}_i(\tau )) - {}^3e\, T_{\perp r}^{(em)}(\tau ,\vec \sigma),\nonumber \\
 &&{}\nonumber \\
 &&{}\nonumber \\
  T^{(em)}_{\perp\perp}(\tau ,\vec{\sigma})&=&
\frac{1}{2\, c\, {}^3e}\, \left(\frac{1}{{}^3e}\, {}^3e_{(a)r}\,
{}^3e_{(a)s}\, \pi^r\, \pi^s + \frac{{}^3e}{2}\, {}^3e^r_{(a)}\,
{}^3e^s_{(a)}\, {}^3e^u_{(b)}\, {}^3e^v_{(b)}\,
F_{ru}\, F_{sv}\right)(\tau ,\vec \sigma),\nonumber\\
&&\nonumber\\
T^{(em)}_{\perp r}(\tau ,\vec{\sigma})&=&  {1\over {c\, {}^3e}}\,
F_{rs}(\tau, \vec{\sigma})\, \pi^s(\tau, \vec{\sigma}).
 \label{3.9}
 \eea

\noindent Let us remark that the mass current density ${\cal
M}_r(\tau ,\vec \sigma)$ does not depend upon the 4-metric.

\bigskip

In Eqs.(\ref{3.8}) $\Gamma (\tau ,\vec \sigma) \approx 0$ is the
electro-magnetic Gauss law, ${\cal H}(\tau ,\vec \sigma ) \approx 0$
and ${\cal H}_{(a)}(\tau ,\vec \sigma ) \approx 0$ are the
super-Hamiltonian and super-momentum constraints respectively, while
${}^3\Theta(\tau ,\vec \sigma ) \approx 0$ are the generators of
3-diffeomorphisms on $\Sigma_{\tau}$. In the constraint ${\cal
H}(\tau ,\vec \sigma) \approx 0$ we have ${}^3e\,
\epsilon_{(a)(b)(c)}\, {}^3e^r_{(a)}\, {}^3e^s_{(b)}\,
\Omega_{rs(c)} = {}^3e\, {}^3R$ with ${}^3R$ the scalar 3-curvature
of the instantaneous 3-space $\Sigma_{\tau}$ and $\Omega_{rs(a)}$
the 3-field strength determined by the 3-spin connection ${}^3{\bar
\omega}_{r(a)}$ given in Eq.(\ref{2.10}).

\bigskip

One can check that the constraints are all first class with the
algebra (see Ref.\cite{35} for the terms containing the Gauss law
constraint)

\begin{eqnarray*}
\lbrace {}^3M_{(a)}(\tau ,\vec \sigma ),{}^3M_{(b)} (\tau ,{\vec
\sigma}^{'})\rbrace &=&\epsilon_{(a)(b)(c)}\, {}^3M_{(c)} (\tau
,\vec \sigma ) \delta^3(\vec \sigma ,{\vec \sigma}^{'}),\nonumber \\
 \lbrace {}^3M_{(a)}(\tau ,\vec \sigma ),{}^3\Theta_r (\tau
,{\vec \sigma}^{'})\rbrace &=&{}^3M_{(a)}(\tau ,{\vec \sigma}
^{'})\, {{\partial \delta^3(\vec \sigma ,{\vec \sigma}^{'})}\over
{\partial \sigma^r}},
 \end{eqnarray*}

\bea
 \lbrace {}^3\Theta_r(\tau ,\vec \sigma ),{}^3\Theta_s (\tau ,{\vec
 \sigma}^{'})\rbrace &=&\Big[{}^3\Theta_r(\tau ,{\vec \sigma} ^{'})
 {{\partial}\over {\partial \sigma^s}} + {}^3\Theta_s(\tau ,\vec
 \sigma ) {{\partial}\over {\partial \sigma^r}}\Big]\, \delta^3(\vec
 \sigma ,{\vec \sigma}^{'}) -\nonumber \\
 &-& \delta_{ru}\, \delta_{sv}\, \Big[{}^3e^u_{(a)}\, {}^3e^m_{(a)}\,
 {}^3e^v_{(b)}\, {}^3e^n_{(b)}\, F_{mn}\Big](\tau ,\vec \sigma)\,
 \Gamma(\tau ,\vec \sigma)\, \delta^3(\vec \sigma, {\vec
 \sigma}^{'}),\nonumber \\
 \lbrace {\cal H}(\tau ,\vec \sigma ),{}^3\Theta_r(\tau ,{\vec
 \sigma}^{'})\rbrace &=& {\cal H}(\tau ,{\vec \sigma}^{'})\, {{\partial
 \delta^3(\vec \sigma ,{\vec \sigma}^{'})}\over {\partial \sigma^r}} +
 \delta_{rs}\, {{\pi^s(\tau ,\vec \sigma)}\over {{}^3e(\tau ,\vec \sigma)}}\,
  \Gamma(\tau ,\vec \sigma)\, \delta^3(\vec \sigma, {\vec
 \sigma}^{'}), \nonumber \\
 \lbrace {\cal H}(\tau ,\vec \sigma ),{\cal
 H}(\tau ,{\vec \sigma}^{'})\rbrace &=& \Big[ {}^3e^r_{(a)}(\tau
 ,\vec \sigma )\, {\cal H} _{(a)}(\tau ,\vec \sigma ) +
  {}^3e^r_{(a)}(\tau ,{\vec \sigma}^{'})\, {\cal
 H}_{(a)}(\tau ,{\vec \sigma}^{'}) \Big] {{\partial \delta^3(\vec
 \sigma ,{\vec \sigma}^{'})}\over {\partial \sigma^r}} =\nonumber \\
 &=&\Big( \Big[{}^3e^r_{(a)}\, {}^3e^s_{(a)}\, [{}^3 \Theta_s +
 {}^3\omega_{s(b)} \, {}^3M_{(b)}]\Big](\tau ,\vec
 \sigma ) + \nonumber \\
 &+&\Big[{}^3e^r_{(a)}\, {}^3e^s_{(a)}\,
 [{}^3\Theta_s + {}^3\omega_{s(b)}\, {}^3M_{(b)}]\Big](\tau ,{\vec
 \sigma}^{'}) \Big) \, {{\partial \delta^3(\vec \sigma ,{\vec
 \sigma}^{'})}\over {\partial \sigma^r}}.
 \label{3.10}
\end{eqnarray}

\bigskip

The energy-momentum tensor of the matter, $T^{AB}(\tau ,\vec \sigma
) = - \Big[ {2\over {\sqrt{-{}^4g}}}\, {{\delta S_{part + em}}\over
{\delta\, {}^4g_{AB}}}\Big](\tau ,\vec \sigma )$, is \footnote{The
Hamilton equations imply ${}^4\nabla_A\, T^{AB} \equiv 0$ in accord
with Einstein's equations and the Bianchi identity.}

 \begin{eqnarray*}
 T^{\tau\tau}(\tau ,\vec \sigma ) &=& {{{\cal M}(\tau ,\vec \sigma )}\over
 {[{}^3e\, (1 + n)^2](\tau ,\vec \sigma )}},\nonumber \\
 T^{\tau r}(\tau ,\vec \sigma ) &=& {{{}^3e^r_{(a)}\, \Big[
 (1 + n)\, {}^3e^s_{(a)}\, {\cal M}_s - n_{(a)}\,
 {\cal M}\Big]}\over {{}^3e\, (1 + n)^2}}(\tau ,\vec \sigma ),
 \nonumber \\
 T^{rs}(\tau ,\vec \sigma ) &=& {{1}\over {[{}^3e\, (1 + n)^2](\tau ,\vec \sigma
 )}}\, \sum_{i=1}^N\, \delta^3(\vec \sigma,{\vec \eta}_i(\tau
 ))\qquad\qquad\qquad M_i\, c = \eta_i\, {\bar M}_i\, c,\nonumber \\
 &&\Big[{{\eta_i\, {}^3e^r_{(a)}\, {}^3e^s_{(b)}}\over
 {{\bar M}_i}}\, \Big((1 + n)\, {}^3e^m_{(a)}\, \Big(\kappa_{im}(\tau ) -
 {{Q_i}\over c}\, A_m \Big) - n_{(a)}\, {\bar M}_i\Big)\nonumber \\
 &&\Big((1 + n)\, {}^3e^n_{(b)}\, \Big(\kappa_{in}(\tau ) -
 {{Q_i}\over c}\, A_n \Big) - n_{(b)}\,
 {\bar M}_i\Big)\Big](\tau ,{\vec \eta}_i(\tau )) + T^{rs}_{(em)}(\tau
 ,\sigma),
 \end{eqnarray*}

  \begin{eqnarray*}
  T^{rs}_{(em)}(\tau ,\vec \sigma) &=&{1\over c}\, \Big[{1\over
  {{}^3e^2}}\, \Big(- \pi^r\, \pi^s + {1\over 2}\,
  \sum_{abcduv}\, {}^3e^r_{(a)}\, {}^3e^s_{(b)}\,
  \Big[\Big(\delta_{ab} + {{n_{(a)}\, n_{(b)}}\over {(1 + n)^2}}\Big)\,
  \delta_{cd} +\nonumber \\
  &+& 4\, {{n_{(a)}\, n_{(b)}\, n_{(c)}\, n_{(d)}}\over {(1 + n)^4}}\Big]\,
  {}^3e_{(c)u}\, {}^3e_{(d)v}\, \pi^u\, \pi^v\Big) +\nonumber \\
  &+& \sum_{ablm}\, {{{}^3e^r_{(a)}\, {}^3e^s_{(b)}}\over {{}^3e}}\, {{n_{(a)}\,
  {}^3e^l_{(b)} + n_{(b)}\, {}^3e^l_{(a)}}\over {1 + n}}\, F_{lm}\, \pi^m +\nonumber \\
  &+& \sum_{abcdlmuv}\, {}^3e^r_{(a)}\, {}^3e^s_{(b)}\, \Big(
  \Big[{}^3e^l_{(a)}\, {}^3e^m_{(b)} - {1\over 4}\, \Big(\delta_{ab} -
  {{n_{(a)}\, n_{(b)}}\over {(1 + n)^2}}\Big)\, \sum_e\,
  {}^3e^l_{(e)}\, {}^3e^m_{(e)}\Big]\, \delta_{cd} -\nonumber \\
  &-& {}^3e^m_{(a)}\, {}^3e^l_{(b)}\, {{n_{(c)}\, n_{(d)}}\over {(1 + n)^2}}
 \Big)\, {}^3e^u_{(c)}\, {}^3e^v_{(d)}\, F_{ul}\, F_{vm}
 \Big](\tau ,\sigma),\nonumber \\
 &&{}\nonumber \\
 &&\Downarrow \qquad {}^4\nabla_A\, T^{AB} \equiv 0,
 \end{eqnarray*}

\bea
  &&\Big[\partial_{\tau} - \sum_{ar}\, {}^3e^r_{(a)}\, n_{(a)}\,
 \partial_r - (1 + n)\, {}^3K\Big]\, (({}^3e)^{-1}\, {\cal M})
 +{{1 + n}\over {{}^3e}}\, \sum_{rs}\, \partial_s\, \Big[\,
 \sum_{ar}\, {}^3e^s_{(a)}\, {}^3e^r_{(a)}\, {\cal M}_r\Big]
 +\nonumber \\
 &&+ 2\, ({}^3e)^{-1}\, \sum_{ars}\, \partial_s n\, {}^3e^s_{(a)}\,
 {}^3e^r_{(a)}\, {\cal M}_r - (1 + n)\, \sum_{abrsuv}\, {}^3K_{rs}\, {}^3e^r_{(a)}\,
 {}^3e^s_{(b)}\, {}^3e^u_{(a)}\, {}^3e^v_{(b)}\,
 T_{uv} \equiv 0,\nonumber \\
 &&\nonumber\\
 &&{}\nonumber \\
 &&\Big[\partial_{\tau} - \sum_{as}\, {}^3e^s_{(a)}\, n_{(a)}\,
 \partial_s\Big]\, (({}^3e)^{-1}\, {\cal M}_r) + \partial_r\, n\, ({}^3e)^{-1}\,
 {\cal M} -\nonumber \\
 &&- \sum_{as}\, \partial_r\, ({}^3e^s_{(a)}\, n_{(a)})\, ({}^3e)^{-1}\,
 {\cal M}_s - (1 + n)\, {}^3K\, ({}^3e)^{-1}\, {\cal M}_r -
 \sum_{auv}\, \partial_u\, n\, {}^3e^u_{(a)}\,{}^3e^v_{(a)}\, T_{rv} +\nonumber \\
 &&+ (1 + n)\, \Big[{1\over {{}^3e}}\, \sum_{auv}\, \partial_u\,
 \Big({}^3e\, {}^3e^u_{(a)}\, {}^3e^v_{(a)}\, T_{rv}\Big) -\nonumber \\
 &&- {1\over 2}\, \sum_{abcuvmn}\, \partial_r\, ({}^3e_{(c)m}\, {}^3e_{(c)n})\,
 {}^3e^m_{(a)}\, {}^3e^n_{(b)}\, \, {}^3e^u_{(a)}\, {}^3e^v_{(b)}\,
 T_{uv}\Big] \equiv 0.\nonumber \\
 &&{}
 \label{3.11}
\eea

\noindent We have used the conventions  ${\cal M} = {}^3e\,
T_{\perp\perp} = {}^3e\, (1 + n)^2\, T^{\tau\tau}$, ${\cal M}_r = -
{}^3e\, T_{\perp r} = {}^3e\, (1 + n)\, \sum_a\, {}^3e_{(a)r}\,
\Big[\sum_s\, {}^3e_{(a)s}\, T^{\tau s} + n_{(a)}\,
T^{\tau\tau}\Big]$ ans $T^{AB} = T_{\perp\perp}\, l^A\, l^B - \sgn\,
T_{\perp r}\, {}^3g^{rs}\, (l^A\, \delta^B_s + l^B\, \delta^A_s) +
T_{rs}\, {}^3g^{ru}\, {}^3g^{sv}\, \delta^A_u\, \delta^B_v$ in
accord with the notations of Ref.\cite{10} \footnote{In the ADM
formalism \cite{a} one uses the decomposition $T^{AB} = E\, l^A\,
l^B + p^A\, l^B + p^B\, l^A + S^{AB}$ with $E= T^{AB}\, l_A\, l_B =
T_{\perp\perp} = ({}^3e)^{-1}\, {\cal M}$, $p^A = - {}^3h^A_B\,
T^{BC}\, l_C = \Big(0; - \sgn\, {\cal M}_r\Big)$, $S^{AB} =
{}^3h^A_C\, {}^3h^B_D\, T^{CD}$ ($S^{\tau\tau} = S^{\tau r} = 0$,
$S^{rs} = T^{rs} + \sum_a\, n_{(a)}\, (T^{\tau r}\, {}^3e^s_{(a)} +
T^{\tau s}\, {}^3e^r_{(a)}) + T^{\tau\tau}\, \sum_{ab}\, n_{(a)}\,
n_{(b)}\, {}^3e^r_{(a)}\, {}^3e^s_{(b)}$).}. To get the explicit
expression of ${}^4\nabla_A\, T^{AB} \equiv 0$ (in accord with
Ref.\cite{a}) we have used Eqs. (\ref{2.5}), (\ref{2.6}) and
(\ref{2.11})

\subsubsection{The Dirac Hamiltonian and the Weak ADM Energy }

In the rest-frame instant form of tetrad gravity \cite{3}, after
making the Legendre transformation and after the addition of the
DeWitt surface, the Dirac Hamiltonian has the following expression
in terms of the weak ADM energy  in the canonical basis (\ref{2.7})

\bea
 H_D&=& {1\over c}\, {\hat E}_{ADM} + \int d^3\sigma\, \Big[ n\,
{\cal H} - n_{(a)}\, {\cal H}_{(a)} - {1\over c}\, A_{\tau}\,
\Gamma\Big](\tau ,\vec \sigma )
+ \lambda_r(\tau )\, {\hat P}^r_{ADM} +\nonumber \\
 &+&\int d^3\sigma\, \Big[\lambda_n\, \pi_n + \lambda_{
n_{(a)}}\, \pi_{n_{(a)}} + \lambda_{\varphi_{(a)}}\,  \pi_{
\varphi_{(a)}} + \mu_{(a)}\, {}^3M_{(a)} + \mu\,
\pi^{\tau}\Big](\tau ,\vec \sigma ),
 \label{3.12}
 \eea

\noindent where $\lambda_n(\tau ,\vec \sigma )$, $\lambda_{n_{(a)}}
(\tau ,\vec \sigma )$, $\lambda_{\varphi_{(a)}}(\tau ,\vec \sigma
)$, $\mu_{(a)}(\tau ,\vec \sigma )$, $\lambda_r(\tau )$, are the
Dirac multipliers in front of the primary constraints and the mass
density ${\cal M}$ is defined in Eqs.(\ref{3.9}). Since we have
${\hat E}_{ADM} = c\, {\hat P}^{\tau}_{ADM}$, $[H_D] = [E/c] = [m\,
l\, t^{-1}]$ and $[\tau ] = [ l]$, Hamilton's equations are written
as $\partial_{\tau}\, F\, \cir\, \{ F, H_D \}$, where the notation
$\cir$ means "evaluated by using the equations of motion". To agree
with the standard ADM equations we need a minus sign in front of the
shift function.
\bigskip

Since the scalar 3-curvature ${}^3e\, {}^3R = {}^3e\,
\epsilon_{(a)(b)(c)}\, {}^3e^r_{(a)}\, {}^3e^s_{(b)}\,
\Omega_{rs(c)}$ of $\Sigma_{\tau}$ can be decomposed in the
following way \cite{3}

\bea
  {}^3\bar e(\tau ,\vec \sigma)\, {}^3R(\tau ,\vec \sigma)
  &{\buildrel {def}\over =}&\, {\cal S}(\tau ,\vec \sigma) + {\cal
 T}(\tau ,\vec \sigma),\nonumber \\
 &&{}\nonumber \\
 {\cal S}(\tau ,\vec \sigma ) &{\buildrel {def}\over =}& \Big[{}^3e\, \sum_{rsuv}\,
 {}^3e^r_{(a)}\, {}^3e^s_{(a)}\, \Big({}^3\Gamma^u_{rv}\, {}^3\Gamma^v_{su} -
 {}^3\Gamma^u_{rs}\, {}^3\Gamma^v_{uv}\Big)\Big](\tau ,\vec
 \sigma),\nonumber \\
 {\cal T}(\tau ,\vec \sigma) &{\buildrel {def}\over =}&
 \sum_r\, \partial_r\, \Big({}^3e\, \sum_{tuv}\, {}^3g^{rt}\,
{}^3g^{uv}\, (\partial_u\, {}^3g_{vt} - \partial_t\,
{}^3g_{uv})\Big),
 \label{3.13}
 \eea

\noindent it follows that the weak ADM energy of Eqs.(\ref{2.22})
and the super-Hamiltonian constraint of Eqs.(\ref{3.8}) can be
written in the following form

 \begin{eqnarray*}
 {\hat E}_{ADM} &=& c\, \int d^3\sigma\, \Big[{\cal M} -
  {{c^3}\over {16\pi\, G}}\, {\cal S} +\nonumber \\
 &+&  {{2\pi\, G}\over {c^3\,\, {}^3e}}\, \sum_{abcd}\,
 {}^3G_{o(a)(b)(c)(d)}\, {}^3e_{(a)r}\, {}^3\pi^r_{(b)}\, {}^3e_{(c)s}\,
 {}^3\pi^r_{(d)}\Big](\tau ,\vec \sigma ),\nonumber \\
 &&{}\nonumber \\
 {\cal H}(\tau ,\vec \sigma) &=& \Big[{\cal M} - {{c^3}\over
 {16\pi\, G}}\, ({\cal S} + {\cal T}) +\nonumber \\
 &+& {{2\pi\, G}\over {c^3\,\, {}^3e}}\, \sum_{abcd}\,
 {}^3G_{o(a)(b)(c)(d)}\, {}^3e_{(a)r}\, {}^3\pi^r_{(b)}\,
 {}^3e_{(c)s}\, {}^3\pi^s_{(d)} \Big](\tau ,\vec \sigma)
 \approx 0,\nonumber \\
 &&{}\nonumber \\
 &&\Downarrow\nonumber \\
 &&{}\nonumber \\
 \end{eqnarray*}

\bea
  {1\over c}\, {\hat E}_{ADM} &+& \int d^3\sigma\, \Big(  n\,
  {\cal H}\Big)(\tau ,\vec \sigma ) =  \nonumber \\
 &&{}\nonumber \\
 &&= \int d^3\sigma\, \Big[(1 + n)\, {\cal M}\Big](\tau ,\vec \sigma )
 -  {{c^3}\over {16\pi\, G}}\, \int d^3\sigma\, \Big[(1 + n)\, {\cal S}
 + n\, {\cal T}\,\, \Big] (\tau ,\vec \sigma ) +\nonumber \\
 &&+  {{2\pi\, G}\over {c^3}}\, \int d^3\sigma\, \Big[{1\over {{}^3e}}\, (1 + n)\,
 \sum_{abcd}\, {}^3G_{o(a)(b)(c)(d)}\,
 {}^3e_{(a)r}\, {}^3\pi^r_{(b)}\, {}^3e_{(c)s}\,
 {}^3\pi^r_{(d)}\Big](\tau ,\vec \sigma ).\nonumber \\
 &&{}
 \label{3.14}
\end{eqnarray}

\subsection{The Hamilton Equations for the Gravitational Field}

In the canonical basis (\ref{2.7}) the non-vanishing gravitational
variables are $\varphi_{(a)}$, $n$, $n_{(a)}$, ${}^3e_{(a)r}$ and
${}^3\pi^r_{(a)}$. The Hamilton equations for the first seven
variables are $\partial_{\tau}\, \varphi_{(a)}(\tau, \vec \sigma)\,
\cir\, \lambda_{\varphi_{(a)}}(\tau, \vec \sigma)$,
$\partial_{\tau}\, n(\tau, \vec \sigma)\, \cir\, \lambda_n(\tau,
\vec \sigma)$, $\partial_{\tau}\, n_{(a)}(\tau, \vec \sigma)\,
\cir\, \lambda_{n_{(a)}}(\tau, \vec \sigma)$. The Hamilton equations
for the cotriads ${}^3e_{(a)r}$ are given in Eq.(A10) of Appendix A
of Ref.\cite{6}, and with some effort one could evaluate the
Hamilton equations for the momenta ${}^3\pi^r_{(a)}$. Instead of
giving these Hamilton equations in the canonical basis (\ref{2.7}),
we will give the Hamilton equations for the quantities in the York
canonical basis (\ref{2.8}) in what follows.

\subsection{The Hamilton Equations for the Particles}

Due to the presence of the Grassmann-valued signs of the energy,
which are constants of motion like the Grassmann-valued electric
charges ($\{ \eta_i(\tau), H_D \} \cir 0$, $\{ Q_i(\tau), H_D \}
\cir 0$), the particle Hamilton equations are defined in the
following way

\bea
 \eta_i\, {\dot \eta}^r_i(\tau ) &\cir&  \{ \eta^r_i(\tau ), H_D\},\nonumber \\
 \eta_i\, {\dot \kappa}_{ir}(\tau ) &=& {\dot {\tilde \kappa}}_i(\tau)\,
 \cir\,  \{\kappa_{ir}(\tau ), H_D\}.
 \label{3.15}
 \eea

\bigskip

The derived particle world-lines are $x^{\mu}_i(\tau) = z^{\mu}(\tau
,{\vec \eta}_i(\tau))$. Looking at the expression of the momenta
$\kappa_{ir}(\tau )$, i=1,..,N, we can define the following derived
4-momenta, corresponding to the ones of the standard manifestly
covariant approach \cite{11,13}, which satisfy the mass shell
constraints for each particle also in the curved space-time $M^4$ in
presence of interactions (however, as shown in Ref.\cite{8}, other
definitions are possible for these derived quantities)

\bea
 p^{(\mu )}_i&=&\Big( p^{(o)}_i = \eta_i\, \sqrt{m^2_i\, c^2 +
{}^3e^r_{(a)}(\tau ,{\vec \eta}_i (\tau ))\, \kappa_{ir}(\tau )\,
{}^3e^s_{(a)}(\tau ,{\vec \eta}_i(\tau ))\, \kappa_{is}(\tau )};
\nonumber \\
&&p^{(a)}_i = {}^3e^s_{(a)}(\tau ,{\vec \eta}_i(\tau
))\, \kappa_{is}(\tau )\, \Big), \nonumber \\
 p^{\mu}_i&=&p_i^{(\alpha )}\,
{}^4E^{\mu}_{(\alpha )},\nonumber \\
 p^A_i&=& p^{(\alpha )}_i\, {}^4E^A_{(\alpha )}= \nonumber \\
 &=& \Big(p^{\tau}_i = {{p^{(o)}_i}\over
{1 + n(\tau ,{\vec \eta}_i(\tau ))}}; p^r_i = - ({{n_{(a)}\,
{}^3e_{(a)}^r\, p^{(o)}_i}\over {1 + n}})(\tau ,{\vec \eta}_i(\tau
)) + {}^3e^r_{(a)}(\tau, {\vec \eta}_i(\tau))\, p^{(a)}_i \Big), \nonumber \\
 &&{}\nonumber \\
 &&p^{\mu}_i\, {}^4g_{\mu\nu}\, p^{\nu}_i = p^{(\alpha )}_i\,
{}^4\eta_{(\alpha )(\beta )}\, p^{(\beta )}_i = p^A_i\, {}^4g_{AB}\,
p^B_i = m^2_i\, c^2.
 \label{3.16}
\eea

\bigskip

The Dirac Hamiltonian (\ref{3.12}) generates the following Hamilton
equations for the particles (see Eq.(\ref{3.3}) for the first one)

\bea
 && \eta_i\, {\dot \eta}_i^r(\tau )\,  =\, \eta_i\, {}^3e^r_{(a)}(\tau ,{\vec
 \eta}_i(\tau ))\, \Big[{{(1 + n)\, {}^3e^s_{(a)}\,
 \Big(\kappa_{is}(\tau ) - {{Q_i}\over c}\, A_r\Big)}\over
 {\sqrt{m^2_i\, c^2 + {}^3e^u_{(b)}\, {}^3e^v_{(b)}\,
 \Big(\kappa_{iu}(\tau ) - {{Q_i}\over c}\, A_u\Big)\,
 \Big(\kappa_{iv}(\tau ) - {{Q_i}\over c}\, A_v\Big)}}} -\nonumber \\
 &&- n_{(a)}\Big](\tau,{\vec \eta}_i(\tau )),\nonumber \\
 &&{}\nonumber \\
 &&\eta_i\, \frac{d}{d\tau}\kappa_{ir}(\tau )\cir\,\,  {{\eta_i\, Q_i}\over c}\,
 \Big[\dot{\eta}^u_i(\tau)\, \frac{\partial\, A_u(\tau ,{\vec \eta}_i(\tau
 ))}{\partial\eta_i^r} +  \frac{\partial\, A_\tau(\tau ,{\vec
 \eta}_i(\tau ))}{\partial \eta^r_i}\Big] + \eta_i\, {\cal
 F}_{ir}(\tau ),\nonumber\\
 &&{}\nonumber \\
 &&{}\nonumber\\
 &&{\cal F}_{ir}(\tau ) = \Big({ { m_i\, c\, }\over
 {\sqrt{\Big(1 + n\Big)^2 - {}^3e_{(a)u}\, {}^3e_{(a)v}\,
 \Big(\dot{\eta}^u_i(\tau ) + {}^3e^u_{(b)}\, n_{(b)}\Big)\,
 \Big(\dot{\eta}^v_i(\tau ) + {}^3e^v_{(c)}\, n_{(c)}\Big)}
 }}\Big)(\tau,{\vec \eta}_i(\tau ))\nonumber \\
 &&{}\nonumber \\
 &&\Big[{1\over 2}\,  \frac{\partial \, [{}^3e_{(a)s}\, {}^3e_{(a)t}](\tau ,{\vec \eta}_i(\tau
 ))}{\partial \eta^r_i}\,\,  \Big(\dot{\eta}^s_i(\tau ) +
 [{}^3e_{(b)}^s\, n_{(b)}](\tau ,{\vec \eta}_i(\tau ))\Big)\,
 \Big(\dot{\eta}^t_i(\tau ) + [{}^3e^t_{(c)}\, n_{(c)}](\tau ,{\vec
 \eta}_i(\tau ))\Big) -\nonumber \\
 &&{}\nonumber \\
 &-& (1 + n)\, \frac{\partial n(\tau
 ,{\vec \eta}_i(\tau ))}{\partial \eta^r_i} +
 \frac{\partial [{}^3e^s_{(a)}\, n_{(a)}](\tau ,{\vec \eta}_i(\tau
 ))}{\partial \eta^r_i}\, [{}^3e_{(b)s}\, {}^3e_{(b)t}](\tau ,{\vec
 \eta}_i(\tau ))\,\, \Big(\dot{\eta}^t_i(\tau ) + [{}^3e^t_{(c)}\,
 n_{(c)}](\tau,{\vec \eta}_i(\tau ))\Big) \,\,\Big],\nonumber \\
 &&{}
 \label{3.17}
 \eea

\noindent where ${\cal F}_{ir}(\tau )$ denotes a set of {\it
relativistic forces}, which in non-inertial frames of Minkowski
space-time would be {\it only inertial effects} \cite{10}.

\medskip

As a consequence of the first line of Eqs.(\ref{3.3}), the second
order form of the particle equations of motion implied by Eqs.
(\ref{3.17})  is

\begin{eqnarray*}
 &&\eta_i\, \frac{d}{d\tau}\left(
\frac{ {}^3e_{(a)r}\, {}^3e_{(a)s}\,\, m_i\, c\,
\Big(\dot{\eta}^s_i(\tau ) +
{}^3e^s_{(b)}\,n_{(b)}\Big)}{\sqrt{\Big(1 + n\Big)^2 -
{}^3e_{(c)u}\, {}^3e_{(c)v}\, \Big(\dot{\eta}^u_i(\tau ) +
{}^3e^u_{(d)}\, n_{(d)}\Big)\, \Big(\dot{\eta}^v_i(\tau )
+ {}^3e_{(e)}^v\, n_{(e)}\Big)}} \right)(\tau ,{\vec \eta}_i(\tau ))\,\cir\nonumber \\
 &&\cir\,\,  {{\eta_i\, Q_i}\over c}\, \left[\dot{\eta}^u_i(\tau )\,\left( \frac{\partial
A_u(\tau ,{\vec \eta}_i(\tau ))}{\partial\eta_i^r} - \frac{\partial
A_r(\tau ,{\vec \eta}_i(\tau ))}{\partial\eta_i^u}\right) + \left(
\frac{\partial A_\tau(\tau ,{\vec \eta}_i(\tau ))}{\partial
\eta^r_i} - \frac{\partial A_r(\tau ,{\vec \eta}_i(\tau
))}{\partial\tau}\right) \right] +\nonumber\\
&&\nonumber\\
&+&\eta_i\, {\cal F}_{ir}(\tau ),
 \end{eqnarray*}

\bea
 &&or\nonumber \\
 &&{}\nonumber \\
 &&\eta_i\, m_i\, c\, \frac{d}{d\tau}\left(
\frac{ \dot{\eta}^s_i(\tau ) + {}^3e^s_{(a)}\, n_{(a)}}{\sqrt{\Big(1
+ n\Big)^2 - {}^3e_{(b)u}\, {}^3e_{(b)v}\, \Big(\dot{\eta}^u_i(\tau
) + {}^3e^u_{(c)}\, n_{(c)}\Big)\, \Big(\dot{\eta}^v_i(\tau ) +
{}^3e^v_{(d)}\, n_{(d)}\Big)}} \right)(\tau
,{\vec \eta}_i(\tau))\,\cir\nonumber \\
 &&\cir\,\,  {{\eta_i\, Q_i}\over c}\, [{}^3e^s_{(a)}\,
 {}^3e^r_{(a)}](\tau ,{\vec \eta}_i(\tau ))\, \Big[
\dot{\eta}^u_i(\tau )\,\left( \frac{\partial A_u(\tau , {\vec
\eta}_i(\tau ))}{\partial\eta_i^r} - \frac{\partial A_r(\tau ,
{\vec \eta}_i(\tau ))}{\partial\eta_i^u}\right) +\nonumber \\
 &+& \left( \frac{\partial A_\tau(\tau ,{\vec \eta}_i(\tau
))}{\partial \eta^r_i} - \frac{\partial A_r(\tau ,{\vec \eta}_i(\tau
))}{\partial\tau}\right)\Big] + \eta_i\, {\tilde {\cal F}}^s_i(\tau ),\nonumber \\
 &&{}\nonumber \\
 &&{}\nonumber \\
 &&{\tilde {\cal F}}^s_i(\tau ) =\nonumber \\
 &=& \Big({{m_i\, c\,\,  {}^3e^s_{(a)}\,
 {}^3e^r_{(a)}}\over {\sqrt{\Big(1 + n\Big)^2 -
{}^3e_{(b)u}\, {}^3e_{(b)v}\, \Big(\dot{\eta}^u_i(\tau ) +
{}^3e^u_{(c)}\, n_{(c)}\Big)\, \Big(\dot{\eta}^v_i(\tau ) +
{}^3e^v_{(d)}\, n_{(d)}\Big)}}}\Big)(\tau ,{\vec
\eta}_i(\tau))\nonumber \\
 &&\Big[{1\over 2}\, \Big(\frac{\partial \, [{}^3e_{(e)s}\,
 {}^3e_{(e)t}](\tau ,{\vec \eta}_i(\tau ))}{\partial
\eta^r_i}\,\,  \Big(\dot{\eta}^s_i(\tau ) + [{}^3e^s_{(f)}\,
n_{(f)}](\tau ,{\vec \eta}_i(\tau ))\Big)\, \Big(\dot{\eta}^t_i(\tau
) + [{}^3e^t_{(g)}\, n_{(g)}](\tau ,{\vec \eta}_i(\tau ))\Big) -\nonumber \\
 &&{}\nonumber \\
 &-& (1 + n)\, \frac{\partial n(\tau, {\vec \eta}_i(\tau ))}{\partial\, \eta^r_i}+
 \nonumber \\
  &+&\frac{\partial [{}^3e^s_{(e)}\, n_{(e)}](\tau ,{\vec \eta}_i(\tau ))}{\partial\,
\eta^r_i}\, [{}^3e_{(f)s}\, {}^3e_{(f)t}](\tau ,{\vec \eta}_i(\tau
))\,\, \Big(\dot{\eta}^t_i(\tau ) + [{}^3e^t_{(g)}\, n_{(g)}](\tau
,{\vec \eta}_i(\tau ))\Big)\,\, \Big) -\nonumber \\
 &-& \Big({{\partial\, {}^3e_{(e)r}\, {}^3e_{(e)u}}\over {\partial\, \tau}} + {\dot
 \eta}^v_i(\tau )\, {{\partial\, {}^3e_{(e)r}\, {}^3e_{(e)u}}\over {\partial\,
 \eta^v_i}}\Big)(\tau ,{\vec \eta}_i(\tau ))\, \Big({\dot \eta}^u_i(\tau )
 + [{}^3e_{(f)}^u\, n_{(f)}](\tau ,{\vec \eta}_i(\tau ))\Big) \Big].
 \label{3.18}
 \eea
\medskip

Here ${\tilde {\cal F}}_{ir}(\tau )$ is the new form of the
relativistic forces.

\bigskip

If, as in Ref.\cite{10},  we {\it define} the non-inertial electric
and magnetic fields in the form

\bea
 E_r &\byd& \left(
\frac{\partial A_\tau}{\partial \eta^r_i}-\frac{\partial
A_r}{\partial\tau}\right)
= - F_{\tau r},\nonumber\\
 &&\nonumber\\
 B_r &\byd& \frac{1}{2}\, \varepsilon_{ruv}\,F_{uv} = \epsilon_{ruv}\,
 \partial_u\, A_{\perp\, v}\,\Rightarrow\, F_{uv} = \varepsilon_{uvr}\,B_r,
 \label{3.19}
 \eea

\noindent  the homogeneous Maxwell equations, allowing the
introduction of the electro-magnetic potentials, have the standard
inertial form $\epsilon_{ruv}\, \partial_u\,B_v = 0$,
$\epsilon_{ruv}\, \partial_u\,E_v + \frac{1}{c}\, \frac{\partial\,
B_r}{\partial \tau} = 0$.\medskip

Then also Eqs.(\ref{3.18}) take the following form

 \bea
 &&\eta_i\, \frac{d}{d\tau}\left( \frac{ {}^3e_{(a)r}\, {}^3e_{(a)s}\,\,
 m_ic\, \Big(\dot{\eta}^s_i(\tau ) + {}^3e^s_{(b)}\, n_{(b)}\Big)}
 {\sqrt{\Big(1 + n\Big)^2 - {}^3e_{(c)u}\, {}^3e_{(c)v}\,
 \Big(\dot{\eta}^u_i(\tau ) + {}^3e^u_{(d)}\, n_{(d)}\Big)\,
 \Big(\dot{\eta}^v_i(\tau ) + {}^3e^v_{(e)}\, n_{(e)}\Big)}}\right)(\tau ,{\vec
\eta}_i(\tau))  \cir \nonumber \\
 &&{}\nonumber \\
 &\cir& {{\eta_i\, Q_i}\over c}\, \left[E_r + \epsilon_{ruv}\, {\dot \eta}^u_i(\tau ) B_v\,
 \right](\tau ,{\vec \eta}_i(\tau )) + \eta_i\, {\cal F}_{ir}(\tau ).
 \label{3.20}
  \eea

\medskip

In the next Section we will give their expression in the York
canonical basis.

\subsection{The Hamilton Equations for  the
Electro-Magnetic Field}

The Hamilton-Dirac equations for the electro-magnetic field are

\bea
 \frac{\partial}{\partial \tau}\, A_\tau(\tau, \vec \sigma) &\cir&
c\, \mu(\tau, \vec \sigma),\nonumber \\
 \frac{\partial}{\partial
\tau}\, A_r(\tau ,\vec \sigma)&\cir& \Big(\frac{\partial}{\partial
\sigma^r}\, A_\tau + \frac{1 + n}{{}^3e}\, {}^3e_{(a)r}\,
{}^3e_{(a)s}\, \pi^s + {}^3e^s_{(a)}\, n_{(a)}\, F_{sr}\Big)(\tau ,\vec \sigma),\nonumber\\
&&\nonumber\\
\frac{\partial}{\partial \tau}\, \pi^r(\tau ,\vec \sigma ) &\cir&
\sum_i\, \eta_i\, Q_i\, \dot{\eta}_i^r(\tau )\, \delta^3(\vec
\sigma, {\vec \eta}_i(\tau )) +\nonumber\\
&+&\Big(\frac{\partial}{\partial \sigma^s}\, \Big[ (1 + n)\, {}^3e\,
{}^3e^r_{(a)}\, {}^3e^s_{(a)}\, {}^3e^u_{(b)}\, {}^3e^v_{(b)} \,
F_{uv} - n_{(a)}\, ({}^3e_{(a)}^s\, \pi^r - {}^3e_{(a)}^r\,
\pi^s)\Big]\Big)(\tau ,\vec \sigma ).\nonumber \\
 &&{}
 \label{3.21}
  \eea

\subsubsection{Maxwell Equations}

As in Ref.\cite{10}, Eqs.(\ref{3.21}) imply

 \bea
\pi^s(\tau, \vec \sigma)&=& -\left[- \frac{{}^3e}{1 + n}\,
{}^3e^s_{(a)}\, {}^3e^r_{(a)}\left( F_{\tau r} - {}^3e^v_{(a)}\,
n_{(a)}\, F_{vr}\right)\right](\tau, \vec \sigma) =\nonumber\\
 &&\nonumber\\
 &=& -\sqrt{- {}^4g(\tau, \vec \sigma)}\, {}^4g^{\tau
A}(\tau, \vec \sigma)\, {}^4g^{s B}(\tau, \vec \sigma)\,
F_{AB}(\tau, \vec \sigma).
 \label{3.22}
 \eea

If we introduce the charge density $\bar \rho$, the charge current
density ${\bar j}^r$ and the total charge $Q_{tot} = \sum_i\, Q_i$
on $\Sigma_{\tau}$

 \bea
  \overline{\rho}(\tau, \vec \sigma)&=&
\frac{1}{{}^3e(\tau, \vec \sigma)} \sum_{i=1}^N\, \eta_i\,
Q_i\,\delta^3(\vec \sigma, {\vec \eta}_i(\tau)),\nonumber\\
 &&\nonumber\\
  \overline{J}^r(\tau, \vec \sigma)&=&
\frac{1}{{}^3e(\tau, \vec \sigma)} \sum_{i=1}^N\, \eta_i\,
Q_i\,\dot{\eta}^r_i(\tau)\, \delta^3(\vec \sigma, {\vec
\eta}_i(\tau)),\nonumber \\
  &&{}\nonumber \\
  \Rightarrow&& Q_{tot} = \int d^3\sigma\, {}^3e(\tau, \vec \sigma)\,
\overline{\rho}(\tau, \vec \sigma) = \sum_{i=1}^N\, \eta_i\, Q_i,
 \label{3.23}
  \eea

\noindent then the last of Eqs.(\ref{3.22}) can be rewritten in form

 \bea
\frac{\partial}{\partial\, \sigma^r}\,\pi^r(\tau, \vec
\sigma)&\approx& - {}^3e(\tau, \vec \sigma)\,
\overline{\rho}(\tau, \vec \sigma),\nonumber\\
 &&\nonumber\\
  \frac{\partial\,\pi^r(\tau, \vec \sigma)} {\partial\, \tau} &\,\cir\,&
\frac{\partial}{\partial\, \sigma^s}\, \left[\sqrt{- {}^4g}\,
{}^3e^s_{(a)}\, {}^3e^v_{(a)}\, {}^3e^r_{(b)}\, {}^3e^u_{(b)}\,
F_{vu} - n_{(a)}\, ({}^3e_{(a)}^s\, \pi^r
- {}^3e_{(a)}^r\,\pi^s)\right](\tau, \vec \sigma) +\nonumber\\
 &&\nonumber\\
&+&{}^3e(\tau, \vec \sigma)\, \overline{J}^r(\tau, \vec \sigma).
 \label{3.24}
  \eea
\medskip

If we introduce the 4-current density  $s^A(\tau ,\vec \sigma)$

 \bea
s^\tau(\tau, \vec \sigma)&=& \frac{1}{\sqrt{- {}^4g(\tau, \vec
\sigma)}}\, \sum_{i=1}^N\, \eta_i\, Q_i\,\delta^3(\vec \sigma,
{\vec \eta}_i(\tau)),\nonumber\\
 &&\nonumber\\
  s^r(\tau, \vec \sigma)&=&
\frac{1}{\sqrt{- {}^4g(\tau, \vec \sigma)}}\, \sum_{i=1}^N\,
\eta_i\, Q_i\,\dot{\eta}^r_i(\tau)\, \delta^3(\vec \sigma, {\vec
\eta}_i(\tau)),
 \label{3.25}
  \eea

\noindent and  we use (\ref{3.22}), then Eqs.(\ref{3.24}) can be
rewritten as the standard manifestly covariant Maxwell equations in
general relativity

 \beq
\frac{1}{\sqrt{- {}^4g(\tau, \vec \sigma)}}\,
\frac{\partial}{\partial\, \sigma^A}\left[\sqrt{- {}^4g(\tau, \vec
\sigma)}\, {}^4g^{AB}(\tau, \vec \sigma)\,  {}^4g^{CD}(\tau, \vec
\sigma)\, F_{BD} (\tau, \vec \sigma) \right]\, \cir\, - s^C(\tau,
\vec \sigma).
 \label{3.26}
  \eeq

Eqs.(\ref{3.26}) imply the following continuity equation due to the
skew-symmetry of $F_{AB}$

 \bea
 &&\frac{1}{\sqrt{- {}^4g(\tau, \vec \sigma)}}\,
\frac{\partial}{\partial\, \sigma^C}\, \left[\sqrt{- {}^4g(\tau,
\vec \sigma)}\, s^C(\tau, \vec \sigma)\right]\, \cir\, 0,\nonumber \\
 &&{}\nonumber \\
 &&or\nonumber \\
 &&{}\nonumber \\
 &&\frac{1}{\sqrt{\gamma(\tau, \vec \sigma)}}\,
\frac{\partial}{\partial\, \tau}\, \left[\sqrt{\gamma(\tau, \vec
\sigma)}\, \overline{\rho}(\tau, \vec \sigma)\right] +
\frac{1}{{}^3e(\tau, \vec \sigma)}\, \frac{\partial}{\partial\,
\sigma^r}\, \left[{}^3e(\tau,
\vec \sigma)\,\overline{J}^r(\tau, \vec \sigma)\right]\, \cir\, 0, \nonumber \\
 &&{}
 \label{3.27}
  \eea
\medskip

\noindent so that consistently we recover $\frac{d}{d\tau}\,
Q_{tot}\, \cir\, 0$.

\subsubsection{The Radiation Gauge for the Electro-Magnetic Field in
Non-Inertial Frames in ADM Tetrad Gravity.}

In Ref.\cite{10} there is a general discussion about the
non-covariant decomposition of the vector potential $\vec A(\tau
,\sigma^u)$ and its conjugate momentum $\vec \pi (\tau ,\sigma^u)$
(the electric field) into longitudinal and transverse parts in
absence of matter. Only with this decomposition we can define a
Shanmugadhasan canonical transformation adapted to the two first
class constraints generating electro-magnetic gauge transformations
and identify the physical degrees of freedom (Dirac observables) of
the electro-magnetic field without sources. This method  identifies
the {\it radiation gauge} as the natural one from the point of view
of constraint theory. When there are charged particles,  this method
allows to find the expression of the mutual Coulomb interaction
among the charges in the admissible non-inertial frames of Minkowski
space-time. As said in Appendix B of the first paper of
Ref.\cite{10} the reduction to the radiation gauge has to be done
after fixing the 3+1 splitting of space-time and the associated
radar 4-coordinates, i.e. after fixing the gauge variables
$\theta^i(\tau, \vec \sigma)$ and ${}^3K(\tau, \vec \sigma)$. In the
next Section the radiation gauge will be used in the restriction of
the Hamilton equations to suitable gauges.
\medskip

Here we extend the results of Refs.\cite{10} to our class of
space-times.
\bigskip

If $\Delta = \sum_r\, \partial^2_r$ is the non-covariant flat
Laplacian, associated to the asymptotic Minkowski metric and acting
in the instantaneous non-Euclidean 3-space $\Sigma_{\tau}$, its
inverse defines the following non-covariant distribution

\beq
 {1\over {\Delta}}\,\, \delta^3(\vec \sigma, {\vec \sigma}^{'}) =
 c(\vec \sigma, {\vec \sigma}^{'}) = - \frac{1}{4\pi}\,\frac{1}{\sqrt{\sum_u\,
(\sigma^u - \sigma^{'\, u})^2} } {\buildrel {def}\over =} - {1\over
{4\pi\, |\vec \sigma - {\vec \sigma}^{'}|}},
 \label{3.28}
 \eeq

\noindent with $\delta^3(\vec \sigma, {\vec \sigma}^{'})$ being the
delta function for $\Sigma_{\tau}$.
\medskip

Then we can define the following non-covariant decomposition of the
vector potential and its conjugate momentum ($\Gamma (\tau ,\vec
\sigma ) \approx 0$ is the Gauss law of Eqs.(\ref{3.8}); $\Delta\,
\eta_{em}(\tau ,\vec \sigma) = \delta^{rs}\, \partial_r\,
 A_s(\tau ,\vec \sigma)$; $\eta_{em}$  describes a Coulomb cloud of
longitudinal photons, see Ref.\cite{36})

\bea
  A_r(\tau ,\vec \sigma) &=& A_{\perp\,r}(\tau ,\vec \sigma) - \partial_r\,
 \eta_{em}(\tau ,\vec \sigma),\nonumber \\
  \pi^r(\tau ,\vec \sigma) &=& \pi^r_\perp(\tau ,\vec \sigma) +
  \delta^{rs}\, \partial_s\, \int d^3\sigma'\, c(\vec \sigma,
  {\vec \sigma}^{'})\, \Big(\Gamma(\tau ,{\vec \sigma}^{'}) - \sum_{i=1}^N\, Q_i\,
 \eta_i\, \delta^3({\vec \sigma}^{'}, {\vec \eta}_i(\tau ))\Big),\nonumber \\
 &&{}\nonumber \\
 \eta_{em}(\tau ,\vec \sigma) &=& - \int d^3\sigma^{'}\, c(\vec
 \sigma, {\vec \sigma}^{'})\, \Big(\delta^{rs}\, \partial_r\,
 A_s(\tau ,{\vec \sigma}^{'})\Big),\qquad \{\eta_{em}(\tau ,\vec \sigma),
 \Gamma(\tau ,{\vec \sigma}^{'})\} = \delta^3(\vec \sigma, {\vec \sigma}^{'}),
 \nonumber \\
 &&{}\nonumber \\
  A_{\perp r}(\tau ,\vec \sigma ) &=&  \delta_{ru}\, P^{us}_{\perp}(
\vec \sigma)\,  A_s(\tau ,\vec \sigma ),\qquad
 \pi^r_\perp(\tau ,\vec \sigma) = \sum_s\, P^{rs}_{\perp}(\vec
 \sigma)\, \pi^s(\tau ,\vec \sigma),
 \label{3.29}
 \eea

\noindent where we introduced the projector $P^{rs}_{\perp}(\vec
\sigma ) = \delta^{rs} - \delta^{ru}\, \delta^{sv}\, {{\partial_u\,
\partial_v}\over {\Delta}}$.

\bigskip

If we introduce the following new Coulomb-dressed momenta for the
particles

 \bea
\check{\kappa}_{ir}(\tau ) &=& \kappa_{ir}(\tau ) + {{Q_i}\over c}\,
\frac{\partial}{\partial\eta^r_i}\, \eta_{em}(\tau
,{\vec \eta}_i(\tau )),\nonumber \\
 &&{}\nonumber \\
 \Rightarrow && \kappa_{ir}(\tau ) - {{Q_i}\over c}\, A_r(\tau ,{\vec \eta}_i(\tau )) =
\check{\kappa}_{ir}(\tau ) - {{Q_i}\over c}\, A_{\perp\,r}(\tau
,{\vec \eta}_i(\tau ))
 \label{3.30}
 \eea

\noindent we arrive at the following non-covariant Shanmugadhasan
canonical transformation

\bea
 &&\begin{array}{|c|c|}
\hline
&\\
A_r(\tau ,\vec \sigma)&\eta^r_i(\tau )\\
&\\
\pi^r(\tau ,\vec \sigma)&\kappa_{ir}(\tau )\\
&\\
\hline
\end{array}
\mapsto
\begin{array}{|cc|c|}
\hline
&&\\
A_{\perp\,r}(\tau ,\vec \sigma)&\eta_{em}(\tau ,\vec \sigma)&\eta^r_i(\tau )\\
&&\\
\pi^r_\perp(\tau ,\vec \sigma)&\Gamma(\tau ,\vec \sigma)
\approx 0&\check{\kappa}_{ir}(\tau )\\
&&\\
\hline
\end{array},\nonumber \\
 &&{}\nonumber \\
 &&\{A_{\perp r}(\tau ,\vec \sigma), \pi_\perp^s(\tau ,{\vec \sigma}^{'})\}
=  c\, \delta_{ru}\,  P^{us}_{\perp}(\vec \sigma )\, \delta^3(\vec
\sigma, {\vec
\sigma}^{'}),\nonumber\\
 &&\nonumber\\
 &&\{\eta^r_i(\tau ), \check{\kappa}_{is}(\tau )\} =  \delta^r_s\,\delta_{ij}.
 \label{3.31}
  \eea
\bigskip

The {\it non-covariant radiation gauge} is defined by adding the
gauge fixing $\eta_{em}(\tau ,\vec \sigma) \approx 0$. As shown in
Ref.\cite{10}, the $\tau$-constancy, $\frac{\partial\eta_{em}(\tau,
\sigma^u)}{\partial\tau} = \{\eta_{em}(\tau ,\vec \sigma ) ,H_D\}
\approx 0$, of this gauge fixing implies the secondary gauge fixing
for the primary constraint $\pi^{\tau}(\tau ,\vec \sigma) \approx 0$

 \bea
&&A_\tau(\tau ,\vec \sigma) \approx  - \int d^3\sigma'\, c(\vec
\sigma, {\vec \sigma}^{'})\,\,
\frac{\partial}{\partial\sigma^{\prime\,r}}\,
\Big[[{}^3e^s_{(a)}\, n_{(a)}](\tau ,{\vec \sigma}^{'})\, F_{sr}(\tau ,{\vec \sigma}^{'}) +\nonumber\\
 &+&\frac{\Big(1 + n(\tau ,{\vec \sigma}^{'})\Big)\,
 [{}^3e_{(a)r}\, {}^3e_{(a)s}](\tau
,{\vec \sigma}^{'})}{{}^3e(\tau ,{\vec \sigma}^{'})}\,
\left(\pi_\perp^s(\tau ,{\vec \sigma}^{'}) - \delta^{sn}\,
\sum_{j=1}^N\, Q_j\, \eta_j\, {{\partial\, c(\vec \sigma, {\vec
\eta}_j(\tau))}\over {\partial\, \sigma^{{'}\, n}}} \right) \Big).\nonumber \\
 &&{}
 \label{3.32}
\eea

\medskip

If we eliminate the electro-magnetic variables $A_{\tau}$,
$\pi^{\tau}$, $\eta_{em}$, $\Gamma$ by going to Dirac Brackets
(still denoted $\{.,.\}$ for simplicity), we remain with only the
transverse fields $A_{\perp\, r}$ and $\pi^r_{\perp}$ ($F_{rs} =
\partial_r\, A_{\perp s} - \partial_s\, A_{\perp r}$).

\medskip

Let us remark that in the radiation gauge the non-inertial magnetic
field of Eqs.(\ref{3.19}) is transverse: $B_r = \epsilon_{ruv}\,
\partial_u\, A_{\perp\, v}$. But the non-inertial electric field $E_r = - F_{\tau
r} = - \partial_{\tau}\, A_{\perp\, r} + \partial_r\, A_{\tau}$ is
not transverse: it has $E_{\perp\, r} = - \partial_{\tau}\,
A_{\perp\, r}$ as a transverse component. Instead the transverse
quantity is $\pi^r_{\perp}$ (it coincides with $\delta^{rs}\,
E_{\perp\, s}$ only in the inertial frames of Minkowski space-time),
whose expression in terms of the electric and magnetic fields is $
\pi^r_{\perp}(\tau ,\vec \sigma) = \Big[{{ {}^3e}\over {1 + n}}\,
{}^3e^r_{(a)}\, {}^3e^s_{(a)}\, (E_s - \epsilon_{suv}\, n^u\,
B_v)\Big](\tau , \vec \sigma) - \delta^{rs}\, \sum_i\, Q_i\,
\eta_i\, \partial_s\, c(\vec \sigma, {\vec \eta}_i(\tau))$.

\bigskip

The electromagnetic part of the Hamiltonian ({3.12}) can be
expressed in terms of the new canonical variables of
Eq.(\ref{3.31}), since from Ref.\cite{10} we have ($n^r =
{}^3e^r_{(a)}\, n_{(a)}$)

\begin{eqnarray}
&&\int d^3\sigma\, \Big({}^3e\, \left[(1 + n)\,
T^{(em)}_{\perp\perp} + n^r\,
T^{(em)}_{\perp r}\right]\Big)(\tau ,\vec \sigma ) = \nonumber\\
&&\nonumber\\
&=&\int d^3\sigma\,  \Big[(1 + n)\, \Big({}^3e\,
\check{T}^{(em)}_{\perp\perp} +  {\cal W}_{(n)}\Big) + n^r\,
\Big({}^3e\, \check{T}^{(em)}_{\perp r} + {\cal W}_r\Big)\Big](\tau,
\vec \sigma),
 \label{3.33}
\end{eqnarray}

\noindent where the new energy-momentum tensor has the form

 \bea
 {}^3e(\tau ,\vec \sigma)\, \check{T}^{(em)}_{\perp\perp}(\tau ,\vec \sigma)&=&
 \Big[\frac{{}^3e_{(a)r}\, {}^3e_{(a)s}}{2\, c\, {}^3e}\, \pi_\perp^r  \, \pi_\perp^s
 + \frac{{}^3e}{4c}\, {}^3e^r_{(a)}\, {}^3e^s_{(a)}\, {}^3e^u_{(b)}\, {}^3e^v_{(b)}\,
F_{ru}\, F_{sv}\Big](\tau ,\vec{\sigma}),\nonumber\\
 &&{}\nonumber \\
 &&{}\nonumber \\
 {}^3e(\tau ,\vec \sigma)\, \check{T}^{(em)}_{\perp r}(\tau ,\vec \sigma) &=& {1\over c}\,
  F_{rs}(\tau, \vec{\sigma})\, \pi_\perp^s(\tau, \vec{\sigma}).
 \label{3.34}
 \eea

\medskip

In Eq.(\ref{3.33}) we rewrote the {\it non-inertial Coulomb
potential} ${\cal W}(\tau)$ (a function of the particle
3-coordinates $\eta^r_i(\tau)$), found in Ref.\cite{9}, in the form

\bea
 {\cal W}(\tau) &=& \int d^3\sigma\,  \Big[(1 +
 n)\, {\cal W}_{(n)} + n^r\, {\cal W}_r\Big](\tau ,\vec
 \sigma),\nonumber \\
 &&{}\nonumber \\
 &&{}\nonumber \\
 {\cal W}_{(n)}(\tau ,\vec \sigma) &=& - \Big[\frac{{}^3e_{(a)r}\, {}^3e_{(a)s}}
 {2\, c\, {}^3e}\, \left(2\, \pi_\perp^r - \delta^{rm}\, \sum_{i=1}^N\, Q_i\, \eta_i\,
 {{\partial\, c(\vec \sigma, {\vec \eta}_i(\tau))}\over {\partial\, \sigma^m}}
 \right)\nonumber \\
 && \delta^{sn}\, \sum_{j=1}^N\, Q_j\, \eta_j\, {{\partial\, c(\vec \sigma,
 {\vec \eta}_j(\tau))}\over {\partial\, \sigma^n}}
 \Big](\tau ,\vec \sigma),\nonumber \\
 &&{}\nonumber \\
 {\cal W}_r(\tau ,\vec \sigma) &=&  - {1\over c}\,
 F_{rs}(\tau ,\vec \sigma)\, \delta^{sn}\, \sum_{i=1}^N\, Q_i\, \eta_i\,
 {{\partial\, c(\vec \sigma, {\vec \eta}_i(\tau))}\over {\partial\, \sigma^n}}.
 \label{3.35}
 \eea

\noindent  In Minkowski space-time  its limit to  inertial frames is
the standard Coulomb potential $\sum_{i \not= j}^{1..N}\, {{Q_i\,
Q_j\, \eta_i\, \eta_j}\over {4\pi |{\vec \eta}_i(\tau) - {\vec
\eta}_j(\tau)|}}$. Let us remark that now this potential depends not
only on the particle positions, but also on the non-inertial
electric and magnetic fields, on the 3-metric on $\Sigma_{\tau}$ and
on the lapse and shift functions.\bigskip

The energy-momentum tensor (\ref{3.11}) can be written in the
radiation gauge by means of the substitutions $A_r\, \mapsto\,
A_{\perp\, r}$ and $\pi^r\, \mapsto\, \pi^r_{\perp} - \sum_i\,
\eta_i\, Q_i\, {{\partial\, c(\vec \sigma, {\vec \eta}_i(\tau))
}\over {\partial\, \sigma^r}}$, $\kappa_{ir}\, \mapsto\, {\check
\kappa}_{ir}$.

\subsubsection{The Dirac Hamiltonian and the Hamilton Equations in
the Radiation Gauge}

After the elimination of the variables $\eta_{em}$, $\Gamma$,
$A_{\tau}$, $\pi^{\tau}$ by going to Dirac brackets, the final form
of the Dirac Hamiltonian (\ref{3.12}) in the radiation gauge is

\bea
 H_D&=& {1\over c}\, {\hat E}_{ADM} + \int d^3\sigma\, \Big[ n\,
{\cal H} - n_{(a)}\, {\cal H}_{(a)}\Big](\tau ,\vec
\sigma ) + \lambda_r(\tau )\, {\hat P}^r_{ADM} +\nonumber \\
 &+&\int d^3\sigma\, \Big[\lambda_n\, \pi_n + \lambda_{
n_{(a)}}\, \pi_{n_{(a)}} + \lambda_{\varphi_{(a)}}\,  \pi_{
\varphi_{(a)}} + \mu_{(a)}\, {}^3M_{(a)}\Big](\tau ,\vec \sigma ),
 \label{3.36}
 \eea

\noindent with ${\hat E}_{ADM}$, ${\cal H}(\tau ,\vec \sigma)$ and
${\cal H}_{(a)}(\tau ,\vec \sigma)$ given by Eqs.(\ref{3.8}) and
(\ref{3.14}) but with ${\cal M}(\tau ,\vec \sigma)$ and ${\cal
M}_r(\tau ,\vec \sigma)$ replaced with the following quantities
(they are already given in the York canonical basis (\ref{2.8}))

\begin{eqnarray*}
 {\check {\cal M}}(\tau ,\vec \sigma )&=& \sum_{i=1}^N\, \delta^3(\vec
\sigma ,{\vec \eta}_i(\tau ))\, {\check M}_i(\tau ,\vec \sigma )\, c
+ \Big({}^3e\, {\check T}_{\perp\perp}^{(em)}\Big)(\tau ,\vec
\sigma) + {\cal W}_{(n)}(\tau ,\vec \sigma ),\nonumber \\
 {\check M}_i(\tau ,\vec \sigma )\, c &=& \eta_i\, \sqrt{m_i^2\, c^2 +
 {\tilde \phi}^{- 2/3}\, \sum_{ars}\, Q_a^{-2}\, V_{ra}\, V_{sa}\,
 \Big({\check \kappa}_{ir}(\tau ) - {{Q_i}\over c}\,
 A_{\perp\, r}\Big)\,  \Big({\check \kappa}_{is}(\tau ) -
 {{Q_i}\over c}\, A_{\perp \, s}\Big)}(\tau ,\vec \sigma),\nonumber \\
 &&{}\nonumber \\
 {\check {\cal M}}_r(\tau ,\vec \sigma )&=& \sum_{i=1}^N\, \eta_i\,
 \Big({\check \kappa}_{ir}(\tau ) - {{Q_i}\over c}\,
 A_{\perp r}(\tau ,\vec \sigma)\Big)\, \delta^3(\vec
\sigma ,{\vec \eta}_i(\tau )) - \Big({}^3e\, {\check T}_{\perp
r}^{(em)}\Big)(\tau ,\vec \sigma) - {\cal W}_r(\tau ,\vec \sigma),
 \end{eqnarray*}

\begin{eqnarray*}
  \Big(\tilde \phi\, {\check T}^{(em)}_{\perp\perp}\Big)(\tau ,\vec \sigma) &=& {\tilde
  \phi}^{-1/3}(\tau ,\vec \sigma)\, \Big[{1\over {2c}}\, \sum_{rsa}\, Q_a^2\,
  V_{ra}\, V_{sa}\, \pi^r_{\perp}\,  \pi^s_{\perp} +\nonumber \\
   &+& {1\over {4c}}\, \sum_{abrsuv}\, Q_a^{-2}\, Q_b^{-2}\, V_{ra}\, V_{sa}\,
  V_{ub}\, V_{vb}\, F_{ru}\, F_{sv}\Big](\tau ,\vec
  \sigma),\nonumber \\
  {\cal W}_{(n)}(\tau ,\vec \sigma) &=& - {1\over {2c}}\,
  \Big[{\tilde \phi}^{-1/3}\, \sum_{rsa}\, Q_a^2\, V_{ra}\, V_{sa}\,
  \Big(2\, \pi^r_{\perp} - \delta^{rm}\, \sum_{i=1}^N\, Q_i\, \eta_i\, {{\partial\,
  c(\vec \sigma , {\vec \eta}_i(\tau))}\over {\partial\, \sigma^m}}\Big)
  \nonumber \\
  &&\delta^{sn}\,\sum_{j=1}^N\, Q_j\, \eta_j\, {{\partial\,
  c(\vec \sigma , {\vec \eta}_j(\tau))}\over {\partial\,
  \sigma^n}}\Big](\tau ,\vec \sigma),
  \end{eqnarray*}

\bea
   \Big(\tilde \phi\, \check{T}^{(em)}_{\perp r}\Big)(\tau ,\vec \sigma) &=& {1\over c}\,
  \sum_s\, F_{rs}(\tau, \vec{\sigma})\, \pi_\perp^s(\tau,
  \vec{\sigma}),\nonumber \\
  {\cal W}_r(\tau ,\vec \sigma) &=& - {1\over c}\, \sum_{sn}\,
 F_{rs}(\tau ,\vec \sigma)\, \delta^{sn}\, \sum_{i=1}^N\, Q_i\, \eta_i\,
 {{\partial\, c(\vec \sigma, {\vec \eta}_i(\tau))}\over {\partial\, \sigma^n}}.
 \nonumber \\
 &&{}
 \label{3.37}
 \eea

\noindent Let us remark that also ${\check {\cal M}}_r(\tau ,\vec
\sigma)$ does not depend upon the 4-metric like ${\cal M}_r(\tau
,\vec \sigma)$ of Eqs.(\ref{3.9}).

\bigskip

In the radiation gauge  the Hamilton-Dirac equations (\ref{3.17})
for the particles are replaced by the following equations

\bea
 \eta_i\, \dot{\eta}^r_i(\tau ) &\cir&\frac{\eta_i\, \Big(1 + n\Big)\,
{}^3e^r_{(a)}\, {}^3e^s_{(a)}\, \Big(\check{\kappa}_{is}(\tau ) -
{{Q_i}\over c}\, A_{\perp\,s}\Big)}{\sqrt{m_i^2\, c^2 + \,
{}^3e^u_{(b)}\, {}^3e_{(b)}^v\, \Big(\check{\kappa}_{iu}(\tau ) -
{{Q_i}\over c}\, A_{\perp\,u}\Big)\, \Big(\check{\kappa}_{iv}(\tau )
- {{Q_i}\over c}\, A_{\perp\,v}\Big)}}
 (\tau , {\vec \eta}_i(\tau ))-\nonumber \\
&-& \eta_i\, [{}^3e^r_{(a)}\, n_{(a)}](\tau ,\eta^u_i(\tau )),\nonumber \\
 &&{}\nonumber \\
 \eta_i\, \frac{d}{d\tau}\check{\kappa}_{ir}(\tau ) &\cir&
 {{\eta_i\, Q_i}\over c}\, \dot{\eta}^u_i(\tau )\,
 \frac{\partial\, A_{\perp\,u}(\tau ,{\vec \eta}_i(\tau ))}{\partial\, \eta_i^r} -
 {1\over c}\, \frac{\partial}{\partial \eta^r_i}{\cal
W}(\tau) + \eta_i\, {\cal F}_{ir}(\tau ).
 \label{3.38}
  \eea

\medskip

As already said, the quantity ${\cal W}(\tau )$ of Eq.(\ref{3.35})
is a functional of the 3-coordinates ${\vec \eta}_i(\tau)$ (and also
of the 4-metric and   the non-inertial electric and magnetic
fields), which replaces the two-body Coulomb potential of the
inertial frames of Minkowski space-time. In Ref.\cite{10} it is
shown that we can write $\eta_i\, Q_i\,E_r(\tau ,{\vec \eta}_i(\tau
)) = - \eta_i\, Q_i \frac{\partial A_{\perp r}(\tau ,{\vec
\eta}_i(\tau ))}{\partial\tau} + \eta_i\, Q_i\left(\frac{\partial
A_\tau(\tau ,\vec \sigma)} {\partial\sigma^r}\right)_{\vec \sigma =
{\vec \eta}_i} \approx - \eta_i\, Q_i \frac{\partial A_{\perp
r}(\tau ,{\vec \eta}_i(\tau ))}{\partial\tau} - \frac{\partial{\cal
W}(\tau)}{\partial\, \eta^r_i} = \eta_i\, Q_i\, E_{\perp r}(\tau
,{\vec \eta}_i(\tau )) - \frac{\partial{\cal W}(\tau)} {\partial\,
\eta^r_i}$.

\bigskip

The first of Eqs.(\ref{3.38}) can be inverted to get

\bea
 \eta_i\, \check{\kappa}_{ir}(\tau )& =& \eta_i\, m_i\, c\,
 \Big(\frac{{}^3e_{(a)r}\, {}^3e_{(a)s}\, \Big(\dot{\eta}^s_i(\tau )
 + {}^3e^s_{(b)}\, n_{(b)}\Big)}{\sqrt{\Big(1 + n\Big)^2 - {}^3e_{(c)u}\, {}^3e_{(c)v}\,
 \Big(\dot{\eta}^u_i(\tau ) + {}^3e^u_{(d)}\, n_{(d)}\Big)\, \Big(\dot{\eta}^v_i(\tau )
 + {}^3e^v_{(e)}\, n_{(e)}\Big)}}\Big)(\tau ,{\vec \eta}_i(\tau ))  + \nonumber \\
 &+& {{\eta_i\, Q_i}\over c}\, A_{\perp\,r}(\tau ,{\vec \eta}_i(\tau )).
 \label{3.39}
  \eea

\bigskip

In the radiation gauge the Hamilton equations for the transverse
electro-magnetic fields $A_{\perp\, r}$ and $\pi^r_{\perp}$ are

\begin{eqnarray*}
 \partial_{\tau}\, A_{\perp\, r}(\tau ,\vec \sigma ) &\cir& \{
 A_{\perp\, r}(\tau ,\vec \sigma), H_D \} =\nonumber \\
 &=& \sum_{nuva}\, \delta_{rn}\, P^{nu}_{\perp}(\vec \sigma)\, \Big[
 {{(1 + n)\, {}^3e_{(a)u}\, {}^3e_{(a)v}}\over {{}^3e}}\, \Big(\pi^v_{\perp} -
 \sum_m\, \delta^{vm}\, \sum_{i=1}^N\, Q_i\, \eta_i\, {{\partial\,
 c(\vec \sigma, {\vec \eta}_i(\tau))}\over {\partial\, \sigma^m}}
 \Big) +\nonumber \\
 &+&  {\bar n}_{(a)}\,  {}^3e^v_{(a)}\, F_{vu}
 \Big](\tau ,\vec \sigma),\nonumber \\
 &&{}\nonumber \\
 \partial_{\tau}\, \pi^r_{\perp}(\tau ,\vec \sigma) &\cir& \{
 \pi^r_{\perp}(\tau ,\vec \sigma), H_D \} =\nonumber \\
 &=& \sum_{wma}\, P^{rw}_{\perp}(\vec \sigma)\, \delta_{wm}\, \Big(
 \sum_{i=1}^N\, \eta_i\, Q_i\, \delta^3(\vec \sigma, {\vec
 \eta}_i(\tau))\, {}^3e^m_{(a)}(\tau ,{\vec \eta}_i(\tau))
 \nonumber \\
 &&\Big[{{(1 + n)\, \sum_s\, {}^3e^s_{(a)}\, {\check \kappa}_{is}(\tau)}\over
 {\sqrt{m_i^2\, c^2 + \sum_{uvb}\,
 {}^3e^u_{(b)}\, \Big({\check \kappa}_{iu}(\tau ) - {{Q_i}\over c}\,
 A_{\perp\, u}\Big)\, {}^3e^v_{(b)}\, \Big({\check \kappa}_{iv}(\tau ) -
 {{Q_i}\over c}\, A_{\perp \, v}\Big)}}} -\nonumber \\
  &-& {\bar n}_{(a)}\Big](\tau ,{\vec \eta}_i(\tau)) +
 \end{eqnarray*}

  \bea
 &+&\Big[(1 + n)\, \sum_{svb}\, \Big({}^3e\, {}^3e^s_{(a)}\, {}^3e^v_{(b)}\,
 ({}^3e^n_{(a)}\, {}^3e^m_{(b)} - {}^3e^m_{(a)}\, {}^3e^n_{(b)})\,
 \partial_n\, F_{sv} +\nonumber \\
 &+& \partial_n\, \Big[{}^3e\,  {}^3e^s_{(a)}\, {}^3e^v_{(b)}\,
 ({}^3e^n_{(a)}\, {}^3e^m_{(b)} - {}^3e^m_{(a)}\, {}^3e^n_{(b)})\,\Big]\,
 F_{sv} \Big) +\nonumber \\
 &+& \partial_n\, n\, \sum_{svb}\, {}^3e\, {}^3e^s_{(a)}\, {}^3e^v_{(b)}\, ({}^3e^n_{(a)}\,
 {}^3e^m_{(b)} - {}^3e^m_{(a)}\, {}^3e^n_{(b)})\, F_{sv} +\nonumber \\
 &+& {\bar n}_{(a)}\, \Big({}^3e^n_{(a)}\, \partial_n\,
 \pi_{\perp}^m + \partial_n\, {}^3e^n_{(a)}\, \pi^m_{\perp} -
 \partial_n\, {}^3e^m_{(a)}\, \pi^n_{\perp} \nonumber \\
 &-& \sum_t\, (\partial_n\, {}^3e^n_{(a)}\, \delta^{mt} - \partial_n\,
 {}^3e^m_{(a)}\, \delta^{nt})\, \sum_{i=1}^N\, \eta_i\, Q_i\,
  {{\partial\, c(\vec \sigma, {\vec \eta}_i(\tau)))}\over {\partial\, \sigma^t}}
 -\nonumber \\
 &-& \sum_t\, ({}^3e^n_{(a)}\, \delta^{mt} - {}^3e^m_{(a)}\, \delta^{nt})\,
 \sum_{i=1}^N\, \eta_i\, Q_i\,  {{\partial^2\, c(\vec \sigma, {\vec \eta}_i(\tau)))}
 \over {\partial\, \sigma^t\, \partial\, \sigma^n}} \Big) -\nonumber \\
 &-& \partial_n\, {\bar n}_{(a)}\, \sum_t\, ({}^3e^n_{(a)}\, \delta^{mt} -
 {}^3e^m_{(a)}\, \delta^{nt})\, \sum_{i=1}^N\, \eta_i\, Q_i\,
  {{\partial\, c(\vec \sigma, {\vec \eta}_i(\tau)))}\over {\partial\, \sigma^t}}
 \Big](\tau ,\vec \sigma)\, \Big).
 \label{3.40}
 \eea

\subsection{The Final Form of the Constraints and of the Weak ADM Energy
in the York Canonical Basis.}

\subsubsection{The Super-Momentum Constraints.}

If ${\bar D}_{r(a)(b)} = \delta_{ab}\, \partial_r + {}^3{\bar
\omega}_{r(a)(b)} = \delta_{ab}\, \partial_r +
\epsilon_{(a)(b)(c)}\, {}^3{\bar \omega}_{r(c)}$ is the covariant
derivative defined in terms of the spin connection given in
Eq.(\ref{2.10}), the super-momentum constraint (\ref{3.8}), with
${\check {\cal M}}$ given in Eq.(\ref{3.37}), has the following
expression \cite{6} in the York canonical basis in the
electro-magnetic radiation gauge

\begin{eqnarray*}
 {\cal H}_{(a)} &=& \sum_c\, R_{(a)(c)}\, {\bar {\cal H}}_{(c)} \approx
 0,\nonumber \\
 &&{}\nonumber  \\
 &&{}\nonumber \\
 {\bar {\cal H}}_{(a)} &{\buildrel {def}\over =}& \sum_r\, \partial_r\, {}^3{\bar
 \pi}^r_{(a)} + \sum_{rb}\, {}^3{\bar \omega}_{r(a)(b)}\, {}^3{\bar
 \pi}^r_{(b)} + \sum_v\, {}^3{\bar e}^v_{(a)}\, {\check {\cal M}}_v
 \approx\nonumber \\
  &\approx& {\tilde {\bar {\cal H}}}_{(a)} \, {\buildrel {def}\over
 =}\,\,  \sum_{rb}\, {\bar D}_{r(a)(b)}\, {}^3{\tilde {\bar \pi}}^r_{(b)}
 + \sum_v\, {}^3{\bar e}^v_{(a)}\, {\check {\cal M}}_v =
 \end{eqnarray*}

\begin{eqnarray*}
 &=& \sum_{rb}\, \Big[\delta_{ab}\, \partial_r +
  \sum_u\,  \Big({1\over 3}\, \Big[Q_a\, Q_b^{-1}\,
  V_{ra}\, V_{ub} - Q_b\, Q_a^{-1}\, V_{rb}\,
  V_{ua}\Big]\,  {\tilde \phi}^{-1}\, \partial_u\, \tilde \phi +\nonumber \\
  &+& \sum_{\bar a}\, \Big[\gamma_{\bar aa}\, Q_a\, Q_b^{-1}\, V_{ra}\,
  V_{ub} - \gamma_{\bar ab}\, Q_b\, Q_a^{-1}\, V_{rb}\,
  V_{ua}\Big]\, \partial_u\, R_{\bar a} +\nonumber \\
  &+& {1\over 2}\, \Big[Q_b\, Q_a^{-1}\, V_{ua}\, (\partial_r\, V_{ub} - \partial_u\,
  V_{rb}) - Q_a\, Q_b^{-1}\,  V_{ub }\, (\partial_r\,
  V_{ua} - \partial_u\, V_{ra})\Big] +\nonumber \\
  &+&{1\over 2}\, \sum_{vw}\, Q_a^{-1}\, Q_b^{-1}\, Q_w^2\, (V_{ua}\,
  V_{vb} - V_{ub}\, V_{va})\,  V_{rw}\, \partial_v\,
  V_{uw}  \Big)   \,\,\,  \Big]
  \end{eqnarray*}

\bea
 &&\Big[{\tilde \phi}^{2/3}\, Q_b^{-1}\, V_{rb}\, \pi_{\tilde \phi} + {\tilde \phi}^{-1/3}\,
 \Big(Q^{-1}_b\, V_{rb}\,  \sum_{\bar c}\, \gamma_{\bar cb}\,
 \Pi_{\bar c} + \sum_f^{f \not= b}\, \sum_{twi}\, Q^{-1}_f\, {{V_{rf}\,
 \epsilon_{bft}\, V_{wt}\, B_{iw}}\over {Q_f\, Q^{-1}_b - Q_b\,
 Q^{-1}_f}}\, \pi^{(\theta )}_i \Big) \Big] +\nonumber \\
 &&{}\nonumber \\
 &+& {\tilde \phi}^{-1/3}\, Q_a^{-1}\, \sum_v\, V_{va}\, {\check {\cal M}}_v
 \quad \approx 0.
 \label{3.41}
 \eea

\noindent See Eq.(\ref{b1}) of Appendix B for its explicit
expression. In terms of the shear the super-momentum constraints
take the following form

\begin{eqnarray*}
  {\tilde {\bar {\cal H}}}_{(a)}(\tau ,\vec \sigma ) \, &{\buildrel {def}\over
 =}&\,\,  \Big(\sum_{rb}\, {\bar D}_{r(a)(b)}\, {}^3{\tilde {\bar \pi}}^r_{(b)}
 + \sum_v\, {}^3{\bar e}^v_{(a)}\, {\check {\cal M}}_v\Big)(\tau ,\vec \sigma ) =\nonumber \\
 &&{}\nonumber \\
 &=& \Big(\sum_{rb}\, \Big[\delta_{ab}\, \partial_r +
   \sum_u\, \Big( {1\over 3}\, \Big[Q_a\, Q_b^{-1}\,
  V_{ra}\, V_{ub} - Q_b\, Q_a^{-1}\, V_{rb}\,
  V_{ua}\Big]\,  {\tilde \phi}^{-1}\, \partial_u\,  \tilde \phi +\nonumber \\
  &+& \sum_{\bar a}\, \Big[\gamma_{\bar aa}\, Q_a\, Q_b^{-1}\, V_{ra}\,
  V_{ub} - \gamma_{\bar ab}\, Q_b\, Q_a^{-1}\, V_{rb}\,
  V_{ua}\Big]\, \partial_u\, R_{\bar a} +\nonumber \\
  &+& {1\over 2}\, \Big[Q_a\, Q_b^{-1}\, V_{ua}\, (\partial_r\, V_{ub} - \partial_u\,
  V_{rb}) - Q_a\, Q_b^{-1}\,  V_{ub }\, (\partial_r\,
  V_{ua} - \partial_u\, V_{ra})\Big] +\nonumber \\
  &+&{1\over 2}\, \sum_{vw}\, Q_a^{-1}\, Q_b^{-1}\, Q_w^2\, (V_{ua}\,
  V_{vb} - V_{ub}\, V_{va})\,  V_{rw}\, \partial_v\,
  V_{uw} \Big)  \,\,\,  \Big]
  \end{eqnarray*}

\bea
 && \Big[{\tilde \phi}^{-1/3}\, Q_b^{-1}\, V_{rb}\, \sum_{\bar c}\, \gamma_{\bar
 cb} \Pi_{\bar c} +
  {\tilde \phi}^{2/3}\, \Big(Q_b^{-1}\, V_{rb}\, \pi_{\tilde \phi} -
 {{c^3}\over {8\pi\, G}}\,  \sum_c^{c \not= b}\, Q^{-1}_c\, V_{rc}\,
 \sigma_{(b)(c)} \Big) \Big] +\nonumber \\
 &&{}\nonumber \\
 &+& {\tilde \phi}^{- 1/3}\,  Q_a^{-1}\, \sum_v\, V_{va}\,  {\check {\cal
 M}}_v\Big)(\tau ,\vec \sigma ) \quad \approx 0.
 \label{3.42}
 \eea

 \bigskip

In order to avoid the complications associated with the denominator
$(Q_f\, Q_b^{-1} - Q_b\, Q_f^{-1} )^{-1}$ appearing in
Eq.(\ref{3.41}), one has to solve Eq.(\ref{3.42}) for
$\sigma_{(a)(b)}{|}_{a \not= b}$ and then use Eq.(\ref{2.18}), i.e.
$\pi^{\theta}_i(\tau, \vec \sigma) = {{c^3}\over {8\pi\, G}}\,
\Big(\tilde \phi\, \sum_{wtab}\, A_{wi}\, V_{wt}\, Q_a\, Q_b^{-1}\,
\epsilon_{tab}\, \sigma_{(a)(b)} \Big)(\tau, \vec \sigma)$. Let us
remark that in Eq.(3.7) of Ref.\cite{6} (see its Appendix D for the
needed Green function) there is a formal solution of the
super-momentum constraints (one has only to add to it the matter
term) and a discussion of the emergence of a problem like the Gribov
ambiguity of Yang-Mills theory also in canonical gravity.

\bigskip

Let us remark that the functional derivatives of the super-momentum
constraint must be done with the expression (\ref{3.41}), before
expressing $\pi_i^{(\theta )}$ in terms of the shear $\tilde \phi\,
\sigma_{(a)(b)}{|}_{a \not= b}$.

\subsubsection{The Super-Hamiltonian Constraint and the Weak ADM
Energy}

The weak ADM energy of Eqs.(\ref{3.14}) has the following form in
the York canonical basis and in the electro-magnetic radiation gauge
with ${\check {\cal M}}$ given in Eq.(\ref{3.37})

\bea
  {\hat E}_{ADM} &=& c\, \int d^3\sigma\, \Big[{\check {\cal M}} -
  {{c^3}\over {16\pi\, G}}\, {\cal S} +
    {{2\pi\, G}\over {c^3}}\, {\tilde \phi}^{-1}\, \Big(
 - 3\, (\tilde \phi\, \pi_{\tilde \phi})^2 + 2\, \sum_{\bar b}\,
 \Pi^2_{\bar b} +\nonumber \\
 &+& 2\, \sum_{abtwiuvj}\, {{\epsilon_{abt}\, \epsilon_{abu}\, V_{wt}\,
 B_{iw}\, V_{vu}\, B_{jv}\, \pi_i^{(\theta )}\,
 \pi_j^{(\theta )}}\over {\Big[Q_a\, Q^{-1}_b - Q_b\,
 Q^{-1}_a \Big]^2}}\Big)\,\, \Big](\tau ,\vec \sigma ),\nonumber \\
 &&{}\nonumber \\
 {\cal S}  &=& \tilde \phi\, {}^3R - {\cal T} = {\tilde \phi}^{1/3}\,
 {}^3\hat R - 8\, {\tilde \phi}^{1/6}\, \hat \triangle\, {\tilde
 \phi}^{1/6} - {\cal T}.
 \label{3.43}
 \eea
\medskip

The functions ${\cal S}(\tau ,\vec \sigma)$ and ${\cal T}(\tau ,\vec
\sigma)$ are given in Eqs.(\ref{b8}) and (\ref{b13}), respectively

\medskip

In Eq.(\ref{3.43}) we have also given the expression \footnote{See
Ref.\cite{25} and Eq.(B4) of Ref. \cite{6}.} of the scalar
3-curvature ${}^3R = {}^3R[\phi, \theta^n, R_{\bar a}]$ in terms of
${}^3\hat R = {}^3\hat R[\theta^n, R_{\bar a}]$ and $\hat
\triangle$, which are the 3-curvature of $\Sigma_{\tau}$ and the
Laplace-Beltrami operator for the conformal 3-metric ${}^3{\hat
g}_{rs}$ ($det\, {\hat g}_{rs} = 1$), respectively, and are given in
Eqs.(\ref{b20}) and (\ref{b12}). This expression is needed to put
\cite{6} the super-Hamiltonian constraint in the form of the
Lichnerowitz equation for the conformal factor $\phi = {\tilde
\phi}^{1/6}$, which has the following expression in the York
canonical basis (${}^3{\hat g}^{rs}$ is the inverse of the 3-metric
with unit determinant)

\bigskip

\bea
 {\cal H}(\tau ,\vec \sigma )&=&
  {{c^3}\over {16\pi\, G}}\, {\tilde \phi}^{1/6} (\tau ,\vec \sigma ) \,
[8\,   {\hat \triangle}\, {\tilde \phi}^{1/6} - {\tilde
\phi}^{1/6}\, {}^3{\hat R}\,](\tau ,\vec \sigma ) +
{\check {\cal M}}(\tau ,\vec \sigma ) +\nonumber \\
 &&{}\nonumber \\
  &+&  {{2\pi\, G}\over {c^3}}\, {\tilde \phi}^{-1}\, \Big[
- 3\, (\tilde \phi\, \pi_{\tilde \phi})^2 + 2\, \sum_{\bar b}\,
 \Pi^2_{\bar b} +\nonumber \\
 &+& 2\, \sum_{abtwiuvj}\, {{\epsilon_{abt}\, \epsilon_{abu}\, V_{wt}\,
 B_{iw}\, V_{vu}\, B_{jv}\, \pi_i^{(\theta )}\,
 \pi_j^{(\theta )}}\over {\Big[Q_a\, Q^{-1}_b - Q_b\,
 Q^{-1}_a \Big]^2}}\Big](\tau ,\vec \sigma )
 \approx 0,\nonumber \\
 &&{}\nonumber \\
    \hat \triangle   &=& \partial_r\, ({}^3{\hat g}^{rs}\, \partial_s)
  = {}^3{\hat g}^{rs}\, {}^3{\hat \nabla}_r\, {}^3{\hat \nabla}_s =
   \partial_r\, \Big(\sum_a\, Q_a^{-2}\,
 V_{ra}\, V_{sa}\,  \partial_s\Big) =\nonumber \\
 &=& \sum_a\, Q_a^{-2}\, \Big[V_{ra}\, V_{sa}\, \partial_r\, \partial_s
 + \Big( 2\, V_{ra}\, V_{sa}\, \sum_{\bar b}\, \gamma_{\bar ba}\, \partial_r\, R_{\bar b}
 - \partial_r\, (V_{ra}\, V_{sa})\Big)\, \partial_s\Big].
 \label{3.44}
 \eea

\bigskip

In terms of the shear the super-Hamiltonian constraint and the weak
ADM energy take the following forms (again these forms cannot be
used for the functional derivatives of these quantities)

\begin{eqnarray*}
 {\cal H}(\tau ,\vec \sigma )&=&
  {{c^3}\over {2\pi\, G}}\, {\tilde \phi}^{-1}(\tau ,\vec \sigma )\, \Big(
 {\tilde \phi}^{7/6}\, (\hat \triangle - {1\over 8}\, {}^3\hat R)\,
 {\tilde \phi}^{1/6} +
 {{2\pi\, G}\over {c^3}}\, \tilde \phi\, {\check {\cal M}} +\nonumber \\
 &+& {{8\, \pi^2\, G^2}\over {c^6}}\, \sum_{\bar a}\, \Pi^2_{\bar a}
 + {1\over 8}\, {\tilde \phi}^{2}\, \sum_{ab, a\not= b}\, \sigma^2_{(a)(b)}
 - {{12\, \pi^2\, G^2}\over {c^6}}\, {\tilde \phi}^{2}\,
 \pi^2_{\tilde \phi} \Big)(\tau ,\vec \sigma ) \approx 0,
 \end{eqnarray*}

 \bea
 {\hat E}_{ADM} &=& c\, \int d^3\sigma\, \Big[{\check {\cal M}} -
  {{c^3}\over {16\pi\, G}}\, {\cal S} + {{4\pi\, G}\over {c^3}}\,
  {\tilde \phi}^{-1}\, \sum_{\bar b}\, \Pi^2_{\bar b} +\nonumber \\
  &+& \tilde \phi\, \Big( {{c^3}\over {16\pi\, G}}\,
 \sum_{ab, a\not= b}\, \sigma^2_{(a)(b)}
 - {{6\pi\, G}\over {c^3}}\, \pi^2_{\tilde \phi}\Big)
 \,\, \Big](\tau ,\vec \sigma ).
 \label{3.45}
 \eea

\bigskip

The super-momentum and super-Hamiltonian constraints are coupled
equations for $\tilde \phi$ and $\pi_i^{(\theta )}$ (or $\tilde
\phi\, \sigma_{(a)(b)}{|}_{a \not= b}$). In the form given in
Eqs.(\ref{3.42}) and (\ref{3.45}) the constraints have to be solved
to get the conformal factor $\tilde \phi$ and the off-diagonal terms
of the shear $\sigma_{(a)(b)}{|}_{a \not= b}$.\medskip

Let us remark that Eqs.(\ref{2.18}) for ${}^3K_{rs}$ and
Eq.(\ref{3.43}) for ${}^3R$ imply that the super-Hamiltonian
constraint can be written in the standard ADM form ${}^3R -
{}^3g^{rs}\, {}^3g^{uv}\, {}^3K_{ru}\, {}^3K_{sv} + ({}^3K)^2\,
\approx\, {{16\pi\, G}\over {c^3}}\, {\tilde \phi}^{-1}\, {\cal M}$
with ${\tilde \phi}^{-1}\, {\cal M} = (1 + n)^2\, T^{\tau\tau} =
T_{\perp\perp}$ from Eq.(\ref{3.11}).

\subsubsection{The Dirac Hamiltonian in the Radiation Gauge}

The Dirac Hamiltonian (\ref{3.12}) takes the following form when
written in the York canonical basis (\ref{2.8}) and restricted to
the electro-magnetic radiation gauge

\bea
  H_D&=& {1\over c}\, {\hat E}_{ADM} + \int d^3\sigma\, \Big[ n\,
{\cal H} - n_{(a)}\, {\cal H}_{(a)}\Big](\tau ,\vec \sigma )
+ \lambda_r(\tau )\, {\hat P}^r_{ADM} +\nonumber \\
 &+&\int d^3\sigma\, \Big[\lambda_n\, \pi_n + \lambda_{
{\bar n}_{(a)}}\, \pi_{{\bar n}_{(a)}} + \lambda_{\varphi_{(a)}}\,
\pi_{ \varphi_{(a)}} + \lambda_{\alpha_(a)}\,
\pi^{(\alpha)}_{(a)}\Big](\tau ,\vec \sigma ),
 \label{3.46}
 \eea

\noindent where $\lambda_{\alpha_{(a)}}(\tau, \vec \sigma)$ are new
Dirac multipliers replacing the $\mu_{(a)}(\tau, \vec \sigma)$
appearing in Eq.(\ref{3.12}).

\subsection{The Asymptotic ADM Poincare' Algebra, the Rest-Frame Conditions
and the Center of Mass in the York Canonical basis}

While the weak ADM energy is given in Eq.(\ref{3.43}), the other
weak ADM generators of Eqs.(\ref{2.22}) have the following form in
the York canonical basis

\begin{eqnarray*}
 {\hat P}^r_{ADM} &=& -2\int d^3\sigma\,\Big\{
-\tilde{\phi}^{-2/3}\sum_{a,v,\bar{b}}\,Q_a^{-2}\Big[
V_{ra}V_{va}\Big(\,\frac{1}{3}\gamma_{\bar{b}a}\tilde{\phi}^{-1}\partial_v\tilde{\phi}+
\sum_{\bar{a}}(\gamma_{\bar{b}a}\gamma_{\bar{a}a}-(1/2)\delta_{\bar{a}\bar{b}})\,\partial_vR_{\bar{a}}\Big)
 -\nonumber\\
&&\nonumber\\
&&-\frac{1}{2}\gamma_{\bar{b}a}\,\partial_v(V_{ra}V_{va})\Big]\,\Pi_{\bar{b}} +\nonumber\\
&&\nonumber\\
&&+\tilde{\phi}^{1/3}\sum_{a,v}\,Q_a^{-2}\Big[
V_{ra}V_{va}\Big(\,\frac{2}{3}\tilde{\phi}^{-1}\partial_v\tilde{\phi}+
\sum_{\bar{a}}\gamma_{\bar{a}a}\,\partial_vR_{\bar{a}}\Big)
-\frac{1}{4}\partial_v(V_{ra}V_{va})\Big]\,\pi_{\tilde{\phi}} +\nonumber\\
&&\nonumber\\
&&+\tilde{\phi}^{-2/3}\sum_{a,d,v}\,Q_a^{-1}Q_d^{-1}\Big[
V_{ra}V_{vd}\Big(\,\frac{1}{3}\tilde{\phi}^{-1}\partial_v\tilde{\phi}+
\sum_{\bar{a}}\gamma_{\bar{a}a}\,\partial_vR_{\bar{a}}\Big)
-\frac{1}{2}V_{rd}\partial_v V_{va} +\nonumber\\
&&+\frac{1}{2}\,\sum_{c,u}Q_c^{-2}Q_a^{2}V_{rc}V_{uc}V_{vd}(\partial_vV_{ua}-\partial_uV_{va})
\Big]\sum_{twi}\frac{\epsilon_{adt}V_{wt}B_{iw}\pi^{(\theta_i)}}{Q_dQ_a^{-1}-Q_aQ_d^{-1}}
-\nonumber\\
&&\nonumber\\
&&-\frac{1}{2}\tilde{\phi}^{-2/3}\sum_{a,s}\,Q_a^{-2}\,V_{ra}V_{sa}\,{\cal
M}_s\,\Big\}  \approx 0,
 \end{eqnarray*}

\begin{eqnarray*}
 {\hat J}^{rs}_{ADM} &=&\int d^3\sigma\,\Big\{\,
\sigma^r\,\Big[
-\tilde{\phi}^{-2/3}\sum_{a,v,\bar{b}}\,Q_a^{-2}\Big[
V_{sa}V_{va}\Big(\,\frac{1}{3}\gamma_{\bar{b}a}\tilde{\phi}^{-1}\partial_v\tilde{\phi}+
\sum_{\bar{a}}(\gamma_{\bar{b}a}\gamma_{\bar{a}a}-(1/2)\delta_{\bar{a}\bar{b}})\,\partial_vR_{\bar{a}}\Big)
-\nonumber\\
&&\nonumber\\
&&-\frac{1}{2}\gamma_{\bar{b}a}\,\partial_v(V_{sa}V_{va})\Big]\,\Pi_{\bar{b}} +\nonumber\\
&&\nonumber\\
&&+\tilde{\phi}^{1/3}\sum_{a,v}\,Q_a^{-2}\Big[
V_{sa}V_{va}\Big(\,\frac{2}{3}\tilde{\phi}^{-1}\partial_v\tilde{\phi}+
\sum_{\bar{a}}\gamma_{\bar{a}a}\,\partial_vR_{\bar{a}}\Big)
-\frac{1}{4}\partial_v(V_{sa}V_{va})\Big]\,\pi_{\tilde{\phi}} +\nonumber\\
&&\nonumber\\
&&+\tilde{\phi}^{-2/3}\sum_{a,d,v}\,Q_a^{-1}Q_d^{-1}\Big[
V_{sa}V_{vd}\Big(\,\frac{1}{3}\tilde{\phi}^{-1}\partial_v\tilde{\phi}+
\sum_{\bar{a}}\gamma_{\bar{a}a}\,\partial_vR_{\bar{a}}\Big)
-\frac{1}{2}V_{sd}\partial_v V_{va} +\nonumber\\
&&+\frac{1}{2}\,\sum_{c,u}Q_c^{-2}Q_a^{2}V_{sc}V_{uc}V_{vd}(\partial_vV_{ua}-\partial_uV_{va})
\Big]\sum_{twi}\frac{\epsilon_{adt}V_{wt}B_{iw}\pi^{(\theta_i)}}{Q_dQ_a^{-1}-Q_aQ_d^{-1}}
-\nonumber\\
&&\nonumber\\
&&-\frac{1}{2}\tilde{\phi}^{-2/3}\sum_{a,u}\,Q_a^{-2}\,V_{sa}V_{ua}\,{\cal
M}_u\,\Big]
-\nonumber\\
&&\nonumber\\
&&-\sigma^s\,\Big[
-\tilde{\phi}^{-2/3}\sum_{a,v,\bar{b}}\,Q_a^{-2}\Big[
V_{ra}V_{va}\Big(\,\frac{1}{3}\gamma_{\bar{b}a}\tilde{\phi}^{-1}\partial_v\tilde{\phi}+
\sum_{\bar{a}}(\gamma_{\bar{b}a}\gamma_{\bar{a}a}-(1/2)\delta_{\bar{a}\bar{b}})\,\partial_vR_{\bar{a}}\Big)
-\nonumber\\
&&\nonumber\\
&&-\frac{1}{2}\gamma_{\bar{b}a}\,\partial_v(V_{sa}V_{va})\Big]\,\Pi_{\bar{b}}
+\end{eqnarray*}

\begin{eqnarray*}
&&+\tilde{\phi}^{1/3}\sum_{a,v}\,Q_a^{-2}\Big[
V_{ra}V_{va}\Big(\,\frac{2}{3}\tilde{\phi}^{-1}\partial_v\tilde{\phi}+
\sum_{\bar{a}}\gamma_{\bar{a}a}\,\partial_vR_{\bar{a}}\Big)
-\frac{1}{4}\partial_v(V_{sa}V_{va})\Big]\,\pi_{\tilde{\phi}} +\nonumber\\
&&\nonumber\\
&&+\tilde{\phi}^{-2/3}\sum_{a,d,v}\,Q_a^{-1}Q_d^{-1}\Big[
V_{ra}V_{vd}\Big(\,\frac{1}{3}\tilde{\phi}^{-1}\partial_v\tilde{\phi}+
\sum_{\bar{a}}\gamma_{\bar{a}a}\,\partial_vR_{\bar{a}}\Big)
-\frac{1}{2}V_{rd}\partial_v V_{va} +\nonumber\\
&&+\frac{1}{2}\,\sum_{c,u}Q_c^{-2}Q_a^{2}V_{rc}V_{uc}V_{vd}(\partial_vV_{ua}-\partial_uV_{va})
\Big]\sum_{twi}\frac{\epsilon_{adt}V_{wt}B_{iw}\pi^{(\theta_i)}}{Q_dQ_a^{-1}-Q_aQ_d^{-1}}
-\nonumber\\
&&\nonumber\\
&&-\frac{1}{2}\tilde{\phi}^{-2/3}\sum_{a,u}\,Q_a^{-2}\,V_{ra}V_{ua}\,{\cal
M}_u\,\Big]\, \Big\},
 \end{eqnarray*}

\bea
 {\hat J}^{\tau r}_{ADM} &=& - \int d^3\sigma\,\Big\{
\sigma^r\,\Big[ \frac{c^3}{16\pi G}\,{\cal S}-{\cal M}-\frac{2\pi
G}{c^3}\tilde{\phi}^{-1}
\Big(-3(\tilde{\phi}\pi_{\tilde{\phi}})^2+2\sum_{\bar{b}}\,\Pi_{\bar{b}}^2
+\nonumber\\
&&\nonumber\\
&&+ 2\sum_{abtwiuvj} \frac{
\epsilon_{abt}\epsilon_{abu}V_{wt}B_{iw}V_{vu}B_{jv}\pi^{(\theta)}_i\pi^{(\theta)}_j
}{\Big[Q_aQ_b^{-1}-Q_bQ_a^{-1}\Big]^2}\Big)\,\Big] +\nonumber\\
&&\nonumber\\
&&+\frac{c^3}{16\pi G}
\tilde{\phi}^{1/3}\sum_{a,v}\,\Big\{Q_a^{-2}\Big[
V_{ra}V_{va}\Big(\,-\frac{2}{3}\tilde{\phi}^{-1}\partial_v\tilde{\phi}+
\sum_{\bar{a}}\gamma_{\bar{a}a}\,\partial_vR_{\bar{a}}\Big)
+\partial_v(V_{ra}V_{va})\Big] +\nonumber\\
&&\nonumber\\
&+&\Big[
V_{ra}V_{va}\Big(\,-\frac{1}{3}\tilde{\phi}^{-1}\partial_v\tilde{\phi}+
4\sum_{\bar{a}}\gamma_{\bar{a}a}\,Q_a^{-4}\partial_vR_{\bar{a}}+
\sum_{\bar{a},c}(\gamma_{\bar{a}c}-2\gamma_{\bar{a}a})\,Q_a^{-2}Q_c^{-2}
\partial_vR_{\bar{a}}\Big) +\nonumber\\
&&\nonumber\\
&&+(Q_a^{-4}-Q_a^{-2}\sum_c\,Q_c^{-2})\partial_v(V_{ra}V_{va})\Big]\Big\}\,\Big\}
  \approx 0.
 \label{3.47}
 \eea

\bigskip

The rest-frame conditions ${\hat P}^r_{ADM} \approx 0$ and the
conditions ${\hat J}^{\tau r}_{ADM} \approx 0$ eliminate the
internal 3-center of mass of the 3-universe $\Sigma_{\tau}$. The
gauge fixings ${\hat J}^{\tau r}_{ADM} \approx 0$ imply
$\lambda_r(\tau ) = 0$ in $H_D$ of Eq.(\ref{3.36}), because their
time preservation implies $\partial_{\tau}\, {\hat J}^{\tau
r}_{ADM}\, \cir\, \{ {\hat J}^{\tau r}_{ADM}, H_D \} = {\hat
P}^r_{ADM} - \lambda_r(\tau )\, {\hat E}_{ADM} \approx -
\lambda_r(\tau )\, {\hat E}_{ADM} \approx 0$ \footnote{The
asymptotic Poincare' charges are assumed gauge invariant, i.e. $\{
{\hat J}^{\tau r}_{ADM}, {\cal H}(\tau ,\vec \sigma ) \} \approx \{
{\hat J}^{\tau r}_{ADM}, {\tilde {\bar {\cal H}}}_{(a)}(\tau ,\vec
\sigma ) \} \approx 0$.}. In this way we identify the Fokker-Pryce
center of inertia of the 3-universe as the origin of the
3-coordinates in the instantaneous 3-spaces $\Sigma_{\tau}$: it
replaces $x^{\mu}(\tau)$ in Eq.(\ref{1.1}).
\medskip

\subsection{The Dirac Hamiltonian in the Schwinger Time Gauges and in the
Electro-Magnetic Radiation Gauge}

In what follows we restrict ourselves to the Schwinger time gauges

\beq
 \alpha_{(a)}(\tau ,\vec \sigma ) \approx 0,\qquad
 \varphi_{(a)}(\tau ,\vec \sigma ) \approx 0,
 \label{s}
 \eeq

\noindent whose $\tau$-preservation implies
$\lambda_{\varphi_{(a)}}(\tau ,\vec \sigma ) =
\lambda_{\alpha_{(a)}}(\tau ,\vec \sigma ) = 0$ in Eq.(\ref{3.46}).
The following results are obtained after the elimination of the
variables $\alpha_{(a)}$, $\pi^{(\alpha)}_{(a)}$ and
$\varphi_{(a)}$, $\pi_{\varphi_{(a)}}$ with Dirac brackets.

\bigskip

By using Eqs.(\ref{3.44}) and (\ref{b1}) for the super-Hamiltonian
and super-momentum constraints with ${\check {\cal M}}$ and ${\check
{\cal M}}_r$ given in Eqs.(\ref{3.37}), the Dirac Hamiltonian
(\ref{3.36}) in Schwinger time gauges  and in the electro-magnetic
radiation gauge is

\begin{eqnarray*}
 H_D&=& {1\over c}\, {\hat E}_{ADM} + \int d^3\sigma\, \Big[ n\, {\cal H}
 - {\bar n}_{(a)}\, {\tilde {\bar {\cal H}}}_{(a)}\Big](\tau ,\vec
\sigma )  + \int d^3\sigma\, \Big[\lambda_n\, \pi_n + \lambda_{{\bar
n}_{(a)}}\, \pi_{{\bar n}_{(a)}}\Big](\tau ,\vec \sigma )
=\nonumber \\
 &&{}\nonumber \\
 &=& \int d^3\sigma\, \Big[(1 + n)\, {\check {\cal M}}\Big](\tau ,\vec
 \sigma ) - {{c^3}\over {16\pi\, G}}\, \int d^3\sigma\, \Big[(1 + n)\,
 {\cal S} + n\, {\cal T} \Big](\tau ,\vec \sigma ) +\nonumber \\
 &+& {{2\pi\, G}\over {c^3}}\, \int d^3\sigma\, \Big[
 (1 + n)\, {\tilde \phi}^{-1}\, \Big(- 3\, (\tilde \phi\, \pi_{\tilde \phi})^2
 + 2\, \sum_{\bar b}\, \Pi^2_{\bar b} +\nonumber \\
 &+& 2\, \sum_{abtwuv}^{a\not= b}\, {{\epsilon_{abt}\, \epsilon_{abu}\, V_{wt}\,
 V_{vu}}\over {\Big[Q_a\, Q^{-1}_b - Q_b\,
 Q^{-1}_a \Big]^2}}\, \sum_i\, B_{iw}\, \pi_i^{(\theta
 )}\, \sum_j\, B_{jv}\, \pi_j^{(\theta )}
 \Big) \Big](\tau ,\vec \sigma ) -
 \end{eqnarray*}

 \bea
 &-& \int\, d^3\sigma\, \sum_a\, {\bar n}_{(a)}(\tau ,\vec \sigma )\,\,
 {\tilde \phi}^{-1/3}(\tau, \vec \sigma)\, \Big(\sum_{b \not= a}\,
 \sum_{rtwi}\, {{ \epsilon_{abt}\, Q_b^{-1}\, V_{rb}\, V_{wt}\, B_{iw}}
 \over {Q_b\, Q_a^{-1} - Q_a\, Q_b^{-1}}}\, \partial_r\, \pi_i^{(\theta)}
 +\nonumber \\
 &+& \sum_{ritw}\, \Big( \sum_{b \not= a}\, {{ \epsilon_{abt}}
 \over {Q_b\, Q_a^{-1} - Q_a\, Q_b^{-1}}}\,
  \Big[Q_b^{-1}\, \partial_r\, (V_{rb}\, V_{wt}\, B_{iw}) +\nonumber \\
  &+& {{2\, Q_a^{-1}}\over {Q_b\, Q_a^{-1} - Q_a\, Q_b^{-1}}}\,
  \sum_{\bar c}\, (\gamma_{\bar ca} - \gamma_{\bar cb})\, \partial_r\,
  R_{\bar c}\, V_{rb}\, V_{tw}\, B_{iw}\Big] +\nonumber \\
 &+&  \sum_{bu}\,  \sum_{c \not= b}\, {{\epsilon_{bct}\,
 Q_a^{-1}\, Q_b\, Q_c^{-1}}\over {Q_c\, Q_b^{-1} - Q_b\,
 Q_c^{-1}}}\,  (V_{rc}\, V_{ua} - V_{ra}\, V_{uc})\, \partial_r\, V_{ub}
 \,\,\, V_{wt}\, B_{iw}\Big)\,  \pi_i^{(\theta)} +\nonumber \\
 &&{}\nonumber \\
 &+&Q_a^{-1}\, \sum_r\, V_{ra}\, \Big(\tilde \phi\, \partial_r\, \pi_{\tilde \phi}
 +  \sum_{\bar b}\, \gamma_{\bar ba}\, \partial_r\, \Pi_{\bar b}\Big)
 +\nonumber \\
 &+&  Q_a^{-1}\, \sum_{r\bar b}\, \Big(\gamma_{\bar ba}\,
 \partial_r\, V_{ra} -  V_{ra}\, \partial_r\, R_{\bar b} +
  \sum_{ub}\, \gamma_{\bar bb}\,  V_{ua}\, V_{rb}\, \partial_r\, V_{ub}\Big)\,
 \Pi_{\bar b} +\nonumber \\
 &&{}\nonumber \\
 &+&  Q_a^{-1}\, \sum_r\, V_{ra}\, {\check {\cal M}}_r
 \Big)(\tau ,\vec \sigma) +\nonumber \\
 &&{}\nonumber \\
 &+& \int d^3\sigma\, \Big[ \lambda_n\, \pi_n + \sum_a\,
 \lambda_{\vec n(a)}\, \pi_{\vec n(a)} \Big](\tau ,\vec \sigma ),
 \label{3.48}
 \eea
\medskip

\noindent where we have used the last expression in Eq.(\ref{b1})
for the super-momentum constraints.

\medskip

$H_D$ depends not only upon the tidal variables $R_{\bar a}$,
$\Pi_{\bar a}$ and the matter, but also upon the gauge variables
$n$, ${\bar n}_{(a)}$, $\theta^n$, $\pi_{\tilde \phi}$, which play
the role of {\it inertial potentials}.

\bigskip

Let us remark that in the Hamiltonian (\ref{3.48}), and in
particular in the weak ADM energy (\ref{3.43}), the kinetic term $-
{{6\pi\, G}\over {c^3}}\, \int d^3\sigma\, \Big[\tilde \phi\, (1 +
n)\, \pi^2_{\tilde \phi}\Big](\tau ,\vec \sigma) = - {{c^3}\over
{24\pi\, G}}\, \int d^3\sigma\, \Big[\tilde \phi\, (1 + n)\,
\Big({}^3K\Big)^2\Big](\tau ,\vec \sigma)$ connected to the momentum
gauge variable $\pi_{\tilde \phi}(\tau ,\vec \sigma) = {{c^3}\over
{12\pi\, G}}\, {}^3K(\tau ,\vec \sigma)$, determining the
instantaneous 3-space, is {\it definite negative} in every gauge.
Therefore it plays the role of a {\it dark energy} and it vanishes
only in the CMC gauges ${}^3K(\tau ,\vec \sigma) \approx 0$.
Instead, the kinetic term connected to the momenta
$\pi_i^{(\theta)}(\tau ,\vec \sigma)$ (or by the off-diagonal terms
of the shear $\sigma_{(a)(b)}{|}_{a \not= b}(\tau ,\vec \sigma)$),
determined by the super-momentum constraints, is always positive
definite due to Eq.(\ref{3.45}).

\vfill\eject

\section{The Equations of Motion in Schwinger Time Gauges.}

In this Section we shall write explicitly the Hamilton equations for
the gravitational and matter variables, generated by the Dirac
Hamiltonian (\ref{3.48}), in an arbitrary Schwinger time gauge and
in the York canonical basis.\medskip

Let us remark that in the evaluation of the Hamilton equations the
constraints can be used only after having done the Poisson brackets
with $H_D$. When, like in Section V, some gauge fixings will be
added, we shall restrict the Hamilton equations of this Section to
the chosen gauge and we will not restrict the Dirac Hamiltonian to
the gauge and evaluate the new Hamilton equations with the new
Hamiltonian (even if the two approaches should be equivalent).
\medskip

By comparison in Appendix A there are the standard ADM equations of
canonical metric gravity and a discussion of the contracted Bianchi
identities. The use of the York canonical basis allows to
disentangle the contracted Bianchi identities from the equations for
the remaining gauge variables (the inertial effects), for the tidal
variables and for the matter.
\bigskip

See Appendix B for the explicit form in the York canonical basis of
many terms appearing in the Hamilton equations.

\subsection{The Contracted Bianchi Identities}

The variables $\tilde \phi(\tau, \vec \sigma)$ and
$\pi^{(\theta)}_i(\tau, \vec \sigma)$ are the quantities determined
by the (non-hyperbolic) partial differential equations corresponding
to the super-Hamiltonian and super-momentum constraints,
respectively. The Hamilton equations for them are the contracted
Bianchi identities ensuring the $\tau$-preservation of the
constraint sub-manifold: they hold independently from the form of
the solution of such partial differential equations.

\bigskip

For $\tilde \phi$ the Hamilton-Dirac equations obtained by using the
Dirac Hamiltonian (\ref{3.48}) in the York canonical basis, in the
electro-magnetic radiation gauge and with the Fokker-Pryce observer
as origin of the 3-coordinates are

\bea
 \partial_{\tau}\, {\tilde \phi} (\tau ,\vec \sigma ) &\cir& \{ \tilde
 \phi(\tau ,\vec \sigma ), H_D \} =  {{\delta\, H_D}\over {\delta\,
 \pi_{\tilde \phi}(\tau ,\vec \sigma )}} =\nonumber \\
 &=& \Big[ - {{12\pi\, G}\over { c^3}}\, (1 + n)\, \tilde \phi\,
 \pi_{\tilde \phi} +
   {\tilde \phi}^{2/3}\,  \sum_{ra}\, Q_a^{-1}\,
 \Big(\partial_r\, {\bar n}_{(a)}\, V_{ra} +\nonumber \\
 &+& {\bar n}_{(a)} \, \Big[V_{ra}\, \Big({2\over 3}\,
 {\tilde \phi}^{-1}\, \partial_r\, \tilde \phi - \sum_{\bar b}\,
 \gamma_{\bar ba}\, \partial_r\, R_{\bar b}\Big) + \partial_r\,
 V_{ra}\Big]\, \Big)\Big](\tau ,\vec \sigma ),\nonumber \\
 &&{}
 \label{4.1}
 \eea

\noindent where Eq.(\ref{b2}) was used.

\bigskip

For $\pi_i^{(\theta )}$ the Hamilton-Dirac equations obtained by
using the Dirac Hamiltonian (\ref{3.48}) in the York canonical basis
are

 \bea
 \partial_{\tau}\, \pi_i^{(\theta )}(\tau ,\vec \sigma ) &\cir& \{
 \pi_i^{(\theta )}(\tau ,\vec \sigma ), H_D \} = - {{\delta\, H_D}\over
 {\delta\, \theta^i(\tau ,\vec \sigma )}} =\nonumber \\
 &&{}\nonumber \\
 &=& - \int d^3\sigma_1\, \Big[(1 + n)(\tau, {\vec \sigma}_1)\,
 \Big( {{\delta\, {\check {\cal M}}(\tau ,{\vec \sigma}_1)}\over {\delta\,
 \theta^i(\tau ,\vec \sigma)}} - {{c^3}\over {16\pi\, G}}\,
 {{\delta\, {\cal S}(\tau ,{\vec \sigma}_1)}\over
 {\delta\, \theta^i(\tau, \vec \sigma)}}\Big) -\nonumber \\
 &-& {{c^3}\over {16\pi\, G}}\, n(\tau, {\vec \sigma}_1)\,
 {{\delta\, {\cal T}(\tau ,{\vec \sigma}_1)}\over
 {\delta\, \theta^i(\tau, \vec \sigma)}} \Big] -\nonumber \\
 &-& {{4\pi\, G}\over {c^3}}\, \Big((1 + n)\, {\tilde \phi}^{-1}\,
 \sum^{a \not= b}_{abtwuvij}\, {{\epsilon_{abt}\, \epsilon_{abu}\,
 \pi_i^{(\theta)}\, \pi_j^{(\theta)}}\over
 {(Q_a\, Q_b^{-1} - Q_b\, Q_a^{-1})^2}}\, {{\partial\,
 [V_{wt}\, V_{vu}\, B_{iw}\, B_{jv}]}\over {\partial\,
 \theta^i}} \Big)(\tau ,\vec \sigma) +\nonumber \\
 &+& \int d^3\sigma_1\, {\bar n}_{(a)}(\tau,{\vec \sigma}_1)\,
 {{\delta\, {\bar {\tilde {\cal H}}}_{(a)}
 (\tau, {\vec \sigma}_1)}\over {\delta\, \theta^i(\tau ,\vec \sigma)}},
 \label{4.2}
 \eea

\noindent where Eq.(\ref{b21}), (\ref{b11}), (\ref{b19}) and
(\ref{b5}) have to be used. With Eq.(\ref{2.18}) we can replace
$\pi_i^{(\theta )}$ with $\sigma_{(a)(b)}{|}_{a \not= b}$, so to
obtain the contracted Bianchi identities for $\partial_{\tau}\,
\sigma_{(a)(b)}{|}_{a \not= b}$.

\subsection{The Equations of Motion for the Gauge Variables}

The equation of motion for the lapse and shift functions identify
their $\tau$-derivatives with the arbitrary Dirac multipliers

\bea
 \partial_{\tau}\, n(\tau ,\vec \sigma ) &\cir& \lambda_n(\tau
 ,\vec \sigma ),\nonumber \\
 \partial_{\tau}\, {\bar n}_{(a)}(\tau ,\vec \sigma ) &\cir&
 \lambda_{{\bar n}_{(a)}}(\tau ,\vec \sigma ).
 \label{4.3}
 \eea

\bigskip

For $\pi_{\tilde \phi}$ the Hamilton-Dirac equations obtained by
using the Dirac Hamiltonian (\ref{3.48}) in the York canonical basis
give the following form of the Raychaudhuri equation

  \bea
 \partial_{\tau}\, \pi_{\tilde \phi}(\tau ,\vec \sigma ) &\cir& \{ \pi_{\tilde
\phi}(\tau ,\vec \sigma ), H_D \}= - {{\delta\, H_D}\over {\delta\,
\tilde \phi (\tau ,\vec \sigma )}} = - {1\over 6}\, \phi^{-5}(\tau
,\vec \sigma )\, {{\delta\, H_D}\over {\delta\,  \phi (\tau ,\vec
\sigma )}} =\nonumber \\
 &&{}\nonumber \\
 &=& - {1\over 6}\, {\tilde \phi}^{-5/6}(\tau ,\vec \sigma)\,
 \int d^3\sigma_1\, \Big[(1 + n)(\tau, {\vec \sigma}_1)\,
 \Big( {{\delta\, {\check {\cal M}}(\tau ,{\vec \sigma}_1)}\over {\delta\,
 \phi(\tau ,\vec \sigma)}} - {{c^3}\over {16\pi\, G}}\,
 {{\delta\, {\cal S}(\tau ,{\vec \sigma}_1)}\over
 {\delta\, \phi(\tau, \vec \sigma)}}\Big) -\nonumber \\
 &-& {{c^3}\over {16\pi\, G}}\, n(\tau, {\vec \sigma}_1)\,
 {{\delta\, {\cal T}(\tau ,{\vec \sigma}_1)}\over
 {\delta\, \phi(\tau, \vec \sigma)}} \Big] +\nonumber \\
 &+& {{2\pi\, G}\over {c^3}}\, (1 + n)(\tau ,\vec \sigma)\, \Big[
 3\, \pi^2_{\tilde \phi} + 2\, {\tilde \phi}^{-2}\,
 \Big(\sum_{\bar b}\, \Pi^2_{\bar b} +\nonumber \\
 &+& \sum_{abtwuv}^{a\not= b}\, {{\epsilon_{abt}\, \epsilon_{abu}\, V_{wt}\,
 V_{vu}}\over {\Big[Q_a\, Q^{-1}_b - Q_b\,
 Q^{-1}_a \Big]^2}}\, \sum_i\, B_{iw}\, \pi_i^{(\theta
 )}\, \sum_j\, B_{jv}\, \pi_j^{(\theta )}
 \Big)\Big](\tau ,\vec \sigma) +\nonumber \\
 &+& \Big({\tilde \phi}^{-1/3}\, \sum_{ra}\, {\bar n}_{(a)}\, Q_a^{-1}\,
 V_{ra}\, \partial_r\, \pi_{\tilde \phi}\Big)(\tau ,\vec \sigma),
 \label{4.4}
 \eea

\noindent where Eq.(\ref{b3}) has been used  and Eqs.(\ref{b22}),
(\ref{b10}) and (\ref{b17}) are needed. With Eq.(\ref{2.18}) we can
replace $\pi_i^{(\theta )}$ with $\sigma_{(a)(b)}{|}_{a \not= b}$.

\bigskip

For $\theta^i$ the Hamilton-Dirac equations obtained by using the
Dirac Hamiltonian (\ref{3.48}) in the York canonical basis are

\bea
 \partial_{\tau}\, \theta^i(\tau ,\vec \sigma ) &\cir& \{
\theta^i(\tau ,\vec \sigma ), H_D \} = {{\delta\, H_D}\over
{\delta\, \pi_i^{(\theta )}(\tau ,\vec \sigma )}} =\nonumber \\
 &&{}\nonumber \\
 &=& \Big[ {{8\pi\, G}\over {c^3}}\,
  (1 + n)\, {\tilde \phi}^{-1}\, \sum_{abtwuvj}\, {{\epsilon_{abt}\,
  \epsilon_{abu}\, V_{wt}\, V_{vu}\,  B_{iw}\,  B_{jv}\, \pi_j^{(\theta )}
  }\over {\Big(Q_a\, Q^{-1}_b -  Q_b\, Q^{-1}_a\Big)^2}} -\nonumber \\
  &-& {\tilde \phi}^{-1/3}(\tau ,\vec \sigma)\, \sum_a\, \Big[ {\bar
  n}_{(a)}\,\,\, \Big(\sum_{b \not= a}\, \sum_{rtw}\, \Big[{1\over 3}\,
  {\tilde \phi}^{-1}\, \partial_r\, \tilde \phi  +\nonumber \\
  &+&  \sum_{\bar c}\, \gamma_{\bar ca}\,
\partial_r\, R_{\bar c}\,  \Big]\, {{\epsilon_{abt}\,
  Q_b^{-1}\, V_{rb}\, V_{wt}\, B_{iw}}\over {Q_b\, Q_a^{-1} - Q_a\, Q_b^{-1}}}
  +\nonumber \\
  &+& \sum_{bu}\,  \sum_{c \not= b}\, {{\epsilon_{bct}\,
 Q_a^{-1}\, Q_b\, Q_c^{-1}}\over {Q_c\, Q_b^{-1} - Q_b\,
 Q_c^{-1}}}\,  (V_{rc}\, V_{ua} - V_{ra}\, V_{uc})\, \partial_r\, V_{ub}
 \, V_{wt}\, B_{iw}  \Big) -\nonumber \\
 &-& \sum_r\, \partial_r\, {\bar n}_{(a)}\, \sum_{b \not= a}\,
 \sum_{tw}\, {{\epsilon_{abt}\, Q_b^{-1}\, V_{rb}\, V_{wt}\, B_{iw}}\over
 {Q_b\, Q_a^{-1} - Q_a\, Q_b^{-1}}}\, \Big](\tau ,\vec \sigma),
 \label{4.5}
  \eea

\noindent where Eq.(\ref{b4}) has been used.

\bigskip

Then one has to add the gauge fixing constraints (satisfying the
orbit condition) to the super-hamiltonian and super-momentum
constraints

\bea
 \chi (\tau ,\vec \sigma ) &\approx& 0,\nonumber \\
 &&{}\nonumber \\
  &&(clock\, synchronization\, convention\, or\nonumber \\
  &&determination\, of\, \pi_{\tilde \phi}\, and\, of\, the\,
  instantaneous\, 3-space)\nonumber \\
 &&{}\nonumber \\
 &&{}\nonumber \\
  \chi_r(\tau ,\vec \sigma ) &\approx& 0,\nonumber \\
 &&{}\nonumber \\
  &&(determination\, of\, \theta^n\, and\, of\, the\,
  3-coordinates).
 \label{4.6}
 \eea
\medskip

Their preservation in time, by using the Dirac Hamiltonian
(\ref{3.48}), generates the equations for the lapse and shift
functions consistently with the clock synchronization convention and
with the 3-coordinates (the partial $\tau$-derivatives act on the
possible explicit $\tau$-dependence of the gauge fixings)

\bea
 \partial_{\tau}\, \chi (\tau ,\vec \sigma ) &\cir& {{\partial\,
 \chi (\tau ,\vec \sigma )}\over {\partial\, \tau}} + \{ \chi (\tau
 ,\vec \sigma ), H_D \} = 0,\nonumber \\
 &&{}\nonumber \\
 &&(determination\, of n),\nonumber \\
 &&{}\nonumber \\
 &&{}\nonumber \\
  \partial_{\tau}\, \chi_r(\tau ,\vec \sigma ) &\cir& {{\partial\,
 \chi_r(\tau ,\vec \sigma )}\over {\partial\, \tau}} + \{ \chi_r(\tau
 ,\vec \sigma ), H_D \} = 0,\nonumber \\
 &&{}\nonumber \\
 &&(determination\, of {\bar n}_{(a)}).
 \label{4.7}
 \eea

\noindent These are coupled elliptic equations for the lapse and
shift functions indeoendent from the Dirac multipliers, which are
determined by the Hamilton equations $\lambda_n(\tau ,\vec \sigma
)\, \cir\, \partial_{\tau}\, n(\tau ,\vec \sigma )$, $\lambda_{{\bar
n}_{(a)}}(\tau ,\vec \sigma )\, \cir\, \partial_{\tau}\, {\bar
n}_{(a)}(\tau ,\vec \sigma )$ once a solution for $n(\tau, \vec
\sigma)$ and ${\bar n}_{(a)}(\tau, \vec \sigma)$ has been found.
\medskip

For gauge fixings of the type $\theta^i(\tau, \vec \sigma) \approx
(numerical\, function)^i$ and $\pi_{\tilde \phi}(\tau, \vec \sigma)
\approx numerical\, function$, Eqs. (\ref{4.7}) are just
Eqs.(\ref{4.4}) and (\ref{4.5}), respectively.

\subsection{The Equations of Motion for the Tidal Variables and the Matter}

Let us now consider the Hamilton-Dirac equations of motion implied
by the Dirac Hamiltonian (\ref{3.48}) for the tidal degrees of
freedom and for the particles.
\medskip

\subsubsection{The Hamilton Equations for the Tidal Variables}

For the tidal variables $R_{\bar a}$ we get the following
kinematical Hamilton equations

\bea
 \partial_{\tau}\, R_{\bar a}(\tau ,\vec \sigma ) &\cir& \{ R_{\bar
 a}(\tau ,\vec \sigma ), H_D \} = {{\delta\, H_D}\over
 {\delta\, \Pi_{\bar a}(\tau ,\vec \sigma )}} =\nonumber \\
 &&{}\nonumber \\
 &=&  \Big[{{8\pi\, G}\over {c^3}}\, {\tilde \phi}^{-1}\, (1 + n)\,
 \Pi_{\bar a} -\nonumber \\
 &-& {\tilde \phi}^{-1/3}\, \sum_{ra}\, Q_a^{-1}\, \Big({\bar n}_{(a)}\,
  \Big[\gamma_{\bar aa}\, V_{ra}\, \Big({1\over 3}\,
  {\tilde \phi}^{-1}\, \partial_r\, \tilde \phi +
  \sum_{\bar b}\, \gamma_{\bar ba}\, \partial_r\, R_{\bar b}\Big) -\nonumber \\
  &-& V_{ra}\, \partial_r\, R_{\bar a} + \sum_{sb}\, \gamma_{\bar
  ab}\, V_{sa}\, V_{rb}\, \partial_r\, V_{sb}  \Big] -
   \gamma_{\bar aa}\, \partial_r\, {\bar n}_{(a)}\, V_{ra}
  \Big)\Big](\tau, \vec \sigma),
 \label{4.8}
 \eea

\noindent where Eq.(\ref{b6}) has been used.
\medskip

Eq.(\ref{4.8}) can be inverted to get the momenta $\Pi_{\bar a}$ in
terms of the velocities $\partial_{\tau}\, R_{\bar a}$

\bea
 \Pi_{\bar a}(\tau ,\vec \sigma ) &\cir& {{c^3}\over {8\pi\, G}}\,
 {{\tilde \phi(\tau ,\vec \sigma )}\over {1 + n(\tau ,\vec \sigma )}}\,
 \Big[\partial_{\tau}\, R_{\bar a} +\nonumber \\
 &+& {\tilde \phi}^{-1/3}\, \sum_{ra}\, Q_a^{-1}\, \Big({\bar n}_{(a)}\,
  \Big[\gamma_{\bar aa}\, V_{ra}\, \Big({1\over 3}\, {\tilde \phi}^{-1}\,
  \partial_r\, \tilde \phi +
  \sum_{\bar b}\, \gamma_{\bar ba}\, \partial_r\, R_{\bar b}\Big) -\nonumber \\
  &-& V_{ra}\, \partial_r\, R_{\bar a} + \sum_{sb}\, \gamma_{\bar
  ab}\, V_{sa}\, V_{rb}\, \partial_r\, V_{sb}  \Big] -
   \gamma_{\bar aa}\, \partial_r\, {\bar n}_{(a)}\, V_{ra}
  \Big)\Big](\tau, \vec \sigma).  \nonumber \\
  &&{}
 \label{4.9}
 \eea

\bigskip

The dynamical Hamilton equations for $\Pi_{\bar a}$ are
 \medskip

 \bea
  \partial_{\tau}\, \Pi_{\bar a}(\tau ,\vec \sigma ) &\cir& \{ \Pi_{\bar
 a}(\tau ,\vec \sigma ), H_D \} = - {{\delta\, H_D}\over
 {\delta\, R_{\bar a}(\tau ,\vec \sigma )}} =\nonumber \\
 &&{}\nonumber \\
  &=& - \int d^3\sigma_1\, \Big[(1 + n)(\tau, {\vec \sigma}_1)\,
 \Big( {{\delta\, {\check {\cal M}}(\tau ,{\vec \sigma}_1)}\over {\delta\,
 R_{\bar a}(\tau ,\vec \sigma)}} - {{c^3}\over {16\pi\, G}}\,
 {{\delta\, {\cal S}(\tau ,{\vec \sigma}_1)}\over
 {\delta\, R_{\bar a}(\tau, \vec \sigma)}}\Big) -\nonumber \\
 &-& {{c^3}\over {16\pi\, G}}\, n(\tau, {\vec \sigma}_1)\,
 {{\delta\, {\cal T}(\tau ,{\vec \sigma}_1)}\over
 {\delta\, R_{\bar a}(\tau, \vec \sigma)}} \Big] -\nonumber \\
 &-&{{8\pi\, G}\over {c^3}}\, \Big({\tilde \phi}^{-1}\, (1 + n)
 \nonumber \\
 && \sum_{abtwuvij}^{a\not= b}\, (\gamma_{\bar aa} - \gamma_{\bar ab})\,
 {{\epsilon_{abt}\, \epsilon_{abu}\, V_{wt}\,
 V_{vu}\, B_{iw}\, B_{jv}\, (Q_b\, Q_a^{-1} + Q_a\, Q_b^{-1})}\over
 {\Big(Q_b\, Q^{-1}_a - Q_a\, Q^{-1}_b \Big)^3}}\,   \pi_i^{(\theta
 )}\,   \pi_j^{(\theta )} \Big)(\tau ,\vec \sigma) +\nonumber \\
 &+& \int d^3\sigma_1\, {\bar n}_{(a)}(\tau,{\vec \sigma}_1)\,
 {{\delta\, {\bar {\tilde {\cal H}}}_{(a)}
 (\tau, {\vec \sigma}_1)}\over {\delta\, R_{\bar a}(\tau ,\vec \sigma)}},
 \label{4.10}
 \eea

\noindent where Eqs.(\ref{b23}), (\ref{b9}), (\ref{b15}) and
(\ref{b7}) have to be used. With Eq.(\ref{2.18}) we can replace
$\pi_i^{(\theta )}$ with $\sigma_{(a)(b)}{|}_{a \not= b}$. \bigskip

If we evaluate the $\tau$-derivative of Eq.(\ref{4.9}) and we equate
it to Eq.(\ref{4.10}), we get the following second order equation
for $R_{\bar a}$

 \begin{eqnarray*}
 \partial^2_{\tau}\, R_{\bar a}(\tau ,\vec \sigma ) &\cir&
 \Big[\,\,\,  {\tilde \phi}^{-1/3}\, \sum_{ra}\, Q_a^{-1}\, V_{ra}\, {\bar
 n}_{(a)}\, \sum_{\bar b}\, (\gamma_{\bar aa}\, \gamma_{\bar ba}\,
 - \delta_{\bar a\bar b})\, \partial_r\,
 \partial_{\tau}\, R_{\bar b} +\nonumber \\
 &+&{\tilde \phi}^{-1/3}\, \sum_{ra}\, Q_a^{-1}\, \Big[\Big(\gamma_{\bar aa}\, V_{ra}\,
 ({1\over 3}\, {\tilde \phi}^{-1}\, \partial_r \tilde \phi + \sum_{\bar c}\,
 \gamma_{\bar ca}\, \partial_r\, R_{\bar c}) - V_{ra}\, \partial_r\,
 R_{\bar a}+\nonumber \\
 &+& \sum_{sb}\, \gamma_{\bar ab}\, V_{sa}\, V_{rb}\, \partial_r\, V_{sb}\Big)\,
 {\bar n}_{(a)} - \gamma_{\bar aa}\, V_{ra}\, \partial_r\,
 {\bar n}_{(a)}\Big]\, \sum_{\bar b}\, \gamma_{\bar
 ba}\, \partial_{\tau}\, R_{\bar b} -\nonumber \\
 &-& {\tilde \phi}^{-1}\, \Big[\partial_{\tau}\, R_{\bar a} + {2\over 3}\,
 {\tilde \phi}^{-1/3}\, \sum_{ra}\, Q_a^{-1}\, \Big(\Big[\gamma_{\bar aa}\,
 V_{ra}\, (- {1\over 6}\, {\tilde \phi}^{-1}\, \partial_r\, \tilde \phi +
 \sum_{\bar b}\, \gamma_{\bar ba}\, \partial_r\, R_{\bar b}) -\nonumber \\
 &-&V_{ra}\, \partial_r\, R_{\bar a} + \sum_{sb}\, \gamma_{\bar ab}\,
 V_{sa}\, V_{rb}\, \partial_r\, V_{sb}\Big]\, {\bar n}_{(a)} - \gamma_{\bar aa}\,
 V_{ra}\, \partial_r\, {\bar n}_{(a)}\Big)\Big]\, \partial_{\tau}\, \tilde
 \phi -\nonumber \\
 &-& {1\over 3}\, {\tilde \phi}^{-4/3}\, \sum_{ra}\, \gamma_{\bar aa}\,
 Q_a^{-1}\, V_{ra}\, {\bar n}_{(a)}\, \partial_r\, \partial_{\tau}\,
 \tilde \phi -
 \end{eqnarray*}

 \begin{eqnarray*}
 &-&{\tilde \phi}^{-1/3}\,  \sum_{ra}\, Q_a^{-1}\, \Big[- {{\partial\, V_{ra}}\over
 {\partial\, \theta^i}}\, \Big({\bar n}_{(a)}\, \Big[\partial_r\, R_{\bar a}
 - \gamma_{\bar aa}\, ({1\over 3}\, {\tilde \phi}^{-1}\, \partial_r\,
 \tilde \phi +\nonumber \\
 &+& \sum_{\bar b}\, \gamma_{\bar bb}\, \partial_r\, R_{\bar b})\Big] + \gamma_{\bar aa}\,
 \partial_r\, {\bar n}_{(a)}\Big) + \sum_{sb}\, \gamma_{\bar ab}\, {{\partial\,
 V_{sa}\, V_{rb}\, \partial_r\, V_{sb}}\over {\partial\, \theta^i}}\,
 {\bar n}_{(a)}\Big]\, \partial_{\tau}\, \theta^i +\nonumber \\
 &+&\Big[ \partial_{\tau}\, R_{\bar a} + {\tilde \phi}^{-1/3}\, \sum_{ra}\,
 Q_a^{-1}\, \Big(\Big[\gamma_{\bar aa}\, V_{ra}\, ({1\over 3}\, {\tilde
 \phi}^{-1}\, \partial_r\, \tilde \phi + \sum_{\bar b}\, \gamma_{\bar ab}\,
 \partial_r\, R_{\bar b}) -\nonumber \\
 &-& V_{ra}\, \partial_r\, R_{\bar a} + \sum_{sb}\, \gamma_{\bar ab}\,
 V_{sa}\, V_{rb}\, \partial_r\, V_{sb}\Big]\, {\bar n}_{(a)} -
 \gamma_{\bar aa}\, V_{ra}\, \partial_r\, {\bar n}_{(a)}\Big)\Big]\,
 {{\partial_{\tau}\, n}\over {1 + n}} -\nonumber \\
 &-&{\tilde \phi}^{-1/3}\,  \sum_{ra}\, Q_a^{-1}\, \Big(\Big[\gamma_{\bar aa}\, V_{ra}\,
 ({1\over 3}\, {\tilde \phi}^{-1}\, \partial_r\, \tilde \phi + \sum_{\bar b}\,
 \gamma_{\bar ba}\, \partial_r\, R_{\bar b}) - V_{ra}\, \partial_r\,
 R_{\bar a} +\nonumber \\
 &+& \sum_{sb}\, \gamma_{\bar ab}\, V_{sa}\, V_{rb}\, \partial_r\,
 V_{sb}\Big]\, \partial_{\tau}\, {\bar n}_{(a)} - \gamma_{\bar aa}\,
 V_{ra}\, \partial_r\, \partial_{\tau}\, {\bar n}_{(a)}
  \Big)\,\,\,  \Big](\tau ,\vec \sigma ) +
  \end{eqnarray*}

\bea
 &+& {1\over 2}\, \Big({\tilde \phi}^{-1}\, (1 + n)\Big)(\tau ,\vec \sigma)\, \int
 d^3\sigma_1\, \Big[(1 + n)(\tau, {\vec \sigma}_1)\,
 {{\delta\, {\cal S}(\tau ,{\vec \sigma}_1)}\over
 {\delta\, R_{\bar a}(\tau, \vec \sigma)}} + n(\tau ,{\vec \sigma}_1)\,
 {{\delta\, {\cal T}(\tau ,{\vec \sigma}_1)}\over
 {\delta\, R_{\bar a}(\tau, \vec \sigma)}}\Big] -\nonumber \\
 &-& {{8\pi\, G}\over {c^3}}\, \Big({\tilde \phi}^{-1}\, (1 +
 n)\Big)(\tau ,\vec \sigma)\, \int
 d^3\sigma_1\, (1 + n)(\tau, {\vec \sigma}_1)\,
 {{\delta\, {\check {\cal M}}(\tau ,{\vec \sigma}_1)}\over
 {\delta\, R_{\bar a}(\tau, \vec \sigma)}} +\nonumber \\
 &+&{{8\pi\, G}\over {c^3}}\, \Big({\tilde \phi}^{-1}\, (1 +
 n)\Big)(\tau ,\vec \sigma)\, \int d^3\sigma_1\, {\bar n}_{(a)}(\tau,{\vec \sigma}_1)\,
 {{\delta\, {\bar {\tilde {\cal H}}}_{(a)}
 (\tau, {\vec \sigma}_1)}\over {\delta\, R_{\bar a}(\tau ,\vec \sigma)}}
 -\nonumber \\
 &-&  \Big({{8\pi\, G}\over {c^3}}\Big)^2\, \Big({\tilde \phi}^{-2}\, (1 +
 n)^2\Big)(\tau ,\vec \sigma)\nonumber \\
 && \Big(\sum_{abtwuvij}^{a\not= b}\, (\gamma_{\bar aa} - \gamma_{\bar ab})\,
 {{\epsilon_{abt}\, \epsilon_{abu}\, V_{wt}\,
 V_{vu}\, B_{iw}\, B_{jv}\, (Q_b\, Q_a^{-1} + Q_a\, Q_b^{-1})}\over
 {\Big(Q_b\, Q^{-1}_a - Q_a\, Q^{-1}_b \Big)^3}}\,   \pi_i^{(\theta
 )}\,   \pi_j^{(\theta )} \Big)(\tau ,\vec \sigma).\nonumber \\
 &&{}
 \label{4.11}
 \eea

\bigskip

This equation depends:\medskip

\noindent a) on the $\tau$-derivatives $\partial_{\tau}\, \tilde
\phi(\tau, \vec \sigma)$ and $\partial_{\tau}\, \theta^i(\tau ,\vec
\sigma)$, given in Eqs.(\ref{4.1}) and (\ref{4.5}), respectively;
\medskip

\noindent b) on the $\tau$-derivatives of the lapse and shift
functions, namely on the arbitrary Dirac multipliers appearing in
Eqs.(\ref{4.3}).

\subsubsection{The Hamilton Equations for the Particles}

By using Eqs.(\ref{3.15}), the Hamilton equations (\ref{3.38}) for
the particles in the electro-magnetic radiation gauge take the
following form in the York canonical basis (Eq.(\ref{3.35}) and
(\ref{3.37}) are used for ${\cal W}$)

 \begin{eqnarray*}
 \eta_i\, {\dot \eta}^r_i(\tau ) &\cir&  \{ \eta^r_i(\tau ), H_D \}
 = \nonumber \\
 &=&  \int d^3\sigma\, \Big(\Big[1 + n(\tau ,\vec \sigma )\Big]\,
 {{\partial\, {\check {\cal M}}(\tau ,\vec \sigma )}\over {\partial\, \kappa_{ir}}} -\nonumber \\
 &-& \sum_a\, {\bar n}_{(a)}(\tau ,\vec \sigma )\, \Big[\phi^{-2}\, \sum_v\, Q^{-1}_v\,
 V_{va}\Big](\tau ,\vec \sigma )\, {{\partial\, {\check {\cal M}}_v(\tau ,\vec \sigma )}\over
 {\partial\, \kappa_{ir}}} \Big) =\nonumber \\
 &&{}\nonumber \\
 &=& \eta_i\, \Big({{\phi^{-4}\, (1 + n)\, \sum_{as}\, Q_a^{-2}\, V_{ra}\,
 V_{sa}\,\, \Big({\check \kappa}_{is}(\tau ) - {{Q_i}\over c}\, A_{\perp\, s}\Big)\,
 }\over {\sqrt{m_i^2\, c^2 + \phi^{-4}\, \sum_{cuv}\,
 Q^{-2}_c\, V_{uc}\, V_{vb}\,\,
 \Big({\check \kappa}_{iu}(\tau ) - {{Q_i}\over c}\, A_{\perp\, u}\Big)\,
 \Big({\check \kappa}_{iv}(\tau ) - {{Q_i}\over c}\, A_{\perp\, v}\Big)\,
 }}} -\nonumber \\
 &-& \phi^{-2}\, \sum_a\, Q_a^{-1}\, V_{ra}\, {\bar n}_{(a)}
 \Big)(\tau ,{\vec \eta}_i(\tau )),
 \end{eqnarray*}

\begin{eqnarray*}
  \eta_i\, {d\over {d\tau}}\, {\check \kappa}_{ir}(\tau ) &\cir&
   \{ {\check \kappa}_{ir}(\tau ), H_D \}
 =\nonumber \\
 &=& -  \int d^3\sigma\, \Big(\Big[1 + n(\tau ,\vec \sigma )\Big]\,
 {{\partial\, {\check {\cal M}}(\tau ,\vec \sigma )}\over {\partial\, \eta^r_i}} -\nonumber \\
 &-& \sum_a\, {\bar n}_{(a)}(\tau ,\vec \sigma )\, \Big[\phi^{-2}\, \sum_v\, Q^{-1}_v\,
 V_{va}\Big](\tau ,\vec \sigma )\, {{\partial\, {\check {\cal M}}_v(\tau ,\vec \sigma )}\over
 {\partial\, \eta^r_i}} \Big) =\nonumber \\
 &&{}\nonumber \\
 &&{}\nonumber \\
 &=& - {{\partial}\over {\partial\, \eta^r_i}}\,
  \int d^3\sigma\, \Big[(1 + n)\, {{\partial\,
 {\cal W}_{(n)}}\over {\partial\, \eta^r_i}} +\nonumber \\
 &+& \phi^{-2}\, \sum_a\, Q_a^{-1}\, {\bar n}_{(a)}\, \sum_v\, V_{va}\,
 {{\partial\, {\cal W}_v}\over {\partial\, \eta^r_i}}\Big](\tau ,\vec \sigma)
 +\nonumber \\
 &+&\eta_i\, \sum_{asv}\, \Big[\Big({{\phi^{-4}\, (1 + n)\, Q_a^{-2}\, V_{sa}\, V_{va}\,
\Big({\check \kappa}_{iv}(\tau) - {{Q_i}\over c}\, A_{\perp\,
v}\Big)}\over {\sqrt{m_i^2\, c^2 + \phi^{-4}\, \sum_{cmn}\,
 Q^{-2}_c\, V_{mc}\, V_{nb}\,\,
 \Big({\check \kappa}_{im}(\tau ) - {{Q_i}\over c}\, A_{\perp\, m}\Big)\,
 \Big({\check \kappa}_{in}(\tau ) - {{Q_i}\over c}\, A_{\perp\,
 n}\Big)}
 }} -\nonumber \\
 &-& \phi^{-2}\, Q_a^{-2}\, V_{sa}\, {\bar n}_{(a)}
 \Big)\, {{Q_i}\over c}\, {{\partial\, A_{\perp\, s}}
 \over {\partial\, \eta^r_i}}\Big](\tau ,{\vec \eta}_i(\tau))
 +\nonumber \\
 &+& \eta_i\, {\check F}_{ir}(\tau, {\vec \eta}_i(\tau)),
 \end{eqnarray*}

 \bea
 {\check F}_{ir}(\tau, {\vec \eta}_i(\tau)) &=&
 \Big(\sqrt{m_i^2\, c^2 + \phi^{-4}\, \sum_{cmn}\,
 Q^{-2}_c\, V_{mc}\, V_{nb}\,\,
 \Big({\check \kappa}_{im}(\tau ) - {{Q_i}\over c}\, A_{\perp\, m}\Big)\,
 \Big({\check \kappa}_{in}(\tau ) - {{Q_i}\over c}\, A_{\perp\,
 n}\Big)}\nonumber \\
 &&\Big[- {{\partial\, n}\over {\partial\, \eta^r_i}} + \phi^{-2}\, \sum_a\,
 Q_a^{-1}\, V_{sa}\, {{\partial\, {\bar n}_{(a)}}\over {\partial\,
 \eta^r_i}} \times\nonumber \\
 &&{{\Big({\check \kappa}_{is}(\tau) - {{Q_i}\over c}\, A_{\perp\,
 s}\Big)}\over {\sqrt{m_i^2\, c^2 + \phi^{-4}\, \sum_{cmn}\,
 Q^{-2}_c\, V_{mc}\, V_{nb}\,\,
 \Big({\check \kappa}_{im}(\tau ) - {{Q_i}\over c}\, A_{\perp\, m}\Big)\,
 \Big({\check \kappa}_{in}(\tau ) - {{Q_i}\over c}\, A_{\perp\,
 n}\Big)}}} +\nonumber \\
 &+&\phi^{-4}\, (1 + n)\, \sum_{aus}\, Q_a^{-2}\, V_{ua}\,
 \Big(V_{sa}\, ({1\over 3}\, {\tilde \phi}^{-1}\, \partial_r\,
 \tilde \phi + \sum_{\bar b}\, \gamma_{\bar bb}\, \partial_r\, R_{\bar b})
 - \partial_r\, V_{sa}\Big)\nonumber \\
 && {{\Big({\check \kappa}_{iu}(\tau ) - {{Q_i}\over c}\, A_{\perp\, u}\Big)\,
 \Big({\check \kappa}_{is}(\tau ) - {{Q_i}\over c}\, A_{\perp\, s}\Big)\,  }\over
 {\Big(\sqrt{m_i^2\, c^2 + \phi^{-4}\, \sum_{cmn}\,
 Q^{-2}_c\, V_{mc}\, V_{nb}\,\,
 \Big({\check \kappa}_{im}(\tau ) - {{Q_i}\over c}\, A_{\perp\, m}\Big)\,
 \Big({\check \kappa}_{in}(\tau ) - {{Q_i}\over c}\, A_{\perp\, n}\Big)\,
 }\Big)^2  }} -\nonumber \\
 &-&\phi^{-2}\, \sum_{as}\, Q_a^{-1}\, {\bar n}_{(a)}\,
 \Big(V_{sa}\, ({1\over 3}\, {\tilde \phi}^{-1}\, \partial_r\,
 \tilde \phi + \sum_{\bar b}\, \gamma_{\bar bb}\, \partial_r\, R_{\bar b})
 - \partial_r\, V_{sa}\Big)\nonumber \\
 && {{ \Big({\check \kappa}_{is}(\tau ) - {{Q_i}\over c}\, A_{\perp\, s}\Big)\,  }\over
 {\sqrt{m_i^2\, c^2 + \phi^{-4}\, \sum_{cmn}\,
 Q^{-2}_c\, V_{mc}\, V_{nb}\,\,
 \Big({\check \kappa}_{im}(\tau ) - {{Q_i}\over c}\, A_{\perp\, m}\Big)\,
 \Big({\check \kappa}_{in}(\tau ) - {{Q_i}\over c}\, A_{\perp\, n}\Big)\,
 }  }} \Big]\Big)(\tau ,{\vec \eta}_i(\tau)).\nonumber \\
 &&{}
 \label{4.12}
 \eea

\bigskip

By inverting the first of Eqs.(\ref{4.12}) we get the following form
of Eq.(\ref{3.39})

\bea
 && \check{\kappa}_{ir}(\tau ) =
 {{Q_i}\over c}\, A_{\perp\,r}(\tau ,{\vec \eta}_i(\tau ))
 +\nonumber \\
 &&{}\nonumber \\
 &+& m_i\, c\,\,\ \Big( {\tilde \phi}^{2/3}\, \sum_{sa}\, Q_a^2\, V_{ra}\,
 V_{sa}\, \Big(\dot{\eta}^s_i(\tau ) + {\tilde \phi}^{-1/3}\, \sum_b\,
 Q_b^{-1}\, V_{sb}\, {\bar n}_{(b)}\Big)\,\, \Big[\Big(1 + n\Big)^2 -
 \nonumber \\
 &-& {\tilde \phi}^{2/3}\, \sum_{uvc}\,
 Q_c^2\, V_{uc}\, V_{vc}\, \Big(\dot{\eta}^u_i(\tau ) + {\tilde \phi}^{-1/3}\,
 \sum_d\, Q_d^{-1}\, V_{ud}\, {\bar n}_{(d)} \Big)\, \Big(\dot{\eta}^v_i(\tau )
 + {\tilde \phi}^{-1/3}\, \sum_e\, Q_e^{-1}\, V_{ve}\, {\bar n}_{(e)}
 \Big)\Big]^{-1/2}\Big)\nonumber \\
 &&(\tau ,{\vec \eta}_i(\tau )).
 \label{4.13}
 \eea

\medskip

If we put Eq.(\ref{4.13}) into the second of Eqs.(\ref{4.12}), we
get the second order equation for $\eta^r_i(\tau)$ (corresponding to
Eq.(\ref{3.17}))

\bea
 &&\eta_i\, {{d\, {\check \kappa}_{ir}}\over {d\, \tau}}\,\, \cir\,\,
 \Big(- {{\partial}\over {\partial\, \eta_i^r}}\,  {\cal W}
 + {{\eta_i\, Q_i}\over c}\, {\dot \eta}^s_i(\tau)\, {{\partial\, A_{\perp\, s}}\over
 {\partial\, \eta_i^r}} + \eta_i\, {\check F}_{ir} \Big)(\tau ,{\vec
 \eta}_i(\tau)), \nonumber \\
 &&{}\nonumber \\
 &&{\cal W}(\tau) = \int d^3\sigma\, \Big[(1 + n)\, {\cal W}_{(n)} + \phi^{-2}\, \sum_a\,
 Q_a^{-1}\, {\bar n}_{(a)}\, \sum_v\, V_{va}\, {\cal W}_v\Big](\tau,
 \vec \sigma),\nonumber \\
  {\cal W}_{(n)}(\tau ,\vec \sigma) &=& - {1\over
  {2c}}\,\Big[\phi^{-2}\, \sum_{ars}\, Q_a^2\, V_{ra}\, V_{sa}\,
 \left(2\, \pi_\perp^r - \delta^{rm}\, \sum_{i=1}^N\, Q_i\, \eta_i\,
 {{\partial\, c(\vec \sigma, {\vec \eta}_i(\tau))}\over {\partial\, \sigma^m}}
 \right)\nonumber \\
 && \delta^{sn}\, \sum_{j=1}^N\, Q_j\, \eta_j\, {{\partial\, c(\vec \sigma,
 {\vec \eta}_j(\tau))}\over {\partial\, \sigma^n}}
 \Big](\tau ,\vec \sigma),\nonumber \\
 &&{}\nonumber \\
 {\cal W}_r(\tau ,\vec \sigma) &=&  - {1\over c}\,
 F_{rs}(\tau ,\vec \sigma)\, \delta^{sn}\, \sum_{i=1}^N\, Q_i\, \eta_i\,
 {{\partial\, c(\vec \sigma, {\vec \eta}_i(\tau))}\over {\partial\,
 \sigma^n}},\nonumber \\
 &&{}\nonumber \\
 &&{\check F}_{ir} = m_i\, c\, \Big[\Big(1 + n\Big)^2 -
  {\tilde \phi}^{2/3}\, \sum_{uvc}\,
 Q_c^2\, V_{uc}\, V_{vc}\, \Big(\dot{\eta}^u_i(\tau ) + {\tilde \phi}^{-1/3}\,
 \sum_d\, Q_d^{-1}\, V_{ud}\, {\bar n}_{(d)} \Big)\nonumber \\
 && \Big(\dot{\eta}^v_i(\tau )
 + {\tilde \phi}^{-1/3}\, \sum_e\, Q_e^{-1}\, V_{ve}\, {\bar n}_{(e)}
 \Big)\Big]^{-1/2}\nonumber \\
 &&\Big[- (1 + n)\, {{\partial\, n}\over {\partial\, \eta_i^r}} +
 \phi^2\, \sum_{acsu}\, Q_a^{-1}\, Q_c^2\, \Big(V_{sa}\, {{\partial\,
 {\bar n}_{(a)}}\over {\partial\, \eta_i^r}} -\nonumber \\
 &-&\Big[({1\over 3}\, {\tilde \phi}^{-1}\, \partial_r\, \tilde \phi +
 \sum_{\bar a}\, \gamma_{\bar aa}\, \partial_r\, R_{\bar a})\, V_{sa} -
 \partial_r\, V_{sa}\Big]\, {\bar n}_{(a)}\Big)\, V_{uc}\,
 V_{sc}\, \Big({\dot \eta}_i^u(\tau) + \phi^{-2}\, \sum_b\, Q_b^{-1}\, V_{ub}\,
 {\bar n}_{(b)}\Big) +\nonumber \\
 &+&\phi^4\, \sum_{auv}\, Q_a^2\, V_{ua}\, \Big(
 ({1\over 3}\, {\tilde \phi}^{-1}\, \partial_r\, \tilde \phi +
 \sum_{\bar a}\, \gamma_{\bar aa}\, \partial_r\, R_{\bar a})\, V_{va}
 + \partial_r\, V_{va} \Big)\nonumber \\
 &&\Big({\dot \eta}_i^u(\tau) + \phi^{-2}\, \sum_b\, Q_b^{-1}\, V_{ub}\,
 {\bar n}_{(b)}\Big) \, \Big({\dot \eta}_i^v(\tau) + \phi^{-2}\, \sum_c\, Q_c^{-1}\, V_{vc}\,
 {\bar n}_{(c)}\Big)\,\, \Big].\nonumber \\
 &&{}
  \label{4.14}
  \eea

\subsubsection{The Hamilton Equations for the Transverse
Electro-Magnetic Field}

Eqs.(\ref{3.40}) for the transverse electro-magnetic fields
$A_{\perp\, r}(\tau ,\vec \sigma)$ and $\pi^r_{\perp}(\tau ,\vec
\sigma)$ in the radiation gauge take the following form in the York
canonical basis

\begin{eqnarray*}
 \partial_{\tau}\, A_{\perp\, r}(\tau ,\vec \sigma ) &\cir& \{
 A_{\perp\, r}(\tau ,\vec \sigma), H_D \} =\nonumber \\
 &=& \sum_{nuva}\, \delta_{rn}\, P^{nu}_{\perp}(\vec \sigma)\, \Big[
 {\tilde \phi}^{-1/3}\, (1 + n)\, Q_a^2\, V_{ua}\, V_{va}\, \Big(\pi^v_{\perp} -
 \sum_m\, \delta^{vm}\, \sum_{i=1}^N\, Q_i\, \eta_i\, {{\partial\,
 c(\vec \sigma, {\vec \eta}_i(\tau))}\over {\partial\, \sigma^m}}
 \Big) +\nonumber \\
 &+&  {\tilde \phi}^{-1/3}\, Q_a^{-1}\, V_{va}\, {\bar n}_{(a)}\, F_{vu}
 \Big](\tau ,\vec \sigma),\nonumber \\
 &&{}\nonumber \\
 \partial_{\tau}\, \pi^r_{\perp}(\tau ,\vec \sigma) &\cir& \{
 \pi^r_{\perp}(\tau ,\vec \sigma), H_D \} =\nonumber \\
 &=& \sum_{wma}\, P^{rw}_{\perp}(\vec \sigma)\, \delta_{wm}\, \Big(
 \sum_{i=1}^N\, \eta_i\, Q_i\, \delta^3(\vec \sigma, {\vec
 \eta}_i(\tau))\, \nonumber \\
 &&\Big[{ { {\tilde \phi}^{-2/3}\, (1 + n)\, Q_a^{-2}\, V_{ma}\, \sum_s\,
 V_{sa}\,  {\check \kappa}_{is}(\tau)}\over
 {\sqrt{m_i^2\, c^2 + {\tilde \phi}^{-2/3}\, \sum_{uvb}\, Q_b^{-2}\,
 V_{ub}\, V_{vb}\, \Big({\check \kappa}_{iu}(\tau ) - {{Q_i}\over c}\,
 A_{\perp\, u}\Big)\,  \Big({\check \kappa}_{iv}(\tau ) -
 {{Q_i}\over c}\, A_{\perp \, v}\Big)}}} -\nonumber \\
  &-& {\tilde \phi}^{-1/3}\, Q_a^{-1}\, V_{ma}\,
  {\bar n}_{(a)}\Big](\tau ,{\vec \eta}_i(\tau)) +
 \end{eqnarray*}

  \begin{eqnarray*}
 &+&\Big[(1 + n)\, \sum_{svbn}\, \Big({\tilde \phi}^{-1/3}\, Q_a^{-2}\,
 Q_b^{-2}\, V_{sa}\, V_{vb}\, (V_{na}\, V_{mb} - V_{nb}\, V_{ma})\,
 \partial_n\, F_{sv} +\nonumber \\
 &+&{\tilde \phi}^{-1/3}\, Q_a^{-2}\, Q_b^{-2}\, \Big[\partial_n\,
 \Big(V_{sa}\, V_{vb}\, (V_{na}\, V_{mb} - V_{nb}\, V_{ma})\Big)
 -\nonumber \\
 &-& V_{sa}\, V_{vb}\, (V_{na}\, V_{mb} - V_{nb}\, V_{ma})\, \Big({1\over 3}\,
 {\tilde \phi}^{-1}\, \partial_n\, \tilde \phi + 2\, \sum_{\bar b}\,
 (\gamma_{\bar ba} + \gamma_{\bar bb})\, \partial_n\, R_{\bar b}\Big) \Big]\,
 F_{sv} \Big) +\nonumber \\
 &+&{\tilde \phi}^{-1/3}\,  \sum_{svbn}\, \partial_n\, n\,  Q_a^{-2}\,
 Q_b^{-2}\, V_{sa}\, V_{vb}\, (V_{na}\, V_{mb} - V_{nb}\, V_{ma})\,
 F_{sv} +\end{eqnarray*}

\bea
 &+& {\tilde \phi}^{-1/3}\, {\bar n}_{(a)}\, Q_a^{-1}\, \sum_n\, \Big(V_{na}\, \partial_n\,
 \pi_{\perp}^m +\nonumber \\
 &+&\Big[\partial_n\, V_{na} - V_{na}\, ({1\over 3}\, {\tilde \phi}^{-1}\,
 \partial_n\, \tilde \phi + \sum_{\bar b}\, \gamma_{\bar ba}\,
 \partial_n\, R_{\bar b})\Big]\, \pi^m_{\perp} -\nonumber \\
 &-& \Big[\partial_n\, V_{ma} - V_{ma}\, ({1\over 3}\, {\tilde \phi}^{-1}\,
 \partial_n\, \tilde \phi + \sum_{\bar b}\, \gamma_{\bar ba}\,
 \partial_n\, R_{\bar b})\Big]\, \pi^n_{\perp} -\nonumber \\
 &-& \sum_t\, \Big(\Big[\partial_n\, V_{na} - V_{na}\, ({1\over 3}\, {\tilde \phi}^{-1}\,
 \partial_n\, \tilde \phi + \sum_{\bar b}\, \gamma_{\bar ba}\,
 \partial_n\, R_{\bar b})\Big]\, \delta^{mt} -\nonumber \\
 &-& \Big[\partial_n\, V_{ma} - V_{ma}\, ({1\over 3}\, {\tilde \phi}^{-1}\,
 \partial_n\, \tilde \phi + \sum_{\bar b}\, \gamma_{\bar ba}\,
 \partial_n\, R_{\bar b})\Big]\, \delta^{nt}\Big)\, \sum_i\, \eta_i\, Q_i\,
  {{\partial\, c(\vec \sigma, {\vec \eta}_i(\tau)))}\over {\partial\, \sigma^t}}
 -\nonumber \\
 &-& \sum_t\, (V_{na}\, \delta^{mt} - V_{ma}\, \delta^{nt})\,
 \sum_{i=1}^N\, \eta_i\, Q_i\,  {{\partial^2\, c(\vec \sigma, {\vec \eta}_i(\tau)))}
 \over {\partial\, \sigma^t\, \partial\, \sigma^n}} \Big) -\nonumber \\
 &-& {\tilde \phi}^{-1/3}\, \sum_n\, \partial_n\, {\bar n}_{(a)}\, Q_a^{-1}\,
 \sum_t\, (V_{na}\, \delta^{mt} - V_{ma}\, \delta^{nt})\, \sum_{i=1}^N\, \eta_i\, Q_i\,
  {{\partial\, c(\vec \sigma, {\vec \eta}_i(\tau)))}\over {\partial\, \sigma^t}}
 \Big](\tau ,\vec \sigma)\, \Big).\nonumber \\
 &&{}
 \label{4.15}
 \eea

\subsubsection{The Constraints to be added to the Hamilton
Equations}

To the previous Hamilton equations we have to add the
super-Hamiltonian constraint (\ref{3.44}) [or (\ref{3.45})] for the
determination of $\tilde \phi(\tau ,\vec \sigma)$ (the Lichnerowicz
equation) and the super-momentum constraints (\ref{3.41}) [or
(\ref{3.42})] for the determination of the momenta $\pi_i^{(\theta
)}(\tau ,\vec \sigma)$ [or $\tilde \phi(\tau ,\vec \sigma)\,
\sigma_{(a)(b)}{|}_{a \not= b}(\tau ,\vec \sigma)$ of
Eq.(\ref{2.18})] or their explicit expression given in Ref.\cite{6}.
\medskip

Also the six rest-frame conditions contained in Eqs.(\ref{3.47})
have to be added.

\vfill\eject

\section{Gauges in ADM Canonical Gravity in the York Canonical Basis}

Let us consider some gauge fixings for the secondary
super-hamiltonian and super-momentum first class constraints
determining the gauge variables $\pi_{\tilde \phi}(\tau ,\vec \sigma
)$ and $\theta^n(\tau ,\vec \sigma )$ inside the family of Schwinger
time gauges.

As shown in Eqs.(\ref{4.7}),their preservation in time by using
$H_D$ will generate four secondary gauge fixings determining the
lapse and shift functions consistently with the chosen definition of
instantaneous 3-space (clock synchronization convention) and of
3-coordinates in it.

\bigskip

\subsection{ ADM 4-Coordinate Gauges}

These are not maximal slicing gauges with ${}^3K(\tau ,\vec \sigma )
\approx 0$, because they require $\sum_r\, {}^3K_{rr}(\tau, \vec
\sigma) \approx 0$. As shown in Refs.\cite{37} they are defined by
the following gauge fixings

\bea
 {}^3K(\tau ,\vec \sigma ) &=& {{12\pi\, G}\over {c^3}}\,
 \pi_{\tilde \phi}(\tau ,\vec \sigma ) \approx {{24\pi\, G}\over
 {c^3}}\, \Big({\tilde \phi}^{-1}\, \sum_{\bar a}\, {{\sum_a\,
 \gamma_{\bar aa}\, Q_a^{-2}}\over {\sum_b\, Q_b^{-2}}}\,
 \Pi_{\bar a} \Big)(\tau ,\vec \sigma ),\nonumber \\
 &&{}\nonumber \\
 \chi_r(\tau ,\vec \sigma ) &=& \Big[\sum_s\, \partial_s\, {}^3g_{rs}
 - {1\over 3}\, \partial_r\, \sum_s\, {}^3g_{ss}\Big](\tau ,\vec \sigma )
 =\nonumber \\
 &=& \Big[{\tilde \phi}^{2/3}\, \sum_a\, Q_a^2\, \Big(\sum_s\,
 \partial_s\, (V_{ra}\, V_{sa}) +\nonumber \\
 &+&2\, \sum_s\, ({1\over 3}\, {\tilde \phi}^{-1}\, \partial_s\, \tilde \phi +
 \sum_{\bar b}\, \gamma_{\bar ba}\, \partial_s\, R_{\bar b})\,
 V_{ra}\, V_{sa} -\nonumber \\
 &-& {2\over 3}\, ({1\over 3}\, {\tilde \phi}^{-1}\, \partial_r\, \tilde \phi +
 \sum_{\bar b}\, \gamma_{\bar ba}\, \partial_r\, R_{\bar b})
 \Big)\Big](\tau ,\vec \sigma)\, \approx 0.
  \label{5.1}
  \eea

\noindent The three equations determining the 3-coordinates have the
form $\sum_{as}\, F_{1a}(\tau ,\vec \sigma)\,\partial_s\,
T_{(rs)a}(\theta^n(\tau ,\vec \sigma)) + \sum_{as}\, F_{2as}(\tau
,\vec \sigma)\, T_{(rs)a}(\theta^n(\tau ,\vec \sigma)) + F_3(\tau
,\vec \sigma) = 0$ with $T_{(rs)a} = V_{ra}(\theta^n)\,
V_{sa}(\theta^n) = T_{(sr)a}$. Once $T_{(rs)a}(\tau ,\vec \sigma)$
has been found, one has to find the compatible  parameters
$\theta^n(\tau ,\vec \sigma )$ of the rotation matrix $V(\theta^n)$.

\subsection{The 3-Harmonic Gauges}

These are gauges in which only the 3-coordinates are fixed but not
the instantaneous 3-space

 \bea
  \pi_{\tilde \phi}(\tau ,\vec \sigma )
  &\approx& not\, specified,\nonumber \\
  &&{}\nonumber \\
  \sum_s\, \partial_s\, \Big({}^3e\, {}^3g^{rs}\Big)(\tau ,\vec \sigma ) &=&
  \sum_s\, \partial_s\, \Big({\tilde \phi}^{1/3}\, \sum_a\,
  Q_a^{-2}\, V_{ra}\, V_{sa}\Big)(\tau ,\vec \sigma )  =\nonumber \\
  &=& \Big[{\tilde \phi}^{1/3}\, \sum_{as}\, Q_a^{-2}\,
  \Big(\partial_s\, (V_{ra}\, V_{sa}) +\nonumber \\
  &+& ({1\over 3}\, {\tilde \phi}^{-1}\, \partial_s\, \tilde \phi -
  2\, \sum_{\bar b}\, \gamma_{\bar ba}\, \partial_s\, R_{\bar b})\, V_{ra}\, V_{sa}
  \Big)\Big](\tau ,\vec \sigma) \approx 0.
  \label{5.2}
  \eea

Again we have a linear partial differential equation for $T_{(rs)a}
= V_{ra}(\theta^n)\, V_{sa}(\theta^n) = T_{(sr)a}$.\medskip

A similar maximal slicing, but not 3-harmonic, gauge with
${}^3K(\tau ,\vec \sigma ) \approx 0$ has Eqs.(\ref{5.2}) replaced
with the gauge fixings $\sum_s\, \partial_s\, \Big[
\Big({}^3e\Big)^{2/3}\, {}^3g^{rs}\Big](\tau ,\vec \sigma ) \approx
0$.

\subsection{The 4-Harmonic Gauges}

The family of 4-harmonic gauges, see Eq.(6.4) of the second paper in
Ref.\cite{6}, is defined by the four gauge fixings $\chi^A(\tau
,\vec \sigma ) = \sum_B\, \partial_B\, \Big(\sqrt{|{}^4g|}\,
{}^4g^{AB}\Big)(\tau ,\vec \sigma ) \approx 0$. By using
Eqs.(\ref{2.6}), (\ref{2.10}), (\ref{4.1}) (the kinematical Hamilton
equation for $\tilde \phi$) and Eqs.(\ref{4.3}) (the
$\tau$-derivatives of the lapse and shift functions are Dirac
multipliers) we get the following expression for $\chi^{\tau}(\tau
,\vec \sigma) \approx 0$ in the York canonical basis

\bea
  \chi^{\tau} &=& \partial_{\tau}\, \Big((1 + n)\, {}^3e\, {}^4g^{\tau\tau}\Big) +
 \sum_s\,  \partial_s\, \Big((1 + n)\, {}^3e\, {}^4g^{\tau s}\Big) =\nonumber \\
 &=&  \sgn\, \Big[\partial_{\tau}\, {{\tilde \phi}\over {1 + n}} - \sum_{sa}\,
 \partial_s\, {{{\tilde \phi}^{2/3}\, Q_a^{-1}\, V_{sa}\, {\bar n}_{(a)}}\over
 {1 + n}}\Big] \approx 0,\nonumber \\
 &&{}\nonumber \\
 &&\Downarrow\nonumber \\
 &&{}\nonumber \\
 {{12\pi\, G}\over {c^3}}\, \pi_{\tilde \phi}(\tau ,\vec \sigma) &=&
 {}^3K(\tau ,\vec \sigma) \approx
 - {1\over {\Big(1 + n(\tau ,\vec \sigma)\Big)^2}}\,
 \Big(\lambda_n + {\tilde \phi}^{-1/3}\, \sum_{ra}\,
 Q_a^{-1}\, V_{ra}\, \partial_r\, n \Big)(\tau ,\vec \sigma).
 \nonumber \\
 &&{}
 \label{5.3}
 \eea

\medskip

The other three gauge fixings are

\bea
  \chi^r &=& \partial_{\tau}\, \Big((1 + n)\, {}^3e\, {}^4g^{\tau r}\Big) +
 \sum_s\, \partial_s\, \Big((1 + n)\, {}^3e\, {}^4g^{rs}\Big) =\nonumber \\
 &=& \sgn\, \Big[- \partial_{\tau}\, {{{}^3e\, n^r}\over {1 + n}} - \sum_s\,
 \partial_s\, \Big((1 + n)\, {}^3e\, ({}^3g^{rs} - {{n^r\, n^s}\over
 {(1 + n)^2}})\Big)\Big] =\nonumber \\
 &=& - \sgn\, \Big[\partial_{\tau}\, {{\phi^6\, \sum_a\, {}^3{\bar e}^r_{(a)}\, {\bar n}_{(a)}}\over
 {1 + n}}\Big] - \sgn\, \sum_s\, \partial_s\, \Big((1 + n)\, \phi^6\, \sum_{ab}\, {}^3{\bar e}^r_{(a)}\,
 {}^3{\bar e}_{(b)}^s\nonumber \\
 && \Big[\delta_{ab} - {{{\bar n}_{(a)}\, {\bar n}_{(b)}}\over {(1 + n)^2}}\Big]\Big)
 =\nonumber \\
 &&{}\nonumber \\
 &=& \Big(- {{{\tilde \phi}^{2/3}}\over {1 + n}}\, \sum_a\, Q_a^{-1}\,
 \Big(V_{ra}\, \lambda_{{\bar n}_{(a)}} +
 {\bar n}_{(a)}\, \Big[{{\partial\, V_{ra}}\over {\partial\, \theta^i}}\,
 \partial_{\tau}\, \theta^i +\nonumber \\
 &+& V_{ra}\, ({2\over 3}\, {\tilde \phi}^{-1}\,
 \partial_{\tau}\, \tilde \phi - \sum_{\bar b}\, \gamma_{\bar ba}\,
 \partial_{\tau}\, R_{\bar b}) - V_{ra}\,
 {{\lambda_n}\over {1 + n}}\Big]\Big) -\nonumber \\
 &-&{{{\tilde \phi}^{1/3}}\over {(1 + n)^2}}\, \sum_{sab}\, Q_a^{-1}\,
 Q_b^{-1}\, \Big(V_{ra}\, V_{sa}\,
 \Big[\Big((1 + n)^2\, \delta_{ab} + {\bar n}_{(a)}\, {\bar n}_{(b)}\Big)\,
 \partial_s\, n -\nonumber \\
 &-& (1 + n)\, \Big({\bar n}_{(a)}\, \partial_s\, {\bar n}_{(b)} + {\bar
 n}_{(b)}\, \partial_s\, {\bar n}_{(a)}\Big)\Big] +\nonumber \\
 &+& (1 + n)\, \Big((1 + n)^2\, \delta_{ab} - {\bar n}_{(a)}\,
 {\bar n}_{(b)}\Big)\, \Big[\partial_s\, (V_{ra}\, V_{sb})
 +\nonumber \\
 &+& V_{ra}\, V_{sb}\, ({1\over 3}\, {\tilde \phi}^{-1}\, \partial_s\, \tilde \phi
 - \sum_{\bar b}\, (\gamma_{\bar ba} + \gamma_{\bar bb})\,
 \partial_s\, R_{\bar b}) \Big]\,\,
 \Big)\,\,\Big)(\tau ,\vec \sigma) \approx 0.\nonumber \\
 &&{}
 \label{5.4}
 \eea

The kinematical Hamilton equations (\ref{4.1}), (\ref{4.5}) and
(\ref{4.8}) are needed to get this final form.

 \medskip

These unconventional Hamiltonian constraints ($\chi^{\tau} \approx
0$ does not define a constant mean curvature (CMC) gauge
${}^3K(\tau, \vec \sigma) = const.$ \footnote{Strictly speaking the
CMC condition with ${}^3K \not= 0$ does not define a space-like
Cauchy 3-surface but a hyperboloidal one, asymptotically tangent to
the future null infinity like the mass hyperboloids in Minkowski
space-time, as shown in Ref.\cite{38}.}) are four coupled equations
for $\pi_{\tilde \phi}$ and $\theta^i$ in terms of $\phi$, $R_{\bar
a}$, $\Pi_{\bar a}$, $n$, $\lambda_n =
\partial_{\tau}\, n$, ${\bar n}_{(a)}$, $\lambda_{\vec n (a)} =
\partial_{\tau}\, {\bar n}_{(a)}$.\bigskip

The stability of these gauge fixings requires to impose
$\partial_{\tau}\, \chi_a(\tau ,\vec \sigma ) \approx 0$ and
$\partial_{\tau}\, \chi_{\tau}(\tau ,\vec \sigma ) \approx 0$. In
this way we get four equations for the determination of $n$ and
${\bar n}_{(a)}$. But these are not equations of the "elliptic" type
like with ordinary gauge fixings. They are coupled equations
depending upon $n$, $\partial_r\, n$, $\partial_{\tau}\, n$,
$\partial^2_{\tau}\, n$ and ${\bar n}_{(a)}$, $\partial_r\, {\bar
n}_{(a)}$, $\partial_{\tau}\, {\bar n}_{(a)}$, $\partial^2_{\tau}\,
{\bar n}_{(a)}$, namely {\it hyperbolic} equations. As a consequence
there is a {\it problem of initial conditions not only for $R_{\bar
a}$ but also for the lapse and shift functions of the harmonic
gauge}. Each possible set of initial values should correspond to a
different completely fixed harmonic gauge, since once we have a
solution for $n$ and ${\bar n}_{(a)}$ the corresponding Dirac
multipliers are determined by taking their $\tau$-derivative.

\bigskip

Instead of reading these constraints as gauge fixings determining
$\pi_{\tilde \phi}(\tau, \vec \sigma)$, Eq.(\ref{5.3}), and
$\theta^n(\tau, \vec \sigma)$, Eq.(\ref{5.4}), let us solve them for
the Dirac multipliers $\lambda_n$ and $\lambda_{{\bar n}_{(a)}}$.
Then it is more natural the following interpretation: the family of
4-harmonic gauges is determined by all the Hamiltonian gauge fixings
of the type (\ref{4.6}) for $\pi_{\tilde \phi}(\tau, \vec \sigma)$
and $\theta^n(\tau, \vec \sigma)$, with induced secondary gauge
fixings (\ref{4.7}) for the lapse and shift functions, such that the
Dirac multipliers $\lambda_n(\tau, \vec \sigma)\, \cir \,
\partial_{\tau}\, n(\tau, \vec \sigma)$, $\lambda_{{\bar n}_{(a)}}(\tau,
\vec \sigma)\, \cir \, \partial_{\tau}\, {\bar n}_{(a)}(\tau, \vec
\sigma)$, have the  form implied by Eqs.(\ref{5.3}) and (\ref{5.4}).
\bigskip

As a consequence, let us remark that in the family of harmonic
gauges there is no natural way to visualize the instantaneous
3-spaces (they cannot be Euclidean due to the equivalence principle)
and to check which properties have an inertial origin due to the
freedom in the clock synchronization convention (the gauge freedom
in the York time). Most of the applications (for instance the IAU
conventions for the solar system \cite{39}) are based on
post-Newtonian expansions inside a non-specified 3-space (with
non-vanishing extrinsic curvature of order $O(c^{-2})$) of a
so-called quasi-inertial frame in Minkowski space-time containing
Newton gravity as the zero order!
\medskip

Let us also remark that the harmonic gauges are a particular case of
the gauges used in numerical relativity, see for instance Refs.
\cite{40,41} and their bibliography. In the 3+1 formulation of
numerical relativity these gauges are chosen as conditions leading
to second order hyperbolic evolution equations for the lapse and
shift functions. Examples of these gauge conditions are
$(\partial_{\tau} - n^r(\tau, \vec \sigma)\, \partial_r)\,
\sqrt{\sgn\, {}^4g_{\tau\tau}(\tau, \vec \sigma)} = (f\, \sgn\,
{}^4g_{\tau\tau}\, {}^3K)(\tau, \vec \sigma)$ (for $f = 1$ it is the
harmonic time condition) and $(\partial_{\tau} - n^s(\tau, \vec
\sigma)\, \partial_s)\, n^r(\tau, \vec \sigma) = (\mu\, {\tilde
\Gamma}^r - \eta\, n^r)(\tau, \vec \sigma)$, where $n^r = {}^3{\bar
e}^r_{(a)}\, {\bar n}_{(a)}$ and ${\tilde \Gamma}^r$ is the
3-dimensional conformal connection in the 3-space  $\Sigma_{\tau}$.
Like the harmonic gauges they are not natural in the Hamiltonian
formulation based upon the York canonical basis.

\subsection{The Synchronous Gauges}

The synchronous gauges are defined by the gauge fixings

\beq
 {\bar n}_{(a)}(\tau ,\vec \sigma ) \approx 0,
 \label{5.5}
 \eeq

\noindent implying ${}^4g_{\tau r}(\tau ,\vec \sigma ) \approx 0$
(no gravito-magnetism) and ${}^4g_{\tau\tau}(\tau ,\vec \sigma )
\approx \sgn\, \Big(1 + n(\tau ,\vec \sigma )\Big)^2$,
$\lambda_{{\bar n}_(a)}(\tau,\vec \sigma ) \approx 0$.

These unconventional gauges have a residual gauge freedom in any
fixation of the 3-coordinates implying $\lambda_{{\bar n}_(a)}(\tau
 ,\vec \sigma ) \approx 0$. If we add the "comoving" condition
$n(\tau, \vec \sigma) \approx 0$, we are also restricting the
freedom in the fixation of the York time.

These gauges are used in cosmology in presence of Killing vectors
implying that the solution of Einstein's equations are homogeneous
and isotropic. However their use in absence of Killing symmetries is
questionable, since gravito-magnetism cannot be eliminated in
general.

\vfill\eject

\section{The 3-Orthogonal Schwinger Time Gauges.}

Let us now consider the most natural family of Schwinger time gauges
in the framework of the York canonical basis, i.e. the 3-orthogonal
gauges

\beq
 \theta^i(\tau ,\vec \sigma )\, \approx\, 0,
 \label{6.1}
 \eeq

\noindent in which the 3-metric is diagonal in each point of the
instantaneous 3-space $\Sigma_{\tau}$ \footnote{We are not aware of
global obstructions forbidding these gauges.} and $V_{ra}(\theta^n)
= \delta_{ar}$. The residual gauge freedom, parametrizing the
family, is the selection of the instantaneous 3-spaces by fixing the
gauge variable $\pi_{\tilde \phi}(\tau, \vec \sigma)$. These gauges
are defined on the 3-space and do not need initial conditions in the
past like the harmonic ones.
\medskip

Since, as said after Eqs.(\ref{2.8}), it is convenient to choose the
parameters $\theta^i(\tau, \vec \sigma)$ as first kind coordinates
for the O(3) group manifold, we get the following results: 1)
$V_{(i)ra} = {{\partial\, V_{ra}(\theta^n)}\over {\partial\,
\theta^i}}{|}_{\theta^i = 0} = \epsilon_{ira}$; 2) $B_{(i)ab} =
{{\partial\, B_{ab}(\theta^n)}\over {\partial\,
\theta^i}}{|}_{\theta^i = 0} =  {1\over 2}\, \epsilon_{iab}$; 3)
$A_{(i)ab} = {{\partial\, A_{ab}(\theta^n)}\over {\partial\,
\theta^i}}{|}_{\theta^i = 0} = - {1\over 2}\, \epsilon_{iab}$; for
the derivatives of $V_{ra}(\theta^n)$ and of the Cartan matrices
$B_{ab}(\theta^n)$ and $A = B^{-1}$ in the 3-orthogonal gauges. By
using these quantities the results of Appendix B can be restricted
to the 3-orthogonal gauges. This restriction is given in Appendix C.

\medskip

In the 3-orthogonal gauges Eqs.(\ref{2.18}) become

\bea
  \Pi_{\bar a} &=& - {{c^3}\over {8\pi\, G}}\, \tilde \phi\, \sum_a\,
 \gamma_{\bar aa}\, \sigma_{(a)(a)},\nonumber \\
 \pi^{(\theta)}_i{|}_{\theta^i = 0} &=& {{c^3}\over {8\pi\, G}}\,
 \sum_{ab}\, \epsilon_{iab}\, \tilde \phi\, \sigma_{(a)(b)}{|}_{a \not=
 b}\, Q_a\, Q_b^{-1},\nonumber \\
 &&\qquad \tilde \phi\, \sigma_{(a)(b)}{|}_{a \not= b} = -  {{8\pi\,
G}\over {c^3}}\, \sum_i\, {{\epsilon_{abi}\, \pi_i^{(\theta)}}\over
{Q_b\, Q_a^{-1} - Q_a\, Q_b^{-1}}},\nonumber \\
 &&{}\nonumber \\
 {}^3K_{rs}{|}_{\theta^i = 0} &=&  {\tilde \phi}^{2/3}\, \Big[Q_r\, Q_s\,
 \sigma_{(r)(s)}{|}_{r \not= s} -
  {{4\pi\, G}\over {c^3}}\, \delta_{rs}\, Q^2_r\, \Big(2\, {\tilde \phi}^{-1}\,
 \sum_{\bar a}\, \gamma_{\bar ar}\, \Pi_{\bar a} - \pi_{\tilde
 \phi}\Big)\Big].
 \label{6.2}
 \eea

\subsection{The Weak ADM Energy}

The weak ADM energy of Eq.(\ref{3.45}) becomes

  \begin{eqnarray*}
 {\hat E}_{ADM}{|}_{\theta^i = 0} &=& c\, \int d^3\sigma\,
 \Big[{\check {\cal M}} {|}_{\theta^i = 0} -
  {{c^3}\over {16\pi\, G}}\, {\cal S}{|}_{\theta^i = 0} +
  \nonumber \\
  &+&  {{2\pi\, G}\over {c^3}}\, {\tilde \phi}^{-1}\, \Big(
 - 3\, (\tilde \phi\, \pi_{\tilde \phi})^2 + 2\, \sum_{\bar b}\,
 \Pi^2_{\bar b} + 2\, \sum_{abij}\, {{\epsilon_{abi}\, \epsilon_{abj}\,
 \pi_i^{(\theta )}\, \pi_j^{(\theta )}}\over {\Big[Q_a\, Q^{-1}_b - Q_b\,
 Q^{-1}_a \Big]^2}}\Big)\,\, \Big](\tau ,\vec \sigma ) =\nonumber \\
 \end{eqnarray*}

\bea
 &=& c\, \int d^3\sigma\,
 \Big[{\check {\cal M}} {|}_{\theta^i = 0} -
  {{c^3}\over {16\pi\, G}}\, {\cal S}{|}_{\theta^i = 0} +
  \nonumber \\
  &+& {{4\pi\, G}\over {c^3}}\, {\tilde \phi}^{-1}\, \sum_{\bar a}\,
  \Pi^2_{\bar a} + {{c^4}\over {16\pi\, G}}\, \tilde \phi\, \sum_{a
  \not= b}\, \sigma^2_{(a)(b)} - {{6\pi\, G}\over {c^3}}\, \tilde
  \phi\, \pi^2_{\tilde \phi} \Big](\tau ,\vec \sigma ),\nonumber \\
 &&{}
 \label{6.3}
 \eea

\noindent where Eq.(\ref{6.2}) has been used and Eq. (\ref{c8}) has
to be used.

\bigskip

The mass density appearing in Eq.(\ref{6.3}) has the following
expression implied by Eqs.(\ref{3.37})

 \bea
 {\check {\cal M}}(\tau ,\vec \sigma ){|}_{\theta^i =0}
 &=& \sum_{i=1}^N\, \delta^3(\vec \sigma
 ,{\vec \eta}_i(\tau ))\,\, \eta_i\, \sqrt{m_i^2\, c^2 +
 \phi^{-4}\, \sum_a\, Q_a^{-2}\,  \Big({\check
 \kappa}_{ia}(\tau ) - {{Q_i}\over c}\, A_{\perp\, a}\Big)^2\,}
 (\tau ,\vec \sigma ) +\nonumber \\
 &+&\tilde \phi\, {\check T}^{(em)}_{\perp\perp}(\tau ,\vec \sigma) + {\cal
 W}_{(n)}(\tau ,\vec \sigma),\nonumber \\
  &&{}\nonumber \\
  &&{}\nonumber \\
  {\check T}^{(em)}_{\perp\perp}(\tau ,\vec \sigma) &=& {\tilde
  \phi}^{-4/3}(\tau ,\vec \sigma)\, \Big[{1\over {2c}}\, \sum_{rsa}\, Q_a^2\,
  \delta_{ra}\, \delta_{sa}\, \pi^r_{\perp}\,  \pi^s_{\perp} +
   {1\over {4c}}\, \sum_{ab}\, Q_a^{-2}\, Q_b^{-2}\,  F_{ab}\, F_{ab}\Big](\tau ,\vec
  \sigma),\nonumber \\
  {\cal W}_{(n)}(\tau ,\vec \sigma) &=& - {1\over {2c}}\,
  \Big[{\tilde \phi}^{-1/3}\, \sum_{rsan}\, Q_a^2\, \delta_{ra}\, \delta_{sa}\,
  \Big(2\, \pi^r_{\perp} - \sum_m\, \delta^{rm}\, \sum_{i=1}^N\, Q_i\, \eta_i\, {{\partial\,
  c(\vec \sigma , {\vec \eta}_i(\tau))}\over {\partial\, \sigma^m}}\Big)
  \nonumber \\
  &&\delta^{sn}\,\sum_{j=1}^N\, Q_j\, \eta_j\, {{\partial\,
  c(\vec \sigma , {\vec \eta}_j(\tau))}\over {\partial\,
  \sigma^n}}\Big](\tau ,\vec \sigma),
   \label{6.4}
 \eea

\subsection{The Super-Hamiltonian and Super-Momentum Constraints}

From Eqs.(\ref{3.44}) (or (\ref{3.45})) and (\ref{3.42}) the
constraints determining $\tilde \phi$ and $\pi_i^{(\theta)}$ are
(Eqs.(\ref{c12}), (\ref{c20}) and (\ref{6.3}) are also needed)

 \bea
 {\cal H}(\tau ,\vec \sigma ){|}_{\theta^i = 0} &=&
  {{c^3}\over {16\pi\, G}}\, {\tilde \phi}^{1/6} (\tau ,\vec \sigma ) \,
 [ 8\,   {\hat \triangle}\, {\tilde \phi}^{1/6} -
  {}^3{\hat R}{|}_{\theta^i = 0}\, {\tilde \phi}^{1/6}](\tau ,\vec \sigma )
 + {\check {\cal M}}{|}_{\theta^i = 0}(\tau ,\vec \sigma ) +\nonumber \\
 &&{}\nonumber \\
  &+&  {{2\pi\, G}\over {c^3}}\, {\tilde \phi}^{-1}\, \Big[
- 3\, (\tilde \phi\, \pi_{\tilde \phi})^2 + 2\, \sum_{\bar b}\,
 \Pi^2_{\bar b} + 2\, \sum_{abij}\, {{\epsilon_{abi}\, \epsilon_{abj}\,
 \pi_i^{(\theta )}\, \pi_j^{(\theta )}}\over {\Big[Q_a\, Q^{-1}_b - Q_b\,
 Q^{-1}_a \Big]^2}}\Big](\tau ,\vec \sigma ) =\nonumber \\
 &&{}\nonumber \\
 &=&  {{c^3}\over {2\pi\, G}}\, \phi^{-6}(\tau ,\vec \sigma )\, \Big(
 \phi^7\, ({\hat \triangle}{|}_{\theta^i = 0} -
 {1\over 8}\, {}^3\hat R{|}_{\theta^i = 0})\, \phi +
 {{2\pi\, G}\over {c^3}}\, \phi^6\, {\check {\cal M}}{|}_{\theta^i = 0}
 +\nonumber \\
 &+& {{8\, \pi^2\, G^2}\over {c^6}}\, \sum_{\bar a}\, \Pi^2_{\bar a}
 + {1\over 8}\, \phi^{12}\, \sum_{ab, a\not= b}\, \sigma_{(a)(b)}\,
 \sigma_{(a)(b)} - {{12\, \pi^2\, G^2}\over {c^6}}\, \phi^{12}\,
 \pi^2_{\tilde \phi} \Big)(\tau ,\vec \sigma ) \approx 0,\nonumber \\
 \label{6.5}
 \eea

 \bea
  {\tilde {\bar {\cal H}}}_{(a)}{|}_{\theta^i = 0}(\tau ,\vec \sigma) &=&
 \phi^{-2}(\tau, \vec \sigma)\, \Big[\sum_{b \not= a}\,
 \sum_i\, {{ \epsilon_{abi}\, Q_b^{-1}}
 \over {Q_b\, Q_a^{-1} - Q_a\, Q_b^{-1}}}\, \partial_b\, \pi_i^{(\theta)}
 +\nonumber \\
 &+&2\, \sum_{b \not= a}\, \sum_i\,  {{ \epsilon_{abi}\, Q_a^{-1}}
 \over {\Big(Q_b\, Q_a^{-1} - Q_a\, Q_b^{-1}\Big)^2}}\,
  \sum_{\bar c}\, (\gamma_{\bar ca} - \gamma_{\bar cb})\, \partial_b\,
  R_{\bar c}\,\, \pi_i^{(\theta)} +\nonumber \\
 &&{}\nonumber \\
 &+&Q_a^{-1}\,  \Big(\phi^6\, \partial_a\, \pi_{\tilde \phi}
 +  \sum_{\bar b}\, (\gamma_{\bar ba}\, \partial_a\, \Pi_{\bar b}
 - \partial_a\, R_{\bar b}\, \Pi_{\bar b}) + {\check {\cal M}}_{a}\Big)
 \Big](\tau ,\vec \sigma) =\nonumber \\
 &&{}\nonumber \\
 &=& - {{c^3}\over {8\pi\, G}}\,  {\tilde \phi}^{2/3}(\tau, \vec
 \sigma)\, \Big(\sum_{b \not= a}\, Q_b^{-1}\, \Big[\partial_b\,
 \sigma_{(a)(b)} +\nonumber \\
 &+& \Big({\tilde \phi}^{-1}\, \partial_b\, \tilde \phi +
 \sum_{\bar b}\, (\gamma_{\bar ba} - \gamma_{\bar bb})\, \partial_b\,
 R_{\bar b}\Big)\, \sigma_{(a)(b)}\Big] -\nonumber \\
 &-& {{8\pi\, G}\over {c^3}}\, {\tilde \phi}^{-1}\,
 Q_a^{-1}\,  \Big[\tilde \phi\, \partial_a\, \pi_{\tilde \phi}
 +  \sum_{\bar b}\, (\gamma_{\bar ba}\, \partial_a\, \Pi_{\bar b}
 - \partial_a\, R_{\bar b}\, \Pi_{\bar b}) + {\check {\cal
 M}}_{a}\Big]
 \Big)(\tau ,\vec \sigma) \quad \approx 0.\nonumber \\
 &&{}
 \label{6.6}
 \eea

\noindent To get the second form of the super-momentum constraints
we used Eq.(\ref{6.2}), which implies $\partial_r\,
\pi_i^{(\theta)}{|}_{\theta^n = 0} = {{c^3}\over {8\pi\, G}}\,
\tilde \phi\, \sum_{a \not= b}\, \epsilon_{iab}\, Q_a\, Q_b^{-1}\,
\Big[\partial_r\, \sigma_{(a)(b)} + \Big({\tilde \phi}^{-1}\,
\partial_r\, \tilde \phi + \sum_{\bar b}\, (\gamma_{\bar ba} -
\gamma_{\bar bb})\, \partial_r\, R_{\bar b}\Big)\,
\sigma_{(a)(b)}\Big]$. For the solution of Eq.(\ref{6.6}) the
results of Ref.\cite{6} are needed. The mass current density
${\check {\cal M}}_r$ is given in Eqs.(\ref{3.37}).

\subsection{The Contracted Bianchi Identities}

From Eqs. (\ref{4.1}) and (\ref{4.2}) the Hamilton equations for the
unknowns in the constraints are (the first is the  Raychaudhuri
equation)

\bea
 \partial_{\tau}\, {\tilde \phi} (\tau ,\vec \sigma ){|}_{\theta^i = 0} &\cir&
  \Big[ - {{12\pi\, G}\over { c^3}}\, (1 + n)\, \tilde \phi\,
 \pi_{\tilde \phi} +   {\tilde \phi}^{2/3}\,  \sum_a\, Q_a^{-1}\,
 \Big(\partial_a\, {\bar n}_{(a)} +\nonumber \\
 &+& {\bar n}_{(a)} \,  \Big({2\over 3}\,
 {\tilde \phi}^{-1}\, \partial_a\, \tilde \phi - \sum_{\bar b}\,
 \gamma_{\bar ba}\, \partial_a\, R_{\bar b}\Big)
  \Big)\Big](\tau ,\vec \sigma ),\nonumber \\
 &&{}
 \label{6.7}
 \eea

\begin{eqnarray*}
 \partial_{\tau}\, \pi_i^{(\theta )}(\tau ,\vec \sigma ){|}_{\theta^i = 0} &\cir&
  - \int d^3\sigma_1\, \Big[(1 + n)(\tau, {\vec \sigma}_1)\,
 \Big( {{\delta\, {\check {\cal M}}(\tau ,{\vec \sigma}_1)}\over {\delta\,
 \theta^i(\tau ,\vec \sigma)}}{|}_{\theta^i = 0} - {{c^3}\over {16\pi\, G}}\,
 {{\delta\, {\cal S}(\tau ,{\vec \sigma}_1)}\over
 {\delta\, \theta^i(\tau, \vec \sigma)}}{|}_{\theta^i = 0}\Big) -\nonumber \\
 &-& {{c^3}\over {16\pi\, G}}\, n(\tau, {\vec \sigma}_1)\,
 {{\delta\, {\cal T}(\tau ,{\vec \sigma}_1)}\over
 {\delta\, \theta^i(\tau, \vec \sigma)}}{|}_{\theta^i = 0} \Big] -\nonumber \\
 &-& {{8\pi\, G}\over {c^3}}\, \Big((1 + n)\, {\tilde \phi}^{-1}\,
 \sum^{a \not= b}_{abtkj}\, {{\epsilon_{abt}\, \epsilon_{abj}\,
 (V_{(i)kt} + B_{(i)kt})\, \pi_k^{(\theta)}\, \pi_j^{(\theta)}}\over
 {(Q_a\, Q_b^{-1} - Q_b\, Q_a^{-1})^2}} \Big)(\tau ,\vec \sigma) +\nonumber \\
 &+& \int d^3\sigma_1\, {\bar n}_{(a)}(\tau,{\vec \sigma}_1)\,
 {{\delta\, {\bar {\tilde {\cal H}}}_{(a)}
 (\tau, {\vec \sigma}_1)}\over {\delta\, \theta^i(\tau ,\vec \sigma)}}{|}_{\theta^i = 0},
 \nonumber \\
 &&{}\nonumber \\
 &&\Downarrow
 \end{eqnarray*}

\bea
 \partial_{\tau}\, \sigma_{(a)(b)}{|}_{a \not= b, \theta^i=0} &\cir&
 - \Big[\Big( {\tilde \phi}^{-1}\, \partial_{\tau}\, \tilde \phi +
 {{Q_b\, Q_a^{-1} + Q_a\, Q_b^{-1}}\over {Q_b\, Q_a^{-1} - Q_a\, Q_b^{-1}}}\,
 \sum_{\bar b}\, (\gamma_{\bar ba} - \gamma_{\bar bb})\, \partial_{\tau}\,
 R_{\bar b} \Big)\, \sigma_{(a)(b)}{|}_{a \not= b}\Big](\tau ,\vec \sigma)
 -\nonumber \\
 &-& {{8\pi\, G}\over {c^3}}\, \Big({\tilde \phi}^{-1}\, \sum_i\,
 {{\epsilon_{abi}}\over {Q_b\, Q_a^{-1} - Q_a\, Q_b^{-1}}}\Big)(\tau ,\vec \sigma)
 \nonumber \\
 &&\Big[ - \int d^3\sigma_1\, \Big[(1 + n)(\tau, {\vec \sigma}_1)\,
 \Big( {{\delta\, {\check {\cal M}}(\tau ,{\vec \sigma}_1)}\over {\delta\,
 \theta^i(\tau ,\vec \sigma)}}{|}_{\theta^i = 0} - {{c^3}\over {16\pi\, G}}\,
 {{\delta\, {\cal S}(\tau ,{\vec \sigma}_1)}\over
 {\delta\, \theta^i(\tau, \vec \sigma)}}{|}_{\theta^i = 0}\Big) -\nonumber \\
 &-& {{c^3}\over {16\pi\, G}}\, n(\tau, {\vec \sigma}_1)\,
 {{\delta\, {\cal T}(\tau ,{\vec \sigma}_1)}\over
 {\delta\, \theta^i(\tau, \vec \sigma)}}{|}_{\theta^i = 0} \Big] -\nonumber \\
 &-&{{c^3}\over {8\pi\, G}}\, \Big((1 + n)\, \tilde \phi\,
 \sum_{m \not= n}\, {{\sigma_{(m)(n)}}\over {Q_m\, Q_n^{-1} - Q_n\, Q_m^{-1}}}
 \nonumber \\
 &&\sum_{c \not= d}\, \sum_{tk}\, \epsilon_{mnt}\, (V_{(i)kt} + B_{(i)kt})\,
 \epsilon_{kcd}\, Q_c\, Q_d^{-1}\, \sigma_{(c)(d)} \Big)(\tau ,\vec \sigma) +\nonumber \\
 &+& \int d^3\sigma_1\, {\bar n}_{(a)}(\tau,{\vec \sigma}_1)\,
 {{\delta\, {\bar {\tilde {\cal H}}}_{(a)}
 (\tau, {\vec \sigma}_1)}\over {\delta\, \theta^i(\tau ,\vec
 \sigma)}}{|}_{\theta^i = 0} \Big].
 \label{6.8}
 \eea

Eqs.(\ref{c21}), (\ref{c11}), (\ref{c19}) and (\ref{c5}) are needed.

\subsection{The Shift Functions}

By using Eq.(\ref{4.5}) for the $\tau$-preservation of the gauge
fixings (\ref{6.1}), we get the following equations for the shift
functions (the second expression uses Eqs.(\ref{6.2}))

\bea
 &&\Big[\sum_{ab}\, {{\epsilon_{abi}}\over {Q_a\, Q_b^{-1} - Q_b\,
 Q_a^{-1}}}\, \Big(- {\tilde \phi}^{-1/3}\, Q_b^{-1}\, \Big[({1\over 3}\, {\tilde
 \phi}^{-1}\, \partial_b\, \tilde \phi + \sum_{\bar b}\, \gamma_{\bar bb}\,
 \partial_b\, R_{\bar b})\, {\bar n}_{(a)} - \partial_b\,
 {\bar n}_{(a)}\Big] -\nonumber \\
 &-& {{8\pi\, G}\over {c^3}}\, (1 + n)\, {\tilde \phi}^{-1}\,
 \sum_j\, {{\epsilon_{abj}\, \pi_j^{(\theta)}}\over {Q_a\, Q_b^{-1} -
 Q_b\, Q_a^{-1}}} \Big) \Big](\tau ,\vec \sigma) =\nonumber \\
 &&{}\nonumber \\
 &=& \Big[\sum_{ab}\, {{\epsilon_{abi}}\over {Q_a\, Q_b^{-1} - Q_b\,
 Q_a^{-1}}}\, \Big(- {\tilde \phi}^{-1/3}\, Q_b^{-1}\, \Big[({1\over 3}\, {\tilde
 \phi}^{-1}\, \partial_b\, \tilde \phi + \sum_{\bar b}\, \gamma_{\bar bb}\,
 \partial_b\, R_{\bar b})\, {\bar n}_{(a)} - \partial_b\,
 {\bar n}_{(a)}\Big] -\nonumber \\
 &-& (1 + n)\, \sigma_{(a)(b)}\Big)\, \Big](\tau ,\vec \sigma)
   \approx 0.\nonumber \\
 &&{}
 \label{6.9}
  \eea

\medskip

By saturating with $\epsilon_{cdi}$ we get for $a \not= b$

\bea
 &&\Big(Q_b^{-1}\, \partial_b\, {\bar n}_{(a)} + Q_a^{-1}\, \partial_a\,
 {\bar n}_{(b)} - \Big[Q_b^{-1}\,  \Big({1\over 3}\, {\tilde \phi}^{-1}\, \partial_b\,
 \tilde \phi + \sum_{\bar a}\, \gamma_{\bar aa}\, \partial_b\, R_{\bar a}\Big)\,
 {\bar n}_{(a)} +\nonumber \\
 &+& Q_a^{-1}\,  \Big({1\over 3}\, {\tilde \phi}^{-1}\, \partial_a\, \tilde \phi +
 \sum_{\bar a}\, \gamma_{\bar ab}\, \partial_a\, R_{\bar a}\Big)\,
 {\bar n}_{(b)}\Big]\Big)(\tau ,\vec \sigma) \approx\nonumber \\
 &\approx&  2\, \Big[{\tilde \phi}^{1/3}\, (1 + n)\,
 \sigma_{(a)(b)}{|}_{a \not= b}\Big](\tau ,\vec \sigma).
 \label{6.10}
 \eea

This is the final form of the equations for the shift functions.

\subsection{The Instantaneous 3-Space and the Lapse Functions}

If we recall from Eq.(\ref{2.9}) the definition $\pi_{\tilde \phi} =
{{c^3}\over {12\pi\, G}}\, {}^3\tilde K$, we can restrict ourselves
to the family of gauges

\beq
 \pi_{\tilde \phi}(\tau ,\vec \sigma ) - {{c^3}\over {12\pi\, G}}\,
 F(\tau ,\vec \sigma )\, \approx\, 0,\qquad
 \Rightarrow \,\, {}^3K(\tau ,\vec \sigma ) \approx F(\tau ,\vec
 \sigma ),
 \label{6.11}
 \eeq

\noindent where $F(\tau ,\vec \sigma )$ is a numerical function
independent from the canonical variables, which describes the
relativistic inertial effects associated with the freedom in the
clock synchronization (they do not exist in Newton theory in Galilei
space-times). For $F(\tau ,\vec \sigma ) = const.$ we have the CMC
gauges.\bigskip

By using Eqs.(\ref{4.4}) and (\ref{6.2}) restricted to the
3-orthogonal gauges, the preservation in $\tau$ of the gauge fixing
constraint (\ref{6.11}) gives the following equation for the
determination of the lapse function

\bea
 && {1\over 6}\, {\tilde \phi}^{-5/6}(\tau ,\vec \sigma)\,
 \int d^3\sigma_1\, \Big[(1 + n)(\tau, {\vec \sigma}_1)\,
 \Big( {{\delta\, {\check {\cal M}}(\tau ,{\vec \sigma}_1)}\over {\delta\,
 \phi(\tau ,\vec \sigma)}}{|}_{\theta^i=0} - {{c^3}\over {16\pi\, G}}\,
 {{\delta\, {\cal S}(\tau ,{\vec \sigma}_1)}\over
 {\delta\, \phi(\tau, \vec \sigma)}}{|}_{\theta^i=0}\Big) -\nonumber \\
 &&\qquad - {{c^3}\over {16\pi\, G}}\, n(\tau, {\vec \sigma}_1)\,
 {{\delta\, {\cal T}(\tau ,{\vec \sigma}_1)}\over
 {\delta\, \phi(\tau, \vec \sigma)}}{|}_{\theta^i=0} \Big] -
 \nonumber \\
 &-&  {{2\pi\, G}\over {c^3}}\, (1 + n)(\tau ,\vec \sigma)\, \Big[
 3\, \pi^2_{\tilde \phi} + 2\, {\tilde \phi}^{-2}\,
  \Big(\sum_{\bar b}\, \Pi^2_{\bar b}
 + \sum_{abij}^{a\not= b}\, {{\epsilon_{abi}\, \epsilon_{abj}\,
 \pi_i^{(\theta)}\, \pi_j^{(\theta)} }\over {\Big[Q_a\, Q^{-1}_b - Q_b\,
 Q^{-1}_a \Big]^2}}\, \Big)\Big](\tau ,\vec \sigma) -\nonumber \\
 &-& \Big({\tilde \phi}^{-1/3}\, \sum_a\, {\bar n}_{(a)}\, Q_a^{-1}\,
  \partial_a\, \pi_{\tilde \phi}\Big)(\tau ,\vec \sigma) =\nonumber  \\
 &&{}\nonumber \\
 &=&{1\over 6}\, {\tilde \phi}^{-5/6}(\tau ,\vec \sigma)\,
 \int d^3\sigma_1\, \Big[(1 + n)(\tau, {\vec \sigma}_1)\,
 \Big( {{\delta\, {\check {\cal M}}(\tau ,{\vec \sigma}_1)}\over {\delta\,
 \phi(\tau ,\vec \sigma)}}{|}_{\theta^i=0} - {{c^3}\over {16\pi\, G}}\,
 {{\delta\, {\cal S}(\tau ,{\vec \sigma}_1)}\over
 {\delta\, \phi(\tau, \vec \sigma)}}{|}_{\theta^i=0}\Big) -\nonumber \\
 &&\qquad - {{c^3}\over {16\pi\, G}}\, n(\tau, {\vec \sigma}_1)\,
 {{\delta\, {\cal T}(\tau ,{\vec \sigma}_1)}\over
 {\delta\, \phi(\tau, \vec \sigma)}}{|}_{\theta^i=0} \Big] -
 \nonumber \\
 &-&  {{2\pi\, G}\over {c^3}}\, (1 + n)(\tau ,\vec \sigma)\, \Big[
 3\, \pi^2_{\tilde \phi} + 2\, \Big({\tilde \phi}^{-2}\,
 \sum_{\bar b}\, \Pi^2_{\bar b}
 + {{c^6}\over {64 \pi^2\, G^2}}\,  \sum_{a \not=
 b}\, \sigma^2_{(a)(b)}\Big)\Big](\tau ,\vec \sigma) -\nonumber \\
 &-& \Big({\tilde \phi}^{-1/3}\, \sum_a\, {\bar n}_{(a)}\, Q_a^{-1}\,
  \partial_a\, \pi_{\tilde \phi}\Big)(\tau ,\vec \sigma)   \approx\nonumber \\
 &\approx&  - {{c^3}\over {12\pi\, G}}\,
 {{\partial\, F(\tau ,\vec \sigma )} \over {\partial\, \tau}}.
 \label{6.12}
 \eea

\noindent where Eqs. (\ref{c22}), (\ref{c10}) and (\ref{c17}) have
to be used.\medskip

Note that this equation depends also upon the shift functions.
\medskip

Eqs.(\ref{6.10}) and (\ref{6.12}) are 4 coupled equations for the
lapse and shift functions.

\subsection{The Equations of Motion}

\subsubsection{The Tidal Variables}

The second order equations of motion(\ref{4.11}) for the tidal
variables $R_{\bar a}$ and Eq.(\ref{4.9}) for the momentum
$\Pi_{\bar a}$ become respectively

 \begin{eqnarray*}
 \partial^2_{\tau}\, R_{\bar a}(\tau ,\vec \sigma ) &\cir&
 \Big[{\tilde \phi}^{-1/3}\, \sum_a\,  Q_a^{-1}\,  {\bar
 n}_{(a)}\, \sum_{\bar b}\, (\gamma_{\bar aa}\, \gamma_{\bar ba}
 - \delta_{\bar a\bar b})\, \partial_a\,
 \partial_{\tau}\, R_{\bar b} +\nonumber \\
 &+&{\tilde \phi}^{-1/3}\, \sum_a\, Q_a^{-1}\, \Big[\Big(\gamma_{\bar aa}\,
 ({1\over 3}\, {\tilde \phi}^{-1}\, \partial_a \tilde \phi + \sum_{\bar c}\,
 \gamma_{\bar ca}\, \partial_a\, R_{\bar c}) -  \partial_a\,
 R_{\bar a} \Big)\, {\bar n}_{(a)} -\nonumber \\
 &-& \gamma_{\bar aa}\,  \partial_a\,
 {\bar n}_{(a)}\Big]\, \sum_{\bar b}\, \gamma_{\bar
 ba}\, \partial_{\tau}\, R_{\bar b} -\nonumber \\
 &-& {\tilde \phi}^{-1}\, \Big[\partial_{\tau}\, R_{\bar a} + {2\over 3}\,
 {\tilde \phi}^{-1/3}\, \sum_a\, Q_a^{-1}\, \Big(\Big[\gamma_{\bar aa}\,
 (- {1\over 6}\, {\tilde \phi}^{-1}\, \partial_a\, \tilde \phi +
 \sum_{\bar b}\, \gamma_{\bar ba}\, \partial_a\, R_{\bar b}) -\nonumber \\
 &-& \partial_a\, R_{\bar a}\Big]\, {\bar n}_{(a)} - \gamma_{\bar aa}\,
  \partial_a\, {\bar n}_{(a)}\Big)\Big]\, \partial_{\tau}\, \tilde
 \phi -\nonumber \\
 &-& {1\over 3}\, {\tilde \phi}^{-4/3}\, \sum_a\, \gamma_{\bar aa}\,
 Q_a^{-1}\,  {\bar n}_{(a)}\, \partial_a\, \partial_{\tau}\,
 \tilde \phi +\nonumber \\
 &+&\Big[ \partial_{\tau}\, R_{\bar a} + {\tilde \phi}^{-1/3}\, \sum_a\,
 Q_a^{-1}\, \Big(\Big[\gamma_{\bar aa}\,  ({1\over 3}\, {\tilde
 \phi}^{-1}\, \partial_a\, \tilde \phi + \sum_{\bar b}\, \gamma_{\bar ab}\,
 \partial_a\, R_{\bar b}) -\nonumber \\
 &-&  \partial_a\, R_{\bar a} \Big]\, {\bar n}_{(a)} -
 \gamma_{\bar aa}\,  \partial_a\, {\bar n}_{(a)}\Big)\Big]\,
 {{\partial_{\tau}\, n}\over {1 + n}} -\nonumber \\
 &-& {\tilde \phi}^{-1/3}\, \sum_a\, Q_a^{-1}\, \Big(\Big[\gamma_{\bar aa}\,
 ({1\over 3}\, {\tilde \phi}^{-1}\, \partial_a\, \tilde \phi + \sum_{\bar b}\,
 \gamma_{\bar ba}\, \partial_a\, R_{\bar b}) -  \partial_a\,
 R_{\bar a}\Big]\, \partial_{\tau}\, {\bar n}_{(a)} -\nonumber \\
 &-&\gamma_{\bar aa}\,  \partial_a\, \partial_{\tau}\, {\bar n}_{(a)}
  \Big)\Big](\tau ,\vec \sigma ) +\nonumber \\
 &&{}\nonumber \\
 &+& {1\over 2}\, \Big({\tilde \phi}^{-1}\, (1 + n)\Big)(\tau ,\vec \sigma)\, \int
 d^3\sigma_1\, \Big[(1 + n)(\tau, {\vec \sigma}_1)\,
 {{\delta\, {\cal S}(\tau ,{\vec \sigma}_1)}\over
 {\delta\, R_{\bar a}(\tau, \vec \sigma)}}{|}_{\theta^i = 0}
  + n(\tau ,{\vec \sigma}_1)\,
 {{\delta\, {\cal T}(\tau ,{\vec \sigma}_1)}\over
 {\delta\, R_{\bar a}(\tau, \vec \sigma)}}{|}_{\theta^i = 0}\Big] -\nonumber \\
 &-& {{8\pi\, G}\over {c^3}}\, \Big({\tilde \phi}^{-1}\, (1 +
 n)\Big)(\tau ,\vec \sigma)\, \int
 d^3\sigma_1\, (1 + n)(\tau, {\vec \sigma}_1)\,
 {{\delta\, {\check {\cal M}}(\tau ,{\vec \sigma}_1)}\over
 {\delta\, R_{\bar a}(\tau, \vec \sigma)}}{|}_{\theta^i = 0} +\nonumber \\
 &+&{{8\pi\, G}\over {c^3}}\, \Big({\tilde \phi}^{-1}\, (1 +
 n)\Big)(\tau ,\vec \sigma)\, \int d^3\sigma_1\, {\bar n}_{(a)}(\tau,{\vec \sigma}_1)\,
 {{\delta\, {\bar {\tilde {\cal H}}}_{(a)}
 (\tau, {\vec \sigma}_1)}\over {\delta\, R_{\bar a}(\tau ,\vec \sigma)}}
 {|}_{\theta^i = 0} -\nonumber \\
 &-&  \Big({{8\pi\, G}\over {c^3}}\Big)^2\, \Big({\tilde \phi}^{-2}\, (1 +
 n)^2\Big)(\tau ,\vec \sigma)\nonumber \\
 && \Big(\sum_{abij}^{a\not= b}\, (\gamma_{\bar aa} - \gamma_{\bar ab})\,
 {{\epsilon_{abi}\, \epsilon_{abj}\, (Q_b\, Q_a^{-1} + Q_a\, Q_b^{-1})}\over
 {\Big(Q_b\, Q^{-1}_a - Q_a\, Q^{-1}_b \Big)^3}}\,   \pi_i^{(\theta
 )}\,   \pi_j^{(\theta )} \Big)(\tau ,\vec \sigma),
 \end{eqnarray*}

 \bea
  \Pi_{\bar a}(\tau, \vec \sigma) &=& - {{c^3}\over {8\pi\, G}}\,
  \Big[\tilde \phi\, \sum_a\, \gamma_{\bar aa}\, \sigma_{(a)(a)}\Big](\tau
  ,\vec \sigma) \cir\nonumber \\
 &\cir& {{c^3}\over {8\pi\, G}}\, {{\tilde \phi(\tau ,\vec \sigma)}\over
 {1 + n(\tau, \vec \sigma)}}\, \Big[\partial_{\tau}\, R_{\bar a} +
 {\tilde \phi}^{-1/3}\, \sum_a\,  Q_a^{-1}\nonumber \\
 && \Big(\Big[\gamma_{\bar aa}\,({1\over 3}\, {\tilde \phi}^{-1}\, \partial_a\,
 \tilde \phi + \sum_{\bar b}\, \gamma_{\bar ba}\, \partial_a\, R_{\bar b})\, -
 \partial_a\, R_{\bar a}\Big]\, {\bar n}_{(a)} - \gamma_{\bar aa}\,
 \partial_a\, {\bar n}_{(a)} \Big)\Big](\tau ,\vec \sigma),\nonumber \\
 &&{}
 \label{6.13}
 \eea

\noindent where Eqs. (\ref{4.3}), (\ref{6.7}), (\ref{c9}),
(\ref{c15}), (\ref{c23}) and (\ref{c7}) have to be used.
\medskip

The last term in the equation for $\partial^2_{\tau}\, R_{\bar a}$
can be written as $- \Big((1 + n)\, \sum_{ab, a\not= b}\,
(\gamma_{\bar aa} - \gamma_{\bar ab})\, {{Q_b\, Q_a^{-1} + Q_b\,
Q_a^{-1}}\over {Q_b\, Q_a^{-1} - Q_a\, Q_b^{-1}}}\,
\sigma^2_{(a)(b)}\Big)(\tau ,\vec \sigma)$.

\subsubsection{The Particles}

For the particles the first of the Hamilton equations (\ref{4.12})
and the second order equation (\ref{4.14}) (obtained from the second
of Eqs.(\ref{4.12}) by using the inversion (\ref{4.13}) giving the
momenta) become

 \begin{eqnarray*}
  \eta_i\, {\dot \eta}^r_i(\tau ) &\cir& \eta_i\,
  \Big({{\phi^{-4}\, (1 + n)\,  Q_r^{-2}\,
  \Big({\check \kappa}_{ir}(\tau ) - {{Q_i}\over c}\, A_{\perp\, r}\Big)\,
 }\over {\sqrt{m_i^2\, c^2 + \phi^{-4}\, \sum_c\,
 Q^{-2}_c\,
 \Big({\check \kappa}_{ic}(\tau ) - {{Q_i}\over c}\, A_{\perp\, c}\Big)^2\,
 }}} -\nonumber \\
 &-& \phi^{-2}\, Q_r^{-1}\,  {\bar n}_{(r)}
 \Big)(\tau ,{\vec \eta}_i(\tau )),
 \end{eqnarray*}

\begin{eqnarray*}
 &&\eta_i\, {d\over {d\tau}}\,\, \Big( {{Q_i}\over c}\, A_{\perp\,r}
 +\nonumber \\
 &&{}\nonumber \\
 &+& m_i\, c\,\,\ \Big( {\tilde \phi}^{2/3}\,  Q_r^2\,
 \Big(\dot{\eta}^r_i(\tau ) + {\tilde \phi}^{-1/3}\,
 Q_r^{-1}\,  {\bar n}_{(r)}\Big)\,\, \Big[\Big(1 + n\Big)^2 -
 \nonumber \\
 &-& {\tilde \phi}^{2/3}\, \sum_{c}\,
 Q_c^2\, \Big(\dot{\eta}^c_i(\tau ) + {\tilde \phi}^{-1/3}\,
 Q_c^{-1}\,  {\bar n}_{(c)} \Big)^2 \Big]^{-1/2}
  \Big)(\tau, {\vec \eta}_i(\tau))\,\, \cir\nonumber \\
 &&{}\nonumber \\
 &\cir&\,\, \Big(- {{\partial}\over {\partial\, \eta_i^r}}\,  {\cal W}
 + {{\eta_i\, Q_i}\over c}\, {\dot \eta}^s_i(\tau)\, {{\partial\, A_{\perp\, s}}\over
 {\partial\, \eta_i^r}} + \eta_i\, {\check F}_{ir} \Big)(\tau ,{\vec
 \eta}_i(\tau)), \nonumber \\
 &&{}\nonumber \\
 &&{}\nonumber \\
 &&{\cal W}(\tau) = \int d^3\sigma\, \Big[(1 + n)\, {\cal W}_{(n)} + \phi^{-2}\, \sum_a\,
 Q_a^{-1}\, {\bar n}_{(a)}\,  {\cal W}_a\big](\tau,
 \vec \sigma),\nonumber \\
  {\cal W}_{(n)}(\tau ,\vec \sigma) &=& - {1\over
  {2c}}\,\Big[\phi^{-2}\, \sum_a\, Q_a^2\,
 \left(2\, \pi_\perp^a - \delta^{am}\, \sum_{i=1}^N\, Q_i\, \eta_i\,
 {{\partial\, c(\vec \sigma, {\vec \eta}_i(\tau))}\over {\partial\, \sigma^m}}
 \right)\nonumber \\
 && \delta^{an}\, \sum_{j=1}^N\, Q_j\, \eta_j\, {{\partial\, c(\vec \sigma,
 {\vec \eta}_j(\tau))}\over {\partial\, \sigma^n}}
 \Big](\tau ,\vec \sigma),\nonumber \\
 &&{}\nonumber \\
 {\cal W}_r(\tau ,\vec \sigma) &=&  - {1\over c}\,
 F_{rs}(\tau ,\vec \sigma)\, \delta^{sn}\, \sum_{i=1}^N\, Q_i\, \eta_i\,
 {{\partial\, c(\vec \sigma, {\vec \eta}_i(\tau))}\over {\partial\,
 \sigma^n}},\end{eqnarray*}

 \bea
 &&{\check F}_{ir} = m_i\, c\, \Big[\Big(1 + n\Big)^2 -
  {\tilde \phi}^{2/3}\, \sum_c\,
 Q_c^2\,  \Big(\dot{\eta}^c_i(\tau ) + {\tilde \phi}^{-1/3}\,
 Q_c^{-1}\,  {\bar n}_{(c)} \Big)^2\Big]^{-1/2}\nonumber \\
 &&\Big[- (1 + n)\, {{\partial\, n}\over {\partial\, \eta_i^r}} +
 \phi^2\, \sum_{a}\,  Q_a\, \Big( {{\partial\,
 {\bar n}_{(a)}}\over {\partial\, \eta_i^r}} -\nonumber \\
 &-&({1\over 3}\, {\tilde \phi}^{-1}\, \partial_r\, \tilde \phi +
 \sum_{\bar a}\, \gamma_{\bar aa}\, \partial_r\, R_{\bar a})\,
 {\bar n}_{(a)}\Big)\, \Big({\dot \eta}_i^a(\tau) + \phi^{-2}\,  Q_a^{-1}\,
 {\bar n}_{(a)}\Big) +\nonumber \\
 &+&\phi^4\, \sum_{a}\, Q_a^2\,
 ({1\over 3}\, {\tilde \phi}^{-1}\, \partial_r\, \tilde \phi +
 \sum_{\bar a}\, \gamma_{\bar aa}\, \partial_r\, R_{\bar a})\,
 \Big({\dot \eta}_i^a(\tau) + \phi^{-2}\,  Q_a^{-1}\,
 {\bar n}_{(a)}\Big)^2 \,\, \Big].\nonumber \\
 &&{}
 \label{6.14}
 \eea

The inversion of the first of Eqs.(\ref{6.11}) gives

\bea
 && \check{\kappa}_{ir}(\tau ) =
 {{Q_i}\over c}\, A_{\perp\,r}(\tau ,{\vec \eta}_i(\tau ))
 + m_i\, c\,\,\ \Big[ {\tilde \phi}^{2/3}\,  Q_r^2\,
  \Big(\dot{\eta}^r_i(\tau ) + {\tilde \phi}^{-1/3}\,
 Q_r^{-1}\,  {\bar n}_{(r)}\Big)\,\, \Big(\Big(1 + n\Big)^2 -
 \nonumber \\
 &-& {\tilde \phi}^{2/3}\, \sum_c\,
 Q_c^2\,  \Big(\dot{\eta}^c_i(\tau ) + {\tilde \phi}^{-1/3}\,
  Q_c^{-1}\,  {\bar n}_{(c)} \Big)\, \Big)^{-1/2}\Big]
 (\tau ,{\vec \eta}_i(\tau )).
 \label{6.15}
 \eea

\subsubsection{The Transverse Electro-Magnetic Field}

Finally the Hamilton equations (\ref{4.15}) for the transverse
electro-magnetic fields in the radiation gauge become

\begin{eqnarray*}
 \partial_{\tau}\, A_{\perp\, r}(\tau ,\vec \sigma ) &\cir&
  \sum_{nua}\, \delta_{rn}\, P^{nu}_{\perp}(\vec \sigma)\, \Big[
 {\tilde \phi}^{-1/3}\, (1 + n)\, Q_a^2\, \delta_{ua}\,  \Big(\pi^a_{\perp} -
 \sum_m\, \delta^{am}\, \sum_{i=1}^N\, Q_i\, \eta_i\, {{\partial\,
 c(\vec \sigma, {\vec \eta}_i(\tau))}\over {\partial\, \sigma^m}}
 \Big) +\nonumber \\
 &+&  {\tilde \phi}^{-1/3}\, Q_a^{-1}\,  {\bar n}_{(a)}\, F_{au}
 \Big](\tau ,\vec \sigma),\nonumber \\
 &&{}\nonumber \\
 \partial_{\tau}\, \pi^r_{\perp}(\tau ,\vec \sigma) &\cir&
  \sum_{m}\, P^{rm}_{\perp}(\vec \sigma)\,  \Big(\sum_a\,
 \delta_{ma}\, \sum_{i=1}^N\, \eta_i\, Q_i\, \delta^3(\vec \sigma, {\vec
 \eta}_i(\tau))\, \nonumber \\
 &&\Big[{ { {\tilde \phi}^{-2/3}\, (1 + n)\, Q_a^{-2}\,
   {\check \kappa}_{ia}(\tau)}\over
 {\sqrt{m_i^2\, c^2 + {\tilde \phi}^{-2/3}\, \sum_b\, Q_b^{-2}\,
  \Big({\check \kappa}_{ib}(\tau ) - {{Q_i}\over c}\,
 A_{\perp\, b}\Big)^2\, }}} -\nonumber \\
  &-& {\tilde \phi}^{-1/3}\, Q_a^{-1}\,
  {\bar n}_{(a)}\Big](\tau ,{\vec \eta}_i(\tau)) -
 \end{eqnarray*}

  \bea
 &-&\Big[2\, {\tilde \phi}^{-1/3}\, (1 + n)\, \sum_{ab}\, Q_a^{-2}\, Q_b^{-2}\,
 \delta_{ma}\, \Big(\partial_b\, F_{ab} - \Big[{1\over 3}\,
 {\tilde \phi}^{-1}\, \partial_b\, \tilde \phi + 2\, \sum_{\bar b}\,
 (\gamma_{\bar ba} + \gamma_{\bar bb})\, \partial_b\, R_{\bar b}\Big]\,
 F_{ab} \Big) +\nonumber \\
 &+&2\, {\tilde \phi}^{-1/3}\, \sum_{ab}\, Q_a^{-2}\, Q_b^{-2}\,
 \delta_{ma}\, \partial_b\, n\, F_{ab} -\nonumber \\
 &-& {\tilde \phi}^{-1/3}\, \sum_a\, {\bar n}_{(a)}\, Q_a^{-1}\,
  \Big(\partial_a\, \pi_{\perp}^m -
   \Big[{1\over 3}\, {\tilde \phi}^{-1}\,
 \partial_a\, \tilde \phi + \sum_{\bar b}\, \gamma_{\bar ba}\,
 \partial_a\, R_{\bar b}\Big]\, \pi^m_{\perp} +\nonumber \\
 &+& \delta_{ma}\, \sum_n\,  \Big[ {1\over 3}\, {\tilde \phi}^{-1}\,
 \partial_n\, \tilde \phi + \sum_{\bar b}\, \gamma_{\bar ba}\,
 \partial_n\, R_{\bar b}\Big]\, \pi^n_{\perp} +\nonumber \\
 &+& \sum_{i=1}^N\, \eta_i\, Q_i\, \Big[
   \Big({1\over 3}\, {\tilde \phi}^{-1}\,
 \partial_a\, \tilde \phi + \sum_{\bar b}\, \gamma_{\bar ba}\,
 \partial_a\, R_{\bar b}\Big)\, {{\partial\, c(\vec \sigma,
 {\vec \eta}_i(\tau)))}\over {\partial\, \sigma^m}} -
 {{\partial^2\, c(\vec \sigma, {\vec \eta}_i(\tau)))}
 \over {\partial\, \sigma^m\, \partial\, \sigma^a}}
 -\nonumber \\
 &-& \delta_{ma}\, \sum_n\, \Big(
 \Big[{1\over 3}\, {\tilde \phi}^{-1}\,
 \partial_n\, \tilde \phi + \sum_{\bar b}\, \gamma_{\bar ba}\,
 \partial_n\, R_{\bar b}\Big]\, {{\partial\, c(\vec \sigma,
 {\vec \eta}_i(\tau)))}\over {\partial\, \sigma^n}} -
 {{\partial^2\, c(\vec \sigma, {\vec \eta}_i(\tau)))}
 \over {\partial\, \sigma^n\, \partial\, \sigma^n}} \Big)
 \Big] \Big) +\nonumber \\
 &+& {\tilde \phi}^{-1/3}\, \sum_a\, Q_a^{-1}\, \sum_{i=1}^N\, \eta_i\, Q_i\,
 \Big(\partial_a\, {\bar n}_{(a)}\,
  {{\partial\, c(\vec \sigma, {\vec \eta}_i(\tau))}\over {\partial\, \sigma^m}}
 -\nonumber \\
 &-& \delta_{ma}\, \sum_n\, \partial_n\, {\bar n}_{(a)}\,
 {{\partial\, c(\vec \sigma, {\vec \eta}_i(\tau))}\over {\partial\, \sigma^n}}
 \Big)\,\, \Big](\tau ,\vec \sigma)\, \Big).\nonumber \\
 &&{}
 \label{6.16}
 \eea

\subsection{The Energy-Momentum Tensor}

In the 3-orthogonal gauges for the gravitational field and in the
radiation gauge for the electro-magnetic field the energy-momentum
tensor of Eqs.(\ref{3.11}) has the following form

\bea
 {\check T}^{\tau\tau}(\tau, \vec \sigma) &=& \Big({\tilde \phi}^{-1}\,
 (1 + n)^{-2}\, {\check {\cal M}}\Big)(\tau, \vec
 \sigma),\nonumber \\
 {\check T}^{\tau r}(\tau, \vec \sigma) &=& \Big({\tilde \phi}^{-4/3}\,
 (1 + n)^{-2}\, Q_r^{-1}\, \Big[{\tilde \phi}^{-1/3}\, (1 + n)\,
 {\check {\cal M}}_r - {\bar n}_{(r)}\, {\check {\cal M}}\Big]
 \Big)(\tau, \vec \sigma),\nonumber \\
  {\check T}^{rs}(\tau, \vec \sigma) &=& \sum_{i=1}^N\, \delta^3(\vec \sigma,
  {\vec \eta}_i(\tau))\,  \Big[{\tilde \phi}^{-5/3}\,
 (1 + n)^{-2}\, {{\eta_i}\over {{\check {\bar M}}_i}}\, Q_r^{-1}\,
 Q_s^{-1}\nonumber \\
 &&\Big({\tilde \phi}^{-1/3}\, (1+n)\, Q_r^{-1}\, ({\check
 \kappa}_{ir}(\tau) - {{Q_i}\over c}\, A_{\perp r}) -
 {\bar n}_{(r)}\, {\check {\bar M}}_i\Big)\nonumber \\
 &&\Big({\tilde \phi}^{-1/3}\, (1+n)\, Q_s^{-1}\, ({\check
 \kappa}_{is}(\tau) - {{Q_i}\over c}\, A_{\perp s}) -
 {\bar n}_{(s)}\, {\check {\bar M}}_i\Big)
 \Big](\tau, {\vec \eta}_i(\tau)) +\nonumber \\
 &+& {1\over c}\, {\tilde \phi}^{-2}(\tau, \vec \sigma)\,
  \Big[- \pi^r_{\perp}\, \pi^s_{\perp} + {1\over 2}\, {\tilde
  \phi}^{-2/3}\, Q_r^{-1}\, Q_s^{-1}\, \nonumber \\
  &&\sum_{uv}\, Q_u^{-1}\, Q_v^{-1}\, \Big((\delta_{rs} +
  {{{\bar n}_{(r)}\, {\bar n}_{(s)}}\over {(1 + n)^2}})\,
  \delta_{uv} + 4\, {{{\bar n}_{(r)}\, {\bar n}_{(s)}\,
  {\bar n}_{(u)}\, {\bar n}_{(v)}}\over {(1 + n)^4}}\Big)\,
  \pi^u_{\perp}\, \pi^v_{\perp} +\nonumber \\
 &+& Q_r^{-1}\, Q_s^{-1}\, \sum_{lm}\, Q_l^{-1}\,
 {{{\bar n}_{(r)}\, \delta_{ls} + {\bar n}_{(s)}\, \delta_{lr}}\over
 {1 + n}}\, F_{lm}\, \pi^m_{\perp} +\nonumber \\
 &+& Q_r^{-1}\, Q_s^{-1}\, \sum_{lmuv}\, Q_l^{-1}\, Q_m^{-1}\,
 Q_u^{-1}\, Q_v^{-1}\, \Big([\delta_{lr}\, \delta_{ms} -\nonumber \\
 &-& {1\over 4}\, (\delta_{rs} - {{{\bar n}_{(r)}\, {\bar n}_{(s)}}
 \over {(1 + n)^2}})\, \delta_{lm}]\, \delta_{uv} - \delta_{ls}\,
 \delta_{mr}\, {{{\bar n}_{(u)}\, {\bar n}_{(v)}}\over {(1 + n)^2}}
 \Big)\, F_{ul}\, F_{vm}\Big](\tau, \vec \sigma),\nonumber \\
 &&{}
 \label{6.17}
 \eea

\noindent with ${\check {\cal M}}$, ${\check {\cal M}}_r$ and
${\check M}_i = \eta_i\, {\check {\bar M}}_i$ given in
Eqs.(\ref{3.37}).
\medskip

The identity ${}^4\nabla_A\, T^{AB}(\tau, \vec \sigma) \equiv 0$ of
Eqs.(\ref{3.11}) assumes the following expression

\bea
 &&\Big(\Big[\partial_{\tau} - {\tilde \phi}^{-1/3}\, \sum_s\, Q_s^{-1}\,
 {\bar n}_{(s)}\, \partial_s - (1 + n)\, {}^3K - {\tilde
 \phi}^{-1}\, \partial_{\tau}\, \tilde \phi -\nonumber \\
 &-& {\tilde \phi}^{-1/3}\, \sum_s\, Q_s^{-1}\, {\bar n}_{(s)}\,
 {\tilde \phi}^{-1}\, \partial_s\, \tilde \phi\Big]\, {\check {\cal
 M}} +\nonumber \\
 &&+ {\tilde \phi}^{-2/3}\, (1 + n)\, \sum_s\, Q_s^{-2}\, \Big[
 \partial_s - 2\, (\partial_s\, \Gamma_s^{(1)} + {1\over 3}\,
 {\tilde \phi}^{-1}\, \partial_s\, \tilde \phi - {{\partial_s\,
 n}\over {1 + n}}) \Big]\, {\check {\cal M}}_s -\nonumber \\
 &&- {\tilde \phi}^{-1}\, (1 + n)\, \sum_{uv}\, Q_u^{-1}\, Q_v^{-1}\,
  ({1\over 3}\, {}^3K\, \delta_{uv} + \sigma_{(u)(v)})\, {\check T}_{uv}
 \Big)(\tau, \vec \sigma) \equiv 0,\nonumber \\
 &&{}\nonumber \\
 &&\Big(\Big[\partial_{\tau} - {\tilde \phi}^{-1/3}\, \sum_s\, Q_s^{-1}\,
 {\bar n}_{(s)}\, \partial_s - (1 + n)\, {}^3K - {\tilde
 \phi}^{-1}\, \partial_{\tau}\, \tilde \phi -\nonumber \\
 &-& {\tilde \phi}^{-1/3}\, \sum_s\, Q_s^{-1}\, {\bar n}_{(s)}\,
 {\tilde \phi}^{-1}\, \partial_s\, \tilde \phi\Big]\, {\check {\cal
 M}}_r +\nonumber \\
 &&+ {\tilde \phi}^{-1/3}\, \sum_s\, Q_s^{-1}\, (\partial_r\,
 \Gamma_s^{(1)} + {1\over 3}\, {\tilde \phi}^{-1}\, \partial_r\,
 \tilde \phi)\, {\check {\cal M}}_s + \partial_r\, n\,
 {\check {\cal M}} +\nonumber \\
 &&+ {\tilde \phi}^{1/3}\, (1 + n)\, \sum_u\, Q_u^{-2}\,
 \Big[\partial_u\, {\check T}_{ru} - (2\, \partial_u\, \Gamma_u^{(1)} -
 {1\over 3}\, {\tilde \phi}^{-1}\, \partial_u\,
 \tilde \phi + {{\partial_u\, n}\over {1 + n}})\, {\check T}_{ru}
 -\nonumber \\
 &&- (\partial_r\, \Gamma_u^{(1)} + {1\over 3}\, {\tilde \phi}^{-1}\,
 \partial_r\, \tilde \phi)\, {\check T}_{uu}
 \Big]\Big)(\tau, \vec \sigma) \equiv 0.\nonumber \\
 &&{}
 \label{6.18}
 \eea

\vfill\eject

\section{Conclusions}

In this paper we gave the Hamilton equations of ADM tetrad gravity
coupled to charged scalar particles and to the electro-magnetic
field in the York canonical basis first in an arbitrary Schwinger
time gauge and then in a sub-family of 3-orthogonal gauges. The
electro-magnetic field has been specialized to the non-covariant
radiation gauge, in which we can eliminate the electro-magnetic
gauge variables and visualize the non-inertial Coulomb potential
among the particles.\bigskip

We gave some refinements of the York canonical basis of
Ref.\cite{6}, connected with the congruence of the Eulerian
observers associated to the 3+1 splitting of space-time.

\bigskip

We emphasized the role of the inertial gauge variable ${}^3K(\tau,
\vec \sigma)$, the York time, connected to the freedom in the choice
of the non-dynamical part of the clock synchronization convention
defining the instantaneous 3-spaces: in the York canonical basis it
is the only gauge variable described by a momentum (a reflex of the
Lorentz signature of space-time) and it gives rise to a negative
kinetic term in the weak ADM energy vanishing only in the gauges
${}^3K(\tau, \vec \sigma) = 0$.

\bigskip

Moreover, in connection with the weak ADM Poincare' charges, we
showed that in these asymptotically Minkowskian space-times one can
introduce an interpretation of the isolated system {\it
gravitational field plus matter} like in the inertial and
non-inertial rest frames of Minkowski space-time \cite{8,10}. The
3-universe contained in an instantaneous 3-space can be described as
a decoupled non-covariant non-observable pseudo-particle carrying a
pole-dipole structure, whose mass and spin are those identifying the
configuration of the "gravitational field plus matter" isolated
system present in the 3-universe.

\bigskip

In the next paper \cite{19}, starting from the results obtained in
the family of non-harmonic 3-orthogonal Schwinger gauges with an
arbitrary numerical function describing the inertial York time,
${}^3K(\tau, \vec \sigma) \approx F(\tau, \vec \sigma)$, we will
define a linearization of ADM canonical tetrad gravity plus matter,
to obtain a formulation of Hamiltonian Post-Minkowskian gravity
(without Post-Newtonian expansions) with non-flat Riemannian
3-spaces and asymptotic Minkowski background: i.e. with a
decomposition of the 4-metric tending to the asymptotic Minkowski
metric at spatial infinity, ${}^4g_{AB}\, \rightarrow
{}^4\eta_{AB}$. We can write ${}^4g_{AB} = {}^4\eta_{AB} +
{}^4h_{AB}$, with the small perturbation ${}^4h_{AB}$ vanishing at
spatial infinity, but this decomposition has no intrinsic meaning in
the bulk differently from what happens in harmonic gauges with a
fixed background (in them one has Euclidean 3-spaces with a
violation of the equivalence principle). Then in a third paper
\cite{19a}, on the Post-Minkowskian 2-body problem in absence of
electro-magnetic field, we will show that a consequence of this
approach is the possibility of describing part (or maybe all) dark
matter as a {\it relativistic inertial effect} determined by the
gauge variable ${}^3K(\tau, \sigma^r)$ (not existing in Newtonian
gravity, where the Euclidean 3-space is an absolute notion): the
rotation curves of galaxies would then experimentally determine a
preferred choice of the instantaneous 3-spaces.
\bigskip

Finally in a future paper we will replace the matter with a perfect
fluid (for instance dust \cite{26}) and we will try to see whether
the York canonical basis can help in developing the back-reaction
\cite{42} approach to dark energy. The main result in this direction
is that all the canonical variables in the York basis (with the
exception of the parameters $\theta^i$) are 3-scalars on the
3-space: therefore we can study the spatial average of nearly all
the Hamilton equations.

\vfill\eject

\appendix

\section{The Comparison with the Standard ADM Equations of Motion,
the Raychaudhuri Equation and the Contracted Bianchi Identities.}

In absence of matter, after the 3+1 splitting of space-time and the
introduction of the radar 4-coordinates, the standard ADM equations
of motion of canonical metric gravity  are (see Eqs.(4.11) of
Ref.\cite{2})

\bea
 \partial_{\tau}\, {}^3g_{rs} &\cir& n_{r|s} + n_{s|r} - 2\,
 (1 + n)\, {}^3K_{rs},\nonumber \\
 &&{}\nonumber \\
 \partial_{\tau}\, {}^3K_{rs} &\cir& (1 + n)\, \Big({}^3R_{rs} + {}^3K\, {}^3K_{rs} -
 2\, {}^3K_{ru}\, {}^3K^u{}_s\Big) -\nonumber \\
 &-& n_{|s|r} + n^u{}_{|s}\, {}^3K_{ur} + n^u{}_{|r}\, {}^3K_{us} +
 n^u\, {}^3K_{rs|u}.
 \label{a1}
 \eea

\medskip

In globally hyperbolic space-times these equations, together with
the equations corresponding to the super-Hamiltonian and
super-momentum constraints, are a first order formulation of
Einstein's equations (see for instance Ref.\cite{a}), employing a
3+1 splitting of space-time and the associated extrinsic curvature
tensor, independent from the existence of the Hamiltonian
formulation. Therefore they can also be used when the standard
canonical formalism cannot be applied, like for instance in
cosmology. However also in this first order formulation the
extrinsic curvature tensor ${}^3K_{rs}$ can be expressed in terms of
its trace ${}^3K$ and of the shear of the Eulerian observers
associated with the 3+1 splitting of space-time as shown in the
first line of Eq.(\ref{2.17}). In the space-times admitting the
Hamiltonian formulation Eqs.(\ref{2.17}) show the connection of the
shear with the canonical momenta: in this case Eqs.(\ref{a1}) and
the Hamilton equations are equivalent.

\bigskip

Since from Eqs.(\ref{2.6}) and (\ref{2.14}) we have $\gamma = det\,
{}^3g_{rs} = ({}^3e)^2 = \phi^{12} = {\tilde \phi}^2$ and ${}^3K = -
\sgn\, \theta = {{12\pi\, G}\over {c^3}}\, \pi_{\tilde \phi}$
($\theta$ is the expansion of the non-geodetic Eulerian observers),
we can deduce from Eqs.(\ref{a1}) the following equations

\bea
 \partial_{\tau}\, \tilde \phi &\cir& \Big[- {{12\pi\, G}\over {c^3}}\, (1 + n)\, \pi_{\tilde \phi}
 +\Big({}^3{\bar e}^u_{(a)}\, {\bar n}_{(a)}\Big)_{|u}\Big]\, \tilde
 \phi,\nonumber \\
 &&{}\nonumber \\
 \partial_{\tau}\, {}^3K &=& {{12\pi\, G}\over {c^3}}\,
 \partial_{\tau}\, \pi_{\tilde \phi}\, = - \sgn\,
 \partial_{\tau}\, \theta \cir\nonumber \\
 &\cir& (1 + n)\, \Big[{}^3R + ({}^3K)^2\Big]
 - n_{|u}{}^{|u} + n^u\, {}^3K_{|u} =\nonumber \\
 &=& (1 + n)\, \Big[{}^3R + \Big({{12\pi\, G}\over {c^3}}\, \pi_{\tilde \phi}\Big)^2
 \Big] - n_{|u}{}^{|u} + {{12\pi\, G}\over {c^3}}\, \pi_{\tilde
 \phi}{}_{ |u} {}^3{\bar e}^u_{(a)}\, {\bar n}_{(a)}.
 \label{a2}
 \eea

\bigskip

The second of Eqs.(\ref{a2}) is an equation  for the time evolution
$\partial_{\tau}\, \pi_{\tilde \phi}$ of the gauge variable
$\pi_{\tilde \phi}$, named Raychaudhuri equation. For geodetic
congruences of time-like curves it is connected with the geodesic
deviation equation and is a fundamental ingredient for the
singularity theorems, see Ref.\cite{43}, where it is shown that in
general (also in Minkowski space-time) in a congruence of "time-like
geodesics" caustics will develop  if convergence ($\theta < 0$)
occurs anywhere. In certain space-times there will be real
space-time singularities if other global assumptions hold \cite{43}.
As shown in Ref.\cite{44} the singularity theorems imply that the
gauge fixing identifying the instantaneous 3-spaces (clock
synchronization convention) have to be divided into inequivalent
classes. Those who satisfy the strong energy conditions and have
non-negative 3-curvature imply a singularity (geodesic
incompleteness) in the past if the spatial average of the expansion
does not vanish. In our class of space-times, including the
Christodoulou-Klainermann ones \cite{45} in absence of matter, the
assumed boundary conditions should eliminate the singularities (at
least for finite intervals of time; but in presence of matter the
problem is open). Moreover the congruence of Eulerian observers,
relevant for the York canonical basis, is not geodesic. However, the
implications for these problems of the gauge nature of the York time
${}^3K$ have still to be clarified.

\bigskip

Instead the first of Eqs.(\ref{a2}) is a contracted Bianchi
identity. Let us clarify this point.\medskip

Due to the {\it Bianchi identities} ${}^4G^{\mu\nu}{}_{;\nu} \equiv
0$, the 10 Einstein's equations $G_{\mu\nu} = {}^4R_{\mu\nu} -
{1\over 2}\, {}^4g_{\mu\nu}\, {}^4R\, \cir\, {{\sgn\, c^3}\over
{8\pi\, G}}\, T_{\mu\nu}$ imply the equations of motion
$T^{\mu\nu}{}_{;\nu} \cir 0$ for matter and 4 contracted Bianchi
identities, which say that 4 Einstein's equations are not
independent from the others and their time-derivatives. Therefore
there are only 6 Einstein's equations functionally independent. But,
since 4 Einstein's equations are the super-Hamiltonian and
super-momentum constraints (restrictions on the Cauchy data), it
turns out that only 2 combinations of the 10 Einstein equations are
dynamical, i.e. depending on the accelerations (they can be put in
normal form; see the Hamilton-Dirac equations for $R_{\bar a}$,
$\Pi_{\bar a}$ in Section IV). The four contracted Bianchi
identities, ${}^4G^{\mu\nu}{}_{;\nu}\equiv 0$, imply \cite{43} that,
if the restrictions of Cauchy data are satisfied initially and the
spatial equations ${}^4G_{ij}\, {\buildrel \circ \over =}\, 0$ are
satisfied everywhere, then the secondary constraints are satisfied
also at later times. Behind these identities there is the gauge
invariance of the Einstein-Hilbert action under 4-diffeomorphisms.
However to get a canonical formulation we need a 3+1 splitting of
space-time and the replacement of the Einstein-Hilbert action with
the ADM action

\bigskip

At the Hamiltonian level, one of the first six equations (\ref{a1})
(the one for $\partial_{\tau}\, \tilde \phi$) and three of the other
six equations (\ref{a1}) (the ones for $\partial_{\tau}\,
\sigma_{(a)(b)}{|}_{a \not= b}$) are  the ADM version of the {\it 4
contracted Bianchi identities}, now implied by the gauge invariances
of the ADM action, which are no more the 4-diffeomorphisms due to
the presence of the 3+1 splitting of space-time. At the Hamiltonian
level these gauge invariances are generated by the 8 (14 in tetrad
gravity \footnote{In canonical tetrad gravity there are 6 extra
primary constraints with the associated Dirac multipliers in the
Dirac Hamiltonian and no extra contracted Bianchi identity.}) first
class constraints of ADM canonical metric gravity defining a
pre-symplectic sub-manifold of the phase space symplectic manifold.
However at each fixed value of the radar time $\tau$ the gauge group
is not a Lie group \footnote{It contains spatial diffeomorphisms
acting on $\Sigma_{\tau}$ and the gauge transformations, replacing
the time diffeomorphisms, generated by the super-Hamiltonian
constraint, which modify the shape of the instantaneous 3-space
$\Sigma{\tau}$ with deformations along the {\it normal} to
$\Sigma_{\tau}$ in every point, proportional to the conjugate gauge
variable $\pi_{\tilde \phi} = {{c^3}\over {12\pi\, G}}\, {}^3K$,
namely to the value of the York time in that point.} and its group
manifold, as a transformation group on the symplectic manifold of
ADM canonical metric gravity, does not have 8 functional gauge
parameters but only 4, the 4 Dirac multipliers \footnote{They
correspond to a special choice of the arbitrary generalized
velocities existing at the Lagrangian level due to the gauge
invariances. The kinematical set of the Hamilton-Dirac equations
identify them as the $\tau$-derivatives of the lapse and shift
functions.} in front of the 4 primary constrains (the vanishing of
the lapse and shift momenta). The secondary constraints (the
super-Hamiltonian and super-momentum constraints) have as gauge
parameters the gauge variables corresponding to the lapse and shift
functions, i.e. 4 elements of a Darboux basis of the symplectic
manifold. As a consequence, it is not clear how to define an
abstract group manifold for this type of symplectic gauge group
\footnote{Due to the Hamilton equations $\partial_{\tau}\, n \,
\cir\, \lambda_n$, $\partial_{\tau}\, n_{(a)}\, \cir\,
\lambda_{n_{(a)}}$, one could say that the the natural gauge
parameters of the symplectic gauge group are the lapse and shift
gauge functions {\it and} their $\tau$-derivatives, so that there
are only 4 arbitrary functions replacing the space-time
4-diffeomorphisms (whose abstract group manifold, depending on 4
arbitrary space-time functions as gauge parameters, is not under
mathematical control in the large), the gauge group of the
Einstein-Hilbert action.}. Moreover, the group manifold is
$\tau$-dependent: we have (a-priori) different group manifolds for
the gauge group on each instantaneous 3-space $\Sigma_{\tau}$.
Therefore Eqs.(\ref{a1}) contain an information on the
$\tau$-dependence of the effective group manifold of the gauge
transformation group of canonical gravity describing its
modification from an instantaneous 3-space $\Sigma_{\tau}$ to a
modified one $\Sigma_{\tau + d\tau}$.

\bigskip

By comparison let us consider electro-magnetism (with its Abelian
gauge group) in the Shanmugadhasan canonical basis (\ref{3.31}). In
the rest-frame instant form of Ref.\cite{8} the canonical variables
and the constraint are  $A_{\tau}$, $\pi_{\tau} \approx 0$,
$\eta_{em} = - {1\over {\triangle}}\, \vec \partial \cdot \vec A$,
$\pi_{\eta_{em}} = \Gamma = \vec\partial \cdot \vec \pi \approx 0$,
${\vec A}_{\perp}$, ${\vec \pi}_{\perp}$. The Dirac Hamiltonian is
$H_D = \int d^3\sigma \, \Big[{\vec \pi}^2_{\perp} + {\vec B}^2 -
A_{\tau}\, \pi_{\eta_{em}} + \lambda\, \pi_{\tau} \Big](\tau ,\vec
\sigma )$ and the gauge variable $A_{\tau}$ plays the role of the
lapse and shift functions ($\lambda \, \cir\, \partial_{\tau}\,
A_{\tau}$). The analogue of the equations (\ref{4.4}), (\ref{4.5}),
for the gauge variables is $\partial_{\tau}\, \eta_{em}\, \cir\, -
A_{\tau}$: this equation determines $A_{\tau}$ once the gauge fixing
for $\eta_{em}$ has been given (the  analogue of Eqs.(\ref{4.7})):
$\eta_{em} - f[..] \approx 0$. The analogue of the Hamiltonian
contracted Bianchi identities (\ref{4.1}), (\ref{4.2}), is
$\partial_{\tau}\, \pi_{\eta_{em}} \, \cir\, 0$ \footnote{ The final
equations for $\tilde \phi$ and $\pi_i^{(\theta )}$ in tetrad
gravity are so complicated because the gauge algebra of metric
gravity is non-Abelian and not a Lie algebra.}.

\bigskip

Let us now look at the Hamiltonian interpretation of the ADM
equations (\ref{a1}) in the York canonical basis of tetrad gravity.
The first half of Eqs.(\ref{a1}), rewritten in the form given in
Eqs.(\ref{2.17}), has the same content as the Hamilton equations for
$\partial_{\tau}\, \tilde \phi$ [Eq.(\ref{4.1}); see also the first
of Eqs.(\ref{a2})], $\partial_{\tau}\, \theta^i$ [Eqs.(\ref{4.5})]
and $\partial_{\tau}\, R_{\bar a}$ [Eqs.(\ref{4.8})]. Analogously
the second half of Eqs.(\ref{a1}) could be decomposed in three sets
of equations for $\partial_{\tau}\, {}^3K$ [see the Hamilton
equation (\ref{4.4}) and the second of Eqs.(\ref{a2})],
$\partial_{\tau}\, \sigma_{(a)(a)}$ and $\partial_{\tau}\,
\sigma_{(a)(b)}{|}_{a \not= b}$. In the canonical formalism the
equations for $\partial_{\tau}\, \sigma_{(a)(a)}$ are replaced by
the Hamilton equations (\ref{4.10}) for $\partial_{\tau}\, \Pi_{\bar
a}$, while the equations for $\partial_{\tau}\,
\sigma_{(a)(b)}{|}_{a \not= b}$ are replaced by the Hamilton
equations (\ref{4.2}) for $\partial_{\tau}\, \pi_i^{(\theta)}$.

\vfill\eject

\section{Calculations.}

In this Appendix there are the calculations needed for the form in
the York canonical basis of the Hamilton equations in arbitrary
Schwinger time gauges, generated by the Dirac Hamiltonian
(\ref{3.48}), of Section IV. The expression given below are in terms
of the variable $\phi = {\tilde \phi}^{1/6}$: to express them in
terms of $\tilde \phi$ one needs $\phi^{-1}\, \partial_r\, \phi =
{1\over 6}\, {\tilde \phi}^{-1}\, \partial_r\, \tilde \phi$ and
$\phi^{-1}\, \partial_r\, \partial_s\, \phi = {1\over 6}\, [{\tilde
\phi}^{-1}\, \partial_r\, \partial_s\, \tilde \phi - {5\over 6}\,
{\tilde \phi}^{-1}\, \partial_r\, \tilde \phi\, {\tilde \phi}^{-1}\,
\partial_s\, \tilde \phi]$.

\subsection{The Functional Derivatives of the Super-Momentum
Constraints.}

\subsubsection{The Super-Momentum Constraints}

For the evaluation of the functional derivatives it is more
convenient to use the following form of the super-momentum
constraints obtained from Eq.(\ref{3.41})

\bea
 {\tilde {\bar {\cal H}}}_{(a)}(\tau ,\vec \sigma)
  &=& \phi^{-2}(\tau, \vec \sigma)\, \Big(\sum_{b \not= a}\,
 \sum_{rtwi}\, {{ \epsilon_{abt}\, Q_b^{-1}\, V_{rb}\, V_{wt}\, B_{iw}}
 \over {Q_b\, Q_a^{-1} - Q_a\, Q_b^{-1}}}\, \partial_r\, \pi_i^{(\theta)}
 +\nonumber \\
 &+& \sum_{ritw}\, \Big( \sum_{b \not= a}\, {{ \epsilon_{abt}}
 \over {Q_b\, Q_a^{-1} - Q_a\, Q_b^{-1}}}\,
  \Big[Q_b^{-1}\, \partial_r\, (V_{rb}\, V_{wt}\, B_{iw}) +\nonumber \\
  &+& {{2\, Q_a^{-1}}\over {Q_b\, Q_a^{-1} - Q_a\, Q_b^{-1}}}\,
  \sum_{\bar c}\, (\gamma_{\bar ca} - \gamma_{\bar cb})\, \partial_r\,
  R_{\bar c}\, V_{rb}\, V_{tw}\, B_{iw}\Big] +\nonumber \\
 &+&  \sum_{bu}\,  \sum_{c \not= b}\, {{\epsilon_{bct}\,
 Q_a^{-1}\, Q_b\, Q_c^{-1}}\over {Q_c\, Q_b^{-1} - Q_b\,
 Q_c^{-1}}}\,  (V_{rc}\, V_{ua} - V_{ra}\, V_{uc})\, \partial_r\, V_{ub}
 \,\,\, V_{wt}\, B_{iw}\Big)\,  \pi_i^{(\theta)} +\nonumber \\
 &&{}\nonumber \\
 &+&Q_a^{-1}\, \sum_r\, V_{ra}\, \Big(\phi^6\, \partial_r\, \pi_{\tilde \phi}
 +  \sum_{\bar b}\, \gamma_{\bar ba}\, \partial_r\, \Pi_{\bar b}\Big)
 +\nonumber \\
 &+&  Q_a^{-1}\, \sum_{r\bar b}\, \Big(\gamma_{\bar ba}\,
 \partial_r\, V_{ra} -  V_{ra}\, \partial_r\, R_{\bar b} +
  \sum_{ub}\, \gamma_{\bar bb}\,  V_{ua}\, V_{rb}\, \partial_r\, V_{ub}\Big)\,
 \Pi_{\bar b} +\nonumber \\
 &&{}\nonumber \\
 &+&  Q_a^{-1}\, \sum_r\, V_{ra}\, {\check {\cal M}}_r
 \Big)(\tau ,\vec \sigma) \quad \approx 0,
 \label{b1}
 \eea

This form of the super-momentum constraints will be used to evaluate
their functional derivatives with respect to the arguments $\tilde
\phi = \phi^6$, $\pi_{\tilde \phi}$, $\theta^i$, $\pi_i^{(\theta
)}$, $R_{\bar a}$, $\Pi_{\bar a}$.

\subsubsection{The Functional Derivative with Respect to $\pi_{\tilde \phi}$}

\bea
    &&\int d^3\sigma_1\, \sum_a\, {\bar n}_{(a)}(\tau ,{\vec \sigma}_1)\,
   {{\delta\, {\tilde {\bar {\cal H}}}_{(a)}(\tau ,{\vec \sigma}_1
  )}\over {\delta\, \pi_{\tilde \phi}(\tau ,\vec \sigma )}}\, =\nonumber \\
  &=&- \Big[\phi^4\, \sum_{ra}\, Q_a^{-1}\, \Big(\partial_r\, {\bar
  n}_{(a)}\, V_{ra} +\nonumber \\
  &+& {\bar n}_{(a)}\, \Big[V_{ra}\, \Big(4\, \phi^{-1}\, \partial_r\, \phi -
  \sum_{\bar b}\, \gamma_{\bar ba}\, \partial_r\, R_{\bar b}\Big) +
  \partial_r\, V_{ra}\Big]\, \Big) \Big](\tau ,\vec \sigma).
 \label{b2}
  \eea

\subsubsection{The Functional Derivative with Respect to $\tilde \phi = \phi^6$}

Since we have ${{\delta}\over {\delta\, \tilde \phi(\tau ,\vec
\sigma)}}\, =\, {1\over 6}\, \phi^{-5}(\tau ,\vec \sigma)\,
{{\delta}\over {\delta\, \phi(\tau ,\vec \sigma)}}$, we get

\beq
    \int d^3\sigma_1\, \sum_a\, {\bar n}_{(a)}(\tau ,{\vec \sigma}_1)\,
   {{\delta\, {\tilde {\bar {\cal H}}}_{(a)}(\tau ,{\vec \sigma}_1
  )}\over {\delta\, \phi(\tau ,\vec \sigma )}}\, \approx
   6\, \Big[\phi^3\, \sum_{ra}\, {\bar n}_{(a)}\, Q_a^{-1}\,
  V_{ra}\, \partial_r\, \pi_{\tilde \phi}\Big](\tau ,\vec \sigma).
  \label{b3}
  \eeq

\subsubsection{The Functional Derivative with Respect to $\pi_i^{(\theta )}$}

By  using the results\medskip

$\partial_r\, {{Q_b^{-1}}\over {Q_b\, Q_a^{-1} - Q_a\, Q_b^{-1}}} =
- {{Q_b^{-1}}\over {Q_b\, Q_a^{-1} - Q_a\, Q_b^{-1}}}\, \sum_{\bar
b}\, \Big[\gamma_{\bar bb} - (\gamma_{\bar ba} - \gamma_{\bar bb})\,
{{Q_b\, Q_a^{-1} + Q_a\, Q_b^{-1}}\over {Q_b\, Q_a^{-1} - Q_a\,
Q_b^{-1}}}\Big]\, \partial_r\, R_{\bar b}$

\noindent and\medskip

$\partial_r\, {{Q_a^{-1}\, Q_b\, Q_c^{-1}}\over {Q_c\, Q_b^{-1} -
Q_b\, Q_c^{-1}}} = - {{Q_a^{-1}\, Q_b\, Q_c^{-1}}\over {Q_c\,
Q_b^{-1} - Q_b\, Q_c^{-1}}}\, \sum_{\bar b}\, \Big[\gamma_{\bar ba}
- \gamma_{\bar bb} + \gamma_{\bar bc} - (\gamma_{\bar bb} -
\gamma){\bar bc})\, {{Q_c\, Q_b^{-1} + Q_b\, Q_c^{-1}}\over {Q_c\,
Q_b^{-1} - Q_b\, Q_c^{-1}}}\Big]\, \partial_r\, R_{\bar b}$,\medskip

\noindent we get

\bea
   &&\int d^3\sigma_1\, \sum_a\, {\bar n}_{(a)}(\tau ,{\vec \sigma}_1)\,
   {{\delta\, {\tilde {\bar {\cal H}}}_{(a)}(\tau ,{\vec \sigma}_1
  )}\over {\delta\, \pi_i^{(\theta )}(\tau ,\vec \sigma )}}\, =\nonumber \\
  &=& \phi^{-2}(\tau ,\vec \sigma)\, \sum_a\, \Big[ {\bar
  n}_{(a)}\nonumber \\
  &&\Big(\sum_{b \not= a}\, \sum_{rtw}\, \Big[2\, \phi^{-1}\, \partial_r\, \phi
  +  \sum_{\bar c}\, \gamma_{\bar ca}\, \partial_r\, R_{\bar c}\,
  \Big]\, {{\epsilon_{abt}\,
  Q_b^{-1}\, V_{rb}\, V_{wt}\, B_{iw}}\over {Q_b\, Q_a^{-1} - Q_a\, Q_b^{-1}}}
  +\nonumber \\
  &+& \sum_{bu}\,  \sum_{c \not= b}\, {{\epsilon_{bct}\,
 Q_a^{-1}\, Q_b\, Q_c^{-1}}\over {Q_c\, Q_b^{-1} - Q_b\,
 Q_c^{-1}}}\,  (V_{rc}\, V_{ua} - V_{ra}\, V_{uc})\, \partial_r\, V_{ub}
 \, V_{wt}\, B_{iw}  \Big) -\nonumber \\
 &-& \sum_r\, \partial_r\, {\bar n}_{(a)}\, \sum_{b \not= a}\,
 \sum_{tv}\, {{\epsilon_{abt}\, Q_b^{-1}\, V_{rb}\, V_{wt}\, B_{iw}}\over
 {Q_b\, Q_a^{-1} - Q_a\, Q_b^{-1}}}\, \Big](\tau ,\vec \sigma).
   \label{b4}
  \eea

\subsubsection{The Functional Derivative with Respect to $\theta^i$}

By using the result ${{\delta\, V_{ra}(\theta^i({\vec
\sigma}_1))}\over {\delta\, \theta^n(\vec \sigma )}} = {{\partial\,
V_{ra}(\theta^i({\vec \sigma}_1))}\over {\partial\, \theta^n({\vec
\sigma}_1 )}}\, \delta^3(\vec \sigma ,{\vec \sigma}_1)$, $\sum_r\,
\partial_{1r}\, {{\delta\, V_{ra}(\theta^i({\vec \sigma}_1))}\over
{\delta\, \theta^n(\vec \sigma )}} = \sum_r\, {{\partial\,
V_{ra}(\theta^i(\vec \sigma))}\over {\partial\, \theta^n(\vec \sigma
)}}\, \partial_{1r}\,\delta^3(\vec \sigma ,{\vec \sigma}_1)$, we get

\begin{eqnarray*}
    &&\int d^3\sigma_1\, \sum_a\, {\bar n}_{(a)}(\tau ,{\vec \sigma}_1)\,
   {{\delta\, {\tilde {\bar {\cal H}}}_{(a)}(\tau ,{\vec \sigma}_1
  )}\over {\delta\, \theta^i(\tau ,\vec \sigma )}}\, =\nonumber \\
  &=& - \phi^{-2}(\tau ,\vec \sigma)\, \sum_{ra}\, \partial_r\,
  {\bar n}_{(a)}(\tau, \vec \sigma)\nonumber \\
  &&\Big[Q_a^{-1}\, \sum_{\bar b}\, \Big(\gamma_{\bar ba}\, {{\partial\,
  V_{ra}}\over  {\partial\, \theta^i}} + \sum_{sb}\, \gamma_{\bar bb}\,
  V_{sa}\, V_{rb}\, {{\partial\, V_{sb}}\over  {\partial\, \theta^i}}
  \Big)\, \Pi_{\bar b}  +\nonumber \\
  &+& \sum_{jtw}\, \Big(\sum_{b \not= a}\,  {{\epsilon_{abt}\,
  Q_b^{-1}}\over {Q_b\, Q_a^{-1} - Q_a\, Q_b^{-1}}}\, {{\partial\,
  V_{rb}\, V_{wt}\, B_{jw}}\over {\partial\, \theta^i}} +
  \nonumber \\
  &+& \sum_{sb}\, \sum_{c \not= b}\, {{\epsilon_{bct}\,
  Q_a^{-1}\, Q_b\, Q_c^{-1}}\over {Q_c\, Q_b^{-1} - Q_b\, Q_c^{-1}}}\,
  (V_{rc}\, V_{sa} - V_{ra}\, V_{sc})\, {{\partial\, V_{sb}}\over
  {\partial\, \theta^i}}\, V_{wt}\, B_{jw}\Big)\, \pi_j^{(\theta)}
  \Big](\tau ,\vec \sigma) +
  \end{eqnarray*}

\bea
  &+& \phi^{-2}(\tau ,\vec \sigma)\, \sum_a\,
  {\bar n}_{(a)}(\tau, \vec \sigma)\,\,
  \Big[Q_a^{-1}\, \sum_r\, \Big({{\partial\, V_{ra}}\over {\partial\,
  \theta^i}}\, \Big[\phi^6\, \partial_r\, \pi_{\tilde \phi}
 +\nonumber \\
  &+& \sum_{\bar b}\,
  \Big(\gamma_{\bar ba}\, (2\, \phi^{-1}\, \partial_r\, \phi + \sum_{\bar c}\,
  \gamma_{\bar ca}\, \partial_r\, R_{\bar c})\, \Pi_{\bar b} -
   \partial_r\, R_{\bar b}\, \Pi_{\bar b}\Big) + {\check {\cal M}}_r\Big]
  -\nonumber \\
  &-& \sum_{sb\bar b}\, {{\partial\, V_{sb}}\over {\partial\,
  \theta^i}}\, \gamma_{\bar bb}\, \Big[V_{sa}\, V_{rb}\, \Big(\partial_r\,
  \Pi_{\bar b} - (2\, \phi^{-1}\, \partial_r\, \phi + \sum_{\bar c}\,
  \gamma_{\bar ca}\, \partial_r\, R_{\bar c})\, \Pi_{\bar b}\Big)
  +\nonumber \\
  &+& \partial_r\, (V_{sa}\, V_{rb})\, \Pi_{\bar b}\Big] +
  \sum_{sb\bar b}\, \gamma_{\bar bb}\, {{\partial\, V_{sa}\, V_{rb}}\over
  {\partial\, \theta^i}}\, \partial_r\,  V_{sb}\, \Pi_{\bar b}  \Big)
  -\nonumber \\
 &-&\sum_{rstwjb}\, \sum_{c \not= b}\, {{\epsilon_{bct}\,
  Q_a^{-1}\, Q_b\, Q_c^{-1}}\over {Q_c\, Q_b^{-1} - Q_b\, Q_c^{-1}}}\,
  (V_{rc}\, V_{sa} - V_{ra}\, V_{sc})\, {{\partial\, V_{sb}}\over
  {\partial\, \theta^i}}\, V_{wt}\, B_{jw}\, \partial_r\, \pi_j^{(\theta)}
  +\nonumber \\
 &+& \sum_{jrtw}\, \Big(\sum_{b \not= a}\, {{\epsilon_{abt}\,
  Q_b^{-1}}\over {Q_b\, Q_a^{-1} - Q_a\, Q_b^{-1}}}\,
 (2\, \phi^{-1}\, \partial_r\, \phi + \sum_{\bar c}\,
  \gamma_{\bar ca}\, \partial_r\, R_{\bar c})\, {{\partial\,
  V_{rb}\, V_{wt}\, B_{jw}}\over {\partial\, \theta^i}}
  +\nonumber \\
 &+& \sum_{sb}\, \sum_{c \not= b}\, {{\epsilon_{bct}\,
  Q_a^{-1}\, Q_b\, Q_c^{-1}}\over {Q_c\, Q_b^{-1} - Q_b\, Q_c^{-1}}}\,
 \Big[{{\partial\, V_{sb}}\over {\partial\, \theta^i}}\, \Big(
 \Big[2\, \phi^{-1}\, \partial_r\, \phi + \sum_{\bar b}\, \Big(\gamma_{\bar ba}
 - \gamma_{\bar bb} + \gamma_{\bar bc} -\nonumber \\
 &-& (\gamma_{\bar bb} - \gamma_{\bar bc})\,
 {{Q_c\, Q_b^{-1} + Q_b\, Q_c^{-1}}\over {Q_c\,
 Q_b^{-1} - Q_b\, Q_c^{-1}}} \Big)\, \partial_r\, R_{\bar b}
 \Big]\, (V_{rc}\, V_{sa} - V_{ra}\, V_{sc})\, V_{wt}\, B_{jw} -\nonumber \\
 &-& \partial_r\, \Big[(V_{rc}\, V_{sa} - V_{ra}\, V_{sc})\, V_{wt}\, B_{jw}\Big]
 \Big) +\nonumber \\
 &+& \partial_r\, V_{sb}\, {{\partial\, (V_{rc}\, V_{sa} - V_{ra}\,
 V_{sc})\, V_{wt}\, B_{jw}}\over {\partial\, \theta^i}} \Big]\,\,
 \Big)\, \pi_j^{(\theta)}  \Big](\tau ,\vec \sigma).
  \label{b5}
  \eea

\subsubsection{The Functional Derivative with Respect to $\Pi_{\bar a}$}

\bea
    &&\int d^3\sigma_1\, \sum_a\, {\bar n}_{(a)}(\tau ,{\vec \sigma}_1)\,
   {{\delta\, {\tilde {\bar {\cal H}}}_{(a)}(\tau ,{\vec \sigma}_1
  )}\over {\delta\, \Pi_{\bar a}(\tau ,\vec \sigma )}}\, =\nonumber \\
  &=&\Big[\phi^{-2}\, \sum_{ra}\, Q_a^{-1}\, \Big({\bar n}_{(a)}\,
  \Big[\gamma_{\bar aa}\, V_{ra}\, \Big(2\, \phi^{-1}\, \partial_r\, \phi +
  \sum_{\bar b}\, \gamma_{\bar ba}\, \partial_r\, R_{\bar b}\Big) -\nonumber \\
  &-& V_{ra}\, \partial_r\, R_{\bar a} + \sum_{sb}\, \gamma_{\bar
  ab}\, V_{sa}\, V_{rb}\, \partial_r\, V_{sb}  \Big] -
   \gamma_{\bar aa}\, \partial_r\, {\bar n}_{(a)}\, V_{ra}
  \Big)\Big](\tau, \vec \sigma).
  \label{b6}
  \eea

\subsubsection{The Functional Derivative with Respect to $R_{\bar a}$}

By using the results \medskip

\noindent ${{\delta\, Q_a^n(\tau ,{\vec \sigma}_1)}\over {\delta\,
R_{\bar a}(\tau, \vec \sigma)}} = n\, \gamma_{\bar aa}\, Q_a^n(\tau
,\vec \sigma)\, \delta^3(\vec \sigma, {\vec \sigma}_1)$,

\noindent ${{\delta}\over
 {\delta\, R_{\bar a}(\tau ,\vec \sigma)}}\, {{Q_b^{-1}}\over {Q_b\, Q_a^{-1} - Q_a\,
Q_b^{-1}}}(\tau, {\vec \sigma}_1) = - \Big({{Q_b^{-1}}\over {Q_b\,
Q_a^{-1} - Q_a\, Q_b^{-1}}}\, \Big[\gamma_{\bar ab} - (\gamma_{\bar
aa} - \gamma_{\bar ab})\, {{Q_b\, Q_a^{-1} + Q_a\, Q_b^{-1}}\over
{Q_b\, Q_a^{-1} - Q_a\, Q_b^{-1}}} \Big]\Big)(\tau ,\vec \sigma)\,$
$\delta^3(\vec \sigma, {\vec \sigma}_1)$,

\noindent ${{\delta}\over
 {\delta\, R_{\bar a}(\tau ,\vec \sigma)}}\, {{Q_a^{-1}}\over {(Q_b\, Q_a^{-1} - Q_a\,
Q_b^{-1})^2}}(\tau, {\vec \sigma}_1) = - \Big({{Q_a^{-1}}\over
{(Q_b\, Q_a^{-1} - Q_a\, Q_b^{-1})^2}}\, \Big[\gamma_{\bar aa} - 2\,
(\gamma_{\bar aa} - \gamma_{\bar ab})\, {{Q_b\, Q_a^{-1} + Q_a\,
Q_b^{-1}}\over {Q_b\, Q_a^{-1} - Q_a\, Q_b^{-1}}} \Big]\Big)(\tau
,\vec \sigma)\,$ $\delta^3(\vec \sigma, {\vec \sigma}_1)$,

\noindent ${{\delta}\over {\delta\, R_{\bar a}(\tau ,\vec
\sigma)}}\, {{Q_a^{-1}\, Q_b\, Q_c^{-1}}\over {Q_c\, Q_b^{-1} -
Q_b\, Q_c^{-1}}}(\tau, {\vec \sigma}_1) = - \Big({{Q_a^{-1}\, Q_b\,
Q_c^{-1}}\over {Q_c\, Q_b^{-1} - Q_b\, Q_c^{-1}}}\,
\Big[\gamma_{\bar aa} - (\gamma_{\bar ab} - \gamma_{\bar ac})\,
{{2\, Q_c\, Q_b^{-1}}\over {Q_c\, Q_b^{-1} - Q_b\, Q_c^{-1}}}
\Big]\Big)(\tau ,\vec \sigma)\,$ $\delta^3(\vec \sigma, {\vec
\sigma}_1)$,

\noindent $\partial_r\, {{Q_a^{-1}}\over {(Q_b\, Q_a^{-1} - Q_a\,
Q_b^{-1})^2}} = - {{Q_a^{-1}}\over {(Q_b\, Q_a^{-1} - Q_a\,
Q_b^{-1})^2}}\, \sum_{\bar b}\, \Big[\gamma_{\bar ba} - 2\,
(\gamma_{\bar ba} - \gamma_{\bar bb})\, {{Q_b\, Q_a^{-1} + Q_a\,
Q_b^{-1}}\over {Q_b\, Q_a^{-1} - Q_a\, Q_b^{-1}}}\Big]\,
\partial_r\, R_{\bar b}$,

\medskip

\noindent we get

\begin{eqnarray*}
    &&\int d^3\sigma_1\, \sum_a\, {\bar n}_{(a)}(\tau ,{\vec \sigma}_1)\,
   {{\delta\, {\tilde {\bar {\cal H}}}_{(a)}(\tau ,{\vec \sigma}_1
  )}\over {\delta\, R_{\bar a}(\tau ,\vec \sigma )}}\, =\nonumber \\
  &=& \phi^{-2}(\tau, \vec \sigma)\, \sum_{ra}\, \partial_r\, {\bar
  n}_{(a)}(\tau ,\vec \sigma)\, \Big[Q_a^{-1}\, V_{ra}\, \Pi_{\bar
  a} -\nonumber \\
  &-& 2\, \sum_{b \not= a}\, (\gamma_{\bar aa} - \gamma_{\bar ab})\,
  \sum_{twi}\, {{\epsilon_{abt}\, Q_a^{-1}}\over
  {(Q_b\, Q_a^{-1} - Q_a\, Q_b^{-1})^2}}\,  V_{rb}\, V_{wt}\, B_{iw}\,
  \pi_i^{(\theta)}\Big](\tau ,\vec \sigma) +\nonumber \\
  &&{}\nonumber \\
  &+& \phi^{-2}(\tau ,\vec \sigma)\, \sum_a\, {\bar n}_{(a)}(\tau
  ,\vec \sigma)\nonumber \\
  &&\Big[Q_a^{-1}\, \sum_r\, \Big(\partial_r\, V_{ra}\, \Pi_{\bar a}
  + V_{ra}\, \Big[\partial_r\, \Pi_{\bar a} - (2\, \phi^{-1}\, \partial_r\, \phi +
  \sum_{\bar b}\, \gamma_{\bar ba}\, \partial_r\, R_{\bar b})\,
  \Pi_{\bar a}\Big] -\nonumber \\
  &-& \gamma_{\bar aa}\, \Big[\sum_{\bar b}\, \Big(\gamma_{\bar ba}\,
  \partial_r\, V_{ra} + \sum_{sb}\, \gamma_{\bar bb}\, V_{sa}\, V_{rb}\,
  \partial_r\, V_{sb}\Big)\, \Pi_{\bar b} +\nonumber \\
  &+& V_{ra}\, \Big(\phi^6\, \partial_r\, \pi_{\tilde \phi} + \sum_{\bar b}\,
  (\gamma_{\bar ba}\, \partial_r\, \Pi_{\bar b} -  \partial_r\,
  R_{\bar b}\, \Pi_{\bar b}) + {\check {\cal M}}_r\Big)\Big]
  \Big) -
  \end{eqnarray*}

  \bea
 &-& \gamma_{\bar aa}\, \sum_{b \not= a}\, \sum_{rtwi}\, {{\epsilon_{abt}\,
 Q_b^{-1}}\over {Q_b\, Q_a^{-1} - Q_a\, Q_b^{-1}}}\, V_{rb}\,
 V_{wt}\, B_{iw}\, \partial_r\, \pi_i^{(\theta)} +\nonumber \\
 &+& \sum_i\, \sum_{rtw}\, \Big(2\, \sum_{b \not= a}\,
 {{\epsilon_{abt}\, Q_a^{-1}}\over {(Q_b\, Q_a^{-1} - Q_a\, Q_b^{-1})^2}}\,
 \Big[2\, (\gamma_{\bar aa} - \gamma_{\bar ab})\, \phi^{-1}\, \partial_r\,
 \phi +\nonumber \\
 &+& \sum_{\bar b}\, (\gamma_{\bar aa}\, \gamma_{\bar bb} - \gamma_{\bar ab}\,
 \gamma_{\bar ba})\, \partial_r\, R_{\bar b}\Big]\, V_{rb}\, V_{wt}\,
 B_{iw} -\nonumber \\
 &-& \gamma_{\bar aa}\, \sum_{b \not= a}\, {{\epsilon_{abt}\,
 Q_b^{-1}}\over {Q_b\, Q_a^{-1} - Q_a\, Q_b^{-1}}}\, \partial_r\,
 (V_{rb}\, V_{wt}\, B_{iw}) -\nonumber \\
 &-& \sum_b\, \sum_{c \not= b}\, {{\epsilon_{bct}\, Q_a^{-1}\, Q_b\,
 Q_c^{-1}}\over {Q_c\, Q_b^{-1} - Q_b\, Q_c^{-1}}}\,
 \Big[\gamma_{\bar aa} - (\gamma_{\bar ab} - \gamma_{\bar ac})\, {{2\,
 Q_a\, Q_b^{-1}}\over {Q_c\, Q_b^{-1} - Q_b\, Q_c^{-1}}}\Big]\nonumber \\
  &&\sum_s\, (V_{rc}\, V_{sa} - V_{ra}\, V_{sc})\, \partial_r\, V_{sb}\,
  V_{wt}\, B_{iw}\Big)\, \pi_i^{(\theta)}\Big](\tau ,\vec \sigma).
  \label{b7}
  \eea

\subsection{The Function ${\cal S}$ and its Functional Derivatives.}

By using Eqs. (B1) and (B2) of Ref.\cite{4} we have the following
expression for the function ${\cal S}(\tau, \vec \sigma)$ defined in
Eq.(\ref{3.13})

\begin{eqnarray*}
 {\cal S} &{\buildrel {def}\over =}&
 {}^3e\, \sum_{rsa}\, {}^3e^r_{(a)}\, {}^3e^s_{(a)}\,
 \sum_{uv}\, \Big({}^3\Gamma^u_{rv}\, {}^3\Gamma^v_{su} - {}^3\Gamma^u_{rs}\,
 {}^3\Gamma^v_{uv}\Big) =\nonumber \\
  &&{}\nonumber \\
  &&{}\nonumber \\
  &=& \phi^2\, \sum_a\, Q^{- 2}_a\nonumber \\
  &&\Big(\sum_{rv}\, \Big[2\, \Big(V_{ra}\, \partial_v\, ln\, \phi +
  \delta_{rv}\, \sum_u\, V_{ua}\, \partial_u\, ln\, \phi \Big)
  -\nonumber \\
  &-& 2\, \sum_{bu}\, V_{rb}\, V_{ub}\,
  V_{va}\, Q^2_a\, Q^{-2}_b\,  \partial_u\, ln\, \phi +\nonumber \\
 &+& \sum_{\bar bb}\, \gamma_{\bar bb}\, V_{rb}\, \Big(\delta_{ab}\,
 \partial_v\, R_{\bar b} + V_{vb}\, \sum_u\, V_{ua}\,
 \partial_u\, R_{\bar b}\Big) -\nonumber \\
  &-& \sum_{\bar bbu}\, \gamma_{\bar ba}\, V_{rb}\, V_{ub}\,
  V_{va}\, Q^2_a\, Q^{-2}_b\, \partial_u\, R_{\bar b} +\nonumber \\
 &+& {1\over 2}\, \sum_{bu}\, V_{rb}\, V_{ua}\, \Big(\partial_u\,
 V_{vb} + \partial_v\, V_{ub}\Big) + \nonumber \\
 &+& {1\over 2}\, \sum_{bcw}\, V_{rb}\, V_{wb}\, Q_c^2\, Q_b^{-2}\, \Big(\delta_{ac}\,
 \Big[\partial_v\, V_{wc} - \partial_w\,
 V_{vc}\Big] +\nonumber \\
 &+&  V_{vc}\, \sum_u\, V_{ua}\, \Big[\partial_u\, V_{wc} -
 \partial_w\, V_{uc}\Big] \Big) \Big]\times
  \end{eqnarray*}

\begin{eqnarray*}
  &&\Big[2\, \Big(V_{va}\, \partial_r\, ln\, \phi +
  \delta_{rv}\, \sum_s\, V_{sa}\, \partial_s\, ln\, \phi \Big)
  -\nonumber \\
  &-& 2\, \sum_{ds}\, V_{vd}\, V_{sd}\,
  V_{ra}\, Q^2_a\, Q^{-2}_d\,  \partial_s\, ln\, \phi +\nonumber \\
 &+& \sum_{\bar cd}\, \gamma_{\bar cd}\, V_{vd}\, \Big(\delta_{ad}\,
 \partial_r\, R_{\bar c} + V_{rd}\, \sum_s\, V_{sa}\,
 \partial_s\, R_{\bar c}\Big) -\nonumber \\
  &-& \sum_{\bar cds}\, \gamma_{\bar ca}\, V_{vd}\, V_{sd}\,
  V_{ra}\, Q^2_a\, Q^{-2}_d\, \partial_s\, R_{\bar c} +\nonumber \\
 &+& {1\over 2}\, \sum_{ds}\, V_{vd}\, V_{sa}\, \Big(\partial_s\,
 V_{rd} + \partial_r\, V_{sd}\Big) + \nonumber \\
 &+& {1\over 2}\, \sum_{des}\, V_{vd}\, V_{sd}\, Q_e^2\, Q_d^{-2}\, \Big(\delta_{ae}\,
 \Big[\partial_r\, V_{se} - \partial_s\,
 V_{re}\Big] +\nonumber \\
  &+& V_{re}\, \sum_t\, V_{ta}\, \Big[\partial_t\, V_{se} -
 \partial_s\, V_{te}\Big] \Big)\Big] +
 \end{eqnarray*}

 \begin{eqnarray*}
 &+& 6\, \phi^2\, \sum_{ars}\, Q_a^{-2}\, \phi^{-1}\, \partial_r\,
 \phi\, \Big[\partial_s\, (V_{ra}\, V_{sa}) + 2\, V_{ra}\, V_{sa}\,
 (\phi^{-1}\, \partial_s\, \phi - \sum_{\bar c}\, \gamma_{\bar ca}\,
 \partial_s\, R_{\bar c})\Big] =
  \end{eqnarray*}

\medskip

 \bea
 &=& \phi^2\, \sum_a\, Q_a^{-2}\, \sum_{rs}\, \Big(2\, \phi^{-1}\,
 \partial_r\, \phi\, \Big[\partial_s\, (V_{ra}\, V_{sa}) + 2\, V_{ra}\,
 V_{sa}\, (2\, \phi^{-1}\, \partial_s\, \phi - \sum_{\bar c}\,
 \gamma_{\bar ca}\, \partial_s\, R_{\bar c})\Big] +\nonumber \\
 &+& V_{ra}\, V_{sa}\, \sum_{\bar b\bar c}\, (2\, \gamma_{\bar ba}\,
 \gamma_{\bar ca} - \delta_{\bar b\bar c})\,  \partial_r\, R_{\bar b}\,
 \partial_s\, R_{\bar c} -\nonumber \\
 &-& 2\, \sum_{\bar b}\, \Big[ \gamma_{\bar ba}\, V_{sa}\, \partial_s\, V_{ra}
 + \sum_{vb}\, \gamma_{\bar bb}\, V_{ra}\, V_{vb}\, V_{sb}\, \partial_v\,
 V_{sa}\Big]\, \partial_r\, R_{\bar b} +\nonumber \\
 &+& {1\over 4}\, \sum_{uvbc}\, \Big[2\, \partial_r\, V_{sc}\, \partial_u\,
 V_{vb}\, \Big(V_{ra}\, V_{sb}\, (V_{uc}\, V_{va} - V_{ua}\, V_{vc})
 + V_{rb}\, V_{sa}\, (V_{ua}\, V_{vc} + V_{uc}\, V_{va})\Big) +
 \nonumber \\
 &+& Q_b^2\, Q_c^{-2}\, \partial_r\, V_{sb}\, \partial_u\, V_{vb}\,
 \Big((V_{ra}\, V_{sc} + V_{rc}\, V_{sa})\, (V_{ua}\, V_{vc} + V_{uc}\, V_{va})\,
 - 4\, V_{rc}\, V_{sa}\, V_{uc}\, V_{va}\Big)\Big]\Big).
 \label{b8}
 \eea

\medskip

${\cal S}(\tau ,\vec \sigma )$ is a function of $\tilde \phi =
\phi^6$, $\theta^i$ and $R_{\bar a}$, which, being
3-coordinate-dependent, plays the role of an inertial potential for
the gravitational field.

\bigskip

Due to Eq.(\ref{3.14}), the function ${\cal S}(\tau ,\vec \sigma )$
appears in the Dirac Hamiltonian in the form $\int d^3\sigma\,
\Big[(1 + n)\, {\cal S}\Big](\tau, \vec \sigma)$, so that we need
the functional derivatives of this quantity.

\subsubsection{The Functional Derivative with Respect to $R_{\bar a}$}

\begin{eqnarray*}
   &&\int d^3\sigma_1\, [1 + n(\tau, {\vec \sigma}_1)]\,
   {{\delta\, {\cal S}(\tau ,{\vec \sigma}_1)}\over
 {\delta\, R_{\bar a}(\tau ,\vec \sigma )}} =\nonumber \\
  &&{}\nonumber \\
  &&{}\nonumber \\
 &=&2\, \Big(\phi^2\, \sum_{rsa}\, Q_a^{-2}\, \nonumber \\
 &&\Big[\partial_r\, n\, \Big(V_{ra}\, V_{sa}\, \Big[2\,
 \gamma_{\bar aa}\, \phi^{-1}\, \partial_s\, \phi - \sum_{\bar b}\,
 (2\, \gamma_{\bar aa}\, \gamma_{\bar ba} - \delta_{\bar a\bar b})\,
 \partial_s\, R_{\bar b}\Big] +\nonumber \\
 &+& \gamma_{\bar aa}\, V_{sa}\, \partial_s\, V_{ra} + \sum_{vb}\,
 \gamma_{\bar ba}\, V_{ra}\, V_{vb}\, V_{sb}\, \partial_v\, V_{sa}\Big)
 -\nonumber \\
 &-& (1 + n)\, \Big(V_{ra}\, V_{sa}\, \Big[2\, \gamma_{\bar aa}\,
 (-\phi^{-1}\, \partial_r\, \partial_s\, \phi + 3\, \phi^{-1}\, \partial_r\,
 \phi\, \phi^{-1}\, \partial_s\, \phi) +\nonumber \\
 &+& 2\, \phi^{-1}\, \partial_r\, \phi\,
 \sum_{\bar b}\, (2\, \gamma_{\bar aa}\, \gamma_{\bar ba} -
 \delta_{\bar a\bar b})\, \partial_s\, R_{\bar b} +\nonumber \\
 &+& \sum_{\bar b}\, (2\, \gamma_{\bar aa}\, \gamma_{\bar ba} -
 \delta_{\bar a\bar b})\, \partial_r\, \partial_s\, R_{\bar b} +
 \sum_{\bar b\bar c}\, [2\, \gamma_{\bar ba}\, (\delta_{\bar a\bar c}
 - \gamma_{\bar aa}\, \gamma_{\bar ca}) - \gamma_{\bar aa}\,
 \delta_{\bar b\bar c}]\, \partial_r\, R_{\bar b}\, \partial_s\,
 R_{\bar c}\Big] -\nonumber \\
 &-& 2\, \phi^{-1}\, \partial_r\, \phi\, \Big[\gamma_{\bar aa}\,
  V_{sa}\, \partial_s\, V_{ra} +  \sum_{vb}\, \gamma_{\bar ab}\, V_{ra}\,
  V_{vb}\, V_{sb}\, \partial_v\, V_{sa}\Big] +\nonumber \\
 &+&\sum_{\bar b}\, \Big[ (2\, \gamma_{\bar aa}\, \gamma_{\bar ba} -
 \delta_{\bar a\bar b})\, \partial_s\, (V_{ra}\, V_{sa})
 + 2\, \sum_{vb}\, (\gamma_{\bar ba}\,  \gamma_{\bar ab} - \gamma_{\bar
 aa}\, \gamma_{\bar bb})\, V_{ra}\,
 V_{vb}\, V_{sb}\, \partial_v\, V_{sa}\Big]\, \partial_r\, R_{\bar b}
 -\end{eqnarray*}

 \bea
 &-&  \partial_r\, \Big(\gamma_{\bar aa}\, V_{sa}\, \partial_s\, V_{ra}
 + \sum_{vb}\, \gamma_{\bar ab}\, V_{ra}\, V_{vb}\, V_{sb}\,
 \partial_v\, V_{sa}\Big) +\nonumber \\
 &+&{1\over 4}\, \sum_{uvbc}\, \Big[2\, \gamma_{\bar aa}\,
 \partial_r\, V_{sc}\, \partial_u\,
 V_{vb}\, \Big(V_{ra}\, V_{sb}\, (V_{uc}\, V_{va} - V_{ua}\, V_{vc})
 + V_{rb}\, V_{sa}\, (V_{ua}\, V_{vc} + V_{uc}\, V_{va})\Big) +
 \nonumber \\
 &+& (\gamma_{\bar aa} - \gamma_{\bar ab} + \gamma_{\bar ac})\,
 Q_b^2\, Q_c^{-2}\, \partial_r\, V_{sb}\, \partial_u\, V_{vb}\,
 \Big((V_{ra}\, V_{sc} + V_{rc}\, V_{sa})\, (V_{ua}\, V_{vc} + V_{uc}\, V_{va})\,
 -\nonumber \\
 &-& 4\, V_{rc}\, V_{sa}\, V_{uc}\, V_{va}\Big)
 \Big]\,\, \Big)\Big]\Big)(\tau, \vec \sigma).
 \label{b9}
 \eea

\subsubsection{The Functional Derivative with Respect to $\tilde \phi = \phi^6$}

\bea
   &&\int d^3\sigma_1\, [1 + n(\tau, {\vec \sigma}_1)]\,
   {{\delta\, {\cal S}(\tau ,{\vec \sigma}_1)}\over
 {\delta\, \phi (\tau ,\vec \sigma )}} =\nonumber \\
  &&{}\nonumber \\
  &&{}\nonumber \\
 &=&2\, \Big(\phi\, (1 + n)\, \sum_{rsa}\, Q_a^{-2}\,
 \Big[- 8\, \Big(\phi^{-1}\, \partial_r\, \phi\, \partial_s\, (V_{ra}\, V_{sa})
 +\nonumber \\
 &+& V_{ra}\, V_{sa}\, (\phi^{-1}\, \partial_r\, \partial_s\,
 \phi - 2\, \phi^{-1}\, \partial_r\, \phi\, \sum_{\bar b}\,
 \gamma_{\bar ba}\, \partial_s\, R_{\bar b})\Big) +\nonumber \\
 &+&V_{ra}\, V_{sa}\, \Big(2\, \sum_{\bar b}\, \gamma_{\bar ba}\,
 \partial_r\, \partial_s\, R_{\bar b} - \sum_{\bar b\bar c}\,
 (2\, \gamma_{\bar ba}\, \gamma_{\bar ca} + \delta_{\bar b\bar c})\,
 \partial_r\, R_{\bar b}\, \partial_s\, R_{\bar c}\Big) -\nonumber \\
 &-&2\, \sum_{\bar b}\, \Big(\gamma_{\bar ba}\, \Big[V_{sa}\,
 \partial_s\, V_{ra} - 2\, \partial_s\, (V_{ra}\, V_{sa})\Big]
 +\nonumber \\
 &+&\sum_{vb}\, \gamma_{\bar bb}\, V_{ra}\, V_{vb}\, V_{sb}\,
 \partial_v\, V_{sa}\Big)\, \partial_r\, R_{\bar b}
 - \partial_r\, \partial_s\, (V_{ra}\, V_{sa}) +\nonumber \\
 &+& {1\over 4}\, \sum_{uvbc}\, \Big[2\, \partial_r\, V_{sc}\, \partial_u\,
 V_{vb}\, \Big(V_{ra}\, V_{sb}\, (V_{uc}\, V_{va} - V_{ua}\, V_{vc})
 + V_{rb}\, V_{sa}\, (V_{ua}\, V_{vc} + V_{uc}\, V_{va})\Big) +
 \nonumber \\
 &+& Q_b^2\, Q_c^{-2}\, \partial_r\, V_{sb}\, \partial_u\, V_{vb}\,
 \Big((V_{ra}\, V_{sc} + V_{rc}\, V_{sa})\, (V_{ua}\, V_{vc} + V_{uc}\, V_{va})\,
 - 4\, V_{rc}\, V_{sa}\, V_{uc}\, V_{va}\Big)\Big]\,\,
 \Big] -\nonumber \\
 &-&\phi\, \sum_{rsa}\, \partial_r\, n\, Q_a^{-2}\, \Big[
 2\, V_{ra}\, V_{sa}\, (4\, \phi^{-1}\, \partial_s\, \phi -
 \sum_{\bar b}\, \gamma_{\bar ba}\, \partial_s\, R_{\bar b})
 + \partial_s\, (V_{ra}\, V_{sa})
 \Big]\Big)(\tau ,\vec \sigma).\nonumber \\
 &&{}
 \label{b10}
  \eea

\subsubsection{The Functional Derivative with Respect to $\theta^i$}

\begin{eqnarray*}
   &&\int d^3\sigma_1\, [1 + n(\tau, {\vec \sigma}_1)]\,
   {{\delta\, {\cal S}(\tau ,{\vec \sigma}_1)}\over
 {\delta\, \theta^i(\tau ,\vec \sigma )}} =\nonumber \\
  &&{}\nonumber \\
  &&{}\nonumber \\
 &=& - \Big[\phi^2\, \sum_{rsa}\, Q_a^{-2}\, \partial_r\, n\, \Big(
 2\, {{\partial\, V_{ra}\, V_{sa}}\over {\partial\, \theta^i}}\,
 \phi^{-1}\, \partial_s\, \phi -\nonumber \\
 &-&2\, {{\partial\, V_{sa}}\over {\partial\, \theta^i}}\, \sum_{v\bar b}\,
 \Big[\gamma_{\bar ba}\, V_{ra}\, \delta_{sv} + \sum_b\, \gamma_{\bar bb}\,
 V_{va}\, V_{rb}\, V_{sb}\Big]\, \partial_v\, R_{\bar b}
 +\nonumber \\
 &+&\sum_{uvbc}\, {{\partial\, V_{sc}}\over {\partial\, \theta^i}}\,
 \partial_u\, V_{vb}\, \Big[V_{ra}\, V_{sb}\, (V_{uc}\, V_{va} - V_{ua}\, V_{vc})
 + V_{rb}\, V_{sa}\, (V_{ua}\, V_{vc} + V_{uc}\, V_{va})\Big]
 +\nonumber \\
 &+&{1\over 2}\, \sum_{uvbc}\, Q_b^2\, Q_c^{-2}\, {{\partial\, V_{sb}}\over
 {\partial\, \theta^i}}\, \partial_u\, V_{vb}\,
 \Big[(V_{ra}\, V_{sc} + V_{rc}\, V_{sa})\, (V_{ua}\, V_{vc} +
V_{uc}\, V_{va})\, - 4\, V_{rc}\, V_{sa}\, V_{uc}\, V_{va}\Big]
 \Big)\Big](\tau ,\vec \sigma) +
 \end{eqnarray*}

\begin{eqnarray*}
 &+& \Big[\phi^2\, (1 + n)\, \sum_{rsa}\, Q_a^{-2}\, \Big(
 {{\partial\, V_{ra}\, V_{sa}}\over {\partial\, \theta^i}}\,
 \Big[- 2\, \phi^{-1}\, \partial_r\, \partial_s\, \phi + 6\, \phi^{-1}\,
 \partial_r\, \phi\, \phi^{-1}\, \partial_s\, \phi +\nonumber \\
 &+& \sum_{\bar b\bar c}\, (2\, \gamma_{\bar ba}\, \gamma_{\bar ca}
 - \delta_{\bar b\bar c})\, \partial_r\, R_{\bar b}\, \partial_s\,
 R_{\bar c}\Big] -\nonumber \\
 &-& 2\, \sum_{sb\bar b}\, \gamma_{\bar bb}\, {{\partial\, V_{ra}\,
 V_{vb}\, V_{sb}}\over {\partial\, \theta^i}}\,
 \partial_v\, V_{sa}\, \partial_r\, R_{\bar b} +\nonumber \\
 &+&2\, {{\partial\, V_{sa}}\over {\partial\, \theta^i}}\, \sum_{\bar b}\,
 \Big[\gamma_{\bar ba}\, \Big(V_{ra}\, \partial_r\, \partial_s\, R_{\bar b}
 + \Big[\partial_r\, V_{ra} +\nonumber \\
 &+& 2\, V_{ra}\, (\phi^{-1}\, \partial_r\, \phi - \sum_{\bar c}\,
 \gamma_{\bar ca}\, \partial_r\, R_{\bar c})\Big]\, \partial_s\,
 R_{\bar b} - \partial_s\, V_{ra}\, \partial_r\, R_{\bar b}\Big)
 +\nonumber \\
 &+& \sum_{vb}\, \gamma_{\bar bb}\, \Big(V_{va}\, V_{rb}\, V_{sb}\,
 \Big[\partial_r\, \partial_v\, R_{\bar b} +\nonumber \\
 &+& 2\, (\phi^{-1}\, \partial_r\, \phi - \sum_{\bar c}\, \gamma_{\bar ca}\,
 \partial_r\, R_{\bar c})\, \partial_v\, R_{\bar b}\Big] + \partial_r\,
 (V_{va}\, V_{rb}\, V_{sb})\, \partial_v\, R_{\bar b}\Big)\Big] +
 \end{eqnarray*}

\bea
 &+& \sum_{uvbc}\, \Big[{1\over 2}\, \partial_r\, V_{sc}\,
 \partial_u\, V_{vb}\nonumber \\
 &&{{\partial}\over {\partial\, \theta^i}}\,
 \Big(V_{ra}\, V_{sb}\, (V_{uc}\, V_{va} - V_{ua}\, V_{vc})
 + V_{rb}\, V_{sa}\, (V_{ua}\, V_{vc} + V_{uc}\, V_{va})\Big)
 -\nonumber \\
 &-&{{\partial\, V_{sc}}\over {\partial\, \theta^i}}\, \Big(2\,
 (\phi^{-1}\, \partial_r\, \phi - \sum_{\bar b}\, \gamma_{\bar ba}\,
 \partial_r\, R_{\bar b})\, \partial_u\, V_{vb}\nonumber \\
 &&\Big[V_{ra}\, V_{sb}\, (V_{uc}\, V_{va} - V_{ua}\, V_{vc})
 + V_{rb}\, V_{sa}\, (V_{ua}\, V_{vc} + V_{uc}\, V_{va})\Big]
 +\nonumber \\
 &+& \partial_r\, \Big[\partial_u\, V_{vb}\,
 \Big(V_{ra}\, V_{sb}\, (V_{uc}\, V_{va} - V_{ua}\, V_{vc})
 + V_{rb}\, V_{sa}\, (V_{ua}\, V_{vc} + V_{uc}\, V_{va})\Big)
\Big] \Big)\Big] +\nonumber \\
 &+&{1\over 2}\, \sum_{uvbc}\, Q_b^2\, Q_c^{-2}\, \Big[
 {1\over 2}\, \partial_r\, V_{sb}\, \partial_u\, V_{vb}\,
 \nonumber \\
 &&{{\partial}\over {\partial\, \theta^i}}\,
 \Big((V_{ra}\, V_{sc} + V_{rc}\, V_{sa})\, (V_{ua}\, V_{vc} +
  V_{uc}\, V_{va})\, - 4\, V_{rc}\, V_{sa}\, V_{uc}\, V_{va}\Big)
  -\nonumber \\
  &-& {{\partial\, V_{sb}}\over {\partial\, \theta^i}}\, \Big(2\,
  (\phi^{-1}\, \partial_r\, \phi - \sum_{\bar b}\, (\gamma_{\bar ba} -
  \gamma_{\bar bb} + \gamma_{\bar bc})\, \partial_r\, R_{\bar b})\,
  \partial_u\, V_{vb}\nonumber \\
  && \Big((V_{ra}\, V_{sc} + V_{rc}\, V_{sa})\, (V_{ua}\, V_{vc} +
 V_{uc}\, V_{va})\, - 4\, V_{rc}\, V_{sa}\, V_{uc}\, V_{va}\Big)
 +\nonumber \\
 &+& \partial_r\, \Big[\partial_u\, V_{vb}\,
 \Big((V_{ra}\, V_{sc} + V_{rc}\, V_{sa})\, (V_{ua}\, V_{vc} +
 V_{uc}\, V_{va})\, - 4\, V_{rc}\, V_{sa}\, V_{uc}\, V_{va}\Big)
 \Big] \Big)\Big]\,\, \Big)\Big](\tau ,\vec \sigma).\nonumber \\
 &&{}
 \label{b11}
  \eea

\subsection{The Laplace-Beltrami operator $\hat \triangle$,
the Function ${\cal T}$ and their Functional Derivatives}

The Laplace-Beltrami operator associated with the 3-metric
${}^3{\hat g}_{rs}$ ($det\, {\hat g}_{rs} = 1$) appearing in
Eq.(\ref{3.43}) is

\bea
   \hat \triangle   &=& \partial_r\, ({}^3{\hat g}^{rs}\, \partial_s)
  = {}^3{\hat g}^{rs}\, {}^3{\hat \nabla}_r\, {}^3{\hat \nabla}_s =
 \partial_r\, \Big(\sum_a\, Q_a^{-2}\,
 V_{ra}\, V_{sa}\,  \partial_s\Big) =\nonumber \\
 &=& \sum_a\, Q_a^{-2}\, \Big[V_{ra}\, V_{sa}\, \partial_r\, \partial_s
 - \Big(2\, V_{ra}\, V_{sa}\, \sum_{\bar b}\, \gamma_{\bar ba}\, \partial_r\, R_{\bar b}
 - \partial_r\, (V_{ra}\, V_{sa})\Big)\, \partial_s\Big].
 \label{b12}
 \eea

From Eqs.(\ref{3.13}) and (\ref{2.10}) we have

{\begin{eqnarray*}
 {\cal T}(\tau ,\vec \sigma ) &=&  \sum_r\, \partial_r\,
 \Big({}^3e\, \sum_{suv}\, {}^3g^{rs}\,
{}^3g^{uv}\, (\partial_u\, {}^3g_{vs} - \partial_s\,
{}^3g_{uv})\Big)(\tau ,\vec \sigma ) =\nonumber \\
 &&{}\nonumber \\
  &=& - \sum_{rsa}\, \partial_r\, \Big(\phi^2\, Q_a^{-2}\, \Big[
 2\, V_{ra}\, V_{sa}\, (4\, \phi^{-1}\, \partial_s\, \phi - \sum_{\bar b}\,
 \gamma_{\bar ba}\, \partial_s\, R_{\bar b}) + \partial_s\, (V_{ra}\, V_{sa})
  \Big]\Big)(\tau  ,\vec \sigma ) =\nonumber \\
 &=&- \Big[\phi^2\, \sum_{rsa}\, Q_a^{-2}\, \Big(2\, V_{ra}\, V_{sa}\,
 \Big[4\, (\phi^{-1}\, \partial_r\, \partial_s\, \phi + \phi^{-1}\,
 \partial_r\, \phi\, \phi^{-1}\, \partial_s\, \phi) -\nonumber \\
 &-& \sum_{\bar b}\, \gamma_{\bar ba}\, \Big( \partial_r\,
 \partial_s\, R_{\bar b} + 2\, (5\, \phi^{-1}\, \partial_s\, \phi -
 \sum_{\bar c}\, \gamma_{\bar ca}\, \partial_s\, R_{\bar c})\,
 \partial_r\, R_{\bar b}\Big)\Big] +\nonumber \\
 &+& 2\, \partial_r\, (V_{ra}\, V_{sa})\, (5\, \phi^{-1}\, \partial_s\,
 \phi - 2\, \sum_{\bar b}\, \gamma_{\bar ba}\, \partial_s\, R_{\bar b}) +
 \partial_r\, \partial_s\, (V_{ra}\, V_{sa}) \Big)\Big](\tau ,\vec \sigma)
  =\nonumber \\
 &&{}\nonumber \\
 &{\buildrel {def}\over =}& \Big(- 8\, \phi\, \hat \triangle\, \phi +
 {\cal T}_1\Big)(\tau ,\vec \sigma ),
 \end{eqnarray*}

 \bea
 {\cal T}_1(\tau ,\vec \sigma) &=&
  - \Big[\phi^2\, \sum_{rsa}\, Q_a^{-2}\, \Big(2\, V_{ra}\, V_{sa}\, \Big[4\,
 \phi^{-1}\, \partial_r\, \phi\, \phi^{-1}\, \partial_s\, \phi
 -\nonumber \\
 &-& \sum_{\bar b}\, \gamma_{\bar ba}\, \Big( \partial_r\, \partial_s\, R_{\bar b} +
 2\, (\phi^{-1}\, \partial_s\, \phi - \sum_{\bar c}\, \gamma_{\bar ca}\,
 \partial_s\, R_{\bar c})\, \partial_r\, R_{\bar b}\Big) \Big] +\nonumber \\
 &+& 2\, \partial_r\, (V_{ra}\, V_{sa})\, (\phi^{-1}\, \partial_s\, \phi -
 2\, \sum_{\bar b}\, \gamma_{\bar ba}\, \partial_s\, R_{\bar b}) +
 \partial_r\, \partial_s\, (V_{ra}\, V_{sa})\Big)\Big](\tau ,\vec \sigma).\nonumber \\
 &&{}
 \label{b13}
 \eea

\bigskip

In the Dirac Hamiltonian (\ref{3.48}) there is the term $-
{{c^3}\over {16\pi\, G}}\, \int d^3\sigma\, \Big[n\, {\cal
T}\Big](\tau ,\vec \sigma)$. To evaluate its functional derivatives
we use the form ${\cal T} = - 8\, \phi\, \hat \triangle\, \phi +
{\cal T}_1$. First we evaluate the functional derivatives of the
term ${{c^3}\over {2\pi\, G}}\, \int d^3\sigma\, \Big[n\, \phi\,
\hat \triangle\, \phi\Big](\tau ,\vec \sigma )$ and then of the
quantity $- {{c^3}\over {16\pi\, G}}\, \int d^3\sigma\, \Big[n\,
{\cal T}\Big](\tau ,\vec \sigma) = - {{c^3}\over {16\pi\, G}}\, \int
d^3\sigma\, \Big[n\, \Big(- 8\, \phi\, \hat \triangle\, \phi + {\cal
T}_1\Big)\Big](\tau ,\vec \sigma)$.

\subsubsection{The Functional Derivative with Respect to $R_{\bar a}$}

\bea
 &&\int d^3\sigma_1\, {{\delta\, \Big[n\, \phi\, \hat
\triangle\, \phi\Big](\tau ,{\vec \sigma}_1 )}\over {\delta\,
R_{\bar a}(\tau ,\vec \sigma)}} = 2\, \Big[\phi^2\, \sum_{rsa}\,
\gamma_{\bar aa}\, Q_a^{-2}\, V_{ra}\, V_{sa}\, \phi^{-1}\,
\partial_s\, \phi\, \Big(n\, \phi^{-1}\, \partial_r\, \phi\,  +
\partial_r\, n\Big)
  \Big](\tau ,\vec \sigma),\nonumber \\
  &&{}
  \label{b14}
  \eea

\bigskip

 \bea
   &&\int d^3\sigma_1\, n(\tau, {\vec \sigma}_1)\,
   {{\delta\, {\cal T}(\tau ,{\vec \sigma}_1)}\over
 {\delta\, R_{\bar a}(\tau ,\vec \sigma )}} =
  - 8\, \int d^3\sigma_1\, {{\delta\, \Big[n\, \phi\, \hat
 \triangle\, \phi\Big](\tau ,{\vec \sigma}_1 )}\over {\delta\,
 R_{\bar a}(\tau ,\vec \sigma)}} +\nonumber \\
 &+&2\, \Big[\phi^2\, \sum_{rsa}\, \gamma_{\bar aa}\, Q_a^{-2}\,
 V_{ra}\, V_{sa}\, (\partial_r\, \partial_s\, n + 2\, \partial_r\,
 n\, \phi^{-1}\, \partial_s\, \phi + 8\, n\, \phi^{-1}\,
 \partial_r\, \phi\, \phi^{-1}\, \partial_s\, \phi  )
 \Big](\tau ,\vec \sigma) =\nonumber \\
 &&{}\nonumber \\
 &=& 2\, \Big[\phi^2\, \sum_{rsa}\, \gamma_{\bar aa}\,
 Q_a^{-2}\, V_{ra}\, V_{sa}\, \Big(\partial_r\, \partial_s\, n -
 6\, \phi^{-1}\, \partial_s\, \phi\, \partial_r\, n  \Big)
 \Big](\tau ,\vec \sigma).
 \label{b15}
 \eea

\subsubsection{The Functional Derivative with Respect to $\tilde \phi = \phi^6$}

\bea
 &&\int d^3\sigma_1\, {{\delta\, \Big[n\, \phi\, \hat
\triangle\, \phi\Big](\tau ,{\vec \sigma}_1 )}\over {\delta\,
\phi(\tau ,\vec \sigma)}} =\nonumber \\
  &=& \Big[2\, n\, \hat \triangle\, \phi + 2\, \sum_a\, Q_a^{-2}\, V_{ra}\,
  V_{sa}\, \partial_r\, n\, \partial_s\, \phi +  \phi\, \hat \triangle\,
  n\Big](\tau ,\vec \sigma).
  \label{b16}
  \eea

\bigskip

 \bea
   &&\int d^3\sigma_1\, n(\tau, {\vec \sigma}_1)\,
   {{\delta\, {\cal T}(\tau ,{\vec \sigma}_1)}\over
 {\delta\, \phi(\tau ,\vec \sigma )}} =
 - 8\, \int d^3\sigma_1\, {{\delta\, \Big[n\, \phi\, \hat
\triangle\, \phi\Big](\tau ,{\vec \sigma}_1 )}\over {\delta\,
\phi(\tau ,\vec \sigma)}} +\nonumber \\
 &+&2\, \Big(8\, n\, \hat \triangle\, \phi + \phi\, Q_a^{-2}\,
 \partial_r\, n\, \Big[\partial_s\, (V_{ra}\, V_{sa}) +\nonumber \\
 &+& 2\, V_{ra}\, V_{sa}\, (4\, \phi^{-1}\, \partial_s\, \phi -
 \sum_{\bar b}\, \gamma_{\bar ba}\, \partial_s\, R_{\bar b})
 \Big]\Big)(\tau ,\vec \sigma)=\nonumber \\
 &&{}\nonumber \\
 &=&  - 2\, \Big(\phi\, \sum_{rsa}\, Q_a^{-2}\,
\Big[ 4\, V_{ra}\, V_{sa}\, \partial_r\, \partial_s\, n -
 \nonumber \\
 &-&3\, \Big(2\, V_{ra}\, V_{sa}\, \sum_{\bar b}\, \gamma_{\bar ba}\,
 \partial_r\, R_{\bar b} - \partial_r\, (V_{ra}\, V_{sa})\Big)\,
 \partial_s\, n \Big]\Big)(\tau ,\vec \sigma).
 \label{b17}
 \eea

\subsubsection{The Functional Derivative with Respect to $\theta^i$}

\bea
 &&\int d^3\sigma_1\, {{\delta\, \Big[n\, \phi\, \hat
\triangle\, \phi\Big](\tau ,{\vec \sigma}_1 )}\over {\delta\,
\theta^i(\tau ,\vec \sigma)}} =\nonumber \\
  &=& - \Big(\phi^2\, \sum_a\, Q_a^{-2}\, {{\partial\, V_{ra}\, V_{sa}}\over
  {\partial\, \theta^i}}\, \Big[n\, \phi^{-1}\, \partial_r\, \phi + \partial_r\, n
  \Big]\, \phi^{-1}\, \partial_s\, \phi\Big)(\tau ,\vec
  \sigma),\nonumber \\
  &&{}
  \label{b18}
  \eea

\bigskip

 \bea
   &&\int d^3\sigma_1\, n(\tau, {\vec \sigma}_1)\,
   {{\delta\, {\cal T}(\tau ,{\vec \sigma}_1)}\over
 {\delta\, \theta^i(\tau ,\vec \sigma )}} =
 - 8\, \int d^3\sigma_1\, {{\delta\, \Big[n\, \phi\, \hat
\triangle\, \phi\Big](\tau ,{\vec \sigma}_1 )}\over {\delta\,
\theta^i(\tau ,\vec \sigma)}} -\nonumber \\
 &-&  \Big[\phi^2\, \sum_{rsa}\, Q_a^{-2}\, {{\partial\, V_{ra}\,
V_{sa}}\over {\partial\, \theta^i}}\, \Big(\partial_r\, \partial_s\,
n + 2\, \partial_r\, n\,
\phi^{-1}\, \partial_s\, \phi -\nonumber \\
&-& 8\, n\,  \phi^{-1}\, \partial_r\, \phi\, \phi^{-1}\,
\partial_s\, \phi\Big)\Big](\tau ,\vec \sigma) =\nonumber \\
 &&{}\nonumber \\
 &=& -  \Big(\phi^2\, \sum_{rsa}\, Q_a^{-2}\, {{\partial\, V_{ra}\,
 V_{sa}}\over {\partial\, \theta^i}}\, \Big[\partial_r\,
 \partial_s\, n - 6\, \partial_r\, n\, \phi^{-1}\, \partial_s\, \phi
  \Big]\Big)(\tau ,\vec \sigma).
 \label{b19}
 \eea

\subsection{The Function ${}^3\hat R$.}

As shown in Ref.\cite{6}, for the evaluation of ${}^3\hat R$ (a
function of $\theta^i$ and $R_{\bar a}$), appearing in Eqs.
(\ref{3.43}) and (\ref{3.44}), we need the following results (see
Eqs.(A25) of Ref.\cite{2} and Eq.(\ref{2.10}) for the spin
connection ${}^3{\bar \omega}_{r(a)}$)

\begin{eqnarray*}
 {}^3R[\theta^n, \phi ,R_{\bar a}]
 &=& \sum_{abrs}\, {}^3e\, \epsilon_{(a)(b)(c)}\, {}^3{\bar e}^r_{(a)}\, {}^3{\bar e}^s_{(b)}\,
 {\bar \Omega}_{rs(c)} = {}^3e\, \Big[\partial_r\, {}^3{\bar \omega}_{s(a)(b)} -
   \partial_s\, {}^3{\bar \omega}_{r(a)(b)} +\nonumber \\
   &+& \sum_e\, \Big({}^3{\bar \omega}_{r(a)(e)}\, {}^3{\bar \omega}_{s(e)(b)} -
   {}^3{\bar \omega}_{s(a)(e)}\, {}^3{\bar
   \omega}_{r(e)(b)}\Big)\Big] =\nonumber \\
 &&{}\nonumber \\
 &=& \phi^{-5}\, \Big(- 8\,  {\hat \triangle}\, \phi +
 {}^3{\hat R}\, \phi \Big) = \phi^{-6}\, \Big({\cal S} + {\cal T}\Big) =
 \phi^{-6}\, \Big({\cal S} + {\cal T}_1 - 8\, \phi\, \hat \triangle\, \phi\Big),
 \end{eqnarray*}

\begin{eqnarray*}
 &&\Downarrow\nonumber \\
 &&{}\nonumber \\
 {}^3\hat R[\theta^n, R_{\bar a}] &=&
  \phi^{-2}\, \Big({\cal S} + {\cal T}_1\Big) =
  \end{eqnarray*}

\bea
 &=& \sum_{rsa}\, Q_a^{-2}\, \Big(- \partial_r\, \partial_s\, (V_{ra}\, V_{sa})
 +\nonumber \\
 &+& 2\, \sum_{\bar b}\, \Big[\gamma_{\bar ba}\, \Big(2\, \partial_s\, (V_{ra}\,
 V_{sa}) - V_{sa}\, \partial_s\, V_{ra}\Big) - \sum_{vb}\, \gamma_{\bar bb}\,
 V_{ra}\, V_{vb}\, V_{sb}\, \partial_v\, V_{sa}\Big]\, \partial_r\, R_{\bar b}
 -\nonumber \\
 &-& V_{ra}\, V_{sa}\, \sum_{\bar b}\, \Big[\partial_r\, R_{\bar
 b}\, \partial_s\, R_{\bar b} + 2\, \gamma_{\bar ba}\, \Big(
   \sum_{\bar c}\, \gamma_{\bar ca}\, \partial_s\, R_{\bar c}\,
 \, \partial_r\, R_{\bar b} - \partial_r\, \partial_s\, R_{\bar b}\Big) \Big]
 +\nonumber \\
 &+& {1\over 4}\, \sum_{uvbc}\, \Big[2\, \partial_r\, V_{sc}\, \partial_u\,
 V_{vb}\, \Big(V_{ra}\, V_{sb}\, (V_{uc}\, V_{va} - V_{ua}\, V_{vc})
 + V_{rb}\, V_{sa}\, (V_{uc}\, V_{vc} + V_{uc}\, V_{va})\Big) +
 \nonumber \\
 &+& Q_b^2\, Q_c^{-2}\, \partial_r\, V_{sb}\, \partial_u\, V_{vb}\,
 \Big((V_{ra}\, V_{sc} + V_{rc}\, V_{sa})\, (V_{ua}\, V_{vc} + V_{uc}\, V_{va})\,
 - 4\, V_{rc}\, V_{sa}\, V_{uc}\, V_{va}\Big)\Big] \Big).\nonumber \\
 &&{}
 \label{b20}
 \eea

\subsection{The Functional Derivatives  of ${\check {\cal M}}$}

We need the derivatives with respect $\theta^i$, $\phi$ and $R_{\bar
a}$ of the mass density ${\check {\cal M}}(\tau ,\vec \sigma)$,
appearing in the combination $\int d^3\sigma\, \Big[(1 + n)\,
{\check {\cal M}}\Big](\tau, \vec \sigma)$ in the Dirac Hamiltonian
(\ref{3.48}),given in Eqs.(\ref{3.37}). They are

\bea
 &&\int d^3\sigma_1\, \Big(1 + n(\tau ,{\vec \sigma}_1)\Big)\,
 {{\delta\, {\check {\cal M}}(\tau ,{\vec \sigma}_1)}\over
 {\delta\, \theta^i(\tau ,\vec \sigma)}} =\nonumber \\
 &&{}\nonumber \\
 &=&{1\over 2}\, \sum_{i=1}^N\, \delta^3(\vec \sigma, {\vec
 \eta}_i(\tau))\, \eta_i\, \Big((1 + n)\nonumber \\
 &&{{\phi^{-4}\, \sum_{rsa}\, Q_a^{-2}\, {{\partial\, V_{ra}\, V_{sa}}
 \over {\partial\, \theta^i}}\,  \Big({\check
 \kappa}_{ir}(\tau ) - {{Q_i}\over c}\, A_{\perp\, r}\Big)\,
 \Big({\check \kappa}_{is}(\tau ) - {{Q_i}\over c}\,
 A_{\perp\, s}\Big)}\over {\sqrt{m_i^2\, c^2 +
 \phi^{-4}\, \sum_{rsa}\, Q_a^{-2}\, V_{ra}\, V_{sa}\, \Big({\check
 \kappa}_{ir}(\tau ) - {{Q_i}\over c}\, A_{\perp\, r}\Big)\,
 \Big({\check \kappa}_{is}(\tau ) - {{Q_i}\over c}\, A_{\perp\,
 s}\Big)}}} \Big)(\tau ,\vec \sigma) +\nonumber \\
 &&{}\nonumber \\
  &+& \Big(1 + n(\tau ,\vec \sigma)\Big)\, \Big({\tilde \phi}^{-1/3}\,
 \nonumber \\
  &&\Big[{1\over {2c}}\, \sum_{rsa}\, Q_a^2\,
  {{\partial\, V_{ra}\, V_{sa}}\over {\partial\, \theta^i}}\,
  \pi^r_{\perp}\, \pi^s_{\perp} +
   {1\over {4c}}\, \sum_{abrsuv}\, Q_a^{-2}\, Q_b^{-2}\, {{\partial\, V_{ra}\, V_{sa}\,
  V_{ub}\, V_{vb}}\over {\partial\, \theta^i}}\, F_{ru}\, F_{sv} -\nonumber \\
  &-&  {1\over {2c}}\,\, \sum_{rsan}\, Q_a^2\,
  {{\partial\, V_{ra}\, V_{sa}}\over {\partial\, \theta^i}}\,
  \Big(2\, \pi^r_{\perp} - \sum_m\, \delta^{rm}\,\sum_{i=1}^N\, Q_i\, \eta_i\, \partial_m\,
  c(\vec \sigma , {\vec \eta}_i(\tau))\Big)\nonumber \\
  && \delta^{sn}\, \sum_{j=1}^N\, Q_j\, \eta_j\, \partial_n\,
  c(\vec \sigma , {\vec \eta}_j(\tau))
  \Big]\Big)(\tau ,\vec \sigma),\nonumber \\
  &&{}
 \label{b21}
 \eea

\bea
  &&\int d^3\sigma_1\, \Big(1 + n(\tau ,{\vec \sigma}_1)\Big)\,
 {{\delta\, {\check {\cal M}}(\tau ,{\vec \sigma}_1)}\over
 {\delta\, \phi(\tau ,\vec \sigma)}} =\nonumber \\
 &&{}\nonumber \\
 &=& - 2\, \sum_{i=1}^N\, \delta^3(\vec \sigma, {\vec
 \eta}_i(\tau))\, \eta_i\, \Big((1 + n)\nonumber \\
 &&{{\phi^{-5}\, \sum_{rsa}\, Q_a^{-2}\, V_{ra}\, V_{sa}\,
 \Big({\check
 \kappa}_{ir}(\tau ) - {{Q_i}\over c}\, A_{\perp\, r}\Big)\,
 \Big({\check \kappa}_{is}(\tau ) - {{Q_i}\over c}\,
 A_{\perp\, s}\Big) }\over {\sqrt{m_i^2\, c^2 +
 \phi^{-4}\, \sum_{rsa}\, Q_a^{-2}\, V_{ra}\, V_{sa}\, \Big({\check
 \kappa}_{ir}(\tau ) - {{Q_i}\over c}\, A_{\perp\, r}\Big)\,
 \Big({\check \kappa}_{is}(\tau ) - {{Q_i}\over c}\, A_{\perp\,
 s}\Big)}}} \Big)(\tau ,\vec \sigma) -\nonumber \\
 &&{}\nonumber \\
  &-& \Big(1 + n(\tau ,\vec \sigma)\Big)\, \Big({\tilde \phi}^{-1/2}\,
 \nonumber \\
 &&\Big[{1\over {c}}\, \sum_{rsa}\, Q_a^2\,
  V_{ra}\, V_{sa}\, \pi^r_{\perp}\, \pi^s_{\perp} +
    {1\over {2c}}\, \sum_{abrsuv}\, Q_a^{-2}\, Q_b^{-2}\, V_{ra}\, V_{sa}\,
  V_{ub}\, V_{vb}\, F_{ru}\, F_{sv} -\nonumber \\
 &-&  {1\over {c}}\, \sum_{rsan}\, Q_a^2\, V_{ra}\, V_{sa}\,
  \Big(2\, \pi^r_{\perp} - \sum_m\, \delta^{rm}\, \sum_{i=1}^N\, Q_i\, \eta_i\, \partial_m\,
  c(\vec \sigma , {\vec \eta}_i(\tau))\Big)\nonumber \\
 &&\delta^{sn}\, \sum_{j=1}^N\, Q_j\, \eta_j\, \partial_n\,
  c(\vec \sigma , {\vec \eta}_j(\tau))
 \Big]\Big)(\tau ,\vec \sigma),\nonumber \\
 &&{}
 \label{b22}
 \eea

\bea
   &&\int d^3\sigma_1\, \Big(1 + n(\tau ,{\vec \sigma}_1)\Big)\,
 {{\delta\, {\check {\cal M}}(\tau ,{\vec \sigma}_1)}\over
 {\delta\, R_{\bar a}(\tau ,\vec \sigma)}} =\nonumber \\
 &&{}\nonumber \\
 &=& - \sum_{i=1}^N\, \delta^3(\vec \sigma, {\vec
 \eta}_i(\tau))\, \eta_i\, \Big((1 + n)\nonumber \\
 && {{\phi^{-4}\, \sum_{rsa}\, \gamma_{\bar aa}\, Q_a^{-2}\, V_{ra}\, V_{sa}\,
 \Big({\check
 \kappa}_{ir}(\tau ) - {{Q_i}\over c}\, A_{\perp\, r}\Big)\,
 \Big({\check \kappa}_{is}(\tau ) - {{Q_i}\over c}\,
 A_{\perp\, s}\Big) }\over {\sqrt{m_i^2\, c^2 +
 \phi^{-4}\, \sum_{rsa}\, Q_a^{-2}\, V_{ra}\, V_{sa}\, \Big({\check
 \kappa}_{ir}(\tau ) - {{Q_i}\over c}\, A_{\perp\, r}\Big)\,
 \Big({\check \kappa}_{is}(\tau ) - {{Q_i}\over c}\, A_{\perp\,
 s}\Big)}}} \Big)(\tau ,\vec \sigma) +\nonumber \\
 &&{}\nonumber \\
 &+& \Big(1 + n(\tau ,\vec \sigma)\Big)\,  \Big({\tilde \phi}^{-1/3}\,
 \nonumber \\
 &&\Big[{1\over {c}}\, \sum_{rsa}\, \gamma_{\bar aa}\, Q_a^2\,
  V_{ra}\, V_{sa}\, \pi^r_{\perp}\, \pi^s_{\perp}  -
   {1\over {2c}}\, \sum_{abrsuv}\, (\gamma_{\bar aa} + \gamma_{\bar ab})\,
   Q_a^{-2}\, Q_b^{-2}\, V_{ra}\, V_{sa}\,
  V_{ub}\, V_{vb}\, F_{ru}\, F_{sv} -\nonumber \\
  &-&   {1\over {c}}\, \sum_{rsan}\, \gamma_{\bar aa}\, Q_a^2\, V_{ra}\, V_{sa}\,
  \Big(2\, \pi^r_{\perp} - \sum_m\, \delta^{rm}\,\sum_{i=1}^N\, Q_i\, \eta_i\, \partial_m\,
  c(\vec \sigma , {\vec \eta}_i(\tau))\Big)\nonumber \\
 && \delta^{sn}\, \sum_{j=1}^N\, Q_j\, \eta_j\, \partial_n\,
  c(\vec \sigma , {\vec \eta}_j(\tau))
 \Big]\Big)(\tau ,\vec \sigma).\nonumber \\
 &&{}
 \label{b23}
 \eea

\vfill\eject

\section{The Quantities of Appendix B in the 3-Orthogonal Gauges.}

In this Appendix we give the restriction of the quantities evaluated
in Appendix B the the 3-orthogonal gauges. Also most of these
quantities are given in terms of $\phi = {\tilde \phi}^{1/6}$ like
in Appendix B.

\bigskip

We shall use the results $V_{ra}(0) = \delta_{ra}$, $V_{(i)ra} =
\epsilon_{ira}$, $B_{(i)ab} = {1\over 2}\, \epsilon_{iab}$
introduced after Eq.(\ref{6.1}) (see the discussion after
Eq.(\ref{2.8})).

\bigskip

The functional derivatives of Eqs.(\ref{b2}) - (\ref{b7}) become

\bea
 {\tilde {\bar {\cal H}}}_{(a)}{|}_{\theta^i = 0}(\tau ,\vec \sigma) &=&
 \phi^{-2}(\tau, \vec \sigma)\, \Big[\sum_{b \not= a}\,
 \sum_i\, {{ \epsilon_{abi}\, Q_b^{-1}}
 \over {Q_b\, Q_a^{-1} - Q_a\, Q_b^{-1}}}\, \partial_b\, \pi_i^{(\theta)}
 +\nonumber \\
 &+&2\, \sum_{b \not= a}\, \sum_i\,  {{ \epsilon_{abi}\, Q_a^{-1}}
 \over {\Big(Q_b\, Q_a^{-1} - Q_a\, Q_b^{-1}\Big)^2}}\,
  \sum_{\bar c}\, (\gamma_{\bar ca} - \gamma_{\bar cb})\, \partial_b\,
  R_{\bar c}\,\, \pi_i^{(\theta)} +\nonumber \\
 &&{}\nonumber \\
 &+&Q_a^{-1}\,  \Big(\phi^6\, \partial_a\, \pi_{\tilde \phi}
 +  \sum_{\bar b}\, (\gamma_{\bar ba}\, \partial_a\, \Pi_{\bar b}
 - \partial_a\, R_{\bar b}\, \Pi_{\bar b}) + {\check {\cal M}}_{(a)}\Big)
 \Big](\tau ,\vec \sigma) \approx 0.\nonumber \\
 &&{}
 \label{c1}
 \eea

\bea
 &&\int d^3\sigma_1\, \sum_a\, {\bar n}_{(a)}(\tau ,{\vec \sigma}_1)\,
   {{\delta\, {\tilde {\bar {\cal H}}}_{(a)}(\tau ,{\vec \sigma}_1
  )}\over {\delta\, \pi_{\tilde \phi}(\tau ,\vec \sigma )}}\,
  {|}_{\theta^i = 0} =\nonumber \\
  &=&- \Big[\phi^4\, \sum_a\, Q_a^{-1}\, \Big(\partial_a\, {\bar
  n}_{(a)} + {\bar n}_{(a)}\,  \Big(4\, \phi^{-1}\, \partial_a\, \phi -
  \sum_{\bar b}\, \gamma_{\bar ba}\, \partial_a\,
  R_{\bar b}\Big)\, \Big) \Big](\tau ,\vec \sigma).\nonumber \\
  &&{}
 \label{c2}
 \eea

\beq
    \int d^3\sigma_1\, \sum_a\, {\bar n}_{(a)}(\tau ,{\vec \sigma}_1)\,
   {{\delta\, {\tilde {\bar {\cal H}}}_{(a)}(\tau ,{\vec \sigma}_1
  )}\over {\delta\, \phi(\tau ,\vec \sigma )}}\,
  {|}_{\theta^i = 0} \approx\,
   6\, \Big[\phi^3\, \sum_a\, {\bar n}_{(a)}\, Q_a^{-1}\,
   \partial_a\, \pi_{\tilde \phi}\Big](\tau ,\vec \sigma).
 \label{c3}
 \eeq

\bea
   &&\int d^3\sigma_1\, \sum_a\, {\bar n}_{(a)}(\tau ,{\vec \sigma}_1)\,
   {{\delta\, {\tilde {\bar {\cal H}}}_{(a)}(\tau ,{\vec \sigma}_1
  )}\over {\delta\, \pi_i^{(\theta )}(\tau ,\vec \sigma )}}\,
  {|}_{\theta^i = 0} =\nonumber \\
  &=& \Big(\phi^{-2}\, \sum_{a \not= b}\,  {{\epsilon_{iab}\, Q_b^{-1}}\over
  {Q_b\, Q_a^{-1} -  Q_a\, Q_b^{-1}}}\, \Big[({1\over 3}\, {\tilde
  \phi}^{-1}\, \partial_b\, \tilde \phi + \sum_{\bar b}\,
  \gamma_{\bar ba}\, \partial_b\, R_{\bar b})\, {\bar n}_{(a)} -
  \partial_b\, {\bar n}_{(a)} \Big]\Big)(\tau ,\vec \sigma ).
  \nonumber \\
  &&{}
 \label{c4}
 \eea

\begin{eqnarray*}
   &&\int d^3\sigma_1\, \sum_a\, {\bar n}_{(a)}(\tau ,{\vec \sigma}_1)\,
   {{\delta\, {\tilde {\bar {\cal H}}}_{(a)}(\tau ,{\vec \sigma}_1
  )}\over {\delta\, \theta^i(\tau ,\vec \sigma )}}\,
  {|}_{\theta^i = 0} =\nonumber \\
  &=& - \phi^{-2}(\tau ,\vec \sigma)\, \sum_{ra}\, \partial_r\,
  {\bar n}_{(a)}(\tau, \vec \sigma)\nonumber \\
  &&\Big[Q_a^{-1}\, \sum_{\bar b}\, \Big(\gamma_{\bar ba}\, V_{(i)ra} +
  \gamma_{\bar br}\, V_{(i)ar}\Big)\, \Pi_{\bar b}  +\nonumber \\
  &+& \sum_j\, \Big(\sum_t\, \sum_{b \not= a}\,  {{\epsilon_{abt}\,
  Q_b^{-1}}\over {Q_b\, Q_a^{-1} - Q_a\, Q_b^{-1}}}\, [\delta_{tj}\,
  V_{(i)rb} + \delta_{rb}\, (V_{(i)jt} + B_{(i)jt})] +  \nonumber \\
  &+& \sum_b\, \sum_{c \not= b}\, {{\epsilon_{bcj\, (\delta_{rc}\, V_{(i)ab}
   - \delta_{ra}\, V_{(i)cb})}\,  Q_a^{-1}\, Q_b\, Q_c^{-1}}\over {Q_c\,
   Q_b^{-1} - Q_b\, Q_c^{-1}}}\,  \Big)\, \pi_j^{(\theta)}
  \Big](\tau ,\vec \sigma) +\nonumber \\
  &&{}\nonumber \\
  &+& \phi^{-2}(\tau ,\vec \sigma)\, \sum_a\,
  {\bar n}_{(a)}(\tau, \vec \sigma)\,
  \Big[Q_a^{-1}\,  \Big(\sum_r\, V_{(i)ra}\, \Big[\phi^6\,
  \partial_r\, \pi_{\tilde \phi} +\nonumber \\
  &+& \sum_{\bar b}\,
  \Big(\gamma_{\bar ba}\, (2\, \phi^{-1}\, \partial_r\, \phi + \sum_{\bar c}\,
  \gamma_{\bar ca}\, \partial_r\, R_{\bar c})\, \Pi_{\bar b} -
  \partial_r\, R_{\bar b}\, \Pi_{\bar b}\Big) + {\check {\cal M}}_r\Big]
  -\nonumber \\
  &-& \sum_{b\bar b}\, V_{(i)ab}\, \gamma_{\bar bb}\,   \Big[\partial_b\,
  \Pi_{\bar b} - (2\, \phi^{-1}\, \partial_b\, \phi + \sum_{\bar c}\,
  \gamma_{\bar ca}\, \partial_b\, R_{\bar c})\, \Pi_{\bar b}\Big]
  \Big)
  -\nonumber \\
 &-&\sum_{rjb}\, \sum_{c \not= b}\, {{\epsilon_{bcj}\,
  Q_a^{-1}\, Q_b\, Q_c^{-1}}\over {Q_c\, Q_b^{-1} - Q_b\, Q_c^{-1}}}\,
  (\delta_{rc}\, V_{(i)ab} - \delta_{ra}\, V_{(i)cb})\,
    \partial_r\, \pi_j^{(\theta)}  +\nonumber \\
 &+& \sum_{jr}\, \Big(\sum_t\, \sum_{b \not= a}\, {{\epsilon_{abt}\,
  Q_b^{-1}}\over {Q_b\, Q_a^{-1} - Q_a\, Q_b^{-1}}}\,
 (2\, \phi^{-1}\, \partial_r\, \phi + \sum_{\bar c}\,
  \gamma_{\bar ca}\, \partial_r\, R_{\bar c})\, [\delta_{tj}\,
  V_{(i)rb} + \delta_{rb}\, (V_{(i)jt} + B_{(i)jt})]  +\nonumber \\
 &+& \sum_b\, \sum_{c \not= b}\, {{\epsilon_{bcj}\,
  Q_a^{-1}\, Q_b\, Q_c^{-1}}\over {Q_c\, Q_b^{-1} - Q_b\, Q_c^{-1}}}\,
  (\delta_{rc}\, V_{(i)ab} - \delta_{ra}\, V_{(i)cb})\,
 \Big[2\, \phi^{-1}\, \partial_r\, \phi +\nonumber \\
 && \sum_{\bar b}\, \Big(\gamma_{\bar ba}
 - \gamma_{\bar bb} + \gamma_{\bar bc} - (\gamma_{\bar bb} - \gamma_{\bar bc})\,
 {{Q_c\, Q_b^{-1} + Q_b\, Q_c^{-1}}\over {Q_c\,
 Q_b^{-1} - Q_b\, Q_c^{-1}}} \Big)\, \partial_r\, R_{\bar b}
 \Big] \Big)   \nonumber \\
 && \pi_j^{(\theta)}  \Big](\tau ,\vec \sigma) =
 \end{eqnarray*}

\bea
 &=& - \Big[{\tilde \phi}^{-1/3}\, \sum_a\, \Big(\sum_r\,
 \partial_r\, {\bar n}_{(a)}\, \Big[Q_a^{-1}\, \sum_{\bar b}\,
 (\gamma_{\bar ba}\, V_{(i)ra}\, + \gamma_{\bar br}\, V_{(i)ar})\,
 \Pi_{\bar b} -\nonumber \\
 &-& {{c^3}\over {8\pi\, G}}\, \tilde \phi\, \Big(\sum_{b\not= a}\,
 Q_b^{-1}\, V_{(i)rb}\, \sigma_{(a)(b)} + Q_a^{-1}\, \sum_{b\not= r}\,
 Q_b\, Q_r^{-1}\, V_{(i)ab}\, \sigma_{(r)(b)} -\nonumber \\
 &-& \sum_{tj}\, \sum_{c \not= d}\, {{\epsilon_{art}\, Q_r^{-1}\,
 (V_{(i)jt} + B_{(i)jt})\, \epsilon_{jcd}\, Q_c\, Q_d^{-1}}\over
 {Q_r\, Q_a^{-1} - Q_a\, Q_r^{-1}}}\, \sigma_{(c)(d)}\Big)\Big] +\nonumber \\
 &+& {{c^3}\over {8\pi\, G}}\, \tilde \phi\, \partial_a\, {\bar
 n}_{(a)}\, Q_a^{-1}\, \sum_{b\not= c}\, Q_b\, Q_c^{-1}\,
 V_{(i)cb}\, \sigma_{(b)(c)} \Big)\Big](\tau ,\vec \sigma) +\nonumber \\
 &&{}\nonumber \\
 &+&\Big[{\tilde \phi}^{-1/3}\, \sum_a\, {\bar n}_{(a)}\,
 \Big(Q_a^{-1}\, \sum_b\, \Big[V_{(i)ba}\, (\tilde \phi\, \partial_b\, \pi_{\tilde \phi}
 - \sum_{\bar b}\, \partial_b\, R_{\bar b}\, \Pi_{\bar b})
 - \gamma_{\bar bb}\, V_{(i)ab}\, \partial_b\, \Pi_{\bar b} +\nonumber \\
 &+& \sum_{\bar b}\, (\gamma_{\bar ba}\, V_{(i)ba} + \gamma_{\bar bb}\, V_{(i)ab})\,
 ({1\over 3}\, {\tilde \phi}^{-1}\, \partial_b\, \tilde \phi + \sum_{\bar c}\,
 \gamma_{\bar ca}\, \partial_b\, R_{\bar c})\, \Pi_{\bar b}
  + V_{(i)ba}\, {\check {\cal M}}_b\Big] +\nonumber \\
 &+& {{c^3}\over {8\pi\, G}}\, \tilde \phi\, \Big[Q_a^{-1}\, \sum_{b,
 c \not= b}\, Q_b\, Q_c^{-1}\, \Big(V_{(i)ab}\, \partial_c\, \sigma_{(b)(c)}
 - V_{(i)cb}\, \partial_a\, \sigma_{(b)(c)} + {2\over 3}\,
 (V_{(i)ab}\, {\tilde \phi}^{-1}\, \partial_c\, \tilde \phi -\nonumber \\
 &-& V_{(i)cb}\, {\tilde \phi}^{-1}\, \partial_a\, \tilde \phi)\, \sigma_{(b)(c)}
 - \sum_{\bar b}\, (\gamma_{\bar ba} - \gamma_{\bar bb} + \gamma_{\bar bc})\,
 (V_{(i)ab}\, \partial_c\, R_{\bar b} - V_{(i)cb}\, \partial_a\, R_{\bar b})\,
 \sigma_{(b)(c)}\Big) +\nonumber \\
 &+& \sum_{tj}\, \sum_{b \not= a,c \not= d}\, {{\epsilon_{abt}\, Q_b^{-1}\,
 (V_{(i)jt} + B_{(i)jt})\, \epsilon_{jcd}\, Q_c\, Q_d^{-1}}\over {Q_b\, Q_a^{-1}
 - Q_a\, Q_b^{-1}}}\, ({1\over 3}\, {\tilde \phi}^{-1}\, \partial_b\, \tilde \phi
 + \sum_{\bar b}\, \gamma_{\bar ba}\, \partial_b\, R_{\bar b})\,
 \sigma_{(c)(d)} \Big]\Big) -\nonumber \\
 &-& \sum_{r\, b \not= a}\, Q_b^{-1}\, \sigma_{(a)(b)}\, V_{(i)rb}\,
 ({1\over 3}\, {\tilde \phi}^{-1}\, \partial_r\, \tilde \phi +
 \sum_{\bar c}\, \gamma_{\bar ca}\, \partial_r\, R_{\bar c} )
 \Big](\tau ,\vec \sigma),\nonumber \\
 &&{}
 \label{c5}
 \eea

\noindent where Eq.(\ref{6.2}) has been used to get the final
expression.

\medskip

\bea
   &&\int d^3\sigma_1\, \sum_a\, {\bar n}_{(a)}(\tau ,{\vec \sigma}_1)\,
   {{\delta\, {\tilde {\bar {\cal H}}}_{(a)}(\tau ,{\vec \sigma}_1
  )}\over {\delta\, \Pi_{\bar a}(\tau ,\vec \sigma )}}\,
  {|}_{\theta^i = 0} =\nonumber \\
  &=&\Big[\phi^{-2}\, \sum_a\, Q_a^{-1}\, \Big({\bar n}_{(a)}\,
  \Big[\gamma_{\bar aa}\,  \Big(2\, \phi^{-1}\, \partial_a\, \phi +
  \sum_{\bar b}\, \gamma_{\bar ba}\, \partial_a\, R_{\bar b}\Big) -
    \partial_a\, R_{\bar a}  \Big] -\nonumber \\
  &-& \gamma_{\bar aa}\, \partial_a\, {\bar n}_{(a)}  \Big)\Big](\tau, \vec \sigma).
 \label{c6}
 \eea

\begin{eqnarray*}
    &&\int d^3\sigma_1\, \sum_a\, {\bar n}_{(a)}(\tau ,{\vec \sigma}_1)\,
   {{\delta\, {\tilde {\bar {\cal H}}}_{(a)}(\tau ,{\vec \sigma}_1
  )}\over {\delta\, R_{\bar a}(\tau ,\vec \sigma )}}\,
  {|}_{\theta^i = 0} =\nonumber \\
  &=& \phi^{-2}(\tau, \vec \sigma)\, \sum_{ra}\, \partial_r\, {\bar
  n}_{(a)}(\tau ,\vec \sigma)\, \Big[Q_a^{-1}\, \delta_{ra}\, \Pi_{\bar
  a} -\nonumber \\
  &-& 2\, \sum_{b \not= a}\, (\gamma_{\bar aa} - \gamma_{\bar ab})\,
  \sum_i\, {{\epsilon_{abi}\, Q_a^{-1}}\over
  {(Q_b\, Q_a^{-1} - Q_a\, Q_b^{-1})^2}}\,  \delta_{rb}\,
  \pi_i^{(\theta)}\Big](\tau ,\vec \sigma) +\nonumber \\
  &&{}\nonumber \\
  &+& \phi^{-2}(\tau ,\vec \sigma)\, \sum_a\, {\bar n}_{(a)}(\tau
  ,\vec \sigma)\nonumber \\
  &&\Big[Q_a^{-1}\,  \Big( \partial_a\, \Pi_{\bar a} - (2\, \phi^{-1}\,
  \partial_a\, \phi + \sum_{\bar b}\, \gamma_{\bar ba}\, \partial_a\,
  R_{\bar b})\, \Pi_{\bar a} -\nonumber \\
  &-& \gamma_{\bar aa}\, \Big[ \phi^6\, \partial_a\, \pi_{\tilde \phi}
  + \sum_{\bar b}\,  (\gamma_{\bar ba}\, \partial_a\, \Pi_{\bar b} -
  \partial_a\,  R_{\bar b}\, \Pi_{\bar b}) + {\check {\cal M}}_a\Big]
   \Big) -\nonumber \\
 &-& \gamma_{\bar aa}\, \sum_{b \not= a}\, \sum_i\, {{\epsilon_{abi}\,
 Q_b^{-1}}\over {Q_b\, Q_a^{-1} - Q_a\, Q_b^{-1}}}\,
  \partial_b\, \pi_i^{(\theta)} +\nonumber \\
 &+& 2\, \sum_i\,  \sum_{b \not= a}\,
 {{\epsilon_{abi}\, Q_a^{-1}}\over {(Q_b\, Q_a^{-1} - Q_a\, Q_b^{-1})^2}}\,
 \Big[2\, (\gamma_{\bar aa} - \gamma_{\bar ab})\, \phi^{-1}\, \partial_b\,
 \phi + \sum_{\bar b}\, (\gamma_{\bar aa}\, \gamma_{\bar bb} - \gamma_{\bar ab}\,
 \gamma_{\bar ba})\, \partial_b\, R_{\bar b}\Big]\nonumber \\
 && \pi_i^{(\theta)}\Big](\tau ,\vec \sigma) =
 \end{eqnarray*}

 \bea
 &=&\Big[{\tilde \phi}^{-1/3}\, \sum_a\, Q_a^{-1}\,
 \Big(\partial_a\, {\bar n}_{(a)}\, \Pi_{\bar a} +  {{c^3}\over
 {4\pi\, G}}\, \tilde \phi\, \sum_{b \not= a}\,
 (\gamma_{\bar aa} - \gamma_{\bar ab})\, {{\partial_b\,
 {\bar n}_{(a)}\, \sigma_{(a)(b)}}\over {Q_b\, Q_a^{-1} - Q_a\, Q_b^{-1}}}
 \Big)\Big](\tau ,\vec \sigma) +\nonumber \\
 &&{}\nonumber \\
 &+& \Big[{\tilde \phi}^{-1/3}\, \sum_a\, {\bar n}_{(a)}\,
 \Big(Q_a^{-1}\, \Big[\partial_a\, \Pi_{\bar a} - ({1\over 3}\,
 {\tilde \phi}^{-1}\, \partial_a\, \tilde \phi + \sum_{\bar b}\,
 \gamma_{\bar ba}\, \partial_a\, R_{\bar b})\, \Pi_{\bar a} -\nonumber \\
 &-& \gamma_{\bar aa}\, \Big(\tilde \phi\, \partial_a\, \pi_{\tilde \phi}
 + \sum_{\bar b}\, (\gamma_{\bar ba}\, \partial_a\, \Pi_{\bar b} -
 \partial_a\, R_{\bar b}\, \Pi_{\bar b}) +
 {\check {\cal M}}_a\Big)\Big] +\nonumber \\
 &+& {{c^3}\over {8\pi\, G}}\, \tilde \phi\, \sum_{b \not= a}\,
 Q_b^{-1}\, \Big[\gamma_{\bar aa}\, \partial_b\, \sigma_{(a)(b)} +
 \nonumber \\
 &+&\Big((\gamma_{\bar aa} - {2\over 3}\, (\gamma_{\bar aa} -
 \gamma_{\bar ab})\, {{Q_b\, Q_a^{-1} }\over
 {Q_b\, Q_a^{-1} - Q_a\, Q_b^{-1}}})\,
 {\tilde \phi}^{-1}\, \partial_b\, \tilde \phi -\nonumber \\
 &-& \sum_{\bar b}\, {{\gamma_{\bar aa}\, (\gamma_{\bar ba} -
 \gamma_{\bar bb})\, [Q_b\, Q_a^{-1} + Q_a\, Q_b^{-1}]  + 2\,
 (\gamma_{\bar aa}\, \gamma_{\bar bb} - \gamma_{\bar ab}\,
 \gamma_{\bar ba})\, Q_b\, Q_a^{-1}}\over {Q_b\, Q_a^{-1} -
 Q_a\, Q_b^{-1}}}\, \partial_b\, R_{\bar b}
 \Big)\, \sigma_{(a)(b)} \Big]\Big)\Big](\tau ,\vec
 \sigma).\nonumber \\
 &&{}
 \label{c7}
 \eea

 \bigskip

The function ${\cal S}(\tau, \vec \sigma)$ of Eqs.(\ref{b8}) becomes

\bea
 {\cal S}{|}_{\theta^i = 0} &=& \phi^2\, \sum_a\, Q_a^{-2}\, \Big(\sum_{\bar b}\,
 \Big[\sum_{\bar c}\, (2\, \gamma_{\bar ba}\, \gamma_{\bar ca} -
 \delta_{\bar b\bar c})\, \partial_a\, R_{\bar b}\, \partial_a\, R_{\bar c}
 -\nonumber \\
 &-& 4\, \gamma_{\bar ba}\, \phi^{-1}\,
 \partial_a\, \phi\, \partial_a\, R_{\bar b}\Big] +
  8\, (\phi^{-1}\, \partial_a\, \phi)^2\Big),
 \label{c8}
 \eea

\noindent and its functional derivatives (\ref{b9}), (\ref{b10}) and
(\ref{b11}) become

\bea
   &&\int d^3\sigma_1\,\, [1 + n(\tau ,{\vec \sigma}_1)]\,
   {{\delta\, {\cal S}(\tau ,{\vec \sigma}_1)}\over
 {\delta\, R_{\bar a}(\tau ,\vec \sigma )}} {|}_{\theta^i = 0} =\nonumber \\
  &&{}\nonumber \\
  &&{}\nonumber \\
 &=&2\, \Big(\phi^2\, \sum_a\, Q_a^{-2}\, \Big[\partial_a\, n\, \Big(2\,
 \gamma_{\bar aa}\, \phi^{-1}\, \partial_a\, \phi - \sum_{\bar b}\,
 (2\, \gamma_{\bar aa}\, \gamma_{\bar ba} - \delta_{\bar a\bar b})\,
 \partial_a\, R_{\bar b}\Big) -\nonumber \\
 &-& (1 + n)\, \Big(2\, \gamma_{\bar aa}\, \Big(- \phi^{-1}\, \partial_a^2\,
 \phi + 3\, (\phi^{-1}\, \partial_a\, \phi)^2\Big) +\nonumber \\
 &+&\sum_{\bar b}\, (2\, \gamma_{\bar aa}\,
 \gamma_{\bar ba} - \delta_{\bar a\bar b})\, (\partial_a^2\,
 R_{\bar b} + 2\, \phi^{-1}\, \partial_a\, \phi\, \partial_a\,
 R_{\bar b} )  +\nonumber \\
 &+& \sum_{\bar b\bar c}\, \Big(2\, \gamma_{\bar ba}\,
 (\delta_{\bar a\bar c} - \gamma_{\bar aa}\, \gamma_{\bar ca}) -
 \gamma_{\bar aa}\, \delta_{\bar b\bar c}\Big)\, \partial_a\, R_{\bar b}\,
 \partial_a\, R_{\bar c} \Big) \Big]\Big)(\tau ,\vec \sigma
 ).\nonumber \\
 &&{}
 \label{c9}
 \eea

\bea
  &&\int d^3\sigma_1\, [1 + n(\tau ,{\vec \sigma}_1)]\,
  {{\delta\, {\cal S}(\tau ,{\vec \sigma}_1
  )}\over {\delta\, \phi (\tau ,\vec \sigma )}}\,
  {|}_{\theta^i = 0} =\nonumber \\
  &=&- 2\, \Big(\phi\, \sum_a\, Q_a^{-2}\, \Big[
 \partial_a\, n\, (4\, \phi^{-1}\, \partial_a\,
 \phi - \sum_{\bar b}\, \gamma_{\bar ba}\, \partial_a\,
 R_{\bar b}) +\nonumber \\
 &+& (1 + n)\, \Big(8\, (\phi^{-1}\, \partial_a^2\, \phi -
 2\, \phi^{-1}\, \partial_a\, \phi\, \sum_{\bar b}\,
 \gamma_{\bar ba}\, \partial_a\, R_{\bar b}) - 2\,
 \sum_{\bar b}\, \gamma_{\bar ba}\, \partial_a^2\,
 R_{\bar b} +\nonumber \\
 &+& \sum_{\bar b\bar c}\, (2\, \gamma_{\bar ba}\,
 \gamma_{\bar ca} + \delta_{\bar b\bar c})\, \partial_a\,
 R_{\bar b}\, \partial_a\, R_{\bar c}
 \Big)\Big]\Big)(\tau ,\vec \sigma).
 \label{c10}
 \eea

\bea
   &&\int d^3\sigma_1\, [1 + n(\tau ,{\vec \sigma}_1)]\,
   {{\delta\, {\cal S}(\tau ,{\vec \sigma}_1
  )}\over {\delta\, \theta^i(\tau ,\vec \sigma )}}\,
  {|}_{\theta^i = 0} =\nonumber \\
  &=&- 2\,\Big[\phi^2\, \sum_{ra}\, Q_a^{-2}\, V_{(i)ra}\, \Big(
 \partial_a\, n\, (\phi^{-1}\, \partial_r\, \phi -
 \sum_{\bar b}\, \gamma_{\bar ba}\, \partial_r\, R_{\bar b}) +
 \partial_r\, n\, (\phi^{-1}\, \partial_a\, \phi - \sum_{\bar b}\,
 \gamma_{\bar br}\, \partial_a\, R_{\bar b}) -\nonumber \\
 &-& (1 + n)\, \Big[2\, (3\, \phi^{-1}\, \partial_a\, \phi\,
 \phi^{-1}\, \partial_r\, \phi - \phi^{-1}\, \partial_a\,
 \partial_r\, \phi) + \sum_{\bar b}\, (\gamma_{\bar ba} +
 \gamma_{\bar br})\, \partial_a\, \partial_r\, R_{\bar b}
 +\nonumber \\
 &+& 2\, \sum_{\bar b}\, (\gamma_{\bar ba}\, \phi^{-1}\, \partial_a\,
 \phi\, \partial_r\, R_{\bar b} + \gamma_{\bar br}\, \phi^{-1}\,
 \partial_r\, \phi\, \partial_a\, R_{\bar b}) -\nonumber \\
 &-& \sum_{\bar b\bar c}\, (2\, \gamma_{\bar br}\, \gamma_{\bar ca}
 + \delta_{\bar b\bar c})\, \partial_a\, R_{\bar b}\,
 \partial_r\, R_{\bar c} \Big]\Big) \Big](\tau ,\vec \sigma).
 \label{c11}
 \eea

\bigskip

The Laplace-Beltrami operator of Eq.(\ref{b12}) becomes

\beq
 \hat \triangle{|}_{\theta^i = 0}\, =\, \sum_a\, Q_a^{-2}\, \Big[ \partial^2_a
 -  2\,  \sum_{\bar b}\, \gamma_{\bar ba}\, \partial_a\, R_{\bar b}
 \, \partial_a\Big].
 \label{c12}
 \eeq
\medskip

The functions ${\cal T}(\tau, \vec \sigma)$ and ${\cal T}_1(\tau,
\vec \sigma)$ of Eq.(\ref{b13}) become

\bea
 {\cal T}(\tau ,\vec \sigma ){|}_{\theta^i = 0} &=& - 2\, \Big[
 \phi^2\, \sum_a\, Q_a^{-2}\, \Big(4\, \Big[\phi^{-1}\, \partial_a^2\,
 \phi + (\phi^{-1}\, \partial_a\, \phi)^2\Big] -\nonumber \\
 &-& \sum_{\bar b}\, \gamma_{\bar ba}\, \Big[\partial_a^2\, R_{\bar
 b} + 2\, (5\, \phi^{-1}\, \partial_a\, \phi - \sum_{\bar c}\,
 \gamma_{\bar ca}\, \partial_a\, R_{\bar c})\, \partial_a\, R_{\bar b}
 \Big]\Big)\Big](\tau ,\vec \sigma),\nonumber \\
 &&{}\nonumber \\
 {\cal T}_1(\tau ,\vec \sigma ){|}_{\theta^i = 0} &=& - 2\,
 \Big[\phi^2\, \sum_a\, Q_a^{-2}\, \Big(4\, (\phi^{-1}\, \partial_a\,
 \phi)^2 -\nonumber \\
 &-& \sum_{\bar b}\, \gamma_{\bar ba}\, \Big[\partial_a^2\, R_{\bar
 b} + 2\, (\phi^{-1}\, \partial_a\, \phi - \sum_{\bar c}\,
 \gamma_{\bar ca}\, \partial_a\, R_{\bar c})\, \partial_a\, R_{\bar b}
 \Big]\Big)\Big](\tau ,\vec \sigma).
 \label{c13}
 \eea

\bigskip

The functional derivatives (\ref{b14})-(\ref{b19}) become

\bea
 &&\int d^3\sigma_1\, {{\delta\, \Big[n\, \phi\, \hat
\triangle\, \phi\Big](\tau ,{\vec \sigma}_1 )}\over {\delta\,
R_{\bar a}(\tau ,\vec \sigma)}} {|}_{\theta^i = 0} =\nonumber \\
  &=& 2\, \phi^2\, \Big( \sum_a\, Q_a^{-2}\, \gamma_{\bar aa}\,
 \phi^{-1}\, \partial_a\, \phi\, (n\, \phi^{-1}\, \partial_a\, \phi
 + \partial_a\, n) \Big)(\tau ,\vec \sigma),
 \label{c14}
 \eea

 \bea
   &&\int d^3\sigma_1\, n(\tau, {\vec \sigma}_1)\,
   {{\delta\, {\cal T}(\tau ,{\vec \sigma}_1)}\over
 {\delta\, R_{\bar a}(\tau ,\vec \sigma )}}
 {|}_{\theta^i = 0} =\nonumber \\
  &&{}\nonumber \\
  &&{}\nonumber \\
 &=& 2\, \Big[\phi^2\, \sum_a\, \gamma_{\bar aa}\, Q_a^{-2}\,
 \Big( \partial^2_a\, n - 6\, \phi^{-1}\, \partial_a\,
 \phi\, \partial_a\, n \Big)\Big](\tau ,\vec \sigma).
 \label{c15}
 \eea

\bea
 &&\int d^3\sigma_1\, {{\delta\, \Big[n\, \phi\, \hat
\triangle\, \phi\Big](\tau ,{\vec \sigma}_1 )}\over {\delta\,
\phi(\tau ,\vec \sigma)}} {|}_{\theta^i = 0} =\nonumber \\
  &=& \Big[2\, n\, \hat \triangle {|}_{\theta^i = 0}\, \phi + 2\, \sum_a\, Q_a^{-2}\,
  \partial_a\, n\, \partial_a\, \phi +  \phi\, \hat \triangle{|}_{\theta^i = 0}\,
  n\Big](\tau ,\vec \sigma).
 \label{c16}
 \eea

 \bea
   &&\int d^3\sigma_1\, n(\tau, {\vec \sigma}_1)\,
   {{\delta\, {\cal T}(\tau ,{\vec \sigma}_1)}\over
 {\delta\, \phi(\tau ,\vec \sigma )}}
 {|}_{\theta^i = 0} =\nonumber \\
  &&{}\nonumber \\
  &&{}\nonumber \\
 &=&- 4\, \Big[\phi\, \sum_a\, Q_a^{-2}\,
  \Big(2\, \partial_a^2\, n - 3\,\sum_{\bar b}\, \gamma_{\bar ba}\,
 \partial_a\, R_{\bar b}\, \partial_a\, n \Big)\Big](\tau
 ,\vec \sigma).
 \label{c17}
 \eea

\bea
 &&\int d^3\sigma_1\, {{\delta\, \Big[n\, \phi\, \hat
\triangle\, \phi\Big](\tau ,{\vec \sigma}_1 )}\over {\delta\,
\theta^i(\tau ,\vec \sigma)}}\,\, {|}_{\theta^i = 0} =\nonumber \\
  &=& - \Big(\phi^2\, \sum_{ar}\, Q_a^{-2}\,
 V_{(i)ra}\, \Big[2\, n\, \phi^{-1}\, \partial_r\, \phi\,
 \phi^{-1}\, \partial_a\, \phi +\nonumber \\
 &+& \partial_r\, n\, \phi^{-1}\, \partial_a\, \phi
 + \partial_a\, n\, \phi^{-1}\, \partial_r\, \phi\Big]
  \Big)(\tau ,\vec \sigma),
 \label{c18}
 \eea

 \bea
   &&\int d^3\sigma_1\, n(\tau, {\vec \sigma}_1)\,
   {{\delta\, {\cal T}(\tau ,{\vec \sigma}_1)}\over
 {\delta\, \theta^i(\tau ,\vec \sigma )}}
 {|}_{\theta^i = 0} =\nonumber \\
  &&{}\nonumber \\
  &&{}\nonumber \\
 &=& - 2\, \Big[\phi^2\, \sum_{ra}\, Q_a^{-2}\, V_{(i)ra}\,
 \Big( \partial_r\, \partial_a\, n - 3\, (\partial_r\, n\,
 \phi^{-1}\, \partial_a\, \phi + \partial_a\, n\, \phi^{-1}\,
 \partial_r\, \phi) \Big)\Big](\tau ,\vec \sigma).\nonumber \\
 &&{}
 \label{c19}
 \eea

\bigskip

The scalar 3-curvature of Eq.(\ref{b20}) becomes

\bea
 {}^3{\hat R}(\tau ,\vec \sigma){|}_{\theta^i=0} &=&
  \sum_a\, \Big(Q_a^{-2}\, \sum_{\bar b}\, \Big[2\, \gamma_{\bar ba}\,
 \partial_a^2\, R_{\bar b} - (\partial_a\, R_{\bar b})^2 -\nonumber \\
 &-&2\, \gamma_{\bar ba}\, \sum_{\bar c}\,  \gamma_{\bar ca}
 \partial_a\, R_{\bar c}\, \partial_a\, R_{\bar b}\Big]
 \Big)(\tau ,\vec \sigma).
 \label{c20}
 \eea

\bigskip

Finally the functional derivatives (\ref{b21}), (\ref{b22}) and
(\ref{b23}) of the mass density become

\bea
  &&\int d^3\sigma_1\, \Big(1 + n(\tau ,{\vec \sigma}_1)\Big)\,
 {{\delta\, {\check {\cal M}}(\tau ,{\vec \sigma}_1)}\over
 {\delta\, \theta^i(\tau ,\vec \sigma)}}{|}_{\theta^i=0} =\nonumber \\
 &&{}\nonumber \\
 &=&{1\over 2}\, \sum_{i=1}^N\, \delta^3(\vec \sigma, {\vec
 \eta}_i(\tau))\, \eta_i\, \Big((1 + n)\nonumber \\
 &&{{\phi^{-4}\, \sum_{rsa}\, Q_a^{-2}\, ( V_{(i)ra}\,
 \delta_{sa} + \delta_{ra}\, V_{(i)sa})\,  \Big({\check
 \kappa}_{ir}(\tau ) - {{Q_i}\over c}\, A_{\perp\, r}\Big)\,
 \Big({\check \kappa}_{is}(\tau ) - {{Q_i}\over c}\,
 A_{\perp\, s}\Big)}\over {\sqrt{m_i^2\, c^2 +
 \phi^{-4}\, \sum_{a}\, Q_a^{-2}\,  \Big({\check
 \kappa}_{ia}(\tau ) - {{Q_i}\over c}\, A_{\perp\, a}\Big)^2\,
 }}} \Big)(\tau ,\vec \sigma) +\nonumber \\
 &&{}\nonumber \\
  &+& \Big(1 + n(\tau ,\vec \sigma)\Big)\,  \Big({\tilde \phi}^{-1/3}\,
 {1\over {c}}\, \Big[\, \sum_{ars}\, Q_a^2\,
  V_{(i)ra}\,  \delta_{sa}
  \pi^r_{\perp}\, \pi^s_{\perp} +  \sum_{abr}\, Q_a^{-2}\, Q_b^{-2}\,
  V_{(i)ra}\, F_{rb}\, F_{ab} -\nonumber \\
  &-&  {1\over 2}\, \sum_{arsn}\, Q_a^2\,  \Big(V_{(i)ra}\,
  \delta_{sa} + V_{(i)sa}\,  \delta_{ra}\Big)\,
  \Big(2\, \pi^r_{\perp} - \sum_m\, \delta^{rm}\, \sum_{i=1}^N\, Q_i\, \eta_i\, \partial_m\,
  c(\vec \sigma , {\vec \eta}_i(\tau))\Big)\nonumber \\
 &&\delta^{sn}\, \sum_{j=1}^N\, Q_j\, \eta_j\, \partial_n\,
  c(\vec \sigma , {\vec \eta}_j(\tau))
  \Big]\Big)(\tau ,\vec \sigma),\nonumber \\
  &&{}
 \label{c21}
 \eea

\bea
   &&\int d^3\sigma_1\, \Big(1 + n(\tau ,{\vec \sigma}_1)\Big)\,
 {{\delta\, {\check {\cal M}}(\tau ,{\vec \sigma}_1)}\over
 {\delta\, \phi(\tau ,\vec \sigma)}} =\nonumber \\
 &&{}\nonumber \\
 &=& - 2\, \sum_{i=1}^N\, \delta^3(\vec \sigma, {\vec
 \eta}_i(\tau))\, \eta_i\, \Big((1 + n)\nonumber \\
 &&{{\phi^{-5}\, \sum_{a}\, Q_a^{-2}\,
 \Big({\check \kappa}_{ia}(\tau ) - {{Q_i}\over c}\,
  A_{\perp\, a}\Big)^2\, }\over {\sqrt{m_i^2\, c^2 +
 \phi^{-4}\, \sum_{a}\, Q_a^{-2}\,  \Big({\check
 \kappa}_{ia}(\tau ) - {{Q_i}\over c}\, A_{\perp\, a}\Big)^2\,
 }}} \Big)(\tau ,\vec \sigma) -\nonumber \\
 &&{}\nonumber \\
  &-& \Big(1 + n(\tau ,\vec \sigma)\Big)\, \Big({\tilde \phi}^{-1/2}\,
 \Big[{1\over {c}}\, \sum_{ars}\, Q_a^2\,
  \delta_{ra}\, \delta_{sa}\, \pi^r_{\perp}\, \pi^s_{\perp} +
   {1\over {2c}}\, \sum_{ab}\, Q_a^{-2}\, Q_b^{-2}\, F_{ab}\, F_{ab}
   -\nonumber \\
 &-&  {1\over {c}}\, \sum_{arsn}\, Q_a^2\, \delta_{ra}\, \delta_{sa}\,
  \Big(2\, \pi^r_{\perp} - \sum_m\, \delta^{rm}\, \sum_{i=1}^N\, Q_i\, \eta_i\, \partial_m\,
  c(\vec \sigma , {\vec \eta}_i(\tau))\Big)\nonumber \\
 &&\delta^{sn}\, \sum_{j=1}^N\, Q_j\, \eta_j\, \partial_n\,
  c(\vec \sigma , {\vec \eta}_j(\tau))
 \Big]\Big)(\tau ,\vec \sigma),\nonumber \\
 &&{}
 \label{c22}
 \eea

\bea
   &&\int d^3\sigma_1\, \Big(1 + n(\tau ,{\vec \sigma}_1)\Big)\,
 {{\delta\, {\check {\cal M}}(\tau ,{\vec \sigma}_1)}\over
 {\delta\, R_{\bar a}(\tau ,\vec \sigma)}} =\nonumber \\
 &&{}\nonumber \\
 &=& - \sum_{i=1}^N\, \delta^3(\vec \sigma, {\vec
 \eta}_i(\tau))\, \eta_i\, \Big((1 + n)\nonumber \\
 && {{\phi^{-4}\, \sum_{a}\, \gamma_{\bar aa}\, Q_a^{-2}\,
 \Big({\check \kappa}_{ia}(\tau ) - {{Q_i}\over c}\,
 A_{\perp\, a}\Big)^2\, }\over {\sqrt{m_i^2\, c^2 +
 \phi^{-4}\, \sum_{a}\, Q_a^{-2}\,  \Big({\check
 \kappa}_{ia}(\tau ) - {{Q_i}\over c}\, A_{\perp\, a}\Big)^2\,
 }}} \Big)(\tau ,\vec \sigma) +\nonumber \\
 &&{}\nonumber \\
 &+& \Big(1 + n(\tau ,\vec \sigma)\Big)\,  \Big({\tilde \phi}^{-1/3}\,
 \Big[{1\over {c}}\, \sum_{ars}\, \gamma_{\bar aa}\, Q_a^2\,
  \delta_{ra}\, \delta_{sa}\, \pi^r_{\perp}\, \pi^s_{\perp}  -
   {1\over {2c}}\, \sum_{ab}\, (\gamma_{\bar aa} + \gamma_{\bar ab})\,
   Q_a^{-2}\, Q_b^{-2}\, F_{ab}\, F_{ab} -\nonumber \\
  &-&   {1\over {c}}\, \sum_{arsn}\, \gamma_{\bar aa}\,
  Q_a^2\, \delta_{ra}\, \delta_{sa}\,
  \Big(2\, \pi^r_{\perp} - \sum_m\, \delta^{rm}\,
  \sum_{i=1}^N\, Q_i\, \eta_i\, \partial_m\,
  c(\vec \sigma , {\vec \eta}_i(\tau))\Big)\nonumber \\
 &&\delta^{sn}\,  \sum_{j=1}^N\, Q_j\, \eta_j\, \partial_n\,
  c(\vec \sigma , {\vec \eta}_j(\tau))
 \Big]\Big)(\tau ,\vec \sigma).\nonumber \\
 &&{}
 \label{c23}
 \eea

\vfill\eject

\end{document}